# Signal Formation Processes in Micromegas Detectors and Quality Control for large size Detector Construction for the ATLAS New Small Wheel

Dissertation zur Erlangung des naturwissenschaftlichen Doktorgrades
der Julius-Maximilians-Universität Würzburg

vorgelegt von

**Fabian Kuger**

aus

Schweinfurt

Würzburg 2017

Eingereicht am 12. April 2017

bei der Fakultät für Physik und Astronomie

1. Gutachter: Prof. Dr. Thomas Trefzger

2. Gutachter: Prof. Dr. Otmar Biebel

3. Gutachter: Prof. Dr. Gregor Herten

der Dissertation

Vorsitzende(r): Prof. Dr. Karl Mannheim

1. Prüfer: Prof. Dr. Thomas Trefzger

2. Prüfer: Prof. Dr. Otmar Biebel

3. Prüfer: Prof. Dr. Werner Porod

im Promotionskolloquium

Tag des Promotionskolloquiums: 31. Juli 2017

Doktorurkunde ausgehändigt am: ....................

# Abstract

The Micromegas technology is one of the most successful modern gaseous detector concepts and widely utilized in nuclear and particle physics experiments. Twenty years of R & D rendered the technology sufficiently mature to be selected as precision tracking detector for the New Small Wheel (NSW) upgrade of the ATLAS Muon spectrometer. This will be the first large scale application of Micromegas in one of the major LHC experiments. However, many of the fundamental microscopic processes in these gaseous detectors are still not fully understood and studies on several detector aspects, like the micromesh geometry, have never been addressed systematically.

The studies on signal formation in Micromegas, presented in the first part of this thesis, focuses on the microscopic signal electron loss mechanisms and the amplification processes in electron gas interaction. Based on a detailed model of detector parameter dependencies, these processes are scrutinized in an iterating comparison between experimental results, theory prediction of the macroscopic observables and process simulation on the microscopic level. Utilizing the specialized detectors developed in the scope of this thesis as well as refined simulation algorithms, an unprecedented level of accuracy in the description of the microscopic processes is reached, deepening the understanding of the fundamental process in gaseous detectors.

The second part is dedicated to the challenges arising with the large scale Micromegas production for the ATLAS NSW. A selection of technological choices, partially influenced or determined by the herein presented studies, are discussed alongside a final report on two production related tasks addressing the detectors' core components: For the industrial production of resistive anode PCBs a detailed quality control (QC) and quality assurance (QA) scheme as well as the therefore required testing tools have been developed. In parallel the study on micromesh parameter optimization and production feasibility resulted in the selection of the proposed mesh by the NSW community and its full scale industrial manufacturing. The successful completion of both tasks were important milestones towards the construction of large size Micromegas detectors clearing the path for NSW series production.

## Kurzdarstellung

Die Micromegas Technologie zählt zu den erfolgreichsten Konzepten moderner Gasdetektoren und findet Anwendung in zahlreichen Experimenten der Kern- und Teilchenphysik. Nach zwanzig Jahren Weiterentwicklung wurde die Micromegas Technologie für hinreichend ausgereift befunden, um als Präzisionsspurdetektor in den New Small Wheels (NSW) des ATLAS Myon Spektrometers verwendet zu werden. Dies stellt den ersten großflächigen Einsatz der Micromegas Technologie in einem LHC Experiment dar. Dennoch blieben einige der grundlegenden Prozesse in Gasdetektoren nach wie vor unzureichend verstanden und ausgewählte Detektoraspekte, wie die Geometrie der Mikrogitter, wurden bisher kaum systematisch untersucht.

Die im ersten Teil dieser Doktorarbeit präsentierten Studien zu Signalenstehungsprozessen in Micromegas richten sich daher auf die mikroskopischen Mechanismen zum Elektronenverlust und die Verstärkungsprozesse in Elektron-Gas-Wechselwirkungen. Diese Prozesse werden auf Basis eines Modells ihrer Abhängigkeiten von den Detektorparametern untersucht, wobei stets der Vergleich zwischen experimentell gemessenen Daten, theoretischen Vorhersagen dieser makroskopischen Größen und der Simulation von Prozessen auf mikroskopischer Ebene gezogen wird. In Verbindung mit den im Rahmen dieser Arbeit entwickelten Detektoren und verbesserten Simulationsalgorithmen lieferten diese iterativen Vergleichsstudien ein vertieftes Verständnis der fundamentalen Prozesse in gasgefüllten Detektoren.

Der zweite Teil widmet sich den mit der Konstruktion der großflächigen ATLAS NSW Micromegas Detektoren einhergehenden Herausforderungen und diskutiert Entscheidungen bezüglich ausgewählter Technologieoptionen, die teilweise substantiell durch diese Arbeit beeinflusst wurden. Darüber hinaus wird abschließend über zwei Tätigkeitsbereiche bezüglich der Produktion zentraler Detektorkomponenten berichtet: Für die industrielle Fertigung der resistiven Anoden-PCBs wurde ein unfangreiches und verlässliches Qualitätssicherungs- und Qualitätskontroll-Schema sowie die hierzu notwendigen Messtechniken und -apparaturen entwickelt. Die parallel vorangetriebene Studie zur Optimierung der Parameter des Mikrogitters unter Berücksichtigung der produktionsbedingten Limitationen führte zu der Bestätigung der vorgeschlagenen Spezifikation durch die NSW Kollaboration und der industriellen Fertigung dieses Gewebes. Der erfolgreiche Abschluss beider Projekte waren essenzielle Meilensteine auf dem Weg zur Serienproduktion der NSW Micromegas Detektoren.

# Contents

# Introduction

In 2013 the CERN Council set the course for European high energy physics, stating that *'Europe's top priority should be the exploitation of the full potential of the LHC, including the high-luminosity upgrade of the machine and the detectors'* [1]. This strategy has recently been underpinned up by the European Commission, approving the high-luminosity HL-LHC as one of the ESFRI Landmarks [2]. In order to continue the line of scientific success achieved at the LHC, foremost the discovery of the Higgs Boson [3] [4], with comprehensive precision studies on Standard Model particles and potential new discoveries, an increase of the collision data to be obtained during LHC operation until 2035 is required. A total integrated luminosity of $3000\,\mathrm{fb}^{-1}$ is envisaged for the HL-LHC, exceeding the LHC design objective by one order of magnitude. Successive upgrades on the LHC during the Long Shutdown 2 (LS2, 2019-2020) and 3 (LS3, 2024-2026) will increase the proton density and consequentially the collision rate. *'These upgrades will turn the LHC facility into a Higgs Factory, narrowing the Higgs properties down to 1 % precision, which will lead to a successful mapping in regions beyond the Standard Model.'* [2]

In parallel with the accelerator, the LHC experiments are facing several modifications in order to adapt to the increased luminosity and maintain their demonstrated excellent performance [5] at higher particle rate and detector occupancy. For the ATLAS detector [6], one of the general purpose experiments at the LHC, an intensive upgrade program is envisaged during the upcoming decade [7, 8]. The exchange of the innermost end-cap regions of the ATLAS Muon spectrometer with the New Small Wheels (NSW) [9] is one of the major upgrade activities scheduled for LS2. With multiple physics analysis channels containing muon end states, the preservation of the reconstruction efficiency and -precision under increased event rate in the muon spectrometers end-cap region is a prerequisite for future success of the ATLAS physics program. The NSW will contain two gaseous detector technologies: strip Thin Gap Chambers (sTGC) and Micromesh gaseous structures (Micromegas).

Gaseous detectors have proven over the last decades to be a versatile and cost effective technology for large volume Muon spectrometers. With the invention and evolution of Micro Pattern Gaseous Detectors (MPGD) during the twenty years, gaseous detectors improved significantly in timing- and spatial resolution. Furthermore, the reduced spatial extent of the required gas volumes in MPGDs permits dead time limitation and renders gaseous detectors suitable for high-rate applications. Consequent R & D resulted in substantial experience in construction and operation of gaseous detectors and broad understanding on the operational parameters of these detector technologies. On a microscopic level, however, some of the basic processes in the interaction between gas constituents and electrons or ions are up to now barely understood and elude experimental confirmation.

The first part of this thesis is, therefore, dedicated to detailed and systematic studies of signal formation processes in Micromegas detectors. Based on the temporal succession of the sub-processes during signal formation, which are common for all gaseous detectors, a factorized approach for the signal strength description is introduced. The dependencies of the sub-processes on the parameters of the detector are systematically accessed and a comprehensive dependency model is introduced. Hereon two processes of signal formation are scrutinized in detail: The electron drift and thereto associated electron losses in the gas and at structure surfaces as well as the amplification processes of single electrons during avalanche formation. For both processes the dependencies to a selected set of detector parameters are studied in detail, applying a systematic parameter variation strategy. Exploiting the increasing availability of computing power, microscopic interactions of particles with the gas are simulated and the macroscopic predictions derived from these Monte Carlo models are compared with the data obtained from experimental measurements with dedicated Micromegas prototypes and setups. Although focused on the Micromegas technology, many of the presented results are applicable or transferable for (all) other types of gaseous detectors.

With about $1280\,\mathrm{m}^2$ of active detection area, the NSW Micromegas will be the largest MPGD based system ever utilized in an LHC experiment. According to the detector dimensions and performance requirements the construction of the NSW poses many challenges on detector design, construction- and assembly methods as well as large scale production of detector components in cooperation with industries. The second part of this thesis explains some of the most important technological choices for the NSW Micromegas and addresses major challenges related to the two most complex detector components: The eponymous fine metallic micromesh and the readout anode PCB. For both products their design and characterization, the technology transfer to the industrial producer as well as the development of a quality control / quality assurance (QC/QA) scheme and the corresponding tooling have been tasks addressed in the scope of this thesis.

# Part I.

# Signal Formation in Micromegas Detectors

# 1. Signal Formation Processes in Gaseous Detectors

**G**aseous **D**etectors (GD) have been essential tools since the very beginning of particle physics and even before in nuclear physics research. They combine a comparatively simple principle of operation with a multitude of tuning and optimization possibilities, leading to a variety of different GD technologies.

A brief overview on the operation principles of gaseous detectors is provided in 1.1. Thereafter, some of the more famous and most successful GD concepts will be reviewed in 1.2, focusing on the similarities in the operation principle as well as the specific benefits and drawbacks of each technology. Based on these similarities the systematic approach towards the signal formation studies in Micromegas detectors conducted in the scope of this thesis is described in 1.3.

## 1.1. Operation Principles of Gaseous Detectors (GD)

While offering a large variety of technologies and implementations all gaseous detectors rely on the same processes for signal formation. This section shall provide a basic overview on these processes and introduce the nomenclature required for a comparison of different GD technologies. In the interest of conciseness it is limited to a phenomenological and rather qualitative discussion, referring the reader to the in-depth studies presented in subsequent chapters when applicable. Comprehensive reports on particle detection can be found in [10–12] and [13].

### 1.1.1. Fundamental Processes of Signal Formation

The detection of a particle requires some sort of energy deposition by the trespassing particle into the detector. In a gaseous volume this energy loss is caused by electromagnetic interaction and results in ionization of the gas along the track of a charged particle, as discussed in more detail in chapter 2. While each ionizing interaction between the particle and the gas yields one primary electron-ion pair, the electrons energy might be sufficient to further ionize the gas before thermalisation. Therefore, the total electron-ion yield of a particle passing the gas volume is typically larger than the number of primary electrons. With the term of ionization being used in different context throughout this thesis, the full process of electron-ion pair creation caused by the trespassing particle, involving secondary processes until electron thermalisation, is in the following referred to as *primary ionization*. The total electron yield from this process can potentially contribute to the signal formation process, thus they are named signal electrons.

A group of these signal electrons created by one primary electron is called a (signal-) cluster.

In the absence of an external electric field electrons and ions are likely to recombine and thus be lost for signal formation. To ensure successful charge separation, the charge carriers are subjected to an external electric field. Thus, they are accelerated along the field lines and acquire momentum. Their energy successively increases until they eventually scatter with other gas constituents. The collision yield depends on the scattering energy and typically includes a randomization of the momentum in elastic or inelastic scattering. It can possible yield an electron-ion pair (ionizing scattering) or result in electron loss due to attachment to or dissociation of the gas constituent. While the processes for electrons and ions are similar, ions move several orders of magnitude slower in the gas, due to their higher mass. Their probability to undergo a scattering process requiring a certain energy threshold is drastically reduced compared to electrons in the same electric field. Therefore, the ions yielded by primary ionization are most commonly simply guided to the cathode and neutralized, while the electrons are utilized to create a signal. Ion movement, however, remains an important process once a larger amount of ions is created, as discussed below.

Although the electron-gas interaction processes remain the same throughout signal formation, they are usually discriminated according to the range of the scattering energy. The *electron drift*, scrutinized in chapter 3, is dominated by scattering energies below the ionization threshold of any gas constituent. The repeating process of acceleration and scattering yields a net movement of the electron without charge multiplication. Characteristic values for the drift process are the mean drift velocity and the transverse and longitudinal diffusion across, respectively along, the drift field lines.

The drifting electrons are commonly guided into a volume of higher field strength, where they accumulate more energy in-between collisions and thus become more likely to cause ionization. Each newly freed electron is similarly subjected to the electric field, accelerated and can cause further ionization leading to the formation of an electron avalanche. This charge *amplification* process, which is discussed in-depth in chapter 4, is stopped once the electrons are either neutralized or the electric field strength is sufficiently reduced to suppress ionizing collisions. The charge gain of the multiplication process determines the operation mode of the detector, which is discussed in section 1.1.2.

Once the fast moving electrons left the amplification region, the remaining *ions drift* slowly towards the cathode. This charge separation and subsequent movement induces a charge on the electrodes. Some (or all) of the electrodes can be used as *readout* structures being connected to electronics measuring and processing the current-, or a charge- or voltage signal. Although the amplification electrons might be collected on a readout electrode, the measured signal primarily depends on the relative movement of the charge within the detector volume and, therefore, the ion drift. The ion drift is to a certain extent similar to the electron drift and, therefore, partially covered in 3. An in-depth assessment of ion drift is given in [13, chap. 4.2-4.3]. The theory of charge signal sensing is not addressed in this thesis but can be found in [12, chap. 5].

### 1.1.2. Avalanche Growth and Gas Gain Operation Modes

Gaseous detectors can be operated at different modes defined by the range of mean gas gain $G$ during the amplification of one electron. The gain describes the mean number of electrons reaching the anode per electron-ion pair created in primary ionization.

- *Recombination region ($G < 1$)* - In very low electric fields, the separation of the primary electron-ion pairs is not reliable and a fraction of the charge-pairs recombine. In conjunction with no amplification this yields an average gain smaller than one.

- *Ionization chamber mode or Unity-gain ($G = 1$)* - In electric fields above recombination threshold combined with the absence of an amplification region, only the electron-ion yield form primary ionization is measured.

- *Proportional region ($G \approx 10^3$ - $10^5$)* - In an electric field above amplification threshold, each signal electron triggers an exponentially growing electron avalanche of stable mean gain and gain fluctuation. The integrated charge remains proportional to the initial number of signal electrons, and thus the energy deposited in primary ionization.

- *Range of limited proportionality ($G \approx 10^5$ - $10^8$)* - At higher amplification the amount of charged ions in the vicinity of the anode increases and their space charge reduces the electric field experienced by following electrons. Thus the strict proportionality between deposited energy and the signal is gradually lost.

- *Saturation- or Geiger mode ($G \approx 10^8$ - $10^9$)* - The space charge driven field strength reduction becomes dominant with further increased gain, resulting in a selfquenching of the avalanche. Therefore, the total charge collected on an anode is limited and the signal strength saturates. Additionally, the very low effective field between the fast electron avalanche and the ion space charge allows for recombination processes under photon emission. These photons travel unhindered of the electric field, can ionize the gas and cause secondary avalanches. Thus the amplification spreads along the anode, increases the anode current and eventually causes a breakdown of the voltage, suppressing further amplification until the field is restored.

- *Discharge region ($G > 10^9$)* - If not contained or interrupted, the photon-triggered growth of the avalanche into a streamer can create a plasma conduit between the electrodes and cause a discharge. If not limited in current by an impedance, these 'sparks' can cause severe damage to the detector. The limit for the occurrence of discharges is usually referred to as Raether-Limit [14].

It should be noted that some of the gain limits are dependent on the geometry of the amplification field. The values above refer to a radial geometry [15] and can differ for other GD technologies.

### 1.1.3. Detector Characteristics

With the wide field of gaseous detector concepts described in 1.2, the optimized choice of a technology for a certain application is rather difficult. Comparisons of different detector types commonly refer to the following characteristics, providing a quantified rating of different aspects of the device's detection capability.

- The *efficiency* $\epsilon$ describes the probability of a trespassing particle to yield the expected signal and if applicable overcome a threshold to have this signal recognized. It is a composed value determined by different sources $i$ of inefficiencies:

$$\epsilon = \prod_i \epsilon_i. \tag{1.1}$$

  These sources include, but are not limited to, the geometrical efficiency $\epsilon_{geo}$, accounting for limited coverage or non-sensitive detector volume, the temporal efficiency $\epsilon_t$, for detectors suffering from dead time, as well as the statistical fluctuations in the primary ionization process $\epsilon_{prim}$.

- The *energy resolution* $\Delta E/E$ is central for GDs working in proportional mode and other devices aiming for a measurement of the deposited energy. It is typically referring to the width of a peak in the measured energy spectrum, which is associated with a monochromatic particle source. $\Delta E$ is thereby defined as the **F**ull **W**idth at **H**alf **M**aximum (FWHM) value and usually quoted as a fraction of the nominal peak energy $\Delta E/E$. For a Gaussian distribution of the measured energy, the peaks standard deviation is $\sigma_E = \text{FWHM}/2.36$.

- The *spatial resolution* $\sigma_x$ of a detector describes how accurately the particle's track, precisely the position of a reference point along the track, can be determined by the detector along a spatial coordinate $x$. Commonly any systematic shift in position reconstruction, for example by positioning of the detector, is removed before determination of the spatial resolution. Thus $\sigma_x$ becomes a measure of the residuals between the reconstructed hit position of the event $i$ and its true position.

$$\sigma_x^2 = \frac{1}{n} \sum_{i=1}^{n} (x_{i,true} - x_{i,reco})^2 \tag{1.2}$$

  With most sources of this deviation being normally distributed $\sigma_x$ can often be interpreted as the width of a Gaussian in the hit position distribution.

- Similarly the *time resolution* $\sigma_t$ defines the detector's typical statistical fluctuation of the signal timing referring to a reference value, after subtraction of any constant delay. The definition of the signal timing varies with the technology, applied readout scheme and electronics, but usually involves a time-over-threshold in the analogue current or voltage signal. Complementary thereto is the device's *response time* $\Delta t$, defined as the delay between the actual trespassing of the particle through the detector and the output of the signal. While $\sigma_t$ describes how accurately the timing is measured, $\Delta t$ assess how fast this measurement can be performed.

- The *recovery-* or *dead time* $\tau$ of the detector after an event can as well limit its application possibilities. It can be included in the detector's efficiency by means of a rate dependent inefficient time fraction

$$\epsilon_t = (1 - R'\tau) \tag{1.3}$$

with $R'$ being the measured rate. Two cases need to be distinguished: If an event renders the detector incapable of causing new signals within its complete recovery time, $\tau$ is fixed after each measured event. In this case the recognized event rate scales with $R' = R/(R\tau + 1)$ to the real rate $R$. If within the dead time new events still cause an amplification, which is simply not recognized, the recovery period is prolonged. The measured event rate is then reduced to $R' = R\exp(-R\tau)$.

- Strongly linked thereto is the *rate capability* $R_{max}$, which is the maximum allowed particle flux under which the detector's other characteristics are met. If the rate limit is due to a detector dead time $\tau$, the maximum allowed rate is determined by this recovery time and the envisaged temporal efficiency $\epsilon_t$. In case of a disabled detector with fixed recovery time, $R_{max}$ calculates as

$$R_{max} \leq \frac{1 - \epsilon_t}{\epsilon_t \tau}. \tag{1.4}$$

In case of a prolongation of the dead time after an event within the downtime, the maximum rate is further reduced according to the transcendental condition

$$1 - \epsilon_t \geq R_{max}\tau \exp(-R_{max}\tau). \tag{1.5}$$

Other limits to the rate are set by accumulation of space charges resulting in distorted drift processes, enhanced recombination rates or non-proportional amplification, all of those possibly effecting the energy resolution or enhancing the response time in case of a significant field reduction due to high ion density.

### 1.1.4. Parameters in a Gaseous Detector Setup

A core advantage of gaseous detectors is their tune-ability towards a wide range of applications. Not only do certain technologies perform better with respect to different challenges, but even for one detector type there is a list of parameters which can be optimized to improve a selected characteristic, often demanding a trade-off in another.

With the huge list of gaseous detector parameters and their specialized working range for each GD technology, providing a complete summary is difficult. Still, the most commonly addressed parameters are hereafter grouped and systematically discussed:

- The *gas parameters* obviously cause huge impact on the detector. They include the species and concentration of the *gas mixture*, which can be further discriminated in *nominal mixture* and *contamination.* Furthermore, the gas conditions like phase, pressure and temperature are associated to this category.

- The *detector geometry* comprises all dimensional aspects. It can be subdivided in the overall *layout*, essentially depending on the technology choice, the utilized *materials*, the *coarse geometry*, comprising all 'large' scale dimensions down to the 0.1 mm-level and the *fine geometry* parameters below that scale.
- The *electric configuration* describes the *potentials* applied to the detector electrodes as well as the *wiring*- and *grounding scheme*.
- Additionally, *environmental parameters* can affect the process within the detectors. While the impact of some parameters on the detector, like the *environmental temperature, pressure* and *humidity*, as well as *electric fields*, can be effectively reduced or shielded, others will inevitably influence the signal formation processes, like for instance an external *magnetic field* permeating the detector.
- A more separated kind of 'parameter' are the *readout electronics*, which are the determining factor for the signal readout, but have very little effect on the other processes in the detector.

In combination with the detector geometry, the electric configuration yields a set of effective parameters, like the field strength in a certain detector volume or capacities between parts of the detector. Similarly a distinction between nominal and effective gas parameters account for the impact of environmental parameters. While potentials, geometry, and so forth are the experimental accessible parameters, the effective parameters yield the impact on the signal formation processes, which is why analytic descriptions or microscopic models usually refer to the effective parameters.

## 1.2. From Gaseous Counting Devices to MPGDs

Literally interpreted a gaseous detector is first and foremost a contained gas volume able to detect the appearance, presence or transition of a particle. This definition comprises as well purely imaging based detectors like the cloud chamber [16], utilized in early particle physics discoveries. More commonly a detection in a GD manifests itself in form of an electric signal as, for example, in a Geiger Counter [17], the first prototype and possibly most famous representative of logical detectors. Nowadays the term of gaseous detectors is usually constrained to logical devices, involving the presence of an electric field in the gaseous volume. The electrode shape and, therefore, the field geometry can be used for a coarse categorization of the different GD technologies.

### 1.2.1. Proportional Counters and (Monitored) Drift Tubes (MDT) - Radial Geometry Detectors

The history of gaseous detectors started in the early 20th century with the radiation counter developed by Ernest Rutherford and Hans Geiger [17]. They utilized a radial electric field between the inner surface of the gas filled cylindrical tube and a concentrically stretched thin anode wire to detect individual ionization trails induced by natural radioactivity. These devices evolved in the 1920 - 40's to proportional counter, capable of measuring the amount of charge freed during the primary ionization [18–20]. Until modern times these geometrically simple detectors have been further refined to be utilized as, for example, streamer tube detectors or, more recently, as **M**onitored **D**rift **T**ube (MDT) arrays installed in the ATLAS experiment.

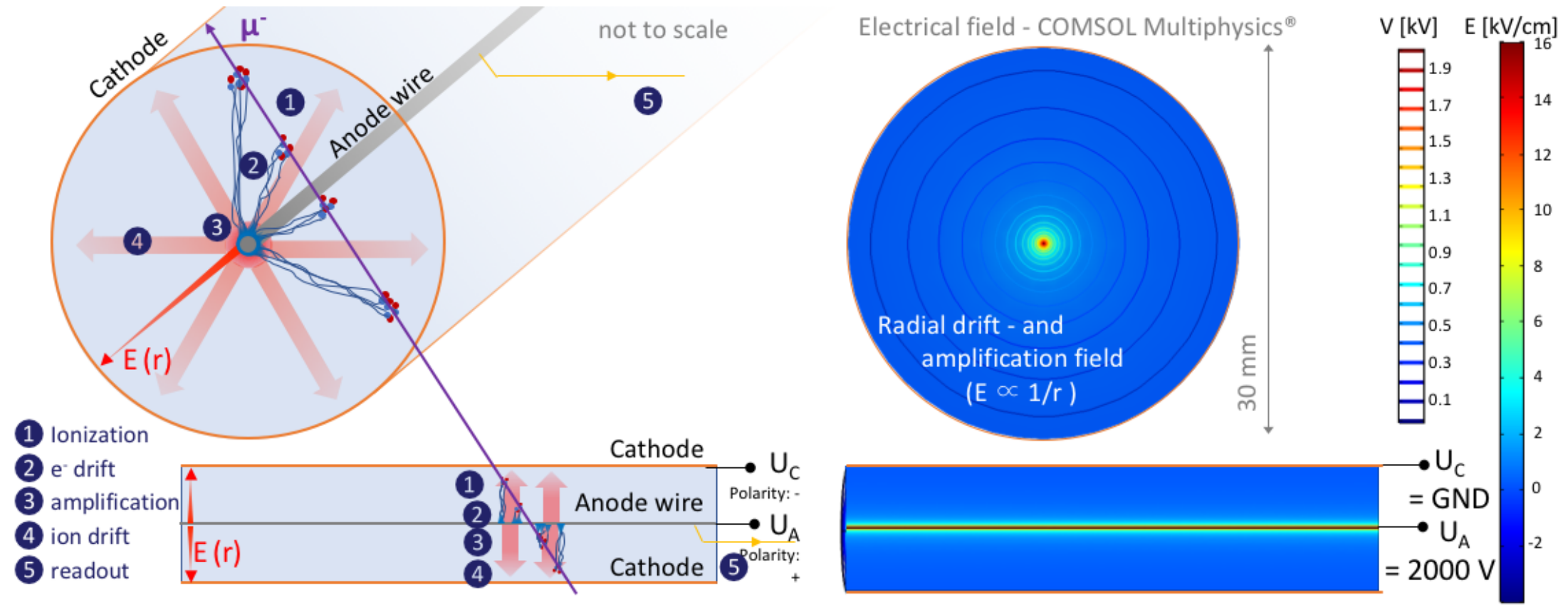

**Figure 1.1.:** Left: Layout of and signal formation processes in a gas filled (shaded blue) drift tube trespassed by a minimum ionizing muon (purple). The charge is in this example read out through the anode wire. The number of depicted electrons (blue), ions (red) and ion density (shaded red) are not to scale. Right: Electric field strength and equipotential lines in this radial geometry detector.

The overall geometry of the detector (figure 1.1) remains similar for all these applications, as do the processes for signal formation: A charged particle traverses the gas and looses energy in form of *primary ionization*, creating electron-ion pairs along its track. Those are separated by the electric field and *drift* towards the anode wire (electrons) or the tube cathode (ions). With the increasing field strength at lower radii, the electrons acquire more and more energy in-between collisions, eventually overcome the ionization threshold and start the *amplification* process. Due to electron diffusion the resulting avalanche is tear-shaped, (partially) surrounding the wire. While the avalanche electrons are almost instantaneously neutralized at the anode wire, the remaining *ions drift* towards the cathode tube, *inducing a signal* on the wire.

The main difference between the above mentioned applications is their selected gain $G$ range[1]. While first Geiger tubes have been used merely as counting devices in saturation- or Geiger mode, most modern gas tube detectors are operated in the proportional regime. This allows for determination of the energy deposited in form of electron-ion pairs during primary ionization, the standard objective for a proportional counter. Taking into account the particle's characteristic energy loss they can be used for particle identification, as done for example in the ATLAS TRT described in 5.1.4. By monitoring the anode arrival time of the different clusters of primary electrons, with the closest arriving first while the one created in vicinity of the tube arrives last, and considering the known drift velocity, the minimal distance of the particle's track through the tube can be determined. This operation mode allows the utilization of stacks of MDTs as precision tracking detectors (5.1.4). Streamer tube detectors, studied in detail in the 1970 - 80's [21, 22] trade this proportionality and particle identification for reduced response time and optical visibility of the event. They usually apply a resistive electrode coating, as already utilized in early Geiger counters [23], to limit the discharge current and are operated in a self-quenching streamer mode.

Comprising a simple to construct geometry, reliable operation and well understood physics processes, drift tubes are still applied in recent HEP experiments. They suffer, however, from a limitation in rate-capability which is owed to a voltage drop induced dead time in streamer- or Geiger mode or the long ion evacuation time in proportional mode. Being determined by the ions drift path and, therefore, the tubes diameter, a newer generation of small (s)MDT chambers with 1.5 cm tube diameter are foreseen as next generation drift tube based muon trackers in ATLAS (chapter 5.2.4).

### 1.2.2. Spark-, Streamer- and Resistive Plate Chambers (RPC) - Parallel Plate Devices

**P**arallel **P**late (PP) devices follow an approach complementary to the radial field detectors. A strong uniform electric field is created in the gas filled volume, enclosed by two parallel orientated flat electrodes. The *primary ionization* electrons freed along a trespassing particle's track immediately start an *amplification process*, once the strong electric field is applied (figure 1.3). While spark- and streamer chambers often rely on the visual detection of the photons emitted during the amplification process, more modern devices, like **R**esisitive **P**late **C**hambers (RPC), use conductive *readout* structures to inductively measure the charge separation and subsequent *ion drift*.

[1] In a radial geometry the gain is, for a given gas mixture, determined primarily by the wire and tube radii and the applied voltages.

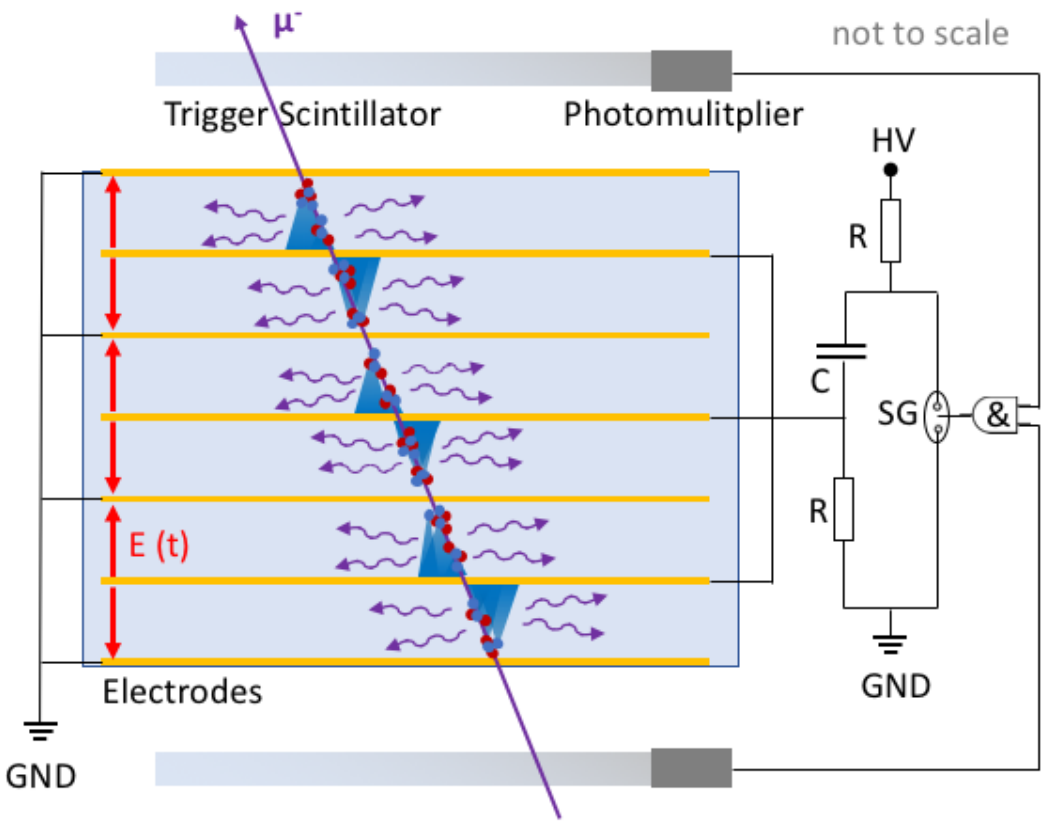

**Figure 1.2.:** Layout, wiring diagram and operation principle of a spark chamber. The trespassing muon induces a signal in both scintillators which triggers a spark gap (SG) and causes the charge up of the electrodes. The free electrons (blue) along the track cause photon (purple) emitting discharges in each gas gap.

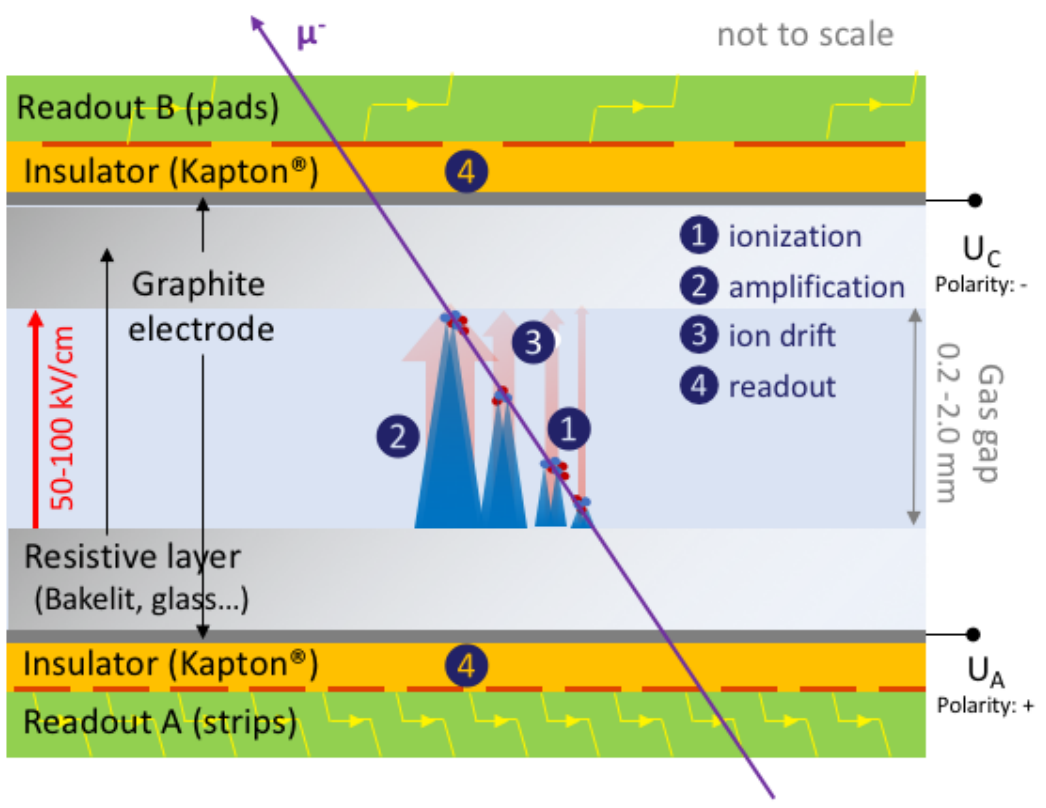

**Figure 1.3.:** Layout of and signal formation processes in a gas filled (shaded blue) RPC trespassed by a minimum ionizing muon. The number of depicted electrons (blue) and ions (red), avalanche size and ion density (shaded red) are not to scale.

The PP layout has first been successfully applied in the spark chamber [24], build in the late 1950's. This detector comprises a stack of parallel plate setups operated in discharge mode (figure 1.2). Since continuous operation in discharge mode is prone to spontaneous discharge and voltage breakdown, a fast trigger mechanism is applied on the top and bottom of the arrangement, enabling the high voltage supply once a particle passes. The charge up time of O(1 μs) is sufficiently short to avoid recombination of the electron-ion pairs before causing the discharge which yields a visible photon flash. The same trigger can be used for image acquisition from different angles, allowing full track reconstruction. An alternative purely electronic readout scheme relies on microphone arrangements recognizing the sound waves emitted by the sparks and their propagation. Apart from its historical importance of being the first detector capable of electronically triggered track imaging, the spark chamber suffers from long dead times of typically O(1 ms). Furthermore, all information on primary ionization density is lost in the discharge.

In a streamer chamber [25], the amplification of the primary charges is triggered similarly to a spark chamber, but interrupted before the streamer can develop into a discharge. This interruption is realized by limiting the exposure time to the high electric field to O(10 ns). During this short period each charge seed develops into a sub-mm long, photon emitting streamer, yielding a visible track. Since the spatial extent of the avalanche is limited, large volumes can be instrumented with a single pair of electrodes. Although the amplification is locally in a saturated mode, the density of streamers and, therefore, charge clusters can be used to determine the particle's characteristic energy loss. Additionally, a multitude of tracks can be visualized simultaneously. Substantial drawbacks are the limitation to a visual readout and the detector dead times in O(10 µs), owed to charge evacuation.

The immediate amplification and absence of a long drift process predestines parallel plate setups as trigger detectors with an intrinsically good time resolution. However, a pulsed operation mode and long recovery time after a discharge forbid trigger application in HEP experiments. These limitations have been overcome with the utilization of resistive electrodes. First proposed with the Pestov Counter in the 1970's [26], this very accurate high pressure detector evolved into the **R**esistive **P**late **C**hamber (RPC), featuring a simplified construction with lower accuracy requirements and operation at atmospheric gas pressure. The key feature of these parallel plate detectors (figure 1.3) are the highly resistive electrodes commonly build from special (Pestov-) glass, Bakelit or thin graphite layers. Operated in the proportional mode or lower streamer regime, the resistive electrodes limit the voltage drop to a 0.1 $\mathrm{cm}^2$ region around the amplification spot. This local voltage drop reduces the field strength and thus quenches the avalanche. An avalanche spread via UV photons is suppressed by an admixture of quenching gases. Furthermore, the resistive electrodes allow for an inductive signal pick up on readout structures outside of the gas volume. Although the recovery time in the region of the voltage drop is still in the order of 10 ms to 0.5 s, depending on the electrodes resistivity, the rest of the detector's active area is unaffected. The rate capability of an RPC can be tuned up to particle rates of 1 $\mathrm{kHz/cm}^2$ by adjusting the mean gain and the surface resistivity. The thickness of the gas gap determines two major detector characteristics, time resolution $\sigma_t$ and detection efficiency $\epsilon$ and, therefore, demands a trade-off between a precise timing detector (thin gap, O(200 µm)) and efficient trigger RPC (larger gap, O(2 mm)). The stacking of several thin gas gaps into a Multigap RPC [27] combines both, but requires a more complicated detector layout.

### 1.2.3. Multi-Wire Proportional Chambers (MWPC) and other Wire Grid Applications

A breakthrough for gaseous detectors has been accomplished by George Charpak in the 1960's with the noble prize rewarded invention of the **M**ulti-**W**ire **P**roportional **C**hamber (MWPC) [28]. This has been the first technique for electronic track reconstruction in uninterrupted large gaseous volumes and combines the advantages of an (almost) uniform drift field with a radial amplification field (figure 1.4 - right).

The MWPC owes its name to the grid of regularly spaced anode wires which are stretched inside an extended gas volume, enclosed by two cathode planes (figure 1.4). Once a particle *ionizes* the gas, the charges are separated and *drift* in the almost uniform electric field. When approaching the wire grid, the parallel field lines evolve into a radial configuration with increasing field strength at small anode distances, where the *amplification* process in proportional mode is triggered. The *ions drift* along the field lines and this charge movement is measured on the wire and/or on *readout* structures at the cathode.

Wire grids have been applied as amplification and readout structures in a variety of detectors and several HEP experiments, among them the huge cylindrical drift chamber of the UA1 experiment of 2.3 m diameter and 5.8 m length, comprising of 17000 field wires and 6125 sensing wires [29]. Wire arrangements are still used as amplification stage in **T**ime **P**rojection **C**hambers (TPC), like the 88 $m^3$ ALICE TPC [30]. Other applications in large volume detectors include signal sensing without amplification, for example in ionization chambers operated at unity gain, or ion back-flow suppression with gating grids used in pulsed operation TPCs. Besides large volume chambers, wire grids are utilized as well in planar detectors like the **C**athode **S**trip **C**hambers (CSC) in ATLAS [6] and CMS [31] or the ATLAS **T**hin **G**ap **C**hambers (TGC), described in more detail in chapter 5.1.4. Both detectors feature segmented readout structures to obtain additional hit position information, either as the cathode (in CSC) or in an RPC-like configuration with a resistive electrode protecting an inductively coupled readout structure (in TGC).

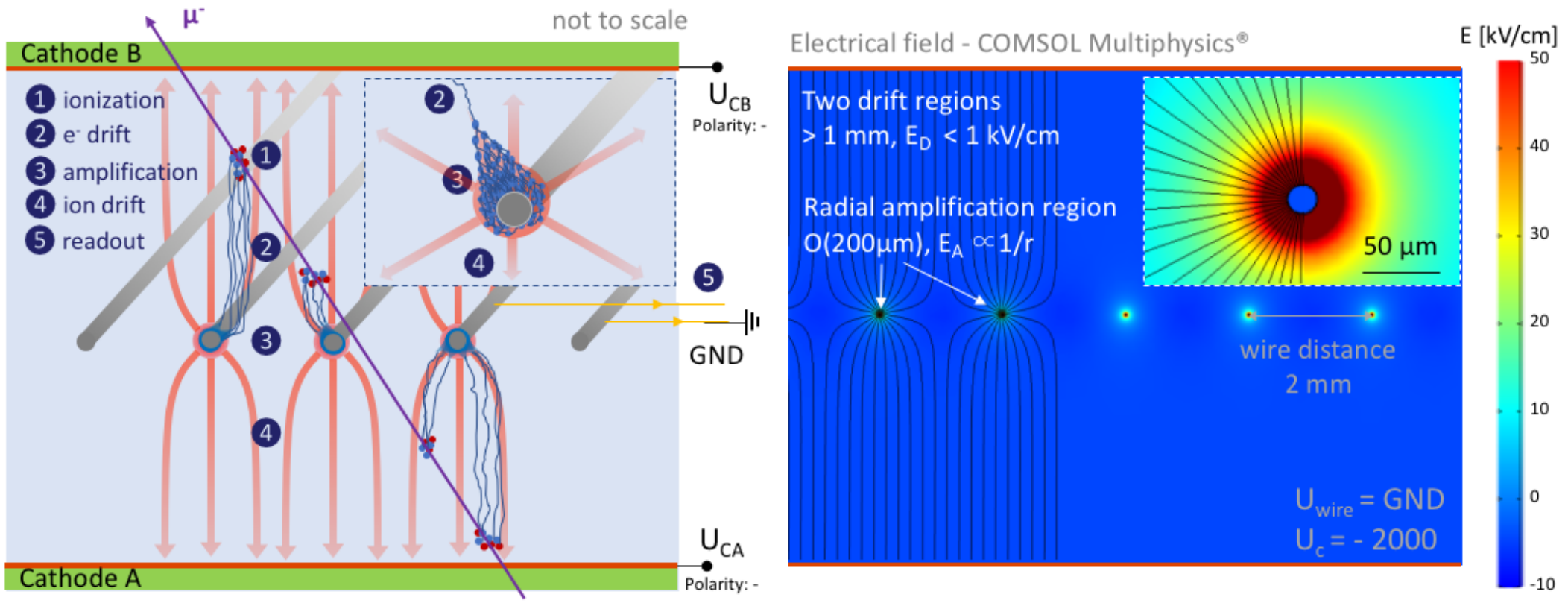

**Figure 1.4.:** Left: Layout of and signal formation processes in a multi-wire proportional chamber. The electrons (blue) freed in primary ionization along a muon track (purple) drift to and are amplified at the anode wires, where the subsequent ion drift (red arrows) is measured. The number of depicted primary clusters and charge densities are not to scale. Right: Electric field configuration in the vicinity to an MWPC anode wire grid with the transition from uniform drift field to radial field amplification regions.

### 1.2.4. Gaseous Electron Multiplier (GEM)

A new approach towards electron amplification has been introduced in the late 1990's by Fabio Sauli with the **G**aseous **E**lectron **M**ultiplier (GEM) [32]. A GEM foil is typically made of 50 μm thin Kapton® with copper layers on both sides. An array of holes, with typical diameter of 50 - 70 μm and 100 - 150 μm spacing, is photo-lithographically etched into this foil. By applying a potential difference of several hundred volts between the two copper electrodes, a strong electric field is generated in these holes (figure 1.5 - right).

A *primary ionization* electron *drifting* along the field lines towards a GEM foil will be focused into the holes and start an *amplification* process within the small volume of the strong electric field. The multiplication is stopped once the avalanche leaves this region, allowing for very well controlled, adjustable gas gain in the $10^1$ - $10^3$-range. The stacking of several GEM foils and their operation at rather small gain per stage became an established concept to reach high gain with low fluctuation, an example of a double stage GEM is shown in figure 1.5.

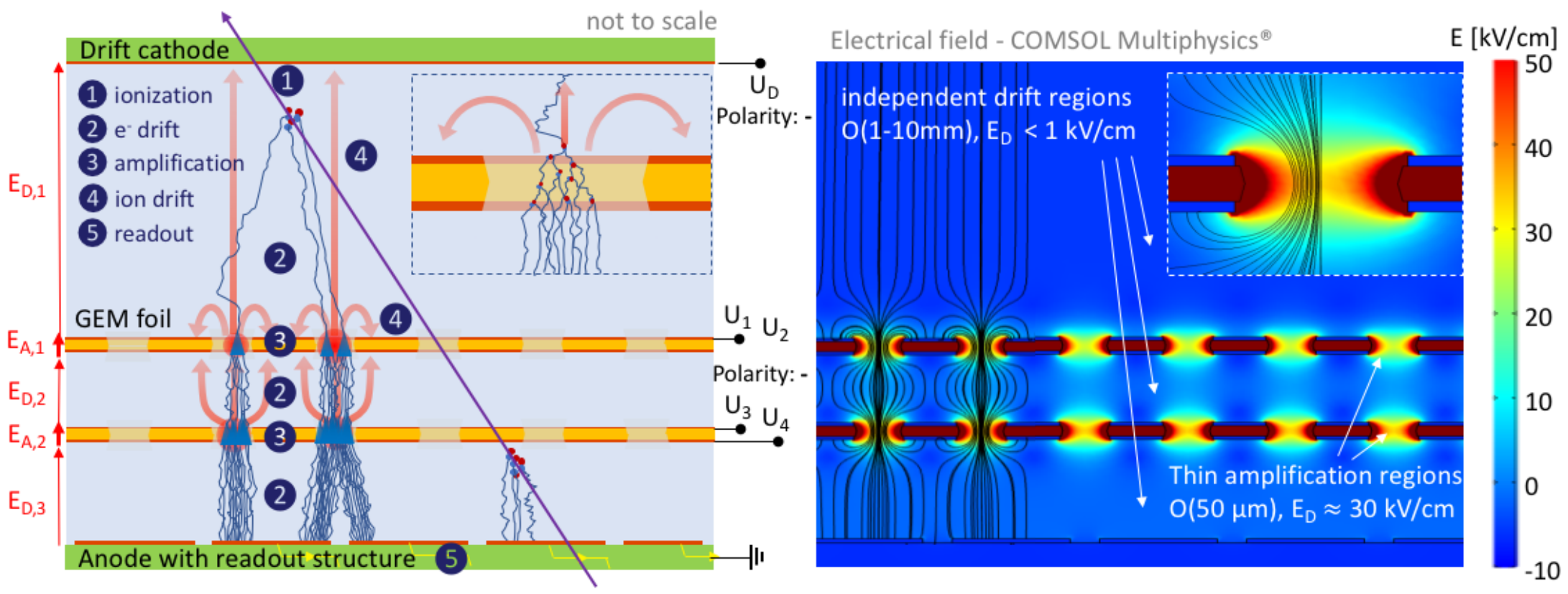

**Figure 1.5.:** Left: Layout of a double stage GEM detector and signal formation processes triggered by a trespassing muon (purple). The electrons (blue) freed along the track are guided into the GEM holes, where they are amplified, extracted on the other side and further amplified in the second GEM. The ion flow (red arrows) is indicated. The primary cluster number and charge densities are not to scale. Right: Electric field configuration in the GEM stack with the extended drift regions, confined amplification volumes and exemplary streamlines to indicate charge drift.

A reliable gain furthermore depends on the collection efficiency of electrons approaching the GEM as well as their extraction efficiency into the next drift- or transfer volume. Both can be optimized by tuning of the potentials and electric fields. This field tuning as well determines the *ion back-flow* (IBF) into the upper gas volume or rather the desirably high fraction of ions neutralized on the GEM electrodes. Since ions produced in lower GEM stages are efficiently collected on the upper GEM electrodes, the first amplification stage and its gain dominates the IBF.

Being independent amplification units GEMs are highly adjustable to the desired application. They can be utilized with different readout techniques, like strips, pads, pixel or wire grids and, if required, be combined with resistive electrodes. GEMs can be applied as thin gap detectors, like the CMS triple-GEM Muon detector upgrade [33], as well as amplification stage in a drift chamber, like the foreseen ALICE TPC upgrade which is based on four layer GEM stacks [34]. Drawbacks of the GEM technology are their complex detector construction, vulnerability to discharges and required trade-off between efficient electron collection and suppressed IBF.

The **th**ick **GEM** (THGEM) [35], also referred to as **L**arge **E**lectron **M**ultiplier (LEM), applies the same mechanism on copper clad fiber glass epoxy boards with a typical thickness and, therefore, amplification range of O(500 µm). These amplification stages are more durable, easy to produce and less prone to discharges, but lack the fine tuning abilities and good energy resolution of a GEM stack.

### 1.2.5. Micromesh Gaseous Structures (Micromegas)

A combined amplification and readout technology has been introduced by Yannis Giomataris with the **Micromesh gaseous** structure (Micromegas, MM) [36]. In a Micromegas detector the gas volume is divided by a thin conductive mesh into a larger drift- or conversion volume of up to several cm and a thin amplification gap of O(100 µm) in front of the readout structure (figure 1.6). Applying potentials on the drift cathode, the mesh and the anode structure electric fields in the two regions are generated independently[2], creating a two stage parallel plate like setup. An *ionizing* particle leaves a trail of electron-ion pairs along its track through the drift volume. The charges are separated and *drift* in the (almost) uniform field to the cathode (ions) or towards the mesh (electrons). Approaching the mesh, due to the much stronger field, they are guided through the mesh openings into the amplification region. Here they start an *amplification* process and the electron avalanche develops in the (ideally) uniform, strong field until it reaches the anode. The remaining *ions drift* towards the mesh where a majority is neutralized and only few are extracted into the drift volume. The charge movement can be measured on a *readout* structure at the anode or at the mesh.

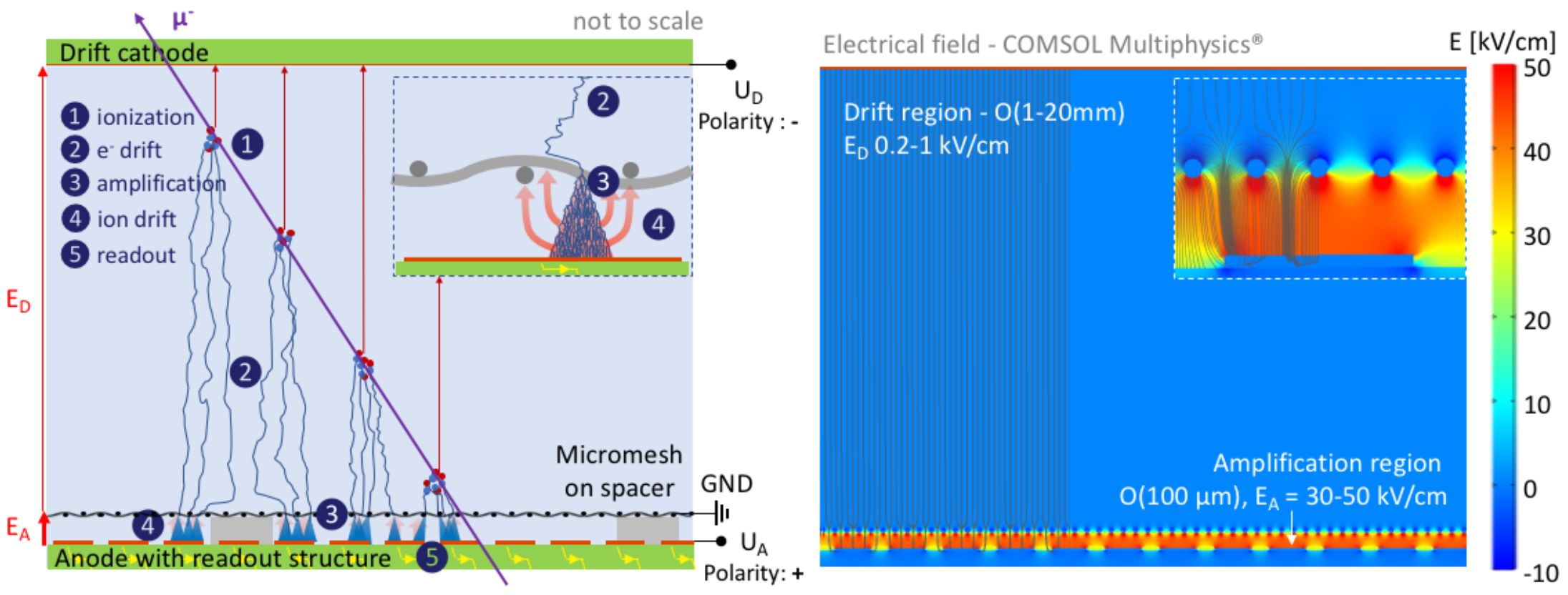

**Figure 1.6.:** Left: Layout of and signal formation processes in a Micromegas detector trespassed by a muon (purple). The freed electrons (blue) drift towards the mesh, enter the amplification region and trigger an avalanche (see magnification). The ions are collected at the mesh. The number and density of charge carriers are not to scale. Right: electric field and streamlines in a Micromegas detector visualizing the double stage parallel plate like configuration. The non-uniformity is due to the mesh structure and anode segmentation (see chapter 8).

The thin amplification gap yields a quick ion evacuation and thence good time resolution and limits the charge spread during amplification, thus not degrading the readout structures spatial resolution. It allows for a fine gain tuning within a typical range of $10^3$ - $10^5$ and features limited gain fluctuations yielding a good energy resolution. This distance between the mesh and the anode is commonly defined by a spacer struc-ture mounted on the anode (more details in chapter 6.3). Since Micromegas detectors are prone to inaccuracies in the amplification gap thickness and contamination of the amplification volume, both increasing the risk of discharges, the construction of these structures requires high precision and is comparatively complex (see chapter 7 for a detailed discussion).

[2]This independence of the fields is limited by the mesh's shielding capability, as discussed in detail in chapter 8.

Micromegas detectors have been used so far in rather specialized applications, such as the CAST detector [?] or in the cylindrical configuration of the CLAS12 [37] experiment. The ATLAS New Small Wheel [9], which will be discussed in part II, will be the first large scale application of this technology.

### 1.2.6. Other Micro Pattern Gaseous Detectors (MPGD)

GEM and Micromegas are the two most utilized technologies of **M**icro **P**attern **G**aseous **D**etectors (MPGD). As the name suggests these devices all utilize very small (O(100 µm) and below) and precise (<10 µm) conductive patterns, usually created with photo-lithography and etching processes. The progress in these production procedures, owed to the worldwide increasing demand on printed circuits, as well as improvements in field simulation techniques allowed for a rapid growth of MPGD concepts, which is still ongoing. With only a few concepts mentioned below, a comprehensive overview can be found in [38].

The Micro Strip Detector [39], featuring an alternating pattern of cathode strips and 10 µm thin anode strips with an O(50 µm) amplification pitch, is generally accounted for as the first MPGD prototype. The well detector [40] and its successors like the resistive µ-well [41] comprise a GEM like hole pattern directly applied on a readout structure. In a similar approach the Integrated Grid (InGrid) [42] technology utilizes a very precise Micromegas directly deployed on a silicon pixel readout chip with one pixel below each mesh opening, forming a GridPix detector [43].

All MPGD layouts share the objective of limiting the amplification process to a narrow and well defined region. Their concepts are usually independent of the drift- or conversion volume and thus they can be applied as planar detectors or as readout for large drift volumes and TPCs. Generally spoken most MPGDs can achieve high rate capability, short response time and spatial accuracy comparable to semiconductor detectors, combined with the possibility to equip large volumes which is characteristic for gaseous detectors. They are, however, sensitive to discharges and prone to damages of their fine metallic structures, a weakness which has been partially overcome by the application of resistive materials, as discussed on the example of the resistive Micromegas in chapter 6.2.

## 1.3. Studies of Signal Formation Processes

Reviewing the operation principle of gaseous detectors it becomes obvious that (A): They all operate on the same fundamental processes. And (B): These processes are temporally ordered, distinct and occur, on a single electron level, one after another.[3]

For their temporal separation, the physics mechanisms occurring in each of the processes are on a microscopic level[4] independent of each other and can be described separately. The full signal formation can thus be composed of the different processes and its outcome can be predicted, provided all mechanisms are sufficiently well understood and the transitions between the stages are accurately described. Conversely it is possible to experimentally study the physics of a single stage by decomposing the signal formation process.

This approach is applied throughout the studies presented in this thesis, which are focused on the mechanisms of electron-gas interaction during drift and amplification and the transition between both stages. Although exercised on the example of Micromegas detectors, with a special emphasis on the New Small Wheel Micromegas configuration, the overall principle and to a certain extent the findings of this study can be applied to any gaseous detector.

### 1.3.1. Factorized Approach on Signal Strength Description

In a detector operated in proportional mode (1.1.2), like a Micromegas (1.2.5), the signal strength $S$ is used to determine the charge deposited in the gas volume by the trespassing particle. Depending on the ion drift behavior, applied readout scheme, involved capacities and utilized electronics, this signal strength can be measured as amplitude of a current peak, voltage drop or as integrated charge. In all cases, the electronics must maintain the proportionality of $S$ to the charge created in the detector's amplification stage and can be, therefore, summarized as a conversion factor $c_{r/o}$. In a Micromegas the charge is separated in the amplification gap and collected on the anode (electrons) and the mesh (majority of ions) or the cathode (small ion fraction). It is proportional to the detectors gain $G$ and the signal electrons arriving in the amplification region and triggering an avalanche. Ideally all signal electrons $n_e$ created in primary ionization should contribute to $S$. In a real Micromegas, however, several processes contribute to a loss of signal electrons before they can trigger an avalanche. This fraction of electrons lost for signal formation $L$ reduces the total charge yield during amplification and thus the signal strength.

$$S = n_e \cdot (1 - L) \cdot G \cdot c_{r/o} \tag{1.6}$$

The main sources of electron losses are recombination $R$ after primary ionization, attachment $A$ to gas constituents during scattering and neutralization $N$ of the electrons at the boundaries of the gas volume. In a Micromegas the latter term is dominated by electron neutralization at the mesh during the transit from the drift- into the amplification region. The fraction of not neutralized electrons $(1 - N)$ is commonly referred to as electron transparency $T$. Thus (1.6) evolves to:

$$S = n_e \cdot (1 - R) \cdot (1 - A) \cdot T \cdot G \cdot c_{r/o} \tag{1.7}$$

[3] Depending on the technology, some process can be omitted or suppressed, like the drift process in an RPC, or occur multiple times, like electron drift and amplification in a GEM stack.

[4] During the reminder of this thesis the term *'microscopic'* is used to describe processes on a per electron, per molecule or per interaction level, while *'macroscopic'* refers to experimental observables which commonly summarize or integrate over several magnitudes of microscopic objects or events.

This factorization allows as well to analytically assess the individual contributions of the sub processes to several detector characteristics. The energy resolution for example is determined by the fluctuations of the signal strength $\sigma_S$ which can be decomposed according to (1.7).

$${\sigma_S}^2 = {\sigma_{n_e}}^2 + {\sigma_{(1-R)}}^2 + {\sigma_{(1-A)}}^2 + {\sigma_T}^2 + {\sigma_G}^2 + {\sigma_{c_{r/o}}}^2 \tag{1.8}$$

In a similar way the timing characteristics can be derived by adding up the average time required for each process to obtain the detector's response time. Herein the dominant contributions are commonly the ion drift and the signal processing time in the electronics. The correctly combined fluctuations in all these contributions pose a lower limit to the time resolution. Assessing the spatial resolution on an analytic level remains difficult for electric fields without an analytic solution like in a GEM or Micromegas. Therefore, these systems are commonly treated numerically utilizing geometrical approximations and Monte Carlo methods.

The validity of this factorization approach is limited, due to the underlying independence assumption of the processes. Although their independence is ensured on a per electron level it is not guaranteed for variables summing over several electrons or processes, like for example the number of signal electrons $n_e$ and the fraction of recombination $R$ or attachment $A$ losses. In this example $n_e$ only represents a number of freed electrons, but neglects their spatial distribution. $R$ on the other hand primarily depends on the external electric field and the local electron-ion density, the latter being determined by primary ionization, but not reflected in $n_e$. Therefore, a careful reflection on the primary ionization processes is required to understand possible dependencies in the experiment. Similarly, the number of collisions between an electron and the gas, and thus the attachment probability of that electron during its whole drift process, is dependent on the actual drift distance and, therefore, the electron's starting position. This can be mitigated by maintaining a statistically stable primary distribution and regarding $A$ as the average attachment probability, implicitly referring to this start point statistic. In a similar way the temporal independence of the summarizing processes can be broken by time delays between, for instance, the amplification of electrons with different drift times. In that case early electrons could yield ion clouds causing space charge effects, sufficiently strong to affect the amplification of electrons arriving later. Space charge effects can be suppressed by keeping the gain and the event rate sufficiently low to limit the ion density.

While process factorization is ideally realized in a single electron response signal, other signals involving multiple signal electrons have to be scrutinized towards possible feedback between the processes. Therefore, a detailed discussion on primary ionization is preceding (chapter 2) the study on drift and amplification processes.

### 1.3.2. Dependency between Parameters, Processes and Characteristics

Many GD studies focus on the causality between a parameter variation and the change in the detector's characteristics, with statements like: 'An increase in gas pressure lead to an improved energy resolution'. While these statements can be perfectly true, they obscure the real chain of causality: The parameter has an impact on the signal formation processes and this change in signal formation yields modified detector characteristics. With each parameter possibly affecting multiple processes, each process being influenced by multiple parameters and influencing in turn several characteristics, this creates a system of dependencies, partially[5] visualized in figure 1.7.

The importance of this remark becomes clear when widening an optimization study from a single parameter-characteristic pair, to a multi-parameter problem with the optimization goal being a trade-off between different detector attributes. While conducting experimental optimization studies on a limited set of parameters is a viable option, the effort scales approximately with the power equal to the number of considered parameters. Additionally a later change of a parameter not taken into account will require a repetition of the study or at least a reinterpretation of its results.

Instead of performing an optimization on predefined parameters, the herein presented study focuses on an improved understanding of the impact of a parameter variation to the signal formation processes. Therefore, a VOTAT [44], short for '**v**ary **o**ne **t**hing at **a** **t**ime', strategy is applied. As the name suggests the key of this procedure is the variation of a single input parameter under preservation of all others. Thus 'VOTAT allows for the direct observation of isolated effects on output variables by manipulating only one input variable at a time' [45] and yields significantly higher success rates and increased understanding for multi-parameter problems, compared to non-VOTAT strategies.

Most detector R & D studies are implicitly utilizing a VOTAT strategy without notice. Therefore, they are often conducted without complete consideration of the full parameter space or lacking a model and, therefore, understanding of the dependencies between the systems in- and output. A conscious, systematic and strict application of the VOTAT strategy, as attempted during the hereafter presented studies, is likely to enhance the insight on the complex dependency system between the parameters of the 'problem' (= detector systems) and its output variables (= characteristics).

The first step thereto was the assessment of the parameter space under study and formation of a dependency-model, utilizing the signal formation processes as a mediator stage between in- and output. The second step is a consideration of all those dependencies in the theoretical model. Therein the correct description of a parameter, and its variation, impact on a process can be difficult to implement, while the preservation of all other parameters is implicit in a theory or simulation model. In the third step, the approach has to be applied on all experimental studies. While several nominal parameters in a GD can be easily changed, like, for example, the electric potentials, others require dedicated detector setups to be modified, like the fine geometry parameters of a Micromegas micromesh. Furthermore, the impact of a variation on the effective parameters has to be well understood in order to avoid second order effects on the process under study. Finally, the preservation of the remaining parameters must be ensured, and the remaining fluctuation be taken into account during data analysis.

[5] The visualized system does not claim completeness. On the contrary some parameters like, for example, a magnetic field and their impact have been suppressed in the interest of clarity.

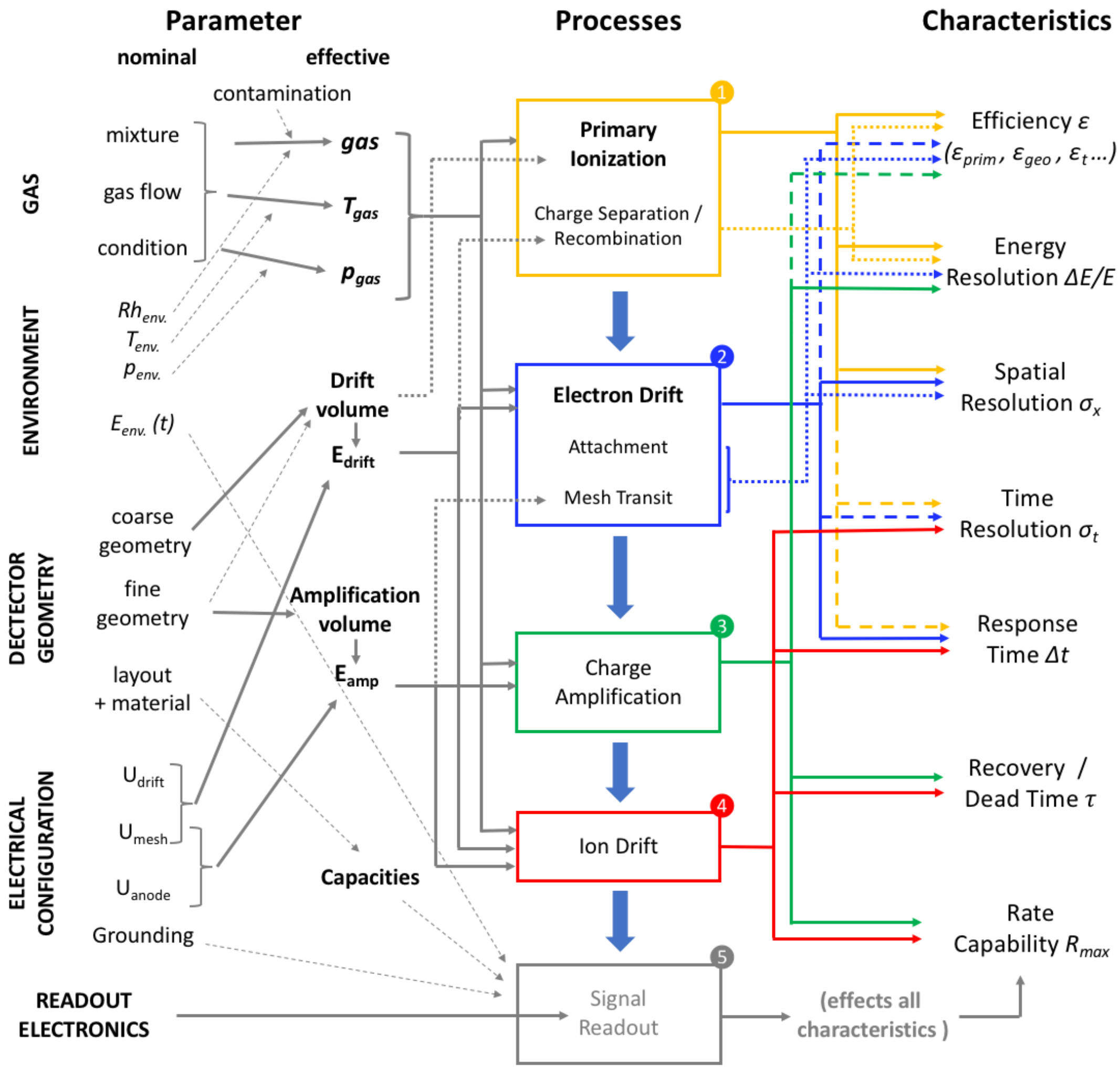

**Figure 1.7.:** Dependency model of parameters (left), processes (center) and characteristics (right) in a Micromegas. Each experimentally controllable nominal parameter determines (grey arrow) or affects (dashed, grey) the effective parameters. They in turn define the signal formation processes (full, grey), or their sub-processes (dotted, grey). The characteristics depend on those processes, or sub-processes (dotted, color), in first (full, color) or higher (dashed, color) order.

### 1.3.3. Analytic Description, Microscopic Model and Experimental Results

As in most fields of physics, this study is based on the comparison of experimental results, reflecting the reality, and a continuously refined theory, comprising our understanding of the reality. In the gaseous detector research the theory can be subdivided into the *analytic description of macroscopic observables* and the *modeling of microscopic processes*.

Analytic descriptions typically refer directly to experimental observables (for example the signal strength), averaged values (like the gain) or statistic parameters of a process (as the electron diffusion during drift). Therefore, they can be easily compared to experimental results, which allows for rapid theory refinement and increasingly accurate phenomenological descriptions. They suffer, however, from two major drawbacks: many of the derived formulas are only applicable to geometries where an analytic solution to the field configuration exists, requiring approximations to simplify more complex systems to an analytically solvable level. Furthermore, they can only grant limited insight on the fundamental microscopic processes on single electron, molecule or interaction level which eventually yield the experimental observable. To measure a mean gain, for example, typically some $10^4 - 10^6$ events are analyzed, each summarizing over O(100) amplification processes involving $10^3 - 10^6$ electrons each, thus reflecting millions of millions of interactions on the microscopic level. Observing only average values and describing them with a limited number of parameters naturally limits the insight to the real fundamental processes. A complete understanding of these microscopic processes is the ultimate goal of fundamental GD R & D.

To enhance this understanding, the experimentally non-accessible microscopic system is mathematically described in a model. Aiming for highly precise process description these models commonly comprise many more parameters than an analytic description. In the above given example the model would include the properties of the electrons and each gas constituent, scattering cross-sections for all (known) electron-gas interaction processes at different energy levels, probabilities for energy transfer mechanisms, an analytic or numeric description of the gas volume and the electric field and much more. Utilizing random number based Monte-Carlo (MC) methods a macroscopic realization of the microscopic system is then simulated. This realization of the model contains the same set of observables and thus allow for comparison with experimental results and, furthermore, grants insight to sub-processes which are possibly hidden on observable level. In understanding differences between the experimental data and the simulation results, the model can be improved in an iterative manner (figure 1.8) and thus the understanding of the fundamental physics processes is enhanced.

Once a model is judged to be valid, based on the degree of agreement to the experimental data, it can be used to refine the analytic description, for example, by adding correction terms or providing parameters. This leads to more precise analytic predictions, with the benefit of comparatively simple to use analytic formulas instead of time consuming model simulation.

Within this study, we put an emphasis on development of simulation tools for microscopic models and model enhancement for two kind of processes: electron drift, including charge loss processes and the electron's transition through structures (chapter 3), as well as the amplification process under consideration of energy transfer mechanisms (chapter 4). In both cases, the experimental setups where optimized towards an unobstructed measurement of the corresponding observables: electron losses during drift and mesh transit, or the gain and gain variance. While a full analytic description of transparency losses is still out of range, the model results will be discussed in comparison to analytically solvable more simple geometries. In the study on avalanche formation our validated model has been used to derive parameters to include energy transfer effects in the analytic description.

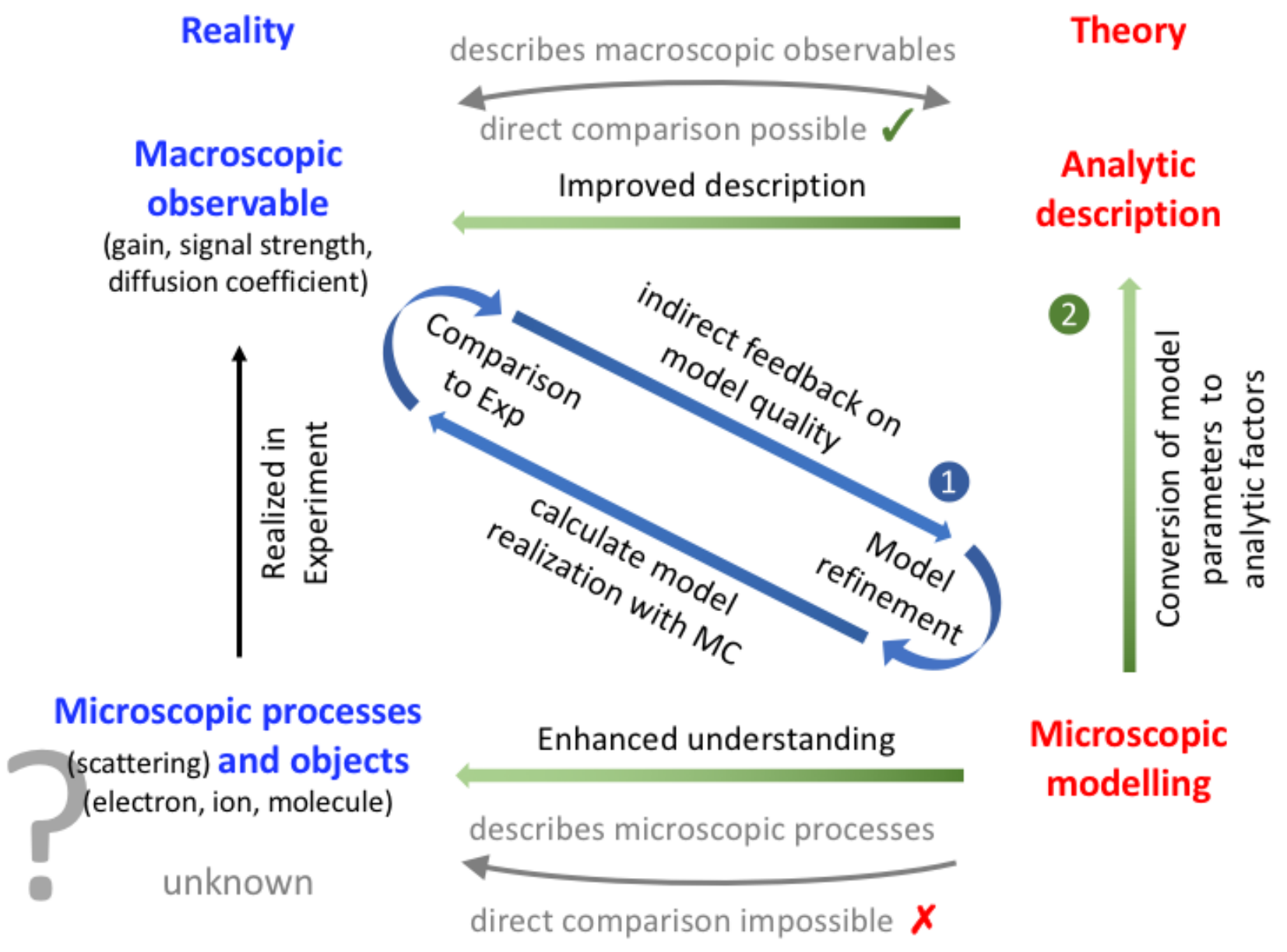

**Figure 1.8.:** Visualization of the interplay between theory (red, right) and reality (blue, left) on macroscopic (top) and microscopic (bottom) scale. (1, blue arrows): Model iteration process under comparison with the experiment. (2, green arrows): Impact of a validated model.

# 2. Primary Ionization in Gases

This chapter covers the mechanisms of energy deposition by a particle in the detector, which is the prerequisite for its detection. The dominant mechanisms for energy deposition are based on electromagnetic interaction between the particle and the detector material. Given the shorter range of weak and strong interactions, their occurrence is suppressed by several orders of magnitude. They remain, however, the only possible channel for the detection of neutral hadrons (strong) or neutrinos (weak interaction).

The electromagnetic interaction mechanisms during matter transit by charged hadrons and muons, electrons and positrons as well as photons are discussed in section 2.1. Thereafter, the primary ionization process of three exemplary event types in gaseous detectors, all of which are utilized in the remainder of the studies, are discussed in detail in section 2.2.

## 2.1. Electromagnetic Interactions of Particles with Matter

The microscopic description of the electromagnetic interaction between charged fermions and photons is provided by the theory of **Q**uantum **E**lectro **D**ynamics (QED).

For charged particles QED describes energy loss processes in a material via scattering, modeled as exchange of virtual photons with the shell electrons of an atom, or by photon radiation. In a scattering process, the energy lost by the charged particle is transferred to the target, typically a shell electron, yielding the atom or molecule in an excited or ionized state.[1] Once a charged particle is thermalized, meaning its energy is reduced to the scale of thermal equilibrium, ionization stops and it becomes invisible to the detector. For photons scattering processes yield an energy transfer to the target, thus an increased wavelength as well as a change in propagation direction defined by a scattering angle. The photoelectric effect leads to excitation or ionization of the target with the photon being absorbed. In a pair production process, the photon converts into an electron-positron pair. While during energy loss processes the original particle is preserved, photons can vanish during absorption interactions or pair production. This fundamental difference is reflected in the penetration depth of a mono-energetic beam of photons or electrons into matter (figure 2.1).

The probability of a single interaction process $i$ is defined by its cross section $\sigma_i$. It can be geometrically interpreted as the effective interaction surface of a target, as seen by the approaching point like projectile particle and is, therefore, measured in units of $\mathrm{cm}^2$ or 'barn', with $1\,\mathrm{barn} = 10^{-24}\,\mathrm{cm}^2$.

[1] Depending on the molecule structure more processes like 3-body attachment, dissociation and excitation of vibration or rotation states are possible, as addressed in chapter 3.

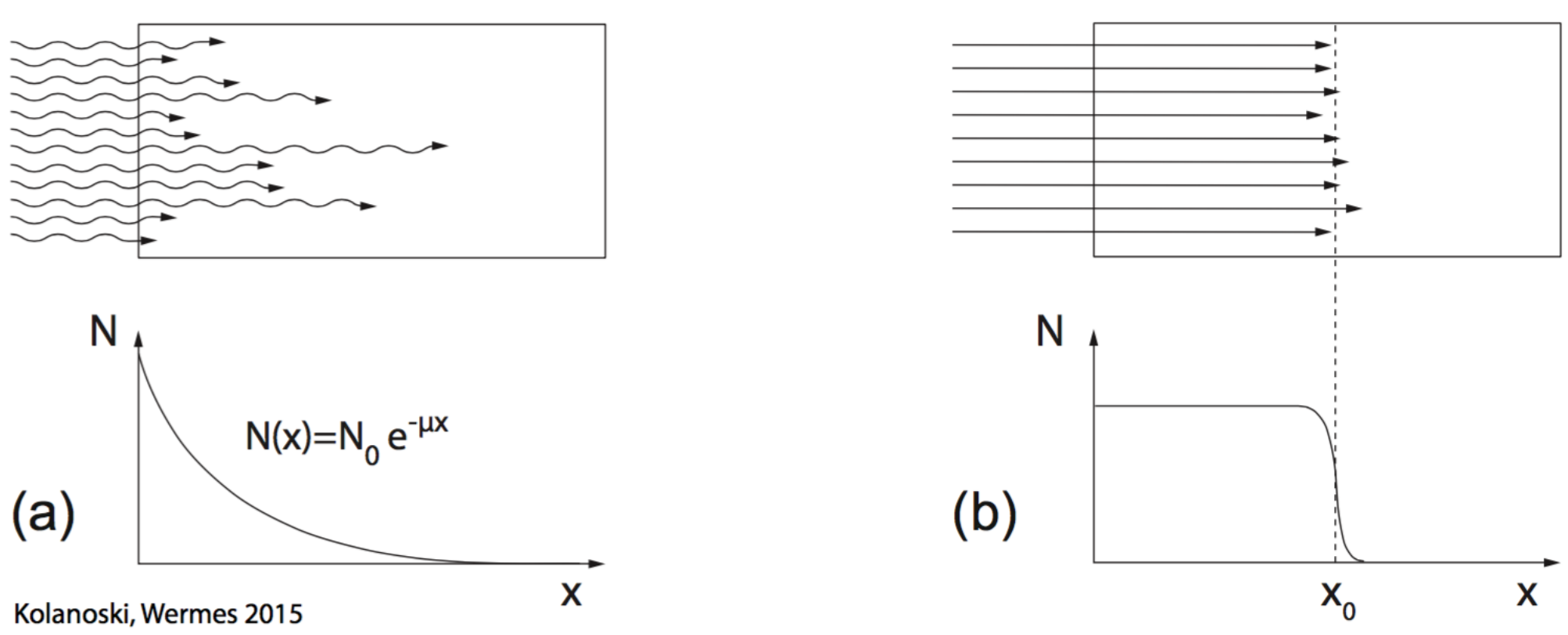

**Figure 2.1.:** Intensity evolution for a beam of mono-energetic photons (left, (a)) and electrons (right, (b)) in matter: the photon intensity decreases exponentially with the path length, but each photon maintains its energy, as long as (Compton-) scattering is suppressed. A charged particle penetrates the material up to a certain characteristic depth, continuously loosing energy along its path until it thermalizes and ionization ends. Thus energy loss yields a flat intensity profile with a localized decrease. [12]

The macroscopic observable counterpart of the cross section is the interaction rate $\dot{N}_{R,i}$ per incoming projectile rate $\dot{N}_{in}$, a measure of the interaction probability. For a sufficiently thin target, the approximation of a non-modified beam during trespassing leads to:

$$\sigma_i = \frac{\dot{N}_{R,i}}{\dot{N}_{in}} \frac{A}{\rho l N_A} \tag{2.1}$$

with the illuminated surface $A$, the material thickness $l$, its density $\rho$ and the Avogadro constant $N_A$.

### 2.1.1. Energy Loss of Charged Hadrons and Muons

For the macroscopic description of primary ionization by energy loss the focus is shifted from a single interaction process to average values summarizing over many interactions. The mean energy loss per path length in a medium $\left\langle -\frac{dE}{dx} \right\rangle$ can be analytically derived from the model (QED) parameters, provided those cross sections and their dependencies are known:

$$\left\langle -\frac{dE}{dx} \right\rangle_i = n \int_{T_{min}}^{T_{max}} T \frac{d\sigma_i}{dT}(m, \beta, T)\, dT \tag{2.2}$$

with differential cross section $\frac{d\sigma_i}{dT}$ for a process i, transferring an energy amount $T$. It is dependent on the projectiles mass $m$ and its velocity $\beta$. The lower limit $T_{min}$ is the ionization or excitation threshold of the target material, and the maximum energy transfer $T_{max}$ is approximated by the scattering with an unbound electron [12]:

$$T_{max} = \frac{2m_e c^2 \beta^2 \gamma^2}{1 + 2\gamma m_e/m + (m_e/m)^2}. \tag{2.3}$$

The mass stopping power, measured as mean energy loss per distance $\langle -\frac{dE}{dx} \rangle$ divided by the materials density $\rho$ (figure 2.2), is determined by the material properties (atomic number $Z$, molar mass $A$, mean ionization energy $I$), the particle properties (charge $z$, mass $m$ by impact on $T_{max}$ (2.3)) and velocity $v$, or rather $\beta = \frac{v}{c}$ with $\gamma = 1/\sqrt{1-\beta^2}$.

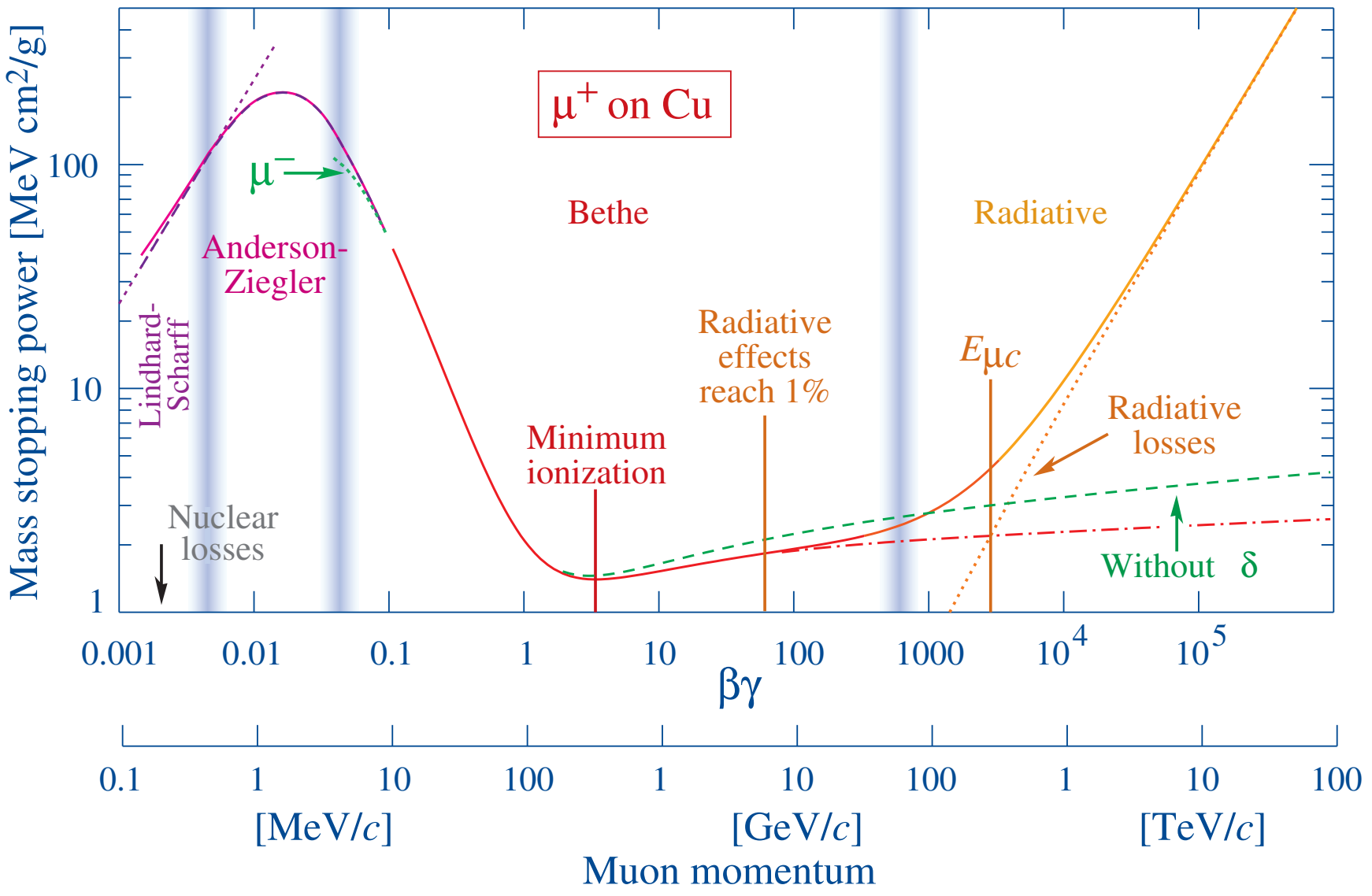

**Figure 2.2.:** Mass stopping power $\langle -\frac{dE}{dx} \rangle$ for positive muons in copper as a function of $\beta\gamma = p/Mc$. Solid curves indicate the total stopping power. The vertical lines indicate regions of validity for different approximations, discussed in the text. [46]

For heavy charged particles in the range of $0.1 \leq \beta\gamma \leq 1000$, the mean energy loss is described by the Bethe-Bloch formula [46]

$$\left\langle -\frac{dE}{dx} \right\rangle_{Ion} = 4Kz^2 \frac{Z}{A} \frac{1}{\beta^2} \left[ \frac{1}{2} \ln \frac{2m_e c^2 \beta^2 \gamma^2 T_{max}}{I^2} - \beta^2 - \frac{\delta(\beta\gamma)}{2} + \frac{C}{Z} \right]. \tag{2.4}$$

Therein $K = \pi N_A r_e^2 m_e c^2$, where $N_A$ is the Avogadro constant, $r_e$ the classical electron radius, $m_e$ the electron mass, $c$ the speed of light in vacuum. $\delta$ corrects for the density effect (see below) and $\frac{C}{Z}$ accounts for shell corrections due to the structure of the atom.

Equation (2.4) gives an accurate estimate on %-level for muons, pions, protons and other charged particles in the same mass range. For heavy projectiles like heavy ions, discrepancies of a factor up to two occur [46].

In the low energy range, the movement of the bound electrons is no longer small compared to the velocity of the projectile and binding energies are not negligible. This is accounted for with the shell correction term $\frac{C}{Z}$ in (2.4). In the range of $0.05 \leq \beta\gamma \leq 0.1$ the discrepancy in energy loss between positively and negatively charged projectiles becomes recognizable (labeled in figure 2.2 with $\mu^-$). For even smaller kinetic energies a theoretical description is missing and phenomenological approximations by Andersen and Ziegler ($0.01 \leq \beta\gamma \leq 0.05$) [47] or Lindhard and Scharff ($\beta\gamma \leq 0.01$) [48, 49] are applied.

In the transition to highly relativistic projectiles a correction to the far interaction of the particle's electric field with the matter must be applied. Referring to the plasma energy $\hbar\omega_p$, this density correction can be calculated as [11]:

$$\delta(\beta\gamma) = \ln(\hbar\omega_p/I) + \ln(\beta\gamma) - 1/2. \tag{2.5}$$

The subtracted correction features a logarithmic increase, which is limited due to the increasing polarization of the material.

In the high relativistic range above the critical energy $E_c$ ($\beta\gamma > 1000$) the energy losses are dominated by radiation of Bremsstrahlung, emitted when the charged particle is decelerated in the coulomb-field of a nucleus. Bremsstrahlung energy losses are increasing proportional to the particle's energy [11]:

$$\left\langle -\frac{dE}{dx} \right\rangle_{Brems} = 4\alpha N_A \frac{Z^2}{A} z^2 \left( \frac{1}{4\pi\epsilon_0} \frac{e^2}{mc^2} \right)^2 \cdot E \ln\left( \frac{183}{Z^{1/3}} \right) \tag{2.6}$$

Charged particles emit Cherenkov radiation when passing through a homogeneous dielectric medium with a velocity greater than the phase velocity of electromagnetic waves in this medium $c/\sqrt{|\epsilon|}$, with $\epsilon$ being the medium's refraction index. In the transition through inhomogeneous media, a charged particle can emit transition radiation photons when passing the boundary between two media with different refraction indices. Both energy loss mechanisms yield photons emitted under a specific angle and of a specific wavelength which can be used for particle ID in specific detectors, but their contribution to the total energy loss is negligible.

Figure 2.2 clearly indicates a minimum of the stopping power at $\beta\gamma \approx 3.5$. Given the small impact of the projectiles mass on (2.4), this minimum occurs for all charged particles in the $3 \leq \beta\gamma \leq 3.5$ range, with the position being determined primarily by the material's atomic number $Z$. Particles with a corresponding energy are referred to as minimal ionizing particles (MIP), as is a large fraction of the muons measured in collider experiments or originating from cosmic radiation. Their interaction with a gaseous volume is discussed in 2.2.1.

### 2.1.2. Ionization Behavior of Electrons and Positrons

For the much lighter electrons and positrons, the critical Energy $E_c$ for radiation dominated energy loss is significantly lower, compared to more heavy particles, as shown in figure 2.3

The energy loss in a specific material, for example lead in figure 2.3, is often referred to in units of the material's radiation length $X_0$, where one radiation length is defined as the thickness of material causing an average energy loss of $\frac{1}{e}$ for a high energetic electron. A good approximation for the radiation length (per density) is given by

$$X_0 = \frac{A}{4\alpha N_A Z(Z+1) r_e^2 \ln(183\, Z^{-1/3})}. \tag{2.7}$$

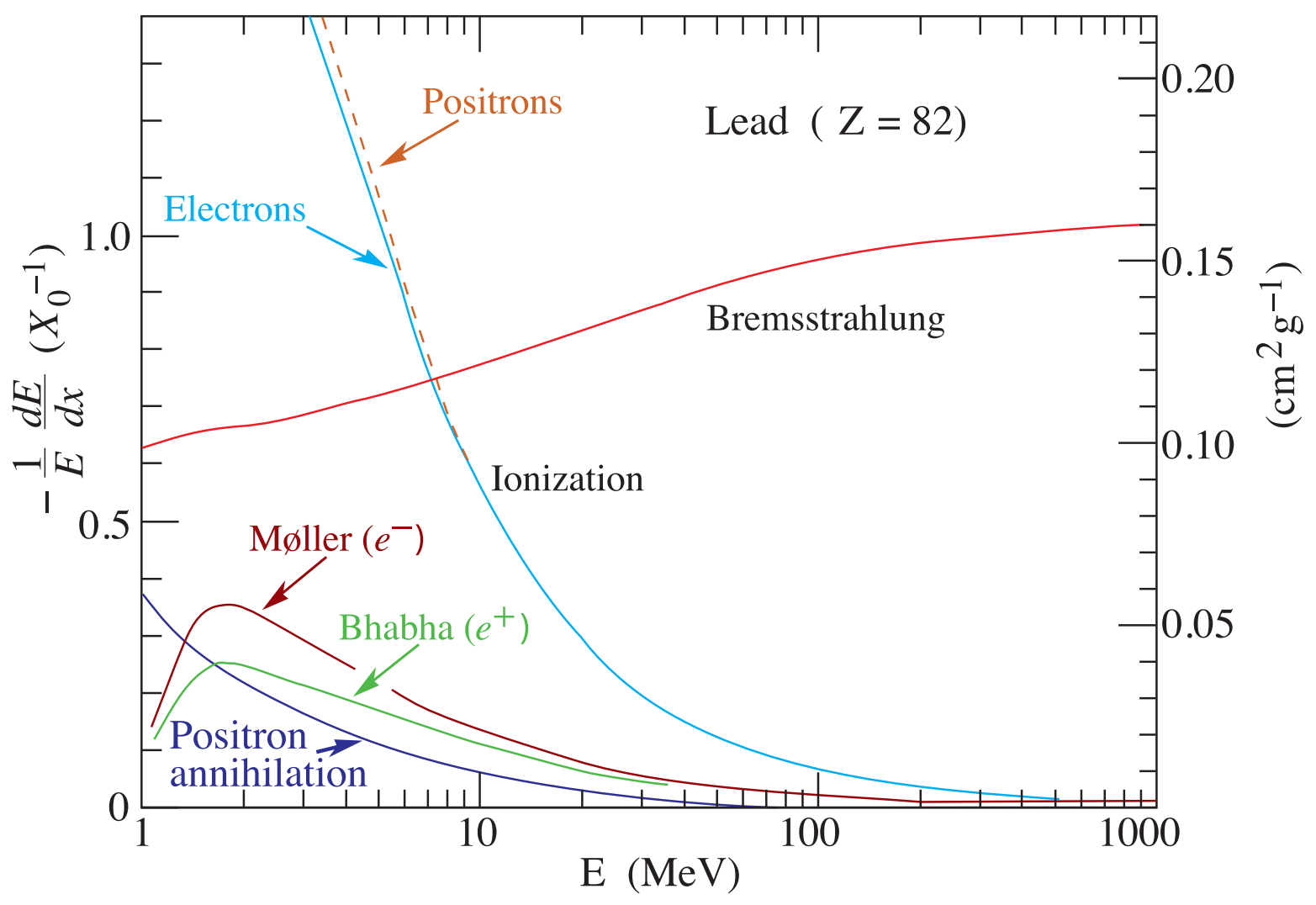

**Figure 2.3.:** Fractional energy loss per radiation length in lead as a function of electron or positron energy [46].

For energies below $E_c$, the energy loss of electrons and positrons is dominated by ionization, with contributions from electron-electron (Møller-) scattering (2.8) and positron-electron (Bhabha-) scattering (2.9):

$$\begin{aligned}\left\langle -\frac{dE}{dx}\right\rangle_{Moller} = &\frac{1}{2}K\frac{Z}{A}\frac{1}{\beta^2}\left[\ln\frac{m_e c^2\beta^2\gamma^2\{m_e c^2(\gamma-1)/2\}}{I^2}\right.\\ &\left. +(1-\beta^2) - \frac{2\gamma-1}{\gamma^2}\ln 2 + \frac{1}{8}\left(\frac{\gamma-1}{\gamma}\right)^2 - \delta\right]\end{aligned} \tag{2.8}$$

$$\begin{aligned}\left\langle -\frac{dE}{dx}\right\rangle_{Bhabha} = &\frac{1}{2}K\frac{Z}{A}\frac{1}{\beta^2}\left[\ln\frac{m_e c^2\beta^2\gamma^2\{m_e c^2(\gamma-1)\}}{2I^2}\right.\\ &\left. + 2\ln 2\frac{\beta^2}{12}\left(23 + \frac{14}{\gamma+1} + \frac{10}{(\gamma+1)^2} + \frac{4}{(\gamma+1)^3}\right) - \delta\right]\end{aligned} \tag{2.9}$$

The logarithmic terms of both processes are comparable to (2.4), substituting $T_{max}$ with $m_e c^2(\gamma-1)/2$ for the symmetrical $e^-$- $e^-$- or $m_e c^2(\gamma-1)$ in $e^+$- $e^-$- scattering. The ionization behavior is slightly different for electrons and positrons, due to their different charge sign. While significant for low positron energies, the annihilation probability rapidly decreases with increasing energy (figure 2.3).

In the $E_{e^-/e^+} > E_c$-region the ionization energy loss is described [10] by

$$\left\langle -\frac{dE}{dx} \right\rangle_{Ion} = 4K\frac{Z}{A}\left[\ln\left(\frac{2m_ec^2\beta^2\gamma^2}{I}\right) - 1\right], \tag{2.10}$$

and, therefore, increases logarithmic with the energy. Radiation losses due to Bremsstrahlung (2.6) per radiation length (2.7) are linear in the particle's energy and thus become dominant for highly relativistic electrons and positrons in the GeV-range.

### 2.1.3. Interaction of Photons with Matter

The primary interaction processes of photons with matter are photoeffect, scattering and pair production, as long as their energy $h\nu$ or rather the reduced photon energy $\epsilon = \frac{h\nu}{m_ec^2}$ is above the discrete absorption levels.

The photoeffect describes the absorption of a photon by an atom under emission of an electron. The K-shell electrons contribute, due to their proximity to the core required as third scattering partner, with about $\sim 80\,\%$ to the total cross sections. The electron is emitted with a kinetic energy of $E_e = h\nu - W$, with $W$ being the binding energy. The photoeffect cross section $\sigma_{p.e.}$ can be estimated in the energy range $E_{K-Abs.}m_ec^2 < \epsilon < 1$ as

$$\sigma_{p.e.} = \frac{32\pi}{3}\sqrt{2}r_e^2Z^5\alpha^4\frac{1}{\epsilon^{7/2}} \tag{2.11}$$

For larger energies $\epsilon >> 1$ one obtains

$$\sigma_{p.e.} = 4\pi r_e^2Z^5\alpha^4\frac{1}{\epsilon}, \tag{2.12}$$

where $\alpha$ is the fine structure constant, $Z$ the atom number and $r_e \approx 2.8\,\text{fm}$ the classic electron radius [10]. Being the dominant process for small photon energies and suppressed with increasing energy, the emitted electrons energy $E_e$ ranges in the highly ionizing regime for electron matter interaction (see 2.1.2) and thermalizes rapidly. The thereby created electron-ion pairs can contribute to signal formation and their amount is proportional to the photo electron's energy. The hole in the inner shell (like the K-shell) is filled by relaxation of an electron from a higher shell under emission of electrons freed by Auger mechanisms or a photon with the specific energy difference. Dependent on the further interaction of this photon with the detector its energy can be measured, for example as a result of a subsequent photoelectric effect, or carried away. The latter effect leads to the occurrence of escape peaks in the energy spectrum.

Scattering processes are distinguished according to the photon's energy and the scattering partner: Rayleigh scattering describes the photon scattering on matter with a small characteristic size compared to the photon wavelength $\lambda$. It occurs due to the polarizability of the material and is a parametric process, hence leaving the scattering target unchanged. Interaction rates of Rayleigh scattering are inversely proportional to the fourth power of the wavelength and thus suppressed for increasing photon energy. Thomson scattering describes the elastic scattering with free charge carriers of mass $M$ and can be seen as the low energy approximation of the Compton scattering for $h\nu \ll Mc^2$. The latter process yields a reduced wavelength of the photon due to an energy transfer to the recoiling electron. The corresponding cross section can be derived from QED and is described in lowest order by the Klein-Nashina equation [10]

$$\sigma_{Compton} = 2\pi r_e^2 \left[ \left( \frac{1+\epsilon}{\epsilon^2} \right) \left( \frac{2(1+\epsilon)}{1+2\epsilon} - \frac{1}{\epsilon} \ln(1+2\epsilon) \right) + \frac{1}{2\epsilon} \ln(1+2\epsilon) - \frac{1+3\epsilon}{(1+2\epsilon)^2} \right] \tag{2.13}$$

At higher photon energies the conversion of a photon in the presence of a coulomb field into an electron positron pair becomes possible. The required energy threshold is

$$h\nu \geq 2m_e c^2 + 2\frac{m_e^2}{M} c^2 \approx 2m_e c^2 = 1.022\,\text{MeV}. \tag{2.14}$$

The approximation is valid in case of the interaction with the Coulomb field of an atom, where $M \gg m_e$. According to (2.14) pair production in an electron field requires $h\nu \geq 4\,m_e c^2$ and is strongly suppressed. The pair production cross section $\sigma_{pair}$ can be approximated for photon energies $h\nu < 137\,Z^{-1/3} m_e c^2$ by

$$\sigma_{pair} = r_e^2 4\alpha Z^2 \left( \frac{7}{9} \ln(2\frac{h\nu}{m_e c^2}) - \frac{109}{54} \right). \tag{2.15}$$

For highly energetic photons, the approximation becomes energy independent:

$$\sigma_{pair} = r_e^2 4\alpha Z^2 \left( \frac{7}{9} \ln\left( \frac{183}{Z^{1/3}} \right) - \frac{1}{54} \right) \approx \frac{7}{9} \frac{A}{X_0 N_A}, \tag{2.16}$$

Hereby the $-1/54$ correction term is neglected and the approximation $Z(Z+1) \approx Z^2$ is used. Equation (2.16) utilizes the same definition for the radiation length as (2.7), thus $X_0$ as well describes the path length corresponding to a probability $p = 1 - \exp(-7/9) \approx 54\,\%$ for pair production by a high energetic photon.

## 2.2. Statistics of Exemplary Events in a Micromegas

Figure 2.4 illustrates three different types of events in a Micromegas detector: Insertion of a single electron (a), trespassing of a minimum ionizing particle along a perpendicular (b) or inclined (c) track and the electron-ion cloud formed by an X-ray absorption (d). The statistics of the ionization yield in each of those events will be discussed in this chapter.

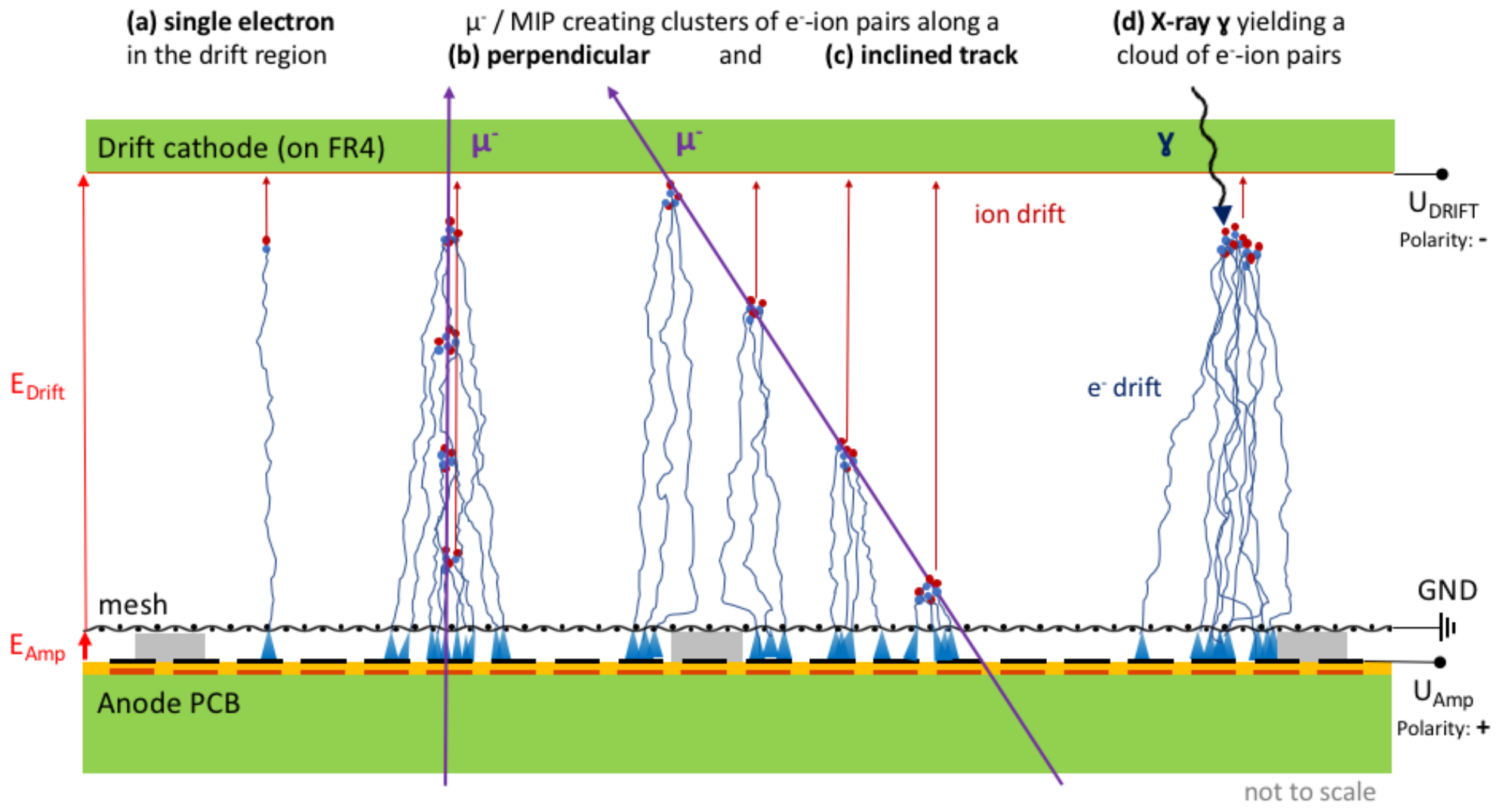

**Figure 2.4.:** Structure of a Micromegas detector, primary ionization and signal formation processes for different event types: (a) a single electron inserted in the detector; (b) perpendicularly or (c) inclined trespassing MIP; (d) electron-ion cloud caused by a X-ray photon. The number of depicted clusters and electrons are for visualization only and smaller compared to a real event.

The ionization yield is primarily determined by the gas mixture. Table 2.1 provides an overview to the gas properties of the noble gases and admixtures utilized during the herein presented experiments.

For the calculation of primary ionization, the interaction of the projectile with a homogeneous gas mixture can be seen as successive transition through thin layers of a pure gas species. The mean energy loss can, therefore, be obtained by a weighted summation of the components $i$ (Bragg additivity), where the weighting factors $w_i$ correspond to their proportion [46]:

$$\left\langle -\frac{dE}{dx} \right\rangle = \sum_i w_i \left\langle -\frac{dE}{dx} \right\rangle_i . \tag{2.17}$$

Similar weighted summation rules can be applied for the average number of primaries $N_P$ and total ionization $N_T$ per path length, as well as for the average energy per electron-ion pair $W_I$.

| Gas | $\rho$ [$\mathrm{mg\,cm^{-3}}$] | $E_X$ [eV] | $E_I$ [eV] | $W_I$ [eV] | $dE/dx\|_{min}$ [$\mathrm{keV\,cm^{-1}}$] | $N_P$ [$\mathrm{cm^{-1}}$] | $N_T$ [$\mathrm{cm^{-1}}$] |
|---|---|---|---|---|---|---|---|
| He | 0.18 | 19.8 | 24.6 | 41 | 0.32 | 3.5 | 7.7 |
| Ne | 0.84 | 16.7 | 21.6 | 37 | 1.45 | 13 | 40 |
| Ar | 1.66 | 11.6 | 15.7 | 26 | 2.53 | 25 | 97 |
| Xe | 5.50 | 8.4 | 12.1 | 22 | 6.87 | 41 | 312 |
| $CO_2$ | 1.84 | 7.0 | 13.8 | 34 | 3.35 | 35 | 100 |
| $iC_4H_{10}$ | 2.49 | 6.5 | 10.6 | 26 | 5.67 | 90 | 220 |

**Table 2.1.:** Properties of noble and molecular gases at normal temperature and pressure (NTP: 20°C, one atm). $E_X$, $E_I$: first excitation, ionization energy; $W_I$ average energy per ion pair; $dE/dx|_{min}$, $N_P$, $N_T$: differential energy loss, primary and total number of electron-ion pairs per cm for a unit charge minimum ionizing particle [46].

### 2.2.1. Ionization along a Minimum Ionization Particle's Track

In collider experiments, like the ATLAS, minimum ionizing particles (MIPs) are the most common particles to be detected by gaseous detectors, for example when applied in a muon system. An abundant source for testing detector response to MIPs are cosmic rays providing $O(1\,\mathrm{Hz/dm^2})$ of high energetic muons. For controlled particle exposure accelerators like the PS or SPS at CERN can provide mono-energetic beams of muons or pions in their test-beam facilities.

For a Micromegas with a 5 mm conversion gap of $Ar:CO_2$ (93 : 7) an average of $12.9 \pm 3.6$ ($13.6 \pm 3.7$) primary interactions with the gas are expected along a perpendicular (20° inclined) track (visualized in figure 2.4 (b) and (c)). The fluctuation in the number of independent primary ionization processes is determined by Poisson statistics with $\sigma(N_P) = \sqrt{N_P}$. In average each of these clusters contains 3.8 electron-ion pairs within a typically $(10\,\mu\mathrm{m})^3$ volume. The fluctuation in cluster size is determined by the differential cross section for a primary ionization with energy transfer $T$ ($E_X < T < T_{max}$) which is proportional to $1/T^2$ [12]. Thus small clusters of one or two electrons are much more likely and large energy transfer in central scatterings to so called 'knock-on' or $\delta$-electrons, occurs scarcely (figure 2.5). These knock-on electrons can reach high energies and, therefore, travel several millimeter in the gas before thermalization, complicating event reconstruction and worsening energy and spatial resolution. The probability density of the energy loss along a fixed path length can in first approximation[2] be described by the Landau distribution (figure 2.6):

$$f_L(\lambda) = \frac{1}{\sqrt{2\pi}} \exp\left(-\frac{1}{2}(\lambda + \exp(-\lambda))\right) \tag{2.18}$$

Here $\lambda$ represents the normalized deviation from the most probable energy loss $\langle \Delta E \rangle$ along a path length $x$:

[2]These approximations include the limit of $T_{max} \to \infty$, negligence of shell effects and assuming a thin material layer.

$$\lambda = \frac{\Delta E - \langle \Delta E \rangle}{\xi} \quad , \quad \xi = \frac{1}{2} K \frac{Z}{A} \frac{z^2}{\beta^2} \rho x. \tag{2.19}$$

For the perpendicular (20° inclined) track through the 0.5 cm Ar : $CO_2$ (93 : 7) gas gap, a mean energy deposit of 1.29 keV (1.37 keV) is expected. This corresponds to an ionization yield of $49 \pm 7$ ($52 \pm 7$), assuming Poisson statistic for the fluctuation. The FWHM of the Landau peak is equal to $4.018\,\xi$ [12] and calculates in our example to 0.26 keV (0.27 keV) corresponding to $\approx 20\,\%$ of the mean energy loss. Thus the (more accurate) Landau description yields slightly higher fluctuation compared to the simple Poisson statistic assumption.

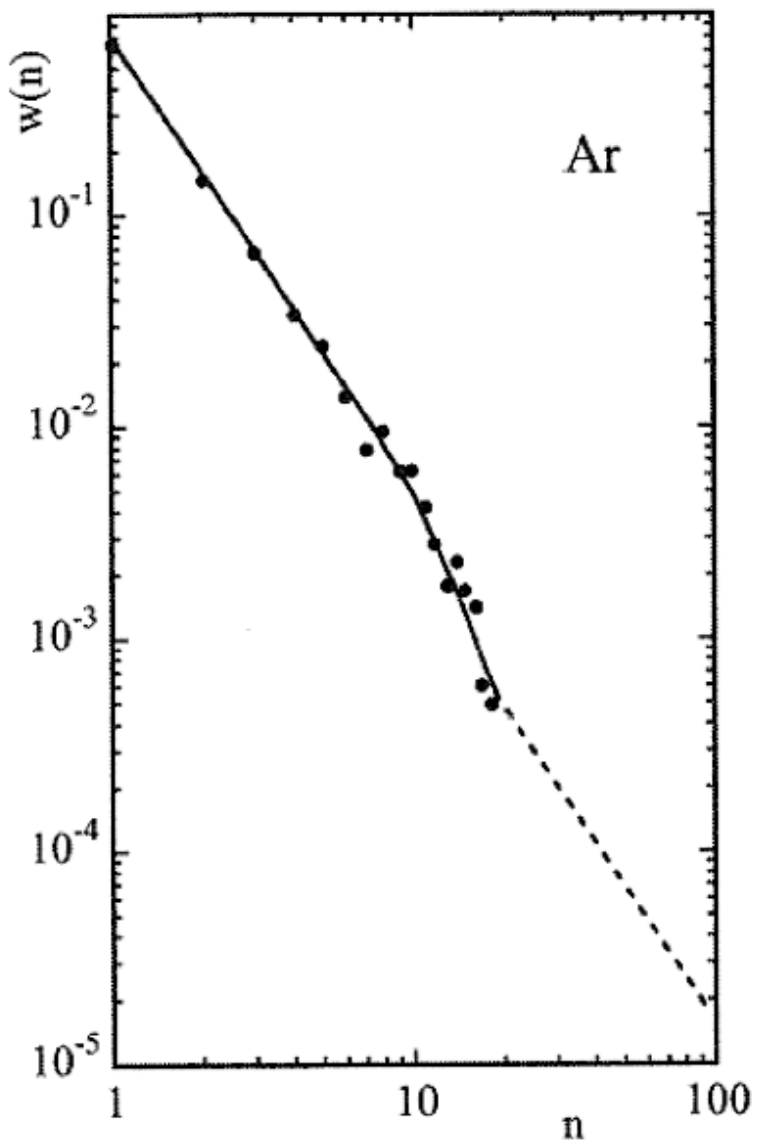

**Figure 2.5.:** Cluster size ($n$) probability ($w(n)$) for minimum ionizing particles in pure Argon. [50]

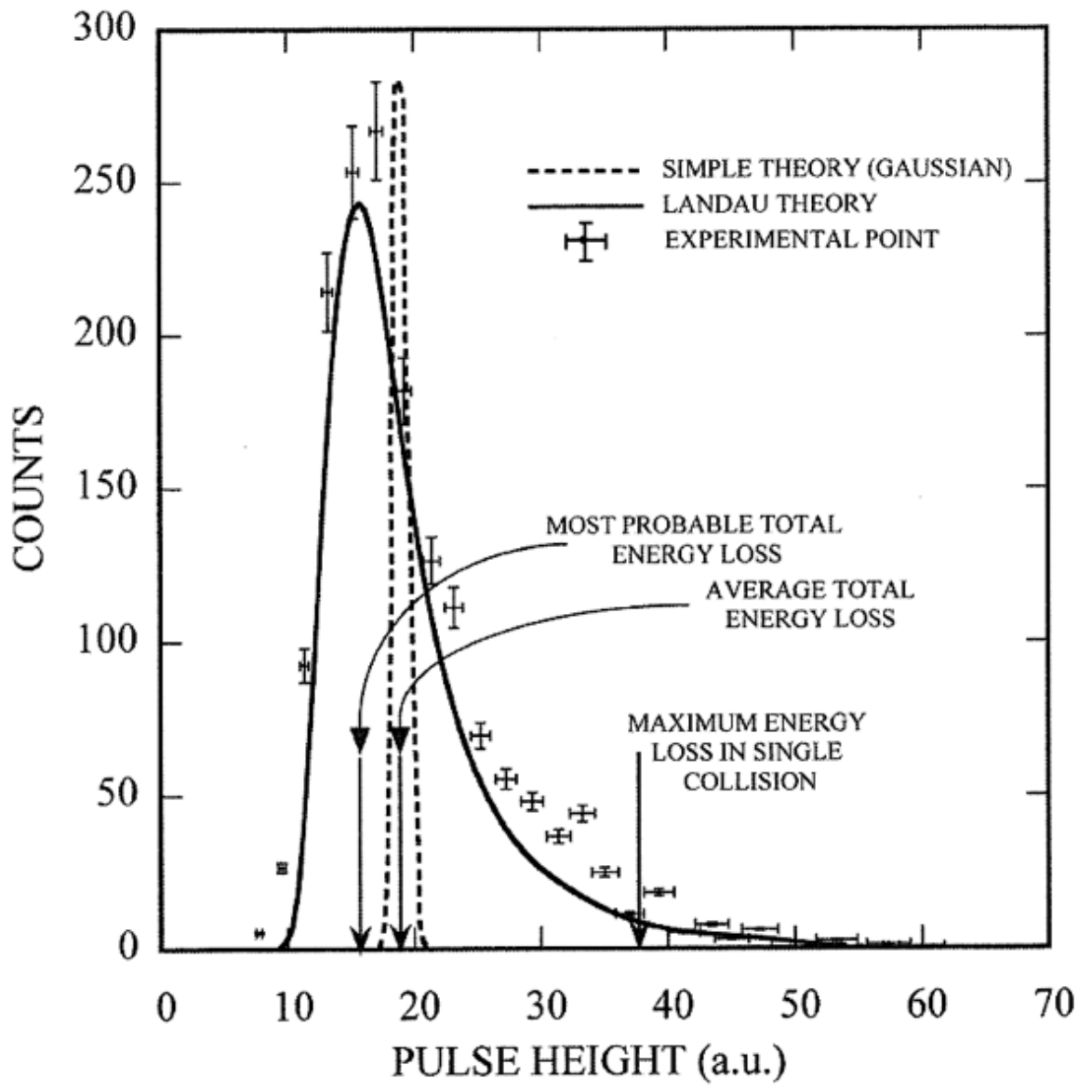

**Figure 2.6.:** Comparison of experimental data and Landau theory calculations of the energy loss in a thin sample of gas. [13]

### 2.2.2. Electron-Ion Clouds formed by X-Rays

For the study of detector characteristics in a laboratory environment X-ray events (figure 2.4 (d)) are often favorable over cosmic muons. X-ray photons can be provided at high rate, spatially collimated and with a well-defined energetic spectrum by using either an X-ray tube or radioactive isotopes. Furthermore, they provide a preferable ionization statistic in thin gaseous volumes due to the 'binary' nature of the absorption processes, compared to a continuous energy loss.

In our experiments a water cooled X-ray tube with a copper anode has been used with a potential difference of 15 kV. The emitted spectrum is composed of continuous Bremsstrahlung with two distinct intensity peaks at $E_{K,\alpha} = 8.0$ keV and $E_{K,\beta} = 8.9$ keV, corresponding to the photon emission during electron relaxation from the L-shell, respectively M-shell, to the K-shell. A 100 µm thin Nickel filter, featuring an absorption edge at 8.3 keV, has been used to significantly reduce the Bremsstrahlung background and suppress the $K_\beta$-peak. The X-ray photons entered the Micromegas detector from the drift-volume side, typically through a thin window of 50 µm Kapton®, Mylar® or a 17 µm copper layer, minimizing absorption losses at the window.

For 8 keV photons, the dominant interaction in Argon is absorption by photoelectric effect and, therefore, the rate $N$ follows an exponential decay (figure 2.1) determined by the absorption coefficient $\mu$:

$$N(x) = N_0 \exp(-\mu x) = N_0 \exp(-\frac{x}{\lambda}). \tag{2.20}$$

The absorption probability in a thin layer and thus the absorption length $\lambda$ (figure 2.8) can be derived from the cross sections (figure 2.7). For the 0.5 cm thin conversion gap filled with Ar : $CO_2$ (93 : 7) the total absorption efficiency for a 8 keV photon is approximately 10 %. The small fraction of $CO_2$ yields a rather small sub-5 % increase of the absorption length compared to pure Argon. Given the low absorption probability in a mm-thick gas layer, the photon rate decreases only little across the subvolumes of the conversion gap and the event rate over position distribution is in first approximation linear.

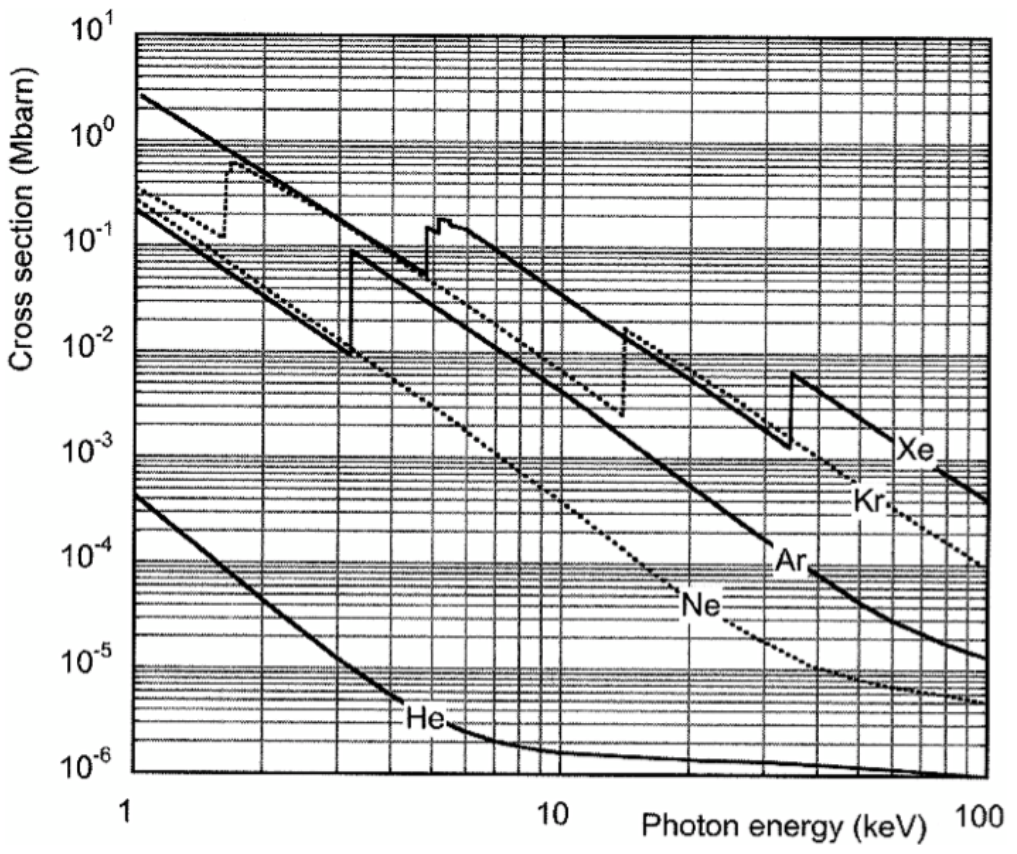

**Figure 2.7.:** Absorption cross section for photons in the keV range in noble gases at STP (T = 0°C, p = 1 atm). [13]

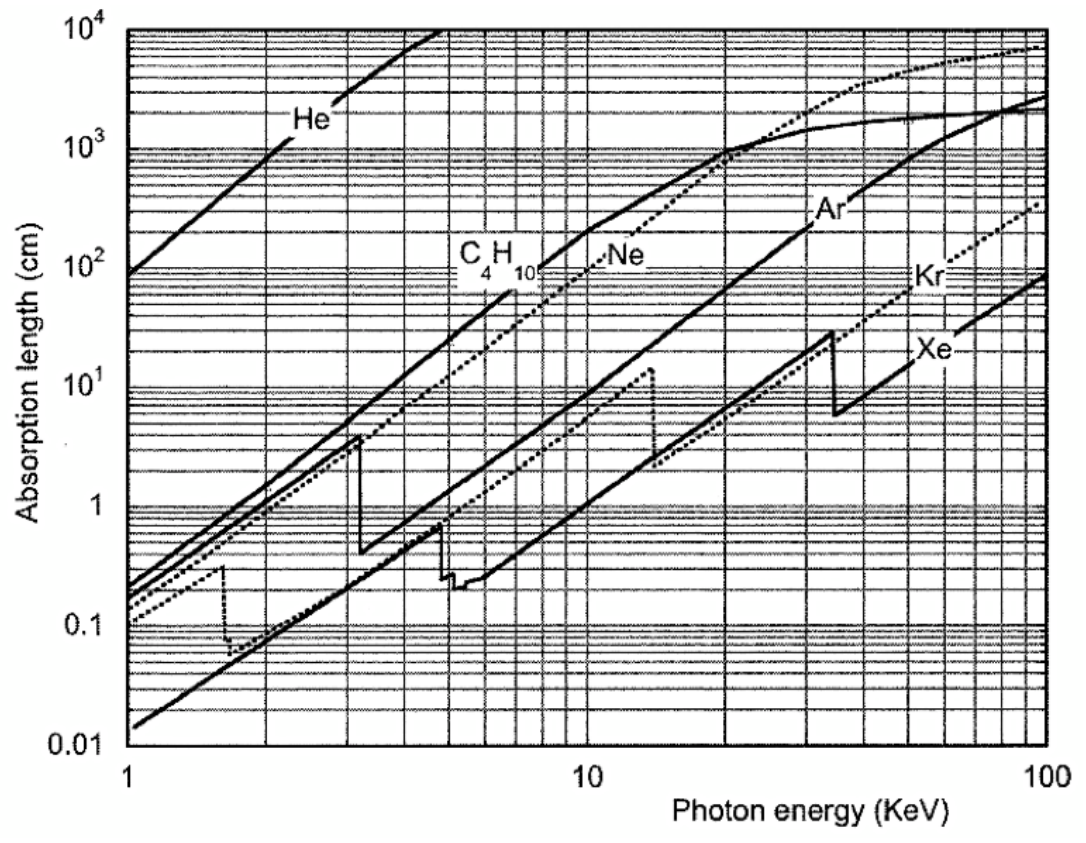

**Figure 2.8.:** Absorption length for photons in the keV range in noble gases at STP (T = 0°C, p = 1 atm). [13]

During photon absorption typically an electron from an Argon K-shell is freed with a kinetic energy of $E_{e^-} = E_{K,\alpha/\beta} - W_{Ar,K}$, where $W_{Ar,K} = 3.2$ keV is the required ionization energy. It is emitted preferably in the plane perpendicular to the photon direction. The 4.8 keV (or 5.7 keV) electron is highly ionizing with a practical range of $\leq 1$ mm according to [13, Figure 2.18]. Therefore, it deposits its energy in form of electron-ion pairs within a sub-mm$^3$ cloud before thermalizing. The vacancy in the Argon K-shell is filled most probably with an electron from the L-shell. In Argon the energy difference of 2.9 keV is released in 92 % of the events in form of one or more electrons freed by Auger mechanism. These electron(s) quickly thermalize in the gas, adding to the total ionization yield, thus the whole $E_{K,\alpha/\beta}$ contributes to the signal. In only 8 % a fluorescence photon is emitted with an energy right below the K-absorption edge. The thence low probability of immediate re-absorption allows the photon to leave the region of the event or possibly the full detector unrecognized. This yields two additional contributions to the measured energy spectrum at $E_{K,\alpha\, esc.} = 4.8\,\mathrm{keV}$ and $E_{K,\beta\, esc.} = 5.7\,\mathrm{keV}$, the so called escape peaks. The photons emitted during subsequent de-excitation of the Argon have a very low energy and, therefore, range and are typically thermalized in the gas.

The average number of electron-ion pairs $n_e$ produced in an $K_\alpha$ event can be calculated utilizing the values from table 2.1 and the Bragg additivity of $W_I$ similar to (2.17). While for charged particles the energy loss fluctuation is described by the Landau distribution, the upper limit of energy loss given by the comparable low kinetic energy modifies the statistic to the more simple dispersion:

$$\sigma(n_e) = \sqrt{F n_e}. \tag{2.21}$$

$F$ is the co called Fano factor, with $F < 1$, which is a measure for the fluctuation of the ionization process [51]. A lower Fano factor is favorable for good energy resolution and can be achieved by optimization of the gas mixture. Especially penning transfer mechanisms, as discussed in chapter 4, contribute significantly to a reduction of $F$. For pure Neon and Argon $F = 0.17$ is estimated by theory, and measurements confirm the range of $0.13 < F < 0.20$ for all noble gases, dependent on the source [52].

In the above given example a $K_{alpha}$ event yields an average of 301 electrons. The fluctuation in primary ionization contributes with $\sigma(n_e)/n_e \approx 2.5\,\%$, or FWHM$\approx 6\,\%$, to the energy resolution limit. Without a change of the gas mixture, the first call for a further reduction of the ionization fluctuation is an increase in photon energy. This leads to an increase of the number of signal electrons and a decrease in relative variance $\sigma(n_e)/n_e \approx 2.5\,\%$. However, the increase in electron energy increases as well its range in the gas, smearing the ion cloud over a larger volume, which is unfavorable for the study of subsequent drift- and amplification processes.

### 2.2.3. Single Electron Insertion

An alternative experimental approach is the insertion of a single electron into the Micromegas drift gap. Experimentally, this is realized by focusing a mono-chromatic photon beam, emitted by a pulsed laser, through a quartz window onto a 0.5 nm thin Nickel-Chromium layer, which is used as the Micromegas cathode. A 337 nm wavelength photon can cause a photoelectric effect in the layer and extract a single electron into the Micromegas drift volume.

The key to single electron response (SER) operation is the adjustment of the laser intensity combined with the pulsed mode of operation: Given the thin absorption layer most of the photons pass through the cathode and, having an energy far below ionization threshold of the gas or the mesh, are absorbed in the readout structure without contributing to the signal. Therefore, a high photon intensity per pulse is required, causing the uncertainty of multiple simultaneous photoelectric effects, yielding multiple electron events. To estimate and suppress the rate of these events, the detector response for each pulse is monitored and the rate is adjusted in a way that only a small fraction $p$ of the pulses causes a signal over noise threshold, while in $(1-p)$ cases a zero-signal is recorded. Assuming negligible loss probability to recombination, attachment and mesh absorption and provided a sufficiently sharp distinction between signal and noise, $p$ is the probability of one or more photo electrons being freed in one laser pulse. With the (unknown) probability of a single electron event $p(1)$, double-, triple- or n-electron events are caused with a probability of $p(2) = p(1)^2$, $p(3) = p(1)^3$ or generalized

$$p(n) = p(1)^n. \tag{2.22}$$

The summation over all possible non-zero events yields the geometric series

$$p = \sum_{k=1}^{\infty} p(k) = \sum_{k=0}^{\infty} p(1)^k - 1 = \frac{1}{1-p(1)} - 1 \tag{2.23}$$

and allows for a determination of the multiple- over single electron rate

$$\frac{p - p(1)}{p(1)} = \frac{p - 1 + \frac{1}{1+p}}{1 - \frac{1}{1+p}} = ... = p. \tag{2.24}$$

The mean ionization yield in the non-zero signals is

$$\langle n_e \rangle = \sum_{k=1}^{\infty} k \frac{p(k)}{p} = \frac{1}{p} \sum_{k=1}^{\infty} k p(1)^k = ... = \frac{p(1)}{p[1-p(1)]^2} = ... = 1 + p \tag{2.25}$$

The variance $\sigma(n_e)^2$ of this electron yield per non-zero signal is

$$\sigma(n_e)^2 = \sum_{k=1}^{\infty} [k - \langle n_e \rangle]^2 \frac{p(k)}{p} = ... = p^2 + p \tag{2.26}$$

The solutions in (2.23) to (2.26) are analytically exact within the convergence radius $p < r = 1$, thus no approximation or constrain is required. With $\langle n_e \rangle \approx 1 + p \approx 1$ for small $p$ the variance around the envisaged electron yield of a single electron per event is similar to $\sigma(n_e)^2$.

With the single electron being extracted in the laser focal point at the drift cathode with a well known energy, this scheme provides ideal conditions for studies of subsequent signal formation processes. However, from a statistical point of view the variation in ionization yield is not negligible unless for very small non-zero event rates, requiring extensive measurement times. Therefore, a mechanism to identify multiple-electron signals or exclude them during data analysis is required, on top of an efficient zero suppression, in order to make best use of this method.

# 3. Electron Drift, Attachment Losses and Transparency of Micromeshes

In most gaseous detectors the electrons freed in primary ionization are required to move towards a gas region where they can be amplified and ultimately induce a signal. This guided electron movement caused by an electric field is called *electron drift*. As described in 1.3.1 and represented in (1.7), two mechanisms contribute to the loss of signal electrons during electron drift in a Micromegas: attachment to gas constituents and absorption at the mesh.

After reviewing the analytic description of the drift process and approaches to predict the electron losses in section 3.1 our method to simulate the macroscopic observables based on the microscopic cross sections is presented in section 3.2. In section 3.3 our setup to experimentally assess electron losses under variation of geometric parameters of the Micromegas according to the VOTAT paradigm is presented and the results are discussed in comparison to the simulation.

The performed simulation study, developed detector concept and the experimental results have been publicly discussed on several occasions during the course of the last three years as oral and poster contributions at the RD51 and MPGD conferences. A compilation of these results has been published in [53]. The latest experimental data set has been acquired in the scope of the bachelor thesis of D. Baur [54], where more details about the applied measurement and analysis procedure can be found.

Annotation on units: for drift processes an appropriate invariant for field dependent variables is the ratio $E_D/N$ where $N$ is the number of molecules per unit volume. Following current customs in detector R & D all results will be presented as a function of the field at NTP conditions (20°C, 1 atm), yielding them easier to interpret for the application of Micromegas in the ATLAS NSW. The results on drift and attachment processes can easily be adopted to non-NTP conditions using the established scaling laws summarized, for example, in [13]. For the interplay of scattering processes with geometry in the transparency simulation, simple scaling laws are not sufficient for an accurate description and a repeated simulation run would be required.

## 3.1. Theory of Low Energy Electron-Gas Interaction

In the absence of an external electric field the movement of thermalized electrons (and ions) follows the Maxwell-Boltzmann law with a mean kinetic energy of $\bar{\epsilon} = kT$, $k$ being the Boltzmann constant and $T$ the temperature. The instantaneous velocity $v$ of this direction randomized movement of a particle with mass $m$ follows the distribution

$$f(v) = 4\pi \left(\frac{m}{2\pi kT}\right)^{\frac{3}{2}} v^2 \exp\left(-\frac{mv^2}{2kT}\right) \quad , \tag{3.1}$$

with the velocities mean $\overline{v}$ and most probable value $v_{MP}$

$$\overline{v} = \int_0^\infty v f(v) dv = \sqrt{\frac{8kT}{\pi m}} \quad , \quad v_{MP} = \sqrt{\frac{2kT}{m}} \quad . \tag{3.2}$$

Moving in the gas they will eventually scatter with the gas constituents, exchanging energy and momentum according to their mass ratio, which is O(1) for ions and O($10^4$) for electrons. Therefore, ions can transfer a significant energy amount to their collision partner, while electron energy is almost preserved for elastic scattering processes[1] and only the direction of the movement is randomized. Due to this repeated process, a point like cloud of electrons (ions) diffuses symmetrically in the gas. The fraction of particles $dN/N$ found in an element $dx$ at a distance $x$ from the origin after a time $t$ follows a Gaussian law

$$\frac{dN}{N} = \frac{1}{\sqrt{4\pi Dt}} \exp\left(-\frac{x^2}{4Dt}\right) dx. \tag{3.3}$$

Herein $D$ is the gas dependent diffusion coefficient and the standard deviation of this linear and volume diffusion process is given by

$$\sigma_x = \sqrt{2Dt} \quad \text{and} \quad \sigma_V = \sqrt{6Dt} \quad . \tag{3.4}$$

Under the influence of an external electric (drift) field $E_D$ a preferred direction is introduced, yielding a net movement of the electrons (ions) towards (in) field line direction. The mean net velocity of this movement, called the drift velocity $v_D$, defines the charge carriers mobility

$$\mu = v_D/E_D \quad . \tag{3.5}$$

For ions, the energy increase between scatterings is largely transferred to the gas, thus the mobility is up to very high fields constant for ions. For electrons the scattering processes are highly dependent on the electrons' characteristic energy $\epsilon_k(E_D)$, which can be much higher than their thermal energy. In both cases, the diffusion can be related to the mobility by

$$\frac{D_{ion}}{\mu_{ion}} = \frac{kT}{e} \quad \text{and} \quad \frac{D_{e^-}}{\mu_{e^-}} = \frac{\epsilon_k}{e} \quad . \tag{3.6}$$

Being macroscopic experimental observables of the drift process, these gas properties ($v_D, \mu, D$) have been extensively studied for a variety of gases and gas mixtures since the 1920's, often with controversial results. A comprehensive recapitulation of measurements can be found in [13, chap. 4.7].

[1] For inelastic scattering processes a part of the energy can be transferred to the gas atom or molecule, as will be discussed in the next section 3.1.1.

### 3.1.1. Low Energy Electron-Gas Scattering Processes

A complementary approach has been started with the studies of the microscopic processes in electron gas interaction. As for higher energies, the probability of a scattering process in a gas is described by its cross section, which is strongly dependent on the energy of the scattering electron $\epsilon$.

For noble gases the effective cross section, shown in figures 3.1 to 3.4, is determined by elastic scatterings for energies up to the first excitation level above 8 eV for Xenon to almost 20 eV for Helium. Surpassing this threshold successively all excitation channels open up and contribute to the cross section. Once the energy surpasses the ionization threshold (see table 2.1) this process becomes the dominant inelastic scattering mode, in Argon and Xenon even exceeding the elastic process at high energies. While the effective cross section for low energies is almost flat in Helium (figure 3.1) and in first order linearly increasing for Neon (figure 3.2), Argon and Xenon (figures 3.3 and 3.4) feature a dip in the elastic cross section. This is a result of the Ramsauer effect [56], occurring when the electron's De-Broglie wavelength is of the order of the atoms' spatial extent. On a macroscopic scale this leads to a local maximum in the drift velocities.

For molecular gases, like Carbon Dioxide or Isobutane, the molecule structure can allow for the excitation of vibration and rotation states for energies as small as 0.1 eV (figure 3.5). Typically, they feature as well a broader range of shell excitation energies due to the different atoms involved. Molecular gases featuring electron affinity or being electro-negative can capture electrons during a scattering process. These attachment cross sections, shown for CO and $CO_2$ in figure 3.6 and for $O_2$ and $H_2O$ in figures 3.7 and 3.8, are typically several orders of magnitude smaller than the respective elastic cross sections and strongly energy dependent. Thus, they contribute little to the overall electron transport properties like drift velocity or diffusion, but cause electron losses during the drift process which will be discussed in section 3.1.3.

The open-access LXCAT project provides a collection of cross sections for electrons and some ion species in a variety of gases. For mixtures of several gas species $g$ with concentration $c_g$ simple composition rules hold for the cross sections:

$$\sigma_{mixture}(\epsilon) = \sum_g c_g \sigma_g(\epsilon) \quad \text{, with} \quad \sum_g c_g = 1 \quad . \tag{3.7}$$

### 3.1.2. Transport Theory of Drift Velocities and Diffusion

With the increasing understanding of electron gas interaction processes on a microscopic level and the availability of their cross sections, algorithms to calculate and predict drift properties became more and more popular. A full recapitulation of the transport theory used in the Magboltz program would exceed the scope of this thesis, therefore, only the basic considerations are discussed following the description in [13].

Similar to the Boltzman theory, the calculations depend on the correct description of the electron energy distribution $F(E_D, gas)$. It is calculated for a given electric field and gas mixture, by equalizing the energy gained from the field in between collisions to the energy lost in the collision process. Assuming that only a negligible fraction of

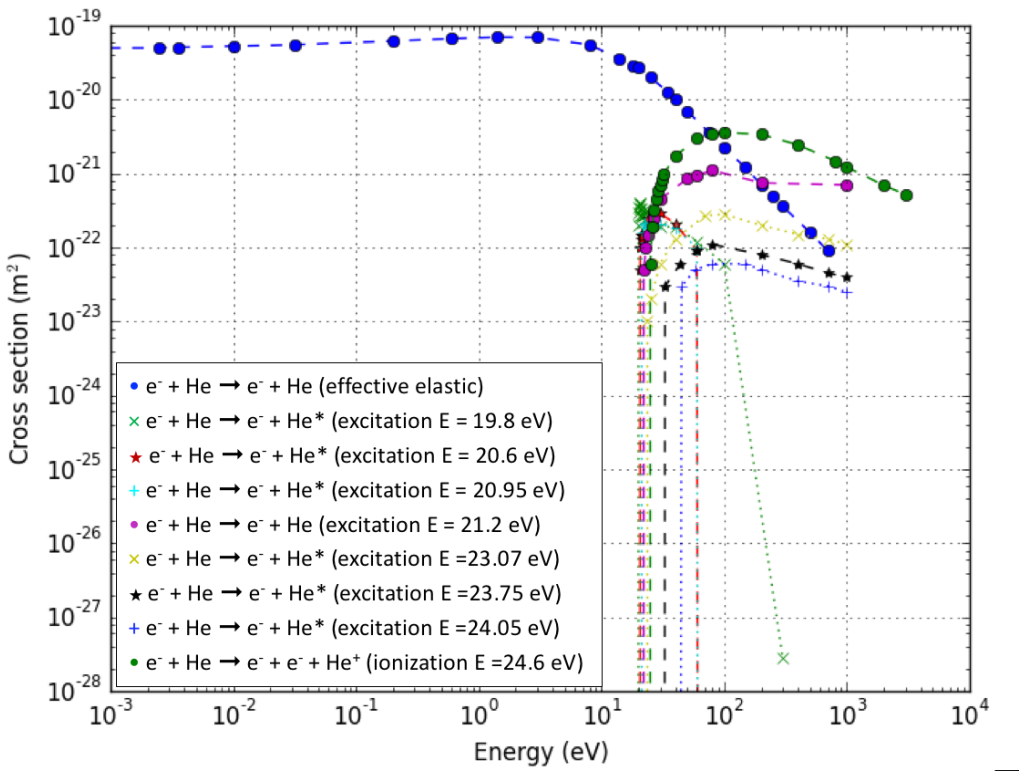

**Figure 3.1.:** Cross sections for different processes in electron-Helium scattering as a function of the electron energy at NTP. Trinity database, retrieved from lxcat.net on February 20, 2017 [55].

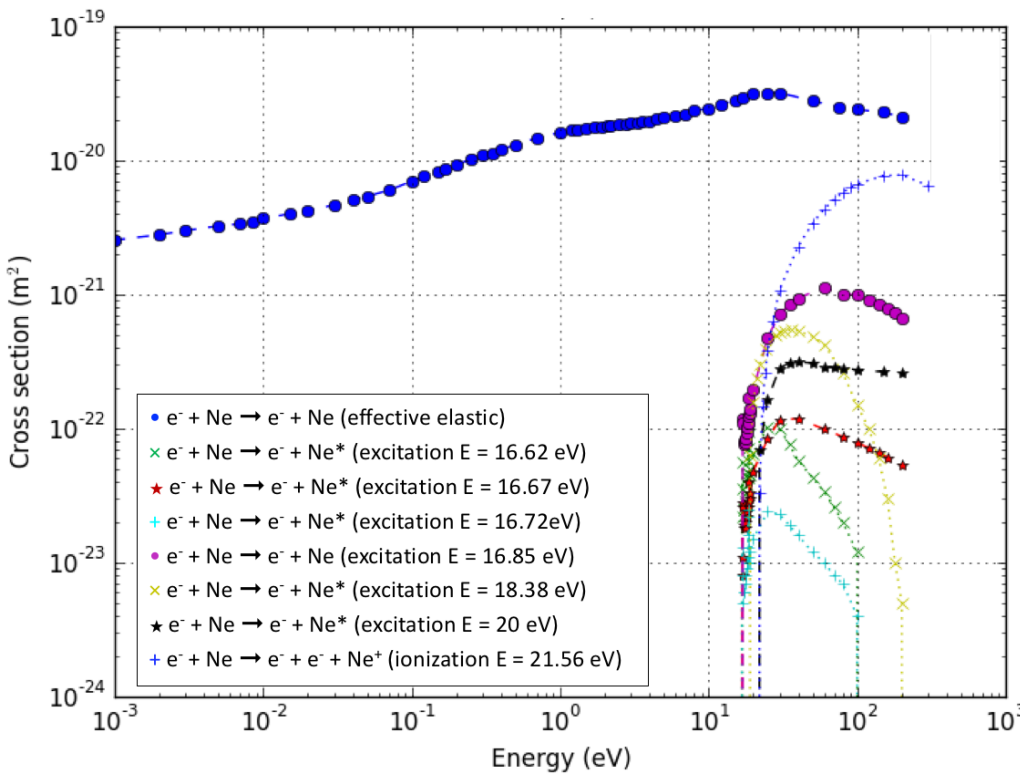

**Figure 3.2.:** Cross sections for different processes in electron-Neon scattering as a function of the electron energy at NTP. Trinity database, retrieved from lxcat.net on February 20, 2017 [55].

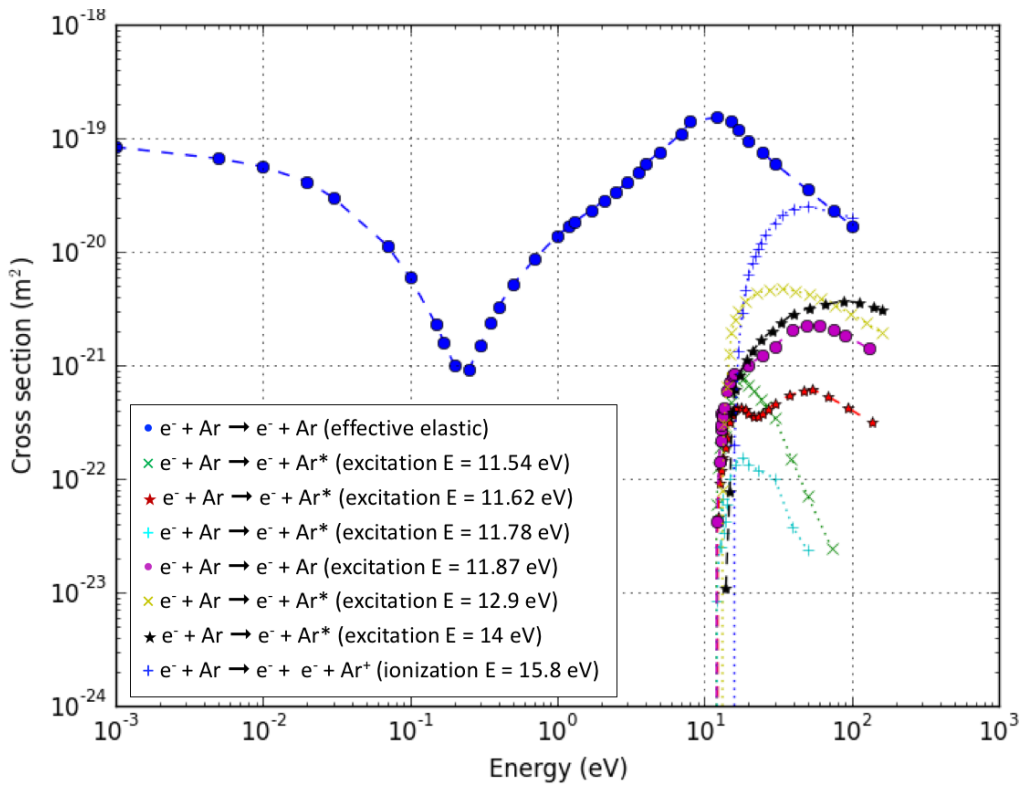

**Figure 3.3.:** Cross sections for different processes in electron-Argon scattering as a function of the electron energy at NTP. Trinity database, retrieved from lxcat.net on February 20, 2017 [55].

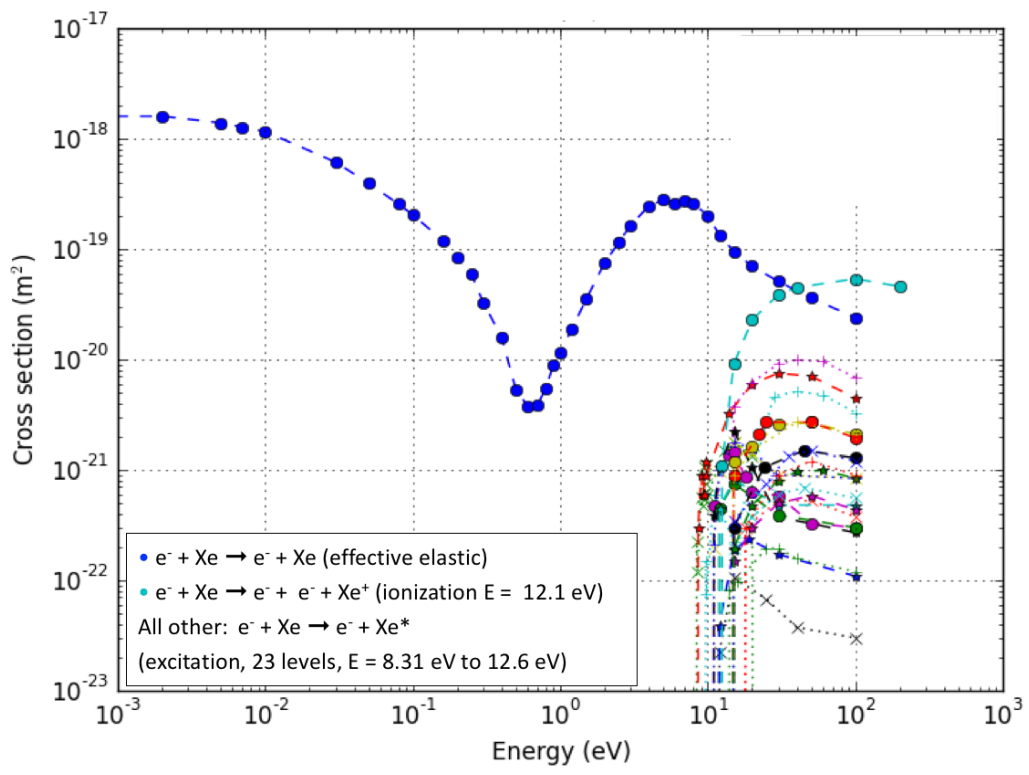

**Figure 3.4.:** Cross sections for different processes in electron-Xenon scattering as a function of the electron energy at NTP. Trinity database, retrieved from lxcat.net on February 20, 2017 [55].

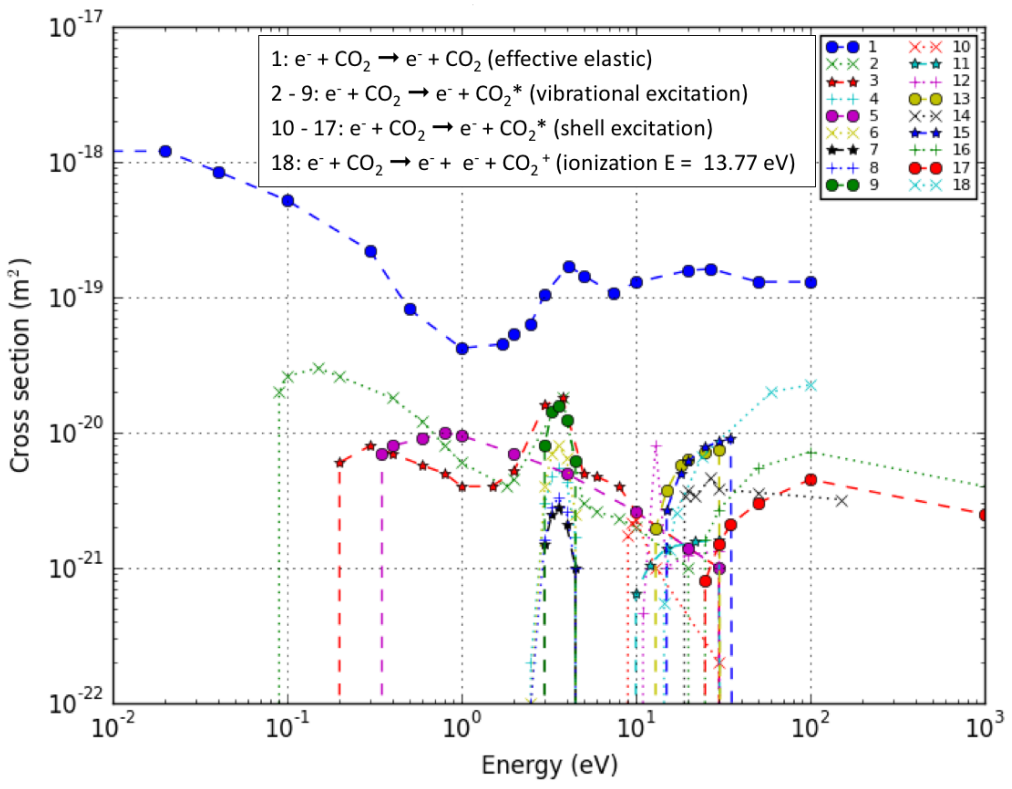

**Figure 3.5.:** Cross sections for different processes in electron-Carbondioxide scattering as a function of the electron energy at NTP. Trinity database, retrieved from lxcat.net on February 20, 2017 [55].

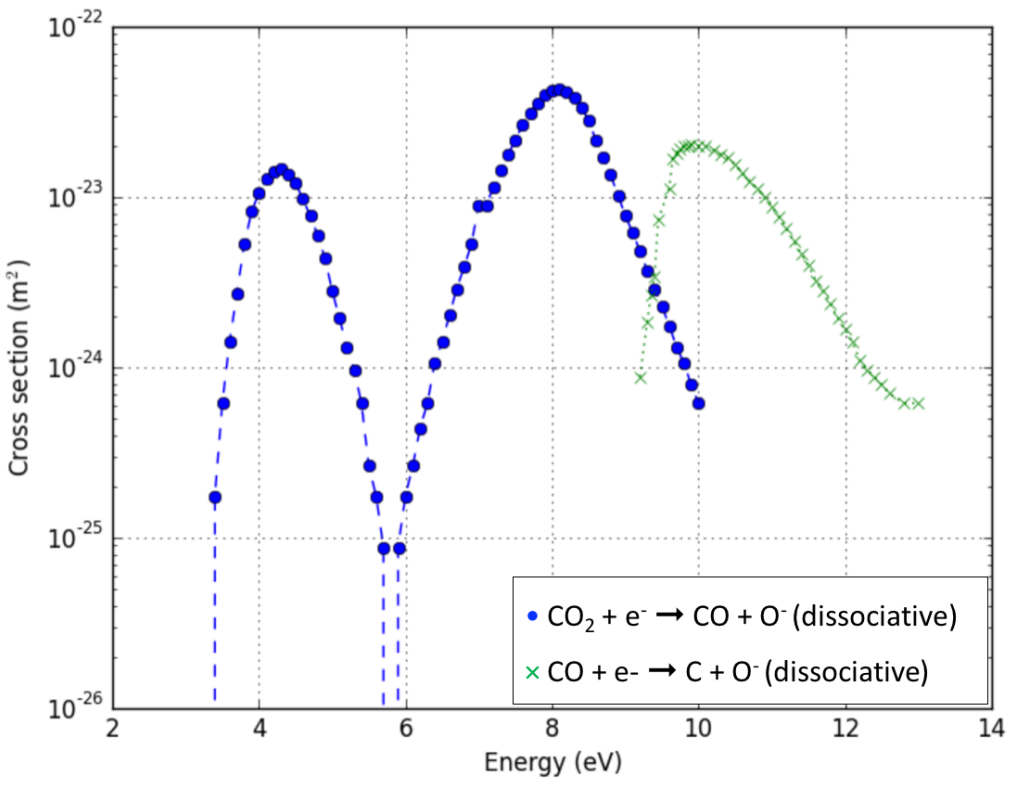

**Figure 3.6.:** Cross sections for attachment of electrons to Carbonmonoxide and Carbondioxide as a function of the electron energy at NTP. Itikawa database, retrieved from lxcat.net on February 20, 2017 [57].

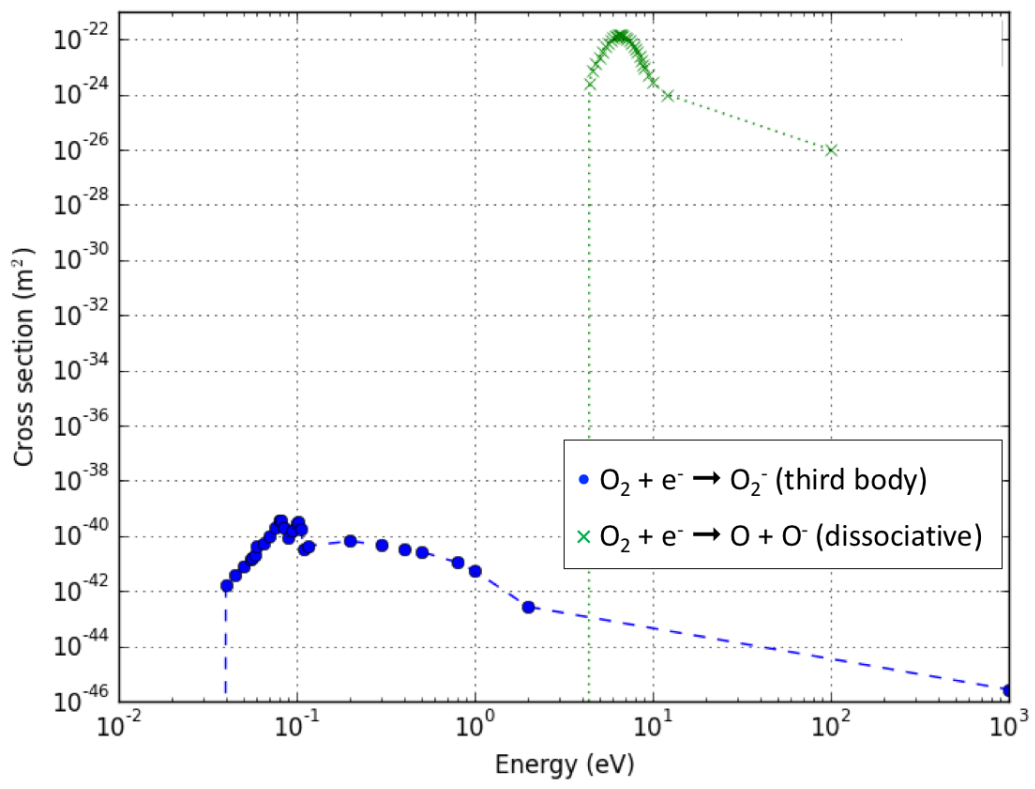

**Figure 3.7.:** Cross sections for attachment of electrons to Oxygen as a function of the electron energy at NTP. Magboltz database, retrieved from lxcat.net on February 20, 2017 [58].

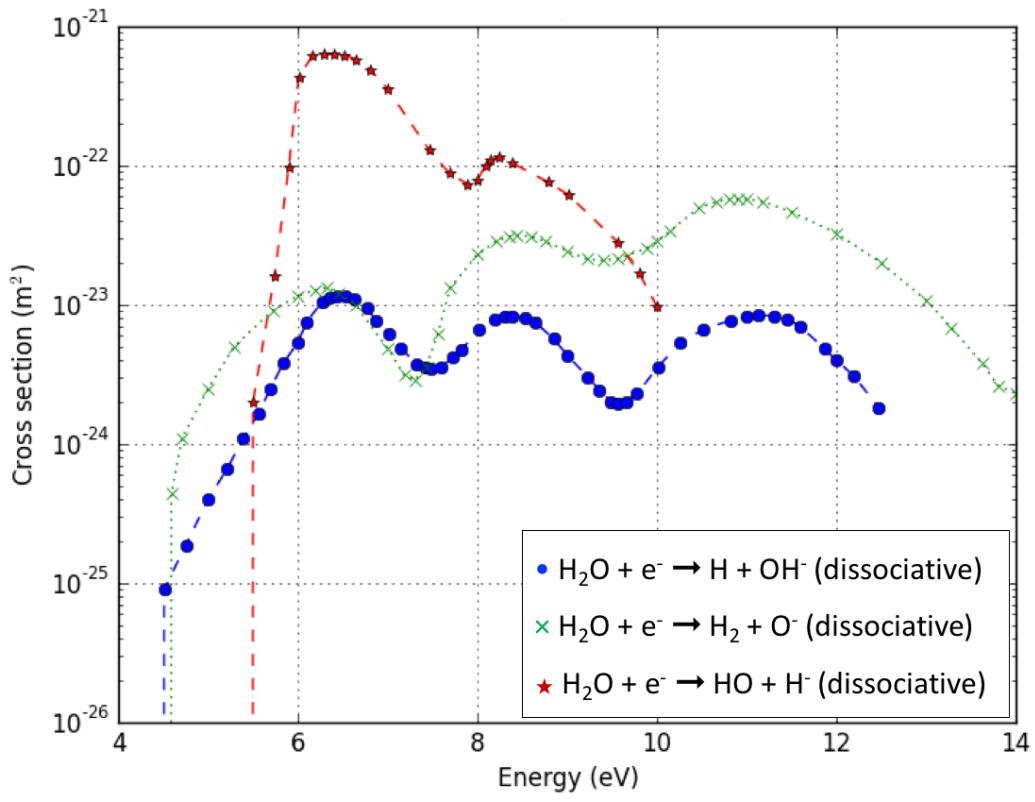

**Figure 3.8.:** Cross sections for attachment of electrons to water molecules as a function of the electron energy at NTP. Itikawa database, retrieved from lxcat.net on February 20, 2017 [57].

collisions are inelastic or ionizing, valid for noble gases and low fields typical for drift processes, the energy distribution can be expressed by

$$F(\epsilon) = C\sqrt{\epsilon}\exp\left(-\int \frac{3\epsilon\Lambda(\epsilon)}{[eE_D\lambda(\epsilon)]^2}\,d\epsilon\right) \quad . \tag{3.8}$$

where $\Lambda(\epsilon)$ is the fractional energy loss in the collisions and $\lambda(\epsilon) = 1/(N\sigma(\epsilon))$ is the mean free path of the exponentially distributed distance between collisions, $N$ is the number of molecules per unit volume and $\sigma(\epsilon)$ the energy dependent total cross section for all contributing processes. Correction terms can be added to (3.8) to account for inelastic collisions, excitation and ionization, provided their cross sections are known. Based on the energy distribution the drift velocity $v_D$ and diffusion coefficient $D$ can be derived by

$$v_D(E_D) = \frac{2e}{3m_e}E_D \int \epsilon\lambda(\epsilon)\frac{d\frac{F(\epsilon)}{\sqrt{2\epsilon/m_e}}}{d\epsilon}\,d\epsilon \tag{3.9}$$

and

$$D(E_D) = \int \frac{\lambda(\epsilon)}{3}\sqrt{2\epsilon/m_e}F(\epsilon)d\epsilon \quad . \tag{3.10}$$

The characteristic electron energy $\epsilon_k$ is customary defined as

$$\epsilon_k = \frac{eE_D D(E_D)}{v_D(E_D)} \tag{3.11}$$

and is a measure for the 'heating' of the gas, since it replaces $kT$ in (3.6).

### 3.1.3. Attachment Losses in (contaminated) Gas-Mixtures

Free electrons can attach to a molecule with electron affinity or being electro-negative. Thus, the attachment probability $h_g$ for a single scattering with a non-noble gas $g$ is finite. This leads to an average attachment time $t_{attach} = 1/hR_{coll}$, with $R_{coll}$ being the collision rate or rather the number of collisions per unit time.

As a consequence of (3.7), the attachment probability of a gas mixture is given by the sum over the gas concentration $c_g$ times its specific attachment probability.

$$h_{mixture} = \sum_g c_g h_g \quad , \text{with} \quad \sum_g c_g = 1 \tag{3.12}$$

With the cross section for different attachment processes being strongly energy dependent (see figures 3.6 to 3.8), the attachment probability becomes a function of the electrical field strength during electron drift $h_g(E_D)$.

Based on the probability of an attachment occurring in a single collision and the mean attachment length $\lambda_{attach} = \lambda/h$, the fraction $A$ of attachment losses after a certain drift length $x$, required for the signal strength calculation in (1.7), can be derived:

$$A(x, E_D) = 1 - \exp\left(-\frac{x}{\lambda_{attach(E_D)}}\right) = 1 - \exp\left(-\frac{xh(E_D)}{\lambda(E_D)}\right) \tag{3.13}$$

While the composition law (3.12) holds for every gas mixture, it does not transfer to the fraction of lost electrons $A$, since $\lambda$ in turn depends on the drift velocity and thus on the gas mixture.

However, small changes in the gas composition, for example ppm-level contamination of $O_2$, do scarcely affect the drift velocity, but significantly change the attachment probability. Assuming a nominal gas mixture (*nom.*), where contamination of a gas species $g$ in concentration $c_g$ has a negligible effect on $\lambda$ and the drift velocity, the total estimated attachment losses $A_{tot.}$ can be composed

$$\begin{aligned} 1 - A(x, E_D)_{tot.} &= (1 - A(x, E_D)_{nom.}) \cdot \prod_g \left(\frac{1 - A(x, E_D)_{nom.+c_g \cdot g}}{1 - A(x, E_D)_{nom.}}\right) \\ &= (1 - A(x, E_D)_{nom.})^{1-|g|} \cdot \prod_g \left(1 - \frac{c_g}{c_{ref}} A(x, E_D)_{nom.+c_{ref} \cdot g}\right) \end{aligned} \tag{3.14}$$

In the latter step $|g|$ is the number of added gas species. Thereby $c_g \leq c_{ref}$ and the reference concentration $c_{ref}$ must be sufficiently small for all contamination species, thus they do not affect $\lambda$. Especially when studying multiple admixtures at different concentrations and drift fields, this condition must be checked before applying the approximation (3.14).

### 3.1.4. Analytic Description of the Transparency of Wire Arrays and Micromeshes

Electrons in contact with the solids confining or interrupting the gas volume can be absorbed and are, therefore, lost for signal formation. In a Micromegas (chapter 1.2.5) each electron must trespass the conductive mesh and is prone to absorption during this process, which is represented with an electron loss probability $(1 - T)$ in (1.7).

If electron scattering and diffusion is negligible compared to the dimensions of the structure, slow electrons can be assumed to follow the electrical field lines. Therefore, the fraction $\Omega$ of field lines terminating on the wire or mesh compared to the total originating from the cathode is commonly used as an estimator for the electron transparency of a structure. For a grid of parallel wires with radius $r$ and wire pitch $p$, $\Omega$ can be analytically derived by reducing the problem to two dimensions [59]. The result depends on the electrical fields on both sides of the wire grid, labeled for a Micromegas according to figure 1.6 with $E_D$ and $E_A$, and a geometrical parameter $\rho = 2\pi r/p$:

$$\begin{aligned} \Omega(E_D, E_A, \rho) =& \frac{1}{\pi E_D}(E_D + E_A)\sqrt{\rho^2 - \left(\frac{E_A - E_D}{E_A + E_D}\right)^2} \\ &- \frac{E_A - E_D}{\pi E_D}\cos^{-1}\left(\frac{E_A - E_D}{E_A + E_D}\frac{1}{\rho}\right). \end{aligned} \tag{3.15}$$

The expression is valid in the range of

$$\frac{1-\rho}{1+\rho} < \frac{E_A}{E_D} < \frac{1+\rho}{1-\rho}. \tag{3.16}$$

Equation (3.15) provides a good estimate for the grids transparency in applications with suppressed scattering, for example high pressure TPCs or liquid gases. In applications beyond these conditions it poses an upper limit, since electron loss probability is increased compared to the zero-scattering case :

$$T_{grid} \leq 1 - \Omega_{grid} \tag{3.17}$$

For a mesh as used in a Micromegas the problem can not be reduced to two dimensions and currently no analytic solution is known. Commonly used is an approximation by two successive grids of crossed wires with the same values for $E_D$ and $E_A$. Although far from being precise, this coarse approximation sets an upper limit on the mesh transparency according to [13]:

$$T_{mesh} \leq (1 - \Omega_{grid})^2 \tag{3.18}$$

For $E_A/E_D$ above the limit given in (3.16), $\Omega$ converges to 0 and thus the limit on the transparency becomes 1 [59].

A lower transparency limit can be derived for very fast electrons approaching the mesh, again under the assumption of zero-diffusion. Having a very high momentum, the electrons' straight movement towards the mesh plane is barely deviated in the comparative short region of bent field lines. Thus, their absorption probability would be equal to the projected surface of the mesh wires on the plane perpendicular to the electrons' approaching direction. It is usually expressed in terms of the mesh's open area $O = ((p-2r)/p)^2$ (discussed in detail in chapter 8.3) or optical transparency $T_{opt}$:

$$T_{mesh} \geq T_{opt} \tag{3.19}$$

These analytic approaches neglect the effect of electron scattering during the drift, which depends on the gas mixture as well as the electrical field. To take these effects into account, a full simulation of the microscopic processes is required.

## 3.2. Simulation of Microscopic Electron Transport

The analytic description of electron transport is reaching its limits in complex field configurations, not analytically describable interference of scattering processes and the absorption at solid structures. Therefore, a different approach on experiment-theory comparison is required to further improve the knowledge on the physics processes. As discussed in chapter 1.3.3 and figure 1.8, the key to this comparison lies in the Monte Carlo based simulation of the microscopic processes yielding macroscopic observables. The simulations performed in this study utilized the Garfield++ program [60] with the algorithms developed during the author's diploma thesis [61], where a detailed description of the method, its realization and its limitations can be found.

The geometry of the Micromegas is modeled with **F**inite **E**lement **M**ethods (FEM) utilizing a woven wire mesh description (figure 3.9), which turned out the be a much more accurate description compared to flat models with intersected cylindrical or rectangular wires. For all Garfield MC simulations the FEM calculations have been performed with the ANSYS® 14.1 program [62]. For a more qualitative discussion on multi-parameter variations in the mesh geometry, as presented in chapter 8, COMSOL Multiphysics® [63] has been used for FEM simulations.

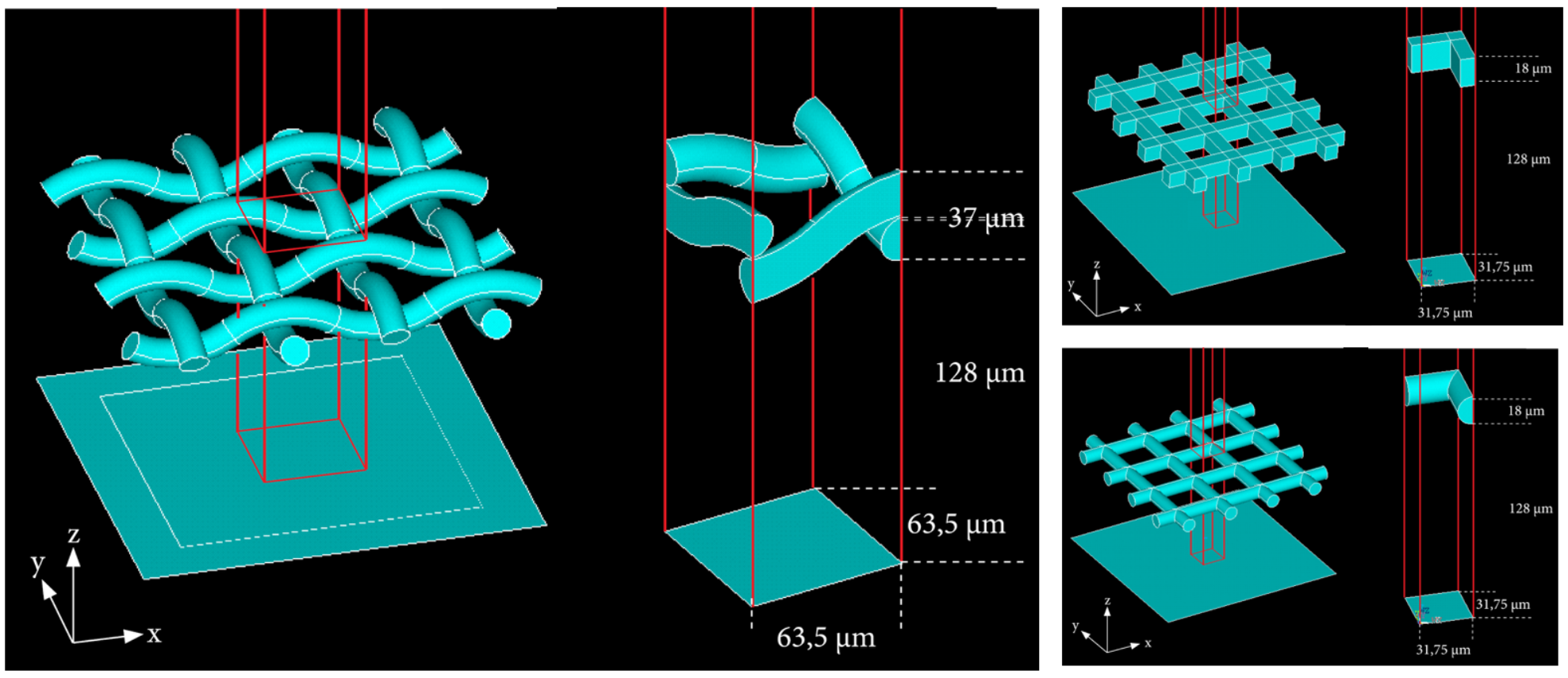

**Figure 3.9.:** Geometry model for FEM field calculation in a Micromegas mesh and amplification region, computed in ANSYS®. The models unit cell is depicted separately with exemplary values. Left: Woven wire mesh model composed of toroid sections. Right: Simplified models with flat geometry using rectangular (top) and cylindrical (bottom) wires [61].

Once the geometry and field configuration is loaded, Garfield++ retrieves the cross sections of the specified gas mixture from the Magboltz [64] database and applies corrections according to the specified pressure and temperature. An electron inserted in the gas volume is then drifted in the *microscopic tracking* mode by repeated iteration of the following steps: The initial condition of the electron and the electrical field in its position is evaluated and the corresponding cross sections and mean free path are derived. The step size is determined by a random number generator, but in relation to the mean free path by the condition of a certain probability of null-collision. The electron's energy gain (or loss) due to the electric field is calculated and the new electron's position and momentum are determined. Based on the updated cross sections it is determined whether scattering occurred. Depending on the selected scattering mechanism a change in the electron's energy and/or momentum as well as a possible excitation or ionization of the scattering target is applied [65]. Eventually each electron will either leave the simulated volume, end up on a boundary surface of the gas volume or be lost to attachment processes. In either case, the electron's track, its endpoint and end condition can be used for further analysis.

While the source mechanism for electron loss during drift can not be directly determined experimentally, attachment losses to the gas and absorption at solids can easily be distinguished in the simulation. Thus, both mechanisms are studied independently and combined according to the factorized approach in (1.7) once they are compared to experimental results.

### 3.2.1. Attachment Losses during Electron Drift

The drift region of a Micromegas is in first order comparable to a parallel plate arrangement and, therefore, the drift field is in a large fraction almost uniform. Deviations of the field occur in the vicinity of the mesh and reach several 100 µm into the drift volume. An in-depth discussion of the range of these non-homogeneity is included in the mesh parameter discussion in chapter 8. For the simulation of drift processes and attachment losses the electric field is assumed to be uniform, a simplification that significantly reduces the simulation time by avoiding an evaluation of the field parameters from an FEM map. Electrons are started at $t_0$=0 in a defined position (0,0,0.5 cm) and drift along the electric field, aligned in z-direction, through the laterally (x-y-plane) not confined gas volume. Their endpoint is registered once they either reach the bottom plane ($z = 0$) of the gas volume and, therefore, leave the sensor or attach to a gas constituent. The fraction of non-attached electrons $(1 - A)$ after a certain drift path $z_1$ along the electric field $E_D$ in a defined *gas* mixture is simply extracted by counting the remaining electrons passing through the $z = z_1$ plane.

$$1 - A(z_1, E_D, gas) = \frac{n_e|_{z=z_1}}{n_e|_{z=z_0}}. \tag{3.20}$$

Additionally, the transverse diffusion can be extracted from the endpoint distribution (figure 3.10 - left) of a sufficiently large number of electron drift events. Therefore, only the non-attached electrons arriving at the $z = 0$ plane are considered, their spatial x-y-distribution is fitted with a Gaussian and its standard deviation yields the diffusion coefficient according to (3.4). The distribution of non-attached electrons over the z-coordinate follows an exponential decrease in agreement with (3.13) and allows for the extraction of the mean attachment length $\lambda_{attach}$. The mean drift velocity $v_D$ can be calculated utilizing the mean end time for electrons arriving at the $z = 0$ plane, extracted from the Gaussian contribution in figure 3.10 - right. The relative width of this Gaussian yields the longitudinal diffusion of the electron cloud along the drift direction. The end time distribution of electrons lost in attachment follow an exponential law and allow for determination of the average attachment time $t_{attach}$.

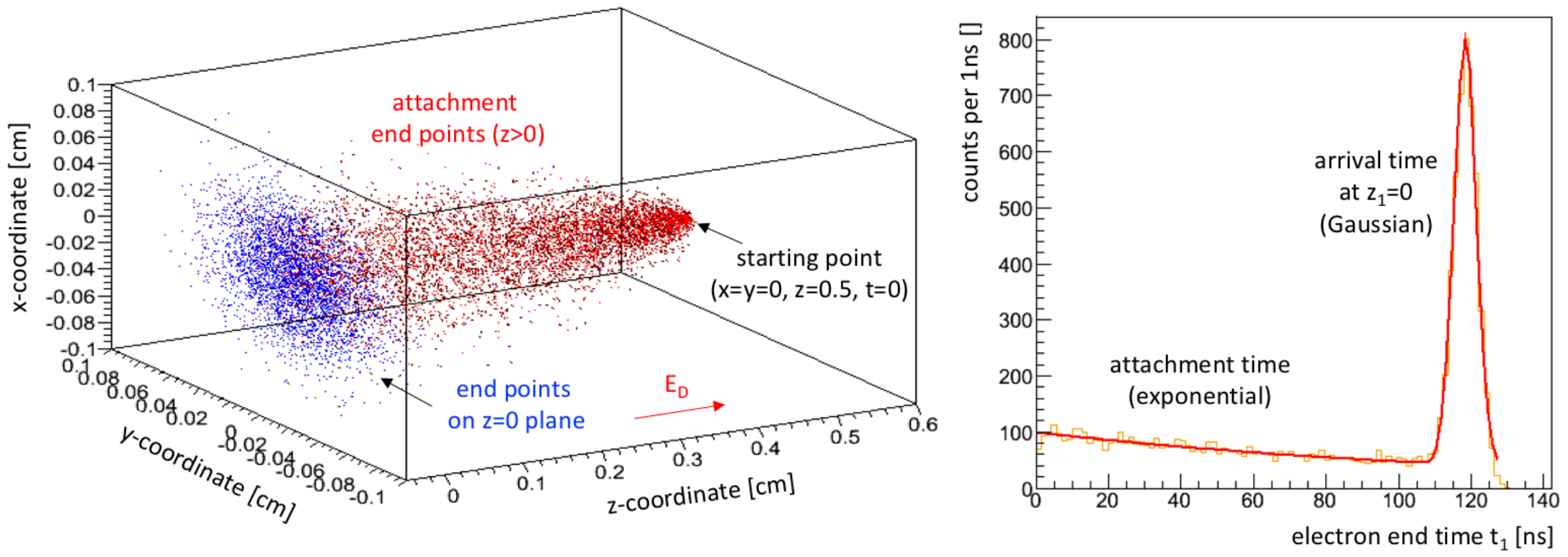

**Figure 3.10.:** End point (left) and end time distribution (right) for the simulated drift of $10^4$ electrons in pure $Ar{:}CO_2$ (93:7) in a homogeneous drift field $E_D$=4 kV/cm.

The drift- and attachment process has been simulated in a $Ar{:}CO_2$ (93:7) gas mixture for a range of $20\,\mathrm{V/cm} \leq E_D \leq 4000\,\mathrm{V/cm}$ with $10^4$ electrons per configuration. Besides the nominal gas mixture, contamination of the gas can have a huge impact on the attachment length. The main source of contamination in our experimental setup is the (humid) atmospheric air. Among its constituents $O_2$ and $H_2O$ yield the largest attachment contribution, since $N_2$ and the noble gases provide no electron attachment mechanism, $CO_2$ is anyhow present in the mixture and small changes in its concentration have a negligible impact, and the fraction of $H_2$ and $CF_4$ in atmospheric air is minuscule. A contribution by attachment to CO produced in a possible dissociative attachment with $CO_2$ is negligible due to the regular gas exchange and, therefore, small concentration and the high dissociation energy required ($\epsilon > 9.2$ eV, see figure 3.6). Different contamination levels of $O_2$ (figure 3.11) and $H_2O$ (figure 3.12) up to percent level in the nominal $Ar{:}CO_2$ (93:7) mixture have been simulated. A reference drift length of 5 mm has been selected, allowing a re-scaling of the expected attachment losses according to (3.13).

For small electron scattering energies of 0.1-1.0 eV attachment losses are dominated by the three body attachment to $O_2$ (figure 3.7). Despite its 17 magnitudes lower cross section, the abundance of scattering processes in this energy range leads to a significant electron loss for $E_D < 500\,\mathrm{V/cm}$ strongly dependent on the Oxygen concentration. At electron scattering energies $> 3$ eV dissociative attachment processes becomes more probable (figures 3.6 to 3.8). While the total cross section maximum for dissociative attachment in $O_2$ and $H_2O$ are higher compared to $CO_2$, these process have a higher energy threshold. Additionally, $O_2$ and $H_2O$ contamination levels are assumed to be small compared to the 7 % $CO_2$. Consequentially increasing drift fields of $E_D \geq 1\,\mathrm{kV/cm}$ yield electron attachment rates dominantly to $CO_2$, only slightly altered by the concentrations of Oxygen up to 0.1 %. A rather large contamination with water vapor up to 1 %-level contributes only for very strong drift fields $E_D \geq 2\,\mathrm{kV/cm}$.

The linear dependence of the attachment coefficient $1/\lambda_{attach}(E_D)$ on the concentration of oxygen (figure 3.13) or water (figure 3.14) underlines the validity of the scaling law introduced in (3.14) in the given concentration range. Therefore, admixtures of both gas species can be calculated based on two reference attachment coefficients extracted from the corresponding linear fit, avoiding the time consuming simulation of all combinations of possible concentrations. For the later applied comparison with experimental data (section 3.3.3) the concentrations of $O_2$ and $H_2O$ as well as a mean drift distance $\overline{x}_{drift}$ must be estimated.

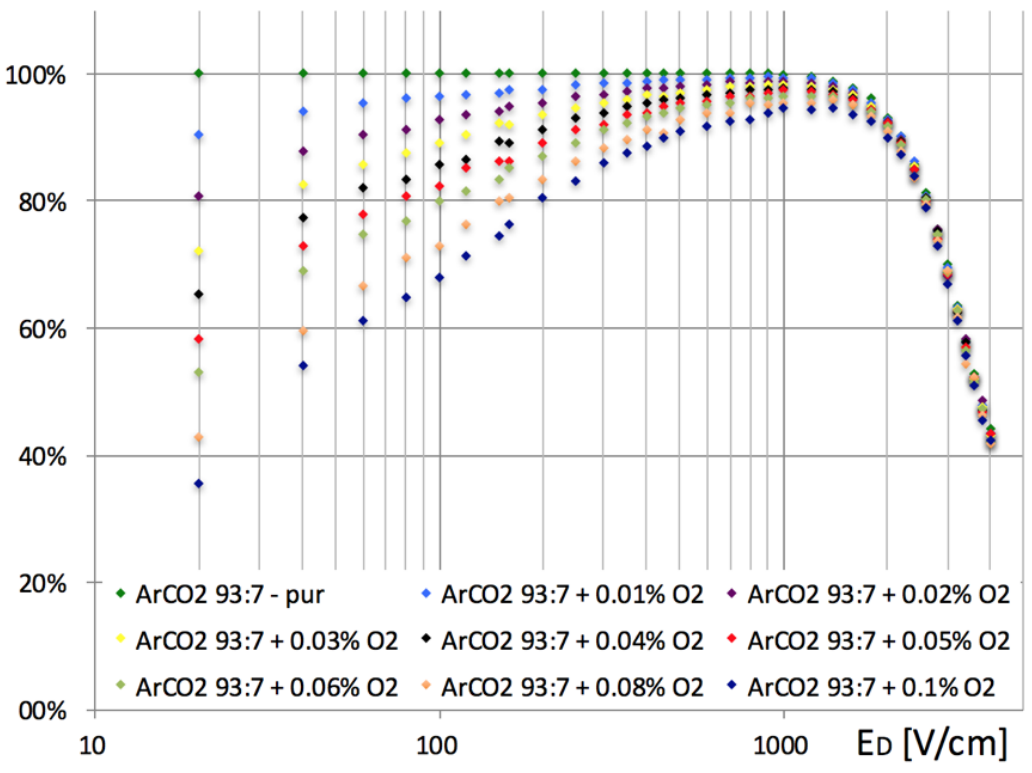

**Figure 3.11.:** Fraction of non-attached electrons after 5 mm drift in Ar:$CO_2$ (93:7) with different levels of $O_2$ contamination.

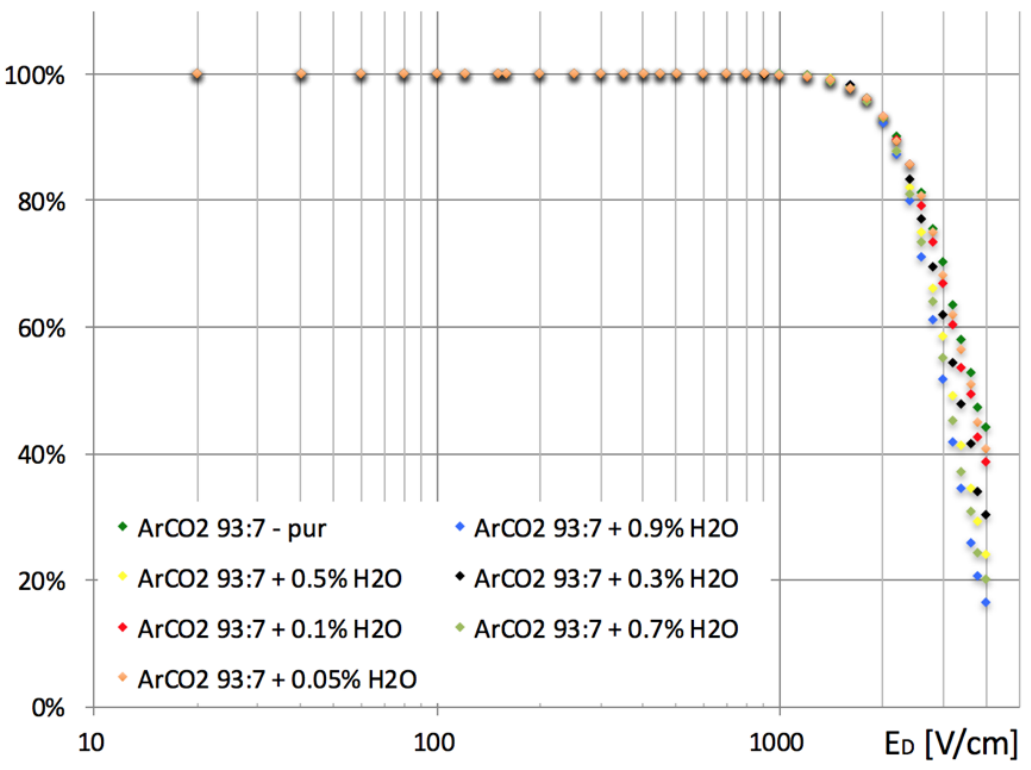

**Figure 3.12.:** Fraction of non-attached electrons after 5 mm drift in Ar:$CO_2$ (93:7) with different levels of $H_2O$ contamination.

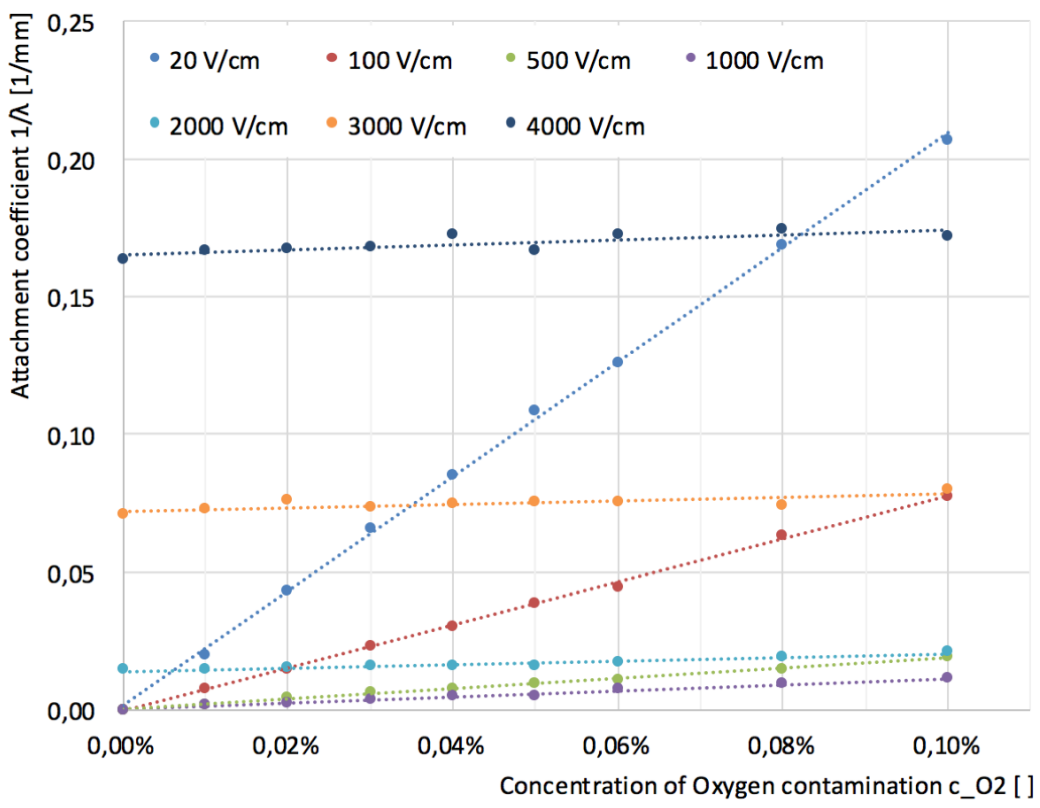

**Figure 3.13.:** Attachment coefficient $1/\lambda_{attach}$ in Ar:$CO_2$ (93:7) with increasing level of Oxygen ($O_2$) contamination up to 0.1 % for various drift field strengths $E_D$. The difference in the offset of the linear fit is owed to the attachment to $CO_2$ in the corresponding drift field.

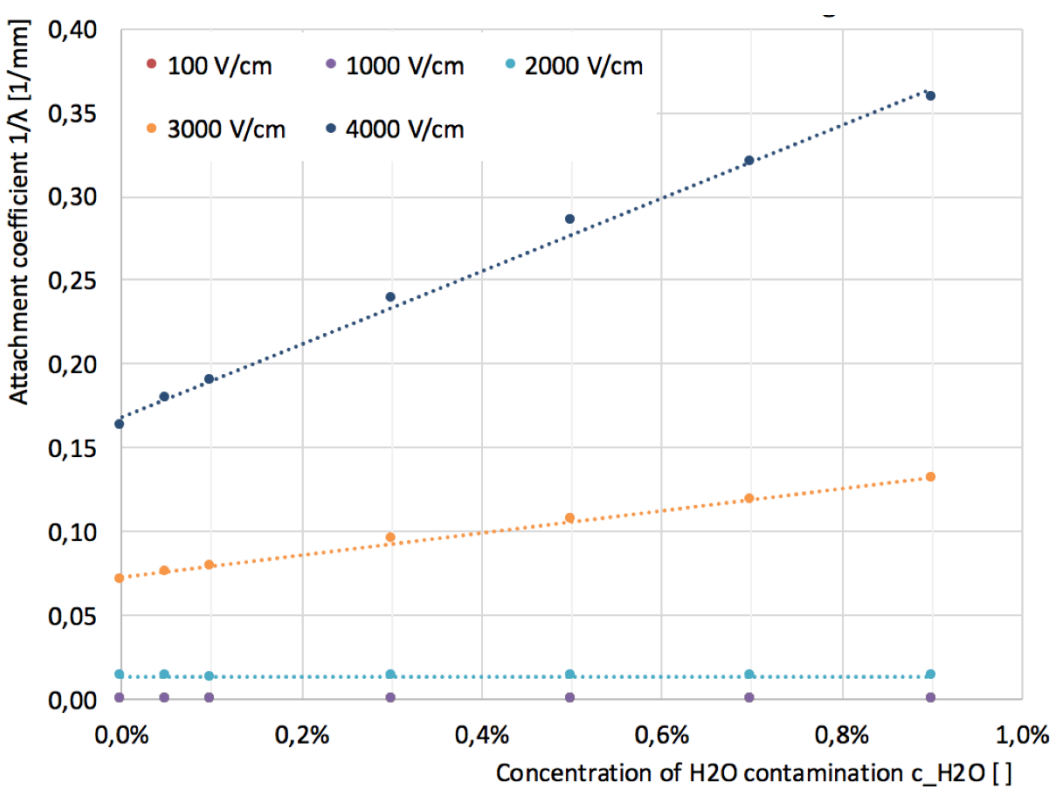

**Figure 3.14.:** Attachment coefficient $1/\lambda_{attach}$ in Ar:$CO_2$ (93:7) with increasing level of water vapor ($H_2O$) contamination up to 0.9 % for various drift field strengths $E_D$. The attachment during drift in lower fields is almost zero and thus suppressed.

For the latter $\overline{x}_{drift}$ slightly larger than 2.5 mm is expected for a 5 mm drift gap, due to the exponential distribution of the photon conversion events in the gas with a mean absorption length in O(5 cm) (figure 2.8). A domination of the starting point by photo electric effect at the copper cathode, instead of the gas, can be ruled out by the frequent occurrence of the escape peak in the spectrum, which is characteristic for photo ionization in Argon. If experimental assessment of the contamination concentrations $c_{\mathrm{O}_2}$ and $c_{\mathrm{H_2O}}$ is not possible, they have to be considered free parameters in a fit. Assuming that the sole source of contamination is the humid atmospheric air entering the detector via backwards diffusion through leaks, their ratio can be estimated to $c_{\mathrm{H_2O}}/c_{\mathrm{O}_2} \sim 11/20$, corresponding to the volume percentages of water vapor and Oxygen at RH= 50 % NTP. However, the validity of this assumption is not secured, since other contamination sources, like the hydroscopic behavior of plastic materials, favor a contamination with water without altering the oxygen content.

### 3.2.2. Electron Transition through Micromeshes and Transparency

For the simulation of the mesh transit, the Micromegas has to be modeled in full geometric detail and with a high granularity of the FEM-mesh to avoid artifacts biasing the results, as we discussed in [61]. The mesh geometry has been modeled for a variety of meshes with 18 µm and 30 µm wire diameters $d$ and different mesh aperture $a$. As described in detail in chapter 8.3, these two parameters completely define the geometry of a symmetrical plain-weave wire mesh and other characteristics, such as the open area $O$ or the pitch $p$, which can be derived by (8.1) and (8.2). Due to the assumed importance of the open area for the mesh transparency the partially redundant notation *'a-d: O'* is used to refer to a mesh geometry.

In the Garfield++ simulation the electrons are started off in a randomized position in the x-y-plane at $z_0 = 400$ µm. Randomization and sufficient distance to the mesh are required to assure a representative approaching behavior of the electrons towards the mesh and avoid a bias on the transparency. Additionally, the mean kinetic energy of the electrons, started with a small momentum in arbitrary direction, must equalize according to the drift field configuration before reaching the region of bent field lines. The limit of 400 µm has been extracted from a fixed-start point transparency study conducted similarly the the one described in [61, chapter 7.1]. Again the endpoint distribution (figure 3.15) of the electrons is used to extract the transparency by counting the fraction of electrons trespassing the mesh.

$$T(mesh, E_D) = \frac{n_e|_{z=z_{mesh-bottom}}}{n_e|_{z=z_{mesh-top}}} \quad . \tag{3.21}$$

Electrons lost to attachment before approaching the mesh ($z_{end} > z_{mesh-top}$), or within the amplification gap after mesh transit ($z_{end} < z_{mesh-bottom}$), do not affect the transparency in this definition. The probability of an electron being lost to attachment within the narrow mesh window is rather small, but a possible bias can be removed by constraining the counting algorithm in (3.21) to electrons with an endstatus $\neq$ 'lost-to-gas-attachment'.

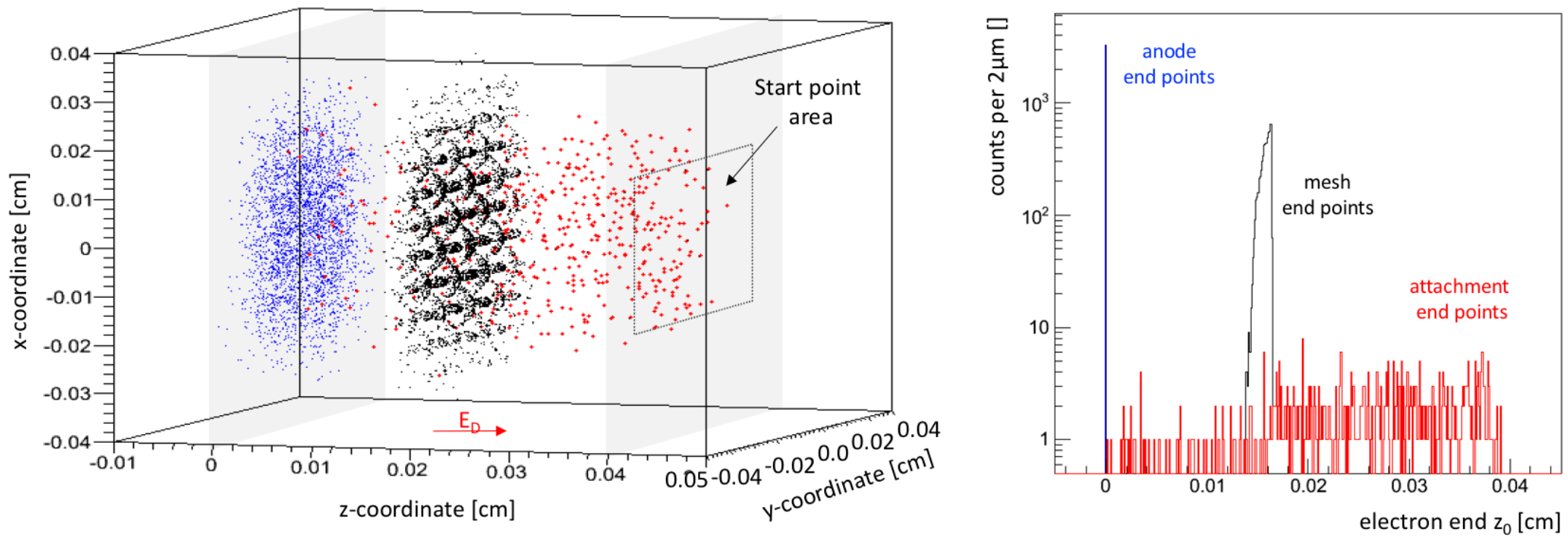

**Figure 3.15.:** End point distribution of $10^4$ electrons approaching the a 45-18: 52.0 % mesh in Ar:$CO_2$ (93:7) with a homogeneous drift field $E_D = 4\,\text{kV/cm}$. The electron endpoints are colored according to their loss mechanism: neutralized at the anode plane $z = 0$ (blue), absorbed at the mesh surface (black) and lost to attachment to the gas (red). (left) 3D endpoint distribution, (right) projection to the z-coordinate.

The fraction of field lines ending on the mesh is determined mainly by the ratio of $E_A/E_D$ (see (3.15)). In a typical Micromegas configuration $E_A \gg E_D$ and thus small changes of $E_A$ within the Micromegas working range of $\sim 40 - 45\,\text{kV/cm}$ are expected to have a negligible small effect on the transparency. This assumption has been cross checked in a preparatory simulation and confirmed by the experimental data (see figure 3.28 in section 3.3.3). Therefore, the mesh and anode potential has been fixed for the majority of the simulation runs at $U_{mesh} = 0$ (GND) and $U_{anode} = +580\,\text{V}$ and only the cathode potential has been varied to cover a drift field range $20\,\text{V/cm} \leq E_D \leq 4000\,\text{V/cm}$. The dependency model introduced in figure 1.7 suggests an impact of the gas mixture on all kind of drift processes and thus on the mesh transit. As discussed in the previous chapter, contamination of $O_2$ up to 0.1 %-level and $H_2O$ below 1 % have a small effect on the overall drift velocities and diffusion coefficients and, therefore, their effect on the transparency is expected to be small. The transparency simulation is, therefore, restricted to the pure Ar:$CO_2$ (93:7) mixture but can be easily adopted for all other gases and mixtures covered by the Magboltz database.

The electron transparency $T$ is shown for different mesh geometries as a function of the drift field strength $E_D$ in figure 3.16. For low drift fields, and hence small mean approaching velocities of the electrons, a transparency close to 100 % is reached for most of the mesh geometries. This observation is in agreement with our argumentation in section 3.1.4, since electrons are expected to closely follow the field lines for low energy drift processes.

With increasing field strength and hence drift velocity the transparency decreases. A descriptive parameter for this decrease is the $E_D$ threshold, above which the transpar-ency drops below a given percentage, like $\max(E_D)|_{T\geq 95\,\%}$ for the $T > 95\,\%$ threshold. Comparing among meshes with equal wire diameter, this threshold field strength increases for meshes with larger aperture, allowing for a wider operation range in drift voltages while maintaining full electron transparency. Beyond $\max(E_D)|_{T\geq 95\,\%}$ the electron transparency continuously decreases until the absorption loss fraction seems to saturate, and transparency converges towards a lower limit. Noticeably this limit is well below the mesh's optical transparency, which is in disagreement with the assumption stated in (3.19). It is furthermore noticeable, that meshes of similar open area do not yield the same transparency for a given field strength and, therefore, the open area is not a good predictor for the mesh transparency, a misconception that is widely believed. The figures 3.16 and 3.17 show clearly that for a given open area, meshes with finer wires yield systematically lower electron transparency.

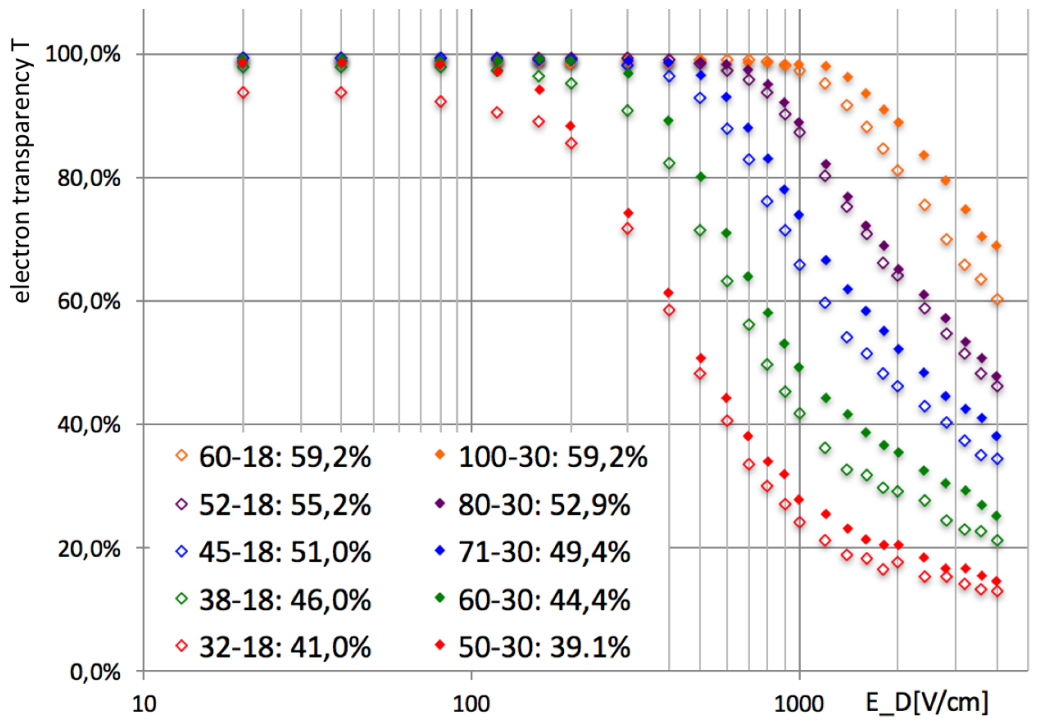

**Figure 3.16.:** Electron transparency $T$ as a function of the drift field strength for different mesh geometries labeled with $a$[µm]-$d$[µm]: $O$[ ].

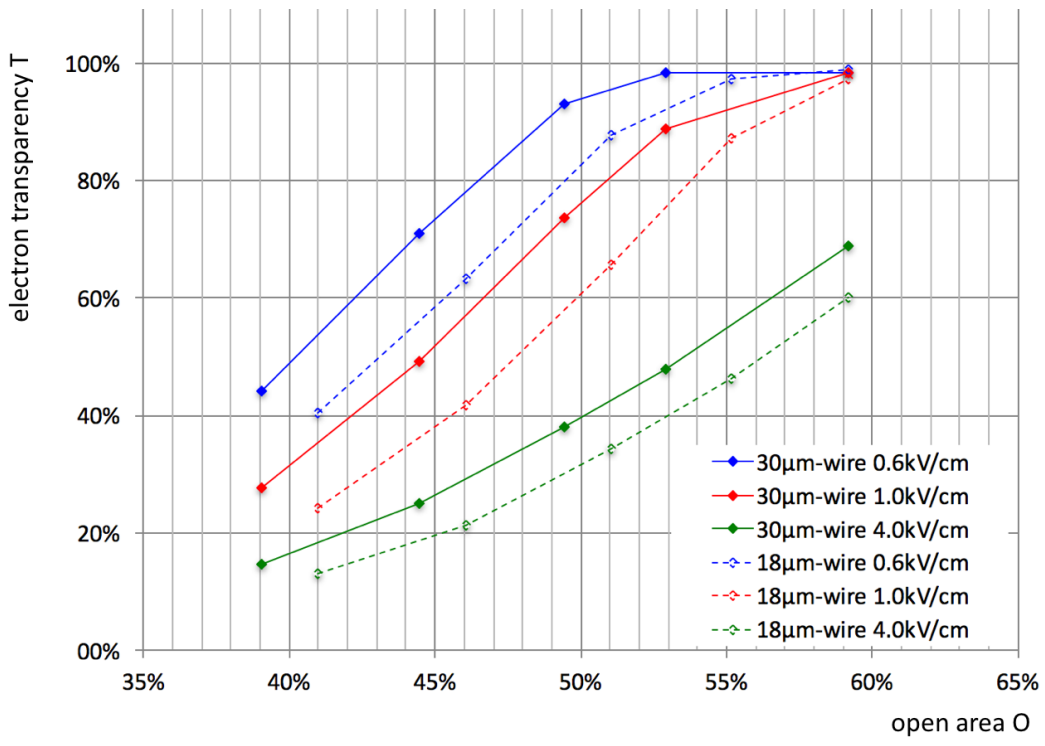

**Figure 3.17.:** Electron transparency $T$ as a function of the open area of meshes with 18 µm and 30 µm wire diameter at three different drift field strengths.

With the high field ratio $E_A/E_D > 10$ in a Micromegas, the fraction of field lines ending on the mesh remains close to 0 % within the studied $E_D$ range. The observed transparency reduction with increasing $E_D$ can, thus, not be explained by the change in the field line configuration, but with the changed drift and scattering behavior of the electrons. Figure 3.18 shows the processes contributing to this continuously increasing loss of electrons with higher $E_D$: while for low electron momenta (figure 3.18 (A)) the electron path can closely follow the field line, the increased momentum of faster electrons (figure 3.18 (B)) leads to a deviation from the field lines due to the electron inertia. This effect always leads to an effective path closer to the wire compared to the field line the electron followed during its approach towards the mesh. Therefore, it allows electrons to reach the wire surface and be absorbed, resulting in a reduction of the transparency with increasing drift field. In its limit of straight tracks, totally undisturbed by the field lines, this inertia effect would yield an electron transparency equal to the optical transparency of the mesh.

An additional deviation from the field lines is caused by the scattering of the electrons with the gas (figure 3.18 (C)). In a collision the electron's direction is randomized and the electron can cross several field lines before being forced into field direction. This leads to a transverse displacement in the direction perpendicular to the field lines. Although this effect is not favoring a certain direction, towards or away from the wires, it allows for electrons to be scattered onto the wire surface and be absorbed, thus contributing to transparency reduction. Creating zones around the wire where absorption after scattering occurs with a certain probability, the effect could be considered to yield an increased effective wire diameter and, therefore, a reduced effective open area of the mesh. With the characteristic dimension being the transverse displacement after a scattering process the corresponding reduction in transparency is larger for fine mesh structures: an effective increase of the wire diameter by a few µm yields a larger reduction of the open area in a 18 µm wire mesh compared to a 30 µm wire mesh of a similar open area. This effect can explain the difference in transparency observed between meshes of comparable open area in figure 3.17.

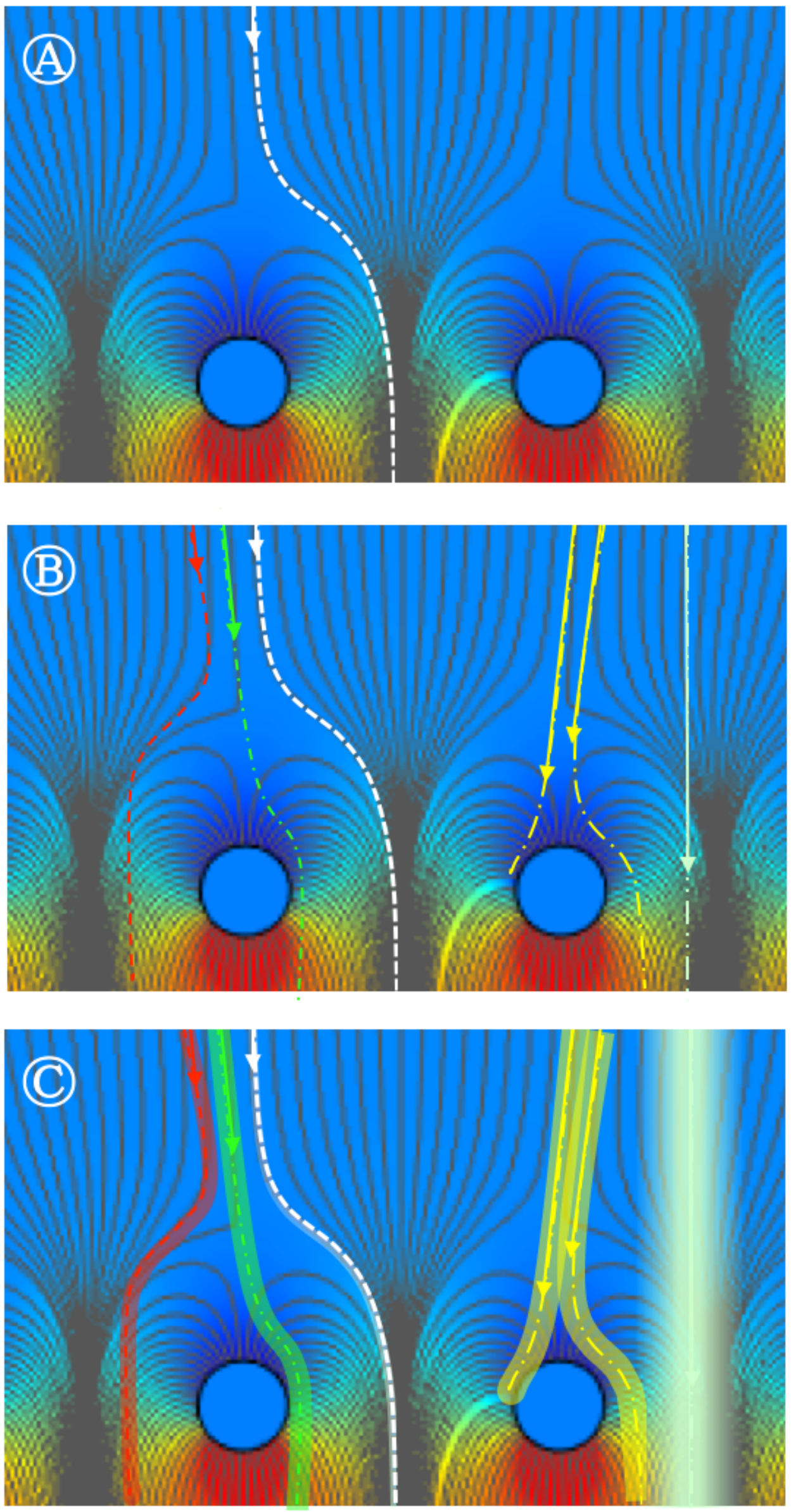

**Figure 3.18.:** Schematic of the electron drift processes and their contribution to absorption losses during mesh transit: (A) electrons with low velocity follow the the field lines closely and absorption losses are dictated by the fraction of field lines ending on the solid, (B) with increasing velocity, the electron's inertia leads to a delayed deflection of the electron and results in a path closer to, or possibly ending on a wire, (C) scattering with the gas results in diffusion along the electron's 'ideal' path and can result in a displacement perpendicular to the track.

## 3.3. Experimental Assessment of Signal Electron Losses

Experimental assessment of signal formation processes relies inevitably on the macroscopic observables, like the signal provided by the detector. It lacks, therefore, the ability to assess single processes as directly and independently of each other as possible in simulation. To study the impact of parameter variations on a particular process we make use of the factorization approach introduced in chapter 1.3.1. Attachment losses $A$ and electron transparency $T$ can be indirectly derived from the measured signal strength $S$ utilizing (1.7). Therefore, it is crucial to avoid, or at least correct for, effects of the intended parameter variation on other signal formation processes. These dependencies have been discussed in chapter 1.3.2 and visualized in figure 1.7. Furthermore, the control of the assumed to be constant parameters, according to a VOTAT strategy, is of great importance to maintain comparability of the data throughout the study.

Among the set of parameters the electric configuration can be precisely controlled and accurately varied, at least on a nominal parameter level. Other parameters prove more difficult to be reproduced reliably: the gas mixture and condition, for example, can be regulated on a nominal level, but effective values can differ due to the impact of contamination from atmospheric air and environmental parameters. The weather dependent environmental pressure remains impossible to control and eventually effects any open exhaust or imperfectly sealed gas setup. While a reproduction of the coarse geometry in-between different detectors is feasible, a controlled variation of the fine geometry parameters, like the dimensions of the micromesh, requires an innovative detector layout.

### 3.3.1. The Exchangeable Mesh Micromegas

Previous approaches to compare the behavior of different meshes in a series of similarly built Micromegas [66] suffer from the repeatability limits in anode and pillar structure dimensions, where deviations on µm-level have significant impact on the detector performance. The **Ex**changeable **Me**sh (ExMe) Micromegas is a novel layout for a Micromegas detector, where the micromesh is mounted on an independent support frame and can be easily exchanged (figure 3.19), first presented in [53]. Utilizing the same readout board and, therefore, identical anode geometry and pillar support structure to operate different meshes, allows for a direct observation of the mesh geometry impact on the detector's performance under strict preservation of all non-mesh related fine geometry parameters.

The design of the ExMe detector follows, to a large extent, the specifications for the ATLAS NSW Micromegas, which are discussed in detail in the chapters 6 and 7.1, with the main difference of the exchangeable mesh feature. The detector is composed, as shown in figure 3.19, of a readout- and a drift panel enclosing the metallic mesh frame and a rubber O-ring for sealing of the created gas volume. Both panels are backed up by FR4-sheet-Al-honeycomb stiffening panels to provide rigidity and maintain the panel's flatness achieved during their face-down gluing on a marble table.

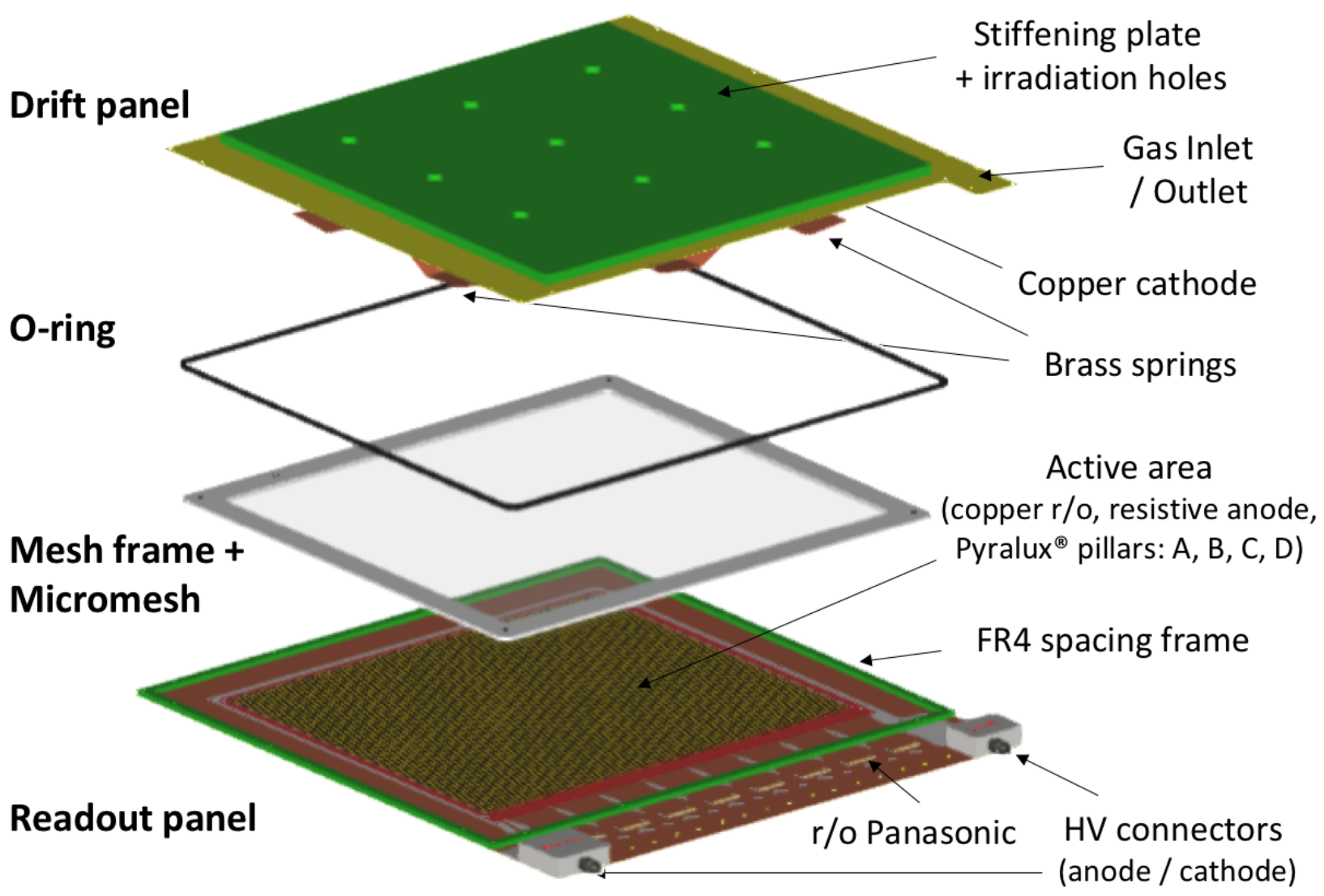

**Figure 3.19.:** Layout of the Exchangeable Mesh (ExMe) Micromegas detector.

The *readout board* comprises a copper structure with 1024 readout strips of 300 µm width, 150 µm spacing and 35 cm length in the active area. They are routed to the rim of the detector and connected with eight 128-channel Panasonic connectors. The resistive protection layer, discussed in chapter 6.2, is forming the detector's anode. Its pattern is a copy of the readout strips with additional interconnections added every 10 mm between alternating neighboring strips (see chapter 6.2.2). It is deposited on an insulating Kapton® foil, which is glued on the copper structure. Two detectors have been built to evaluate two different types of resistive material and deposition procedures (chapter 6.2.3): ExMe1 with a sub-0.1 µm thin carbon sputtered anode and ExMe2 with a screen-printed pattern of 12 µm central strip thickness.

The support structure for the floating mesh (see chapter 6.3) consists of a pattern of 300 µm diameter pillars formed out of a double layer of 64 µm Pyralux® photoresist. The pillars are arranged in a triangular lattice with different inter pillar spacing in the four sectors of the detector: A = 5 mm, B = 7 mm, C = 8.5 mm and D = 10 mm (magnification in figure 3.22). This allows for a study of the pillar distance impact on the Micromegas performance. Dependent on the thickness profile of the resistive anode, the pillars create amplification gaps of slightly different effective heights. The maintained quantity is the volume of the Pyralux®, which is laminated under high temperature and pressure on the readout board. On a flat sputtered anode surface (ExMe1) it, therefore, creates precise 128 µm high pillars. On the 'bumpy' surface of the screen printed anode (ExMe2) the gaps between the resistive strips are filled during the lamination and the material is redistributed. The resulting effective distances are visualized in figure 3.20 and for the ExMe2 detector one records:

$$\text{pillar top} \leftrightarrow \text{strip surface} < 128\,\mu\text{m} < \text{pillar top} \leftrightarrow \text{Kapton}^{®}\,\text{surface} \tag{3.22}$$

Along its circumference the readout panel carries a 5 mm high FR4 frame (figure 3.22) defining the distance between the panels and, hence, the gas gap.

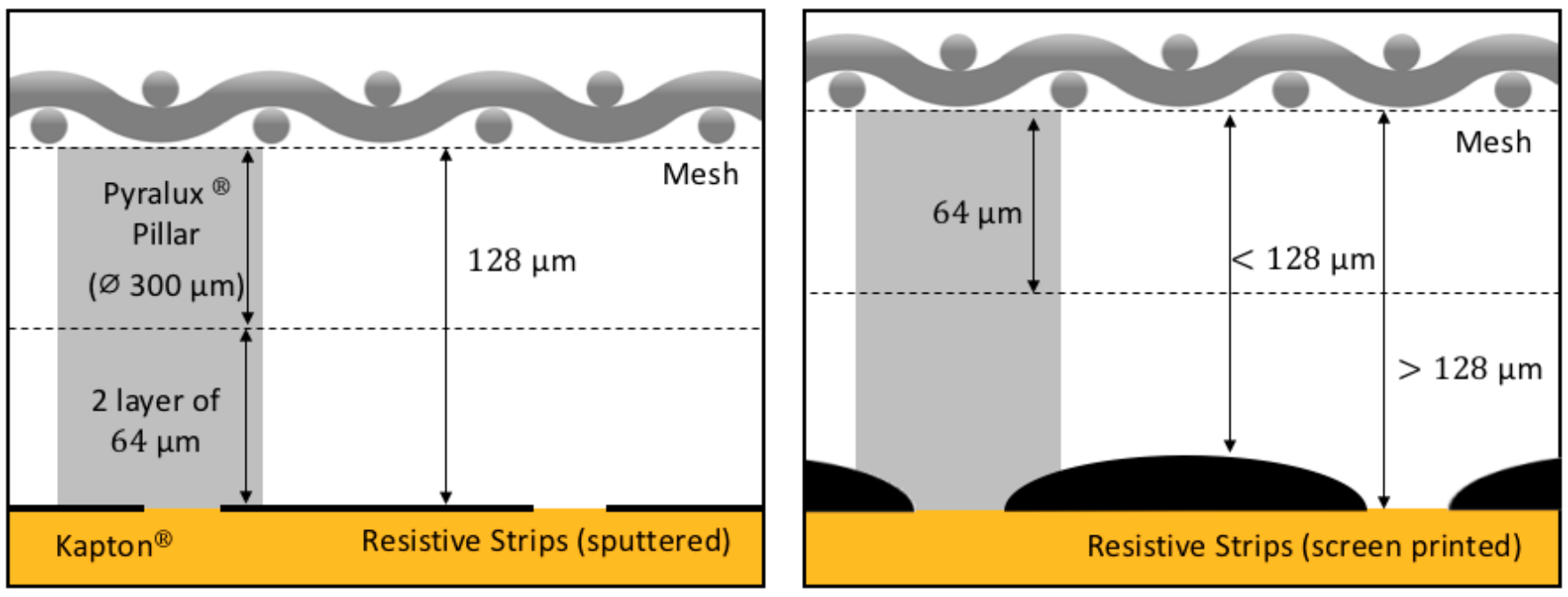

**Figure 3.20.:** Schematic of the ExMe1 (left) and ExMe2 (right) amplification gap cross section formed by the flat / bumpy structure of the resistive anode strips and the mesh resting on pillars of equal material volume. Not to scale. Modified after [54]

The *drift panel* is a multilayer PCB with a flat copper cathode and internal conductive channels. The gas distribution lines are embedded in the panel featuring a series of inlet holes along one side of the cathode and exhaust holes along the opposite side, creating an almost uniform gas flow through the volume. Contrary to the continuous stiff-back of the readout plane, the upper stiffening is pierced with nine holes corresponding to the positions where the drift panels' FR4 is removed and only a thin copper layer is enclosing the gas. These spots are utilized to irradiate the chamber providing on the one hand minimized obstructive material and, therefore, increased rate in the detector, as well as additional guidance for source re-positioning. A set of brass springs is mounted along each side of the cathode. Once the chamber is assembled they press the mesh frame onto the readout panel and assure a strong contact of the mesh on the supporting pillar structure.

The *mesh frames* are 4 mm thick solid iron to withstand deformation under the mesh tension. They are aligned in the chamber by pins in the corners fitting in corresponding holes in the readout- and the drift panel. The reduced frame thickness, compared to the total gas gap leads to a 1 mm clearance housing the brass springs. Being pressed onto the readout panel, the metallic mesh is brought in direct contact with a copper layer connected to the detector ground. Starting with a rather small selection of available meshes (50-30: 39.1 %, 70-30: 49.4 % and 45-18: 51.0 %) others have been successively added during the course of the study, trying to cover the 40 - 60 % range in open area for 18 µm and 30 µm-wire meshes.

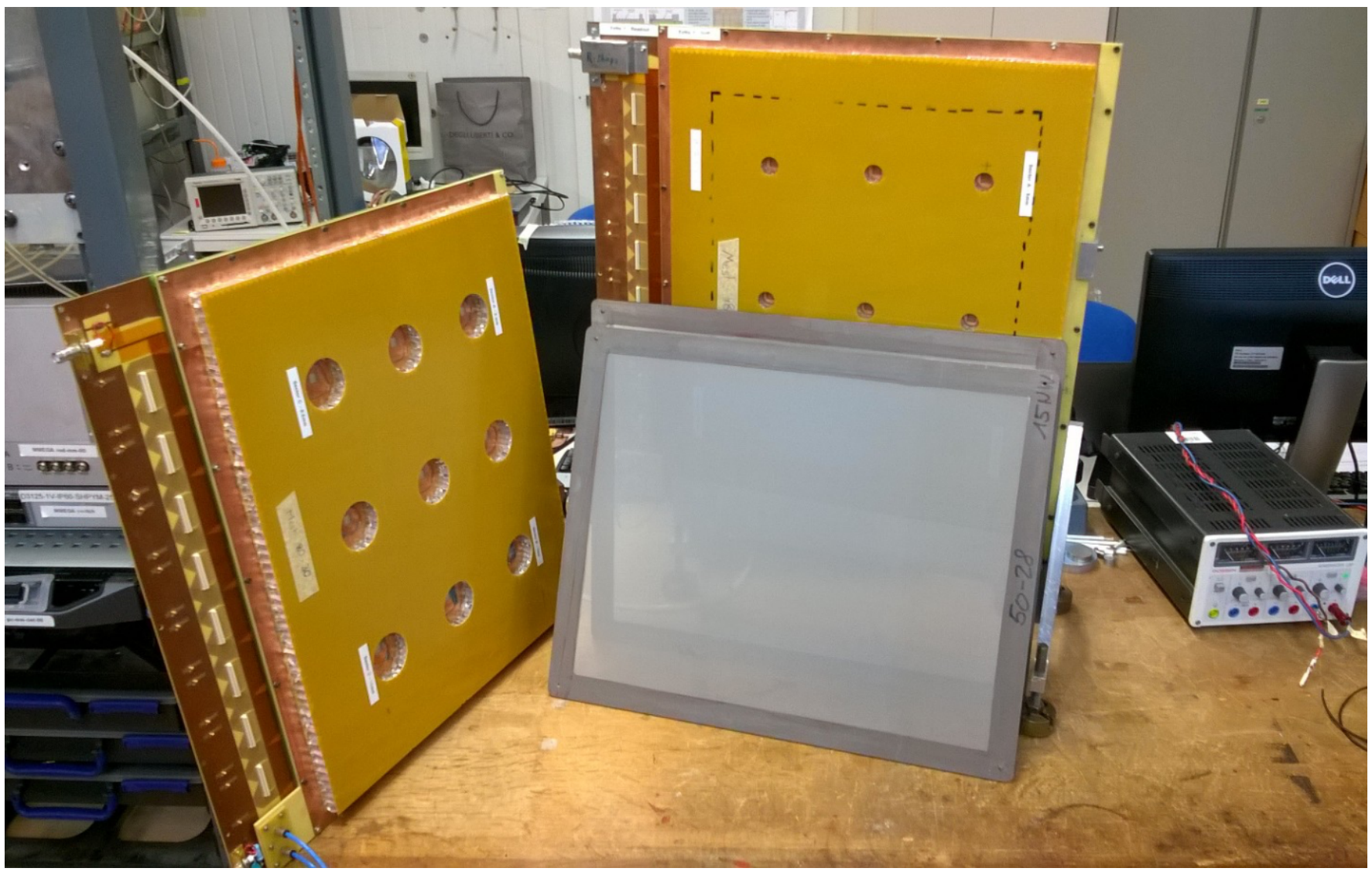

**Figure 3.21.:** Photograph of the two ExMe detectors and two additional mesh frames in the RD51 gaseous detector development laboratory at CERN.

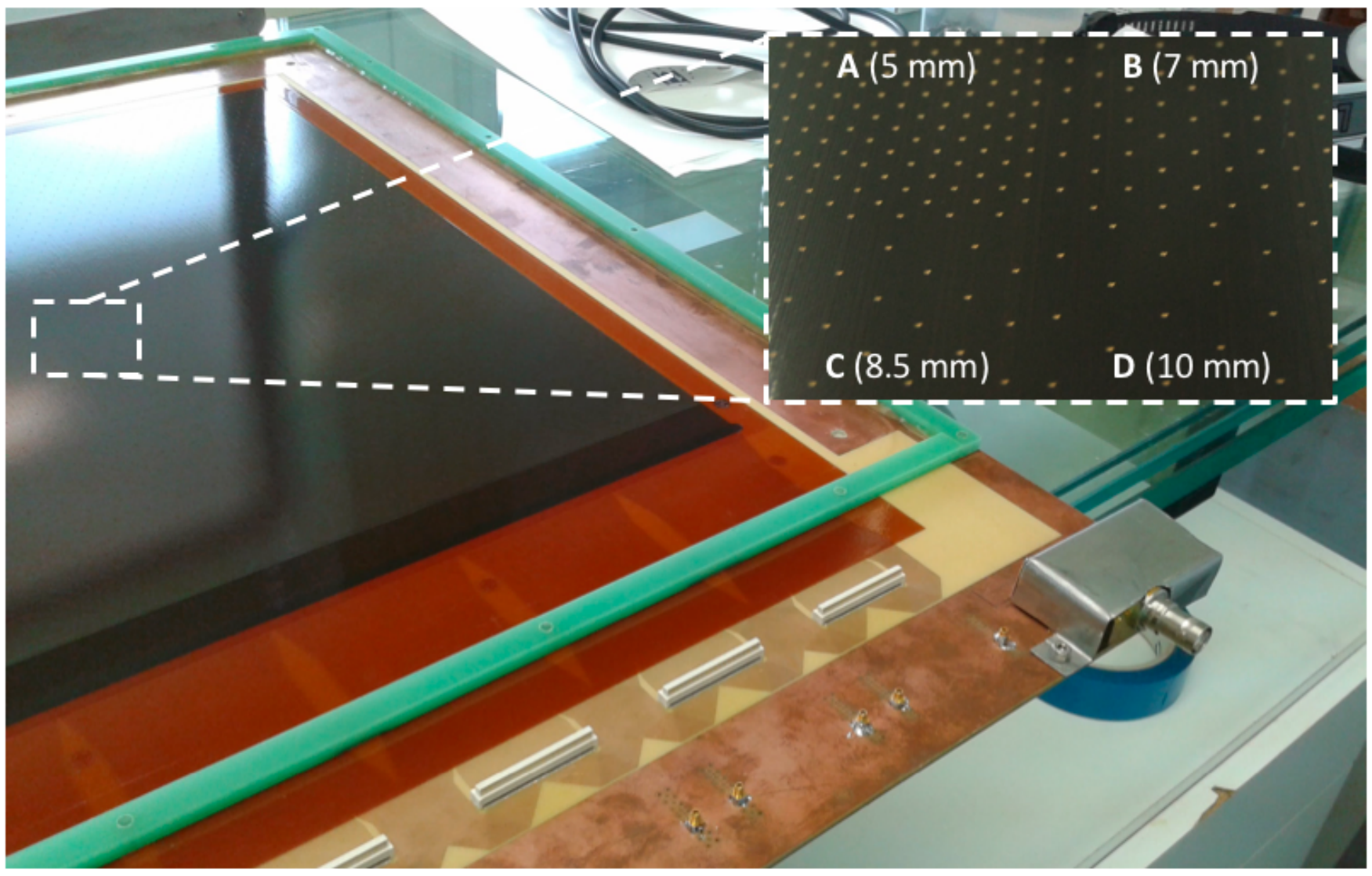

**Figure 3.22.:** Photograph of the ExMe1 readout panel. The magnification shows the central joining region of the four sectors A, B, C and D, corresponding to the four different inter-pillar distances.

### 3.3.2. Experimental Setup and Data Analysis

We studied the response of these two detectors to X-ray irradiation with the experimental setup, schematically shown in figure 3.23, located in the RD51 gaseous detector development laboratory at CERN.

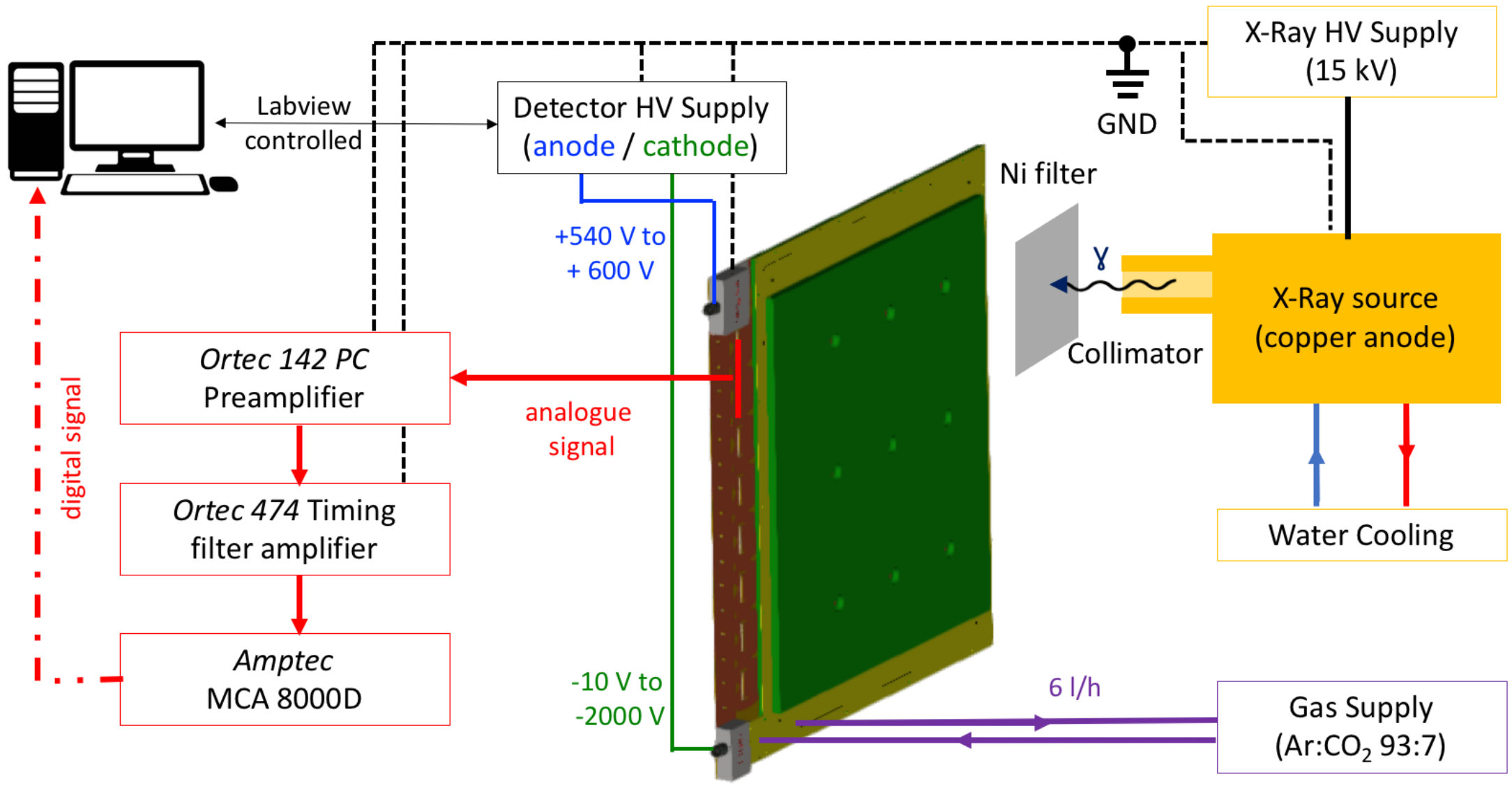

**Figure 3.23.:** Schematic of the experimental setup for X-ray source spectrum measurement with the ExMe detector.

A water cooled X-ray source with a copper anode at $U_{X-Ray} = 15\,\text{kV}$ provides a high intensity $\gamma$-beam, which is collimated and targeted onto the chamber irradiation spots through a nickel filter. The emitted spectrum (figure 3.24) is composed of the continuous Bremsstrahlung and the two distinct peaks corresponding to the $K_\alpha$ and $K_\beta$ transition in copper, as discussed in chapter 2.2.2. The nickel absorber features an absorption edge in between these two energy levels and is, therefore, ideal to suppress the $K_\beta$ peak and, to a certain extent, the Bremsstrahlung contribution.

The detector's gas gap is flushed with a premixed Ar:$CO_2$ (93:7) gas mixture with a flow of $\simeq 6\,\text{l/h}$ resulting in a slight 2-3 mbar over-pressure in the chamber with respect to the atmosphere. The gas flow is controlled at the inlet and cross checked at the gas outlet to detect larger gas leaks. During the course of this study, the originally used plastic pipes have been replaced by copper pipes to reduce water contamination due to the hydroscopic material.

Voltages are applied to the detector anode and cathode with a *CAEN* N1471A dual channel power supply [68], which is remotely controlled by a LabVIEW [69] based interface. Thanks to the adaptation of P. Thuiner's 'GEM monkey' software package, voltage variation according to the data acquisition status has been largely automatized. Great care has been taken to improve ground connection of the detector and all involved electrical components to a common ground, reducing the noise threshold below 3 % of the signal range.

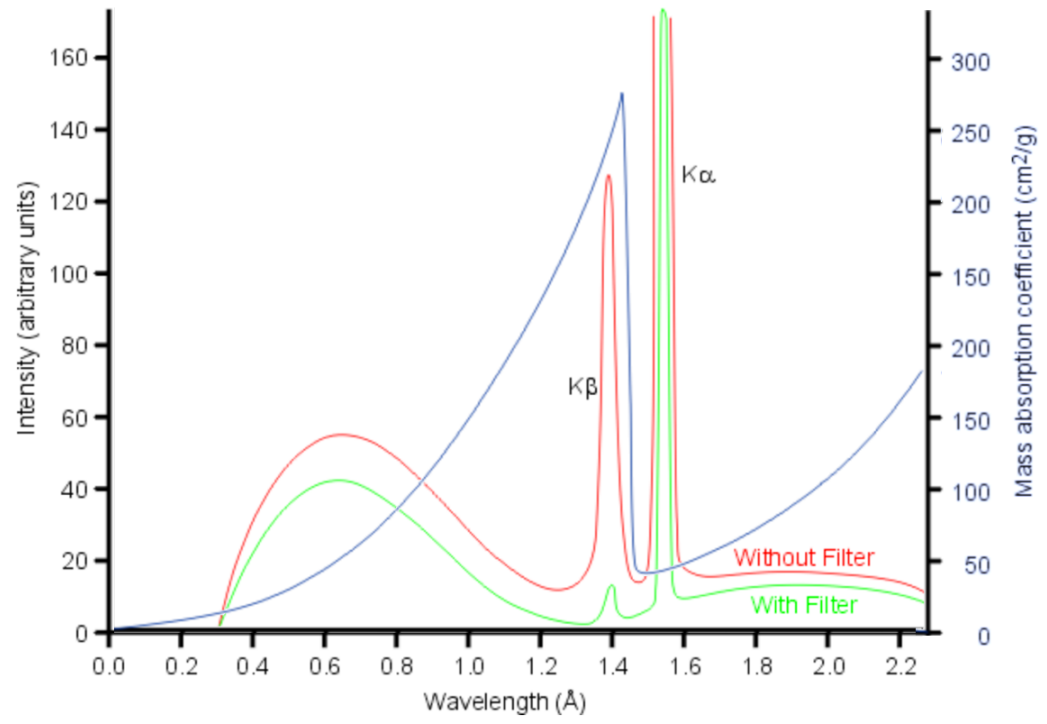

**Figure 3.24.:** Wavelength spectrum of a copper X-ray source before (red) and after (green) trespassing a nickel filter with the corresponding mass absorption coefficient (blue, right axis). [67]

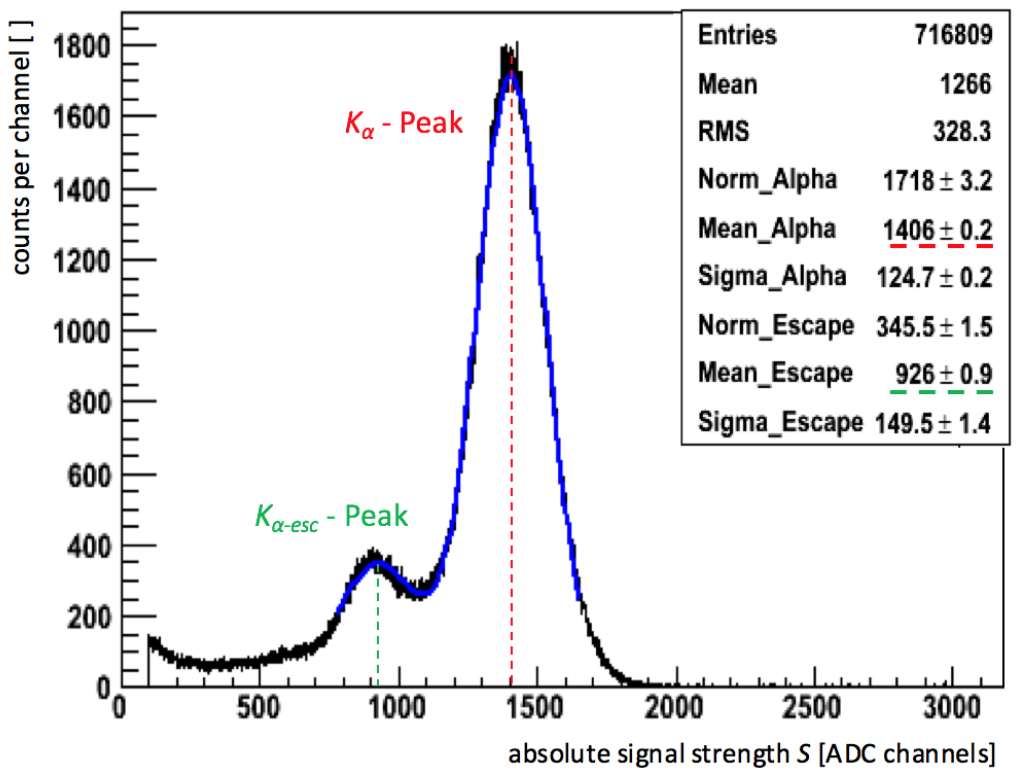

**Figure 3.25.:** Measured ADC spectrum for $7 \cdot 10^5$ X-ray events in an ExMe Micromegas detector, fitted with two Gaussian curves to determine the $K_\alpha$ (red) and $K_{\alpha-esc}$ (green) peak positions.

The charge signal induced on the readout strips is collected on two shortened Panasonic connectors and converted into a voltage signal in an *Ortec* 142 PC preamplifier [70] with 4 V/pC. The settings of the *Ortec* 474 timing filter [71] amplifier have been optimized to allow measurements within a wide range of amplification voltages and, therefore, gas gain and remained unchanged during all measurements to ensure comparability of the results. [2] An *Amptec* 8000D multi-channel-analyzer [72] is used to convert the 0-10 V signal into a digital signal and pass on the charge equivalent in channels (0-4096 ADC) in a self triggered mode. The resulting spectra were acquired by the *Amptec* DppMCA program and exported to Root for further analysis. Figure 3.25 shows a typical spectrum after 3 min measurement time, fitted with two Gaussian curves to model the $K_\alpha$ and the corresponding escape peak $K_{\alpha-esc}$ (see chapter 2.2.2). The $K_\beta$ and $K_{\beta-esc}$ contributions are suppressed by the nickel filter and hidden in the superimposed peaks.

Under the variation of the drift voltage $U_D$ a shift in the $K_\alpha$ peak position is observed as shown in figure 3.26. In terms of the signal strength $S$ in (1.7) the mean position of the $K_\alpha$ peak corresponds to a characteristic number of signal electrons $n_e$ freed by the 8.0 keV photon in the gas. All parameters effecting the gain $G$ and readout $c_{r/o}$ are kept constant in agreement with the VOTAT paradigm. Thus, the observed change in signal strength $S$ reflects a reduction of the number of electrons triggering an avalanche and, therefore, the combination of all electron losses before amplification. Operating above the recombination threshold where $R \sim 0$ the relative shift in the peak position shown in figure 3.27 can be associated with a change in the fraction of non-lost electrons $(1 - A) \cdot T$.

This drift voltage sweep in the range of 10 - 2000 V and the resulting relative electron loss curve corresponds to one run during data acquisition. They were repeated for different amplification voltages, different meshes and in the different sectors of both ExMe detectors. The full list of all measurements performed during the latest comprehensive data taking period is presented in [54].

[2] A comprehensive list of these parameters can be found in [54, Table A1]

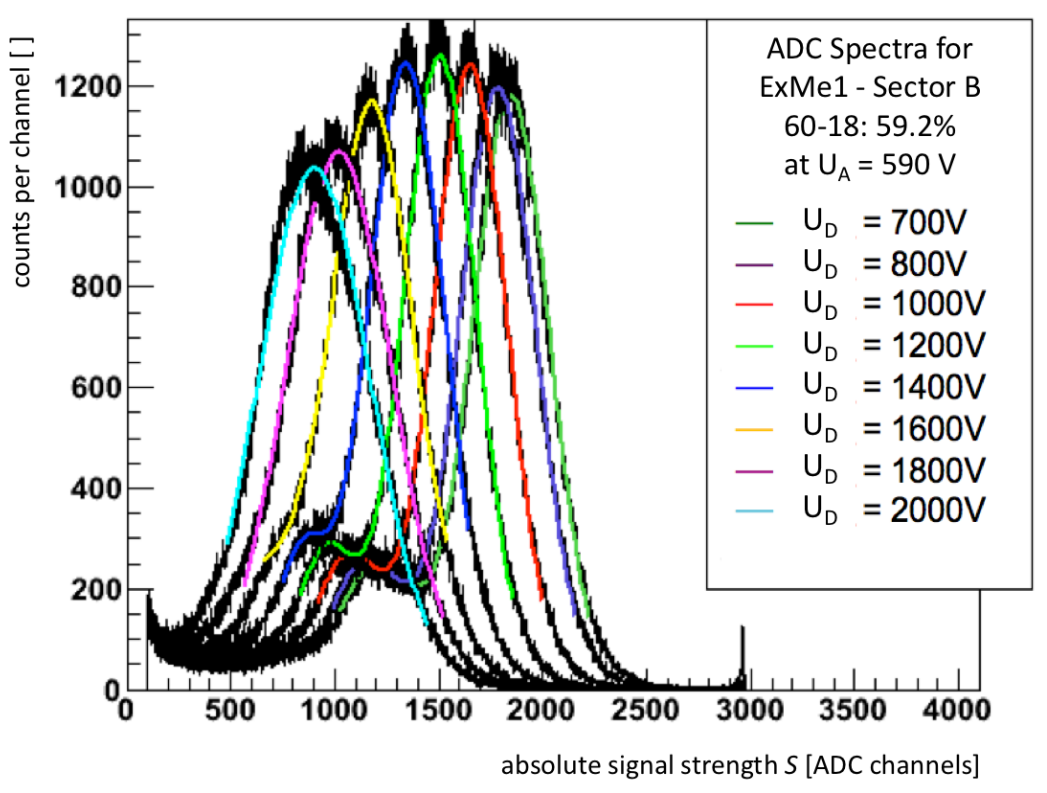

**Figure 3.26.:** Measured ADC spectra for different drift voltages in the ExMe1 detector equipped with a 60-18: 59.2 % mesh, operated at $U_A = 590\,\mathrm{V}$.

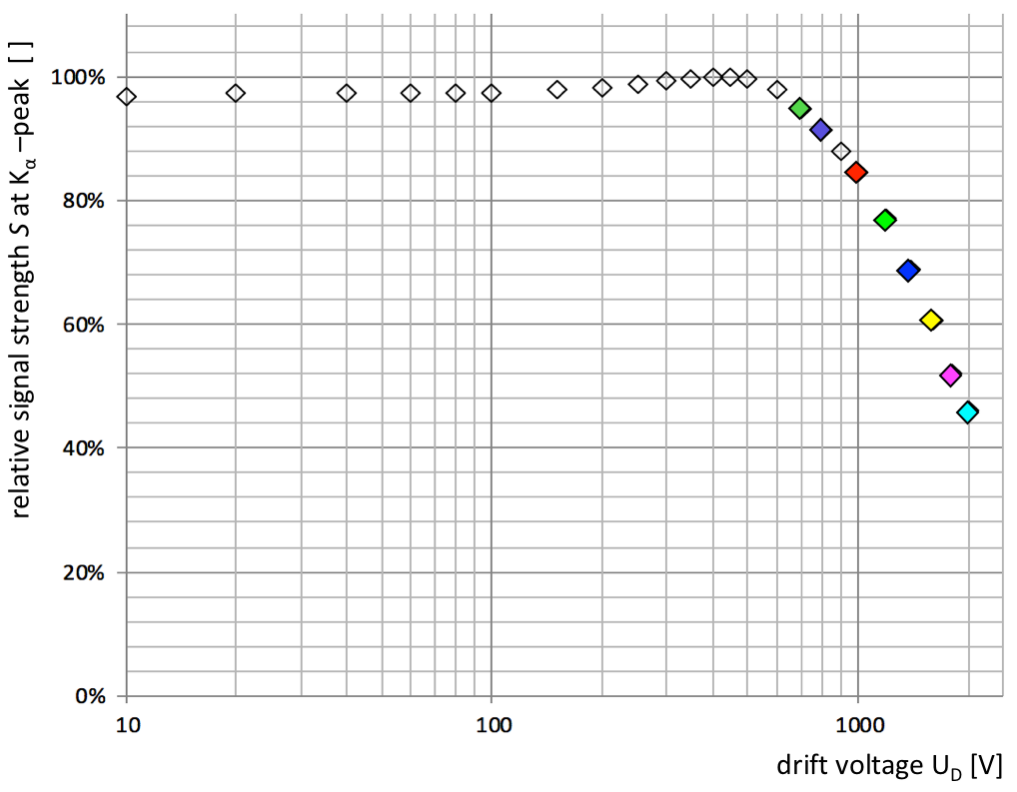

**Figure 3.27.:** Relative signal strength at the $K_\alpha$ peak as a function of the drift voltage $E_D$. Colors corresponding to figure 3.26.

### 3.3.3. Experimental Results and Comparison to Simulation

According to the dependency model introduced in figure 1.7 electron losses are assumed to be to in first order independent of the electric field in the amplification gap within the Micromegas working range. Figure 3.28 shows the experimental test of this assumption for a 45-18: 51.0 % mesh in the ExMe1 chamber. The absolute signal strength $S$ varies as expected for different amplification voltages $U_A$, which is mainly determining the gas gain $G$ (figure 3.28 - left). With the gain being kept constant within one run, normalization of the curves corrects for its impact. The normalized curves of relative signal strength as a function of the drift field show a convincing agreement with each other (figure 3.28 - right). A slight tendency towards a larger number of contributing signal electrons with increasing amplification voltage can be seen. The effect is limited to O($\pm 1.5\,\%$) and is most pronounced for stronger drift fields, where one approaches the limit of the $E_A \gg E_D$ assumption, and thus changes in $E_A$ can cause a more pronounced effect.

A comparison between the two detectors ExMe1 and ExMe2 (figure 3.29) with their different anode structures confirms the independence of the meshes' electron transparency $T$ from the fine geometry parameters of the anode. Similarly, no variation of electron losses between the four detector sectors, corresponding to a variation in the pillar distance, has been observed. All these three experimental observations support the assumption of an independence of the signal electron losses from the amplification field $E_A$ and justify the restriction of the simulation to $U_A = 580\,\mathrm{V}$ as discussed in section 3.2.2.

The experimentally measured relative signal strength, corresponding to the fraction of primary electrons contributing to the signal, is compared to the simulated predic-tion of electron losses for each mesh geometry. Therefore, the signal strength curve is normalized such that its maximum equals the maximum of the simulation, which is typically 100 % if only the mesh's transparency is considered. Already this simple comparison yields a quite satisfactory agreement for measurements in pure gas conditions, as shown in figure 3.30 - left. However, it becomes increasingly inaccurate for larger gas contamination levels and geometries with a lower transparency breakdown threshold $\max(E_D)|_{T\geq 95\,\%}$.

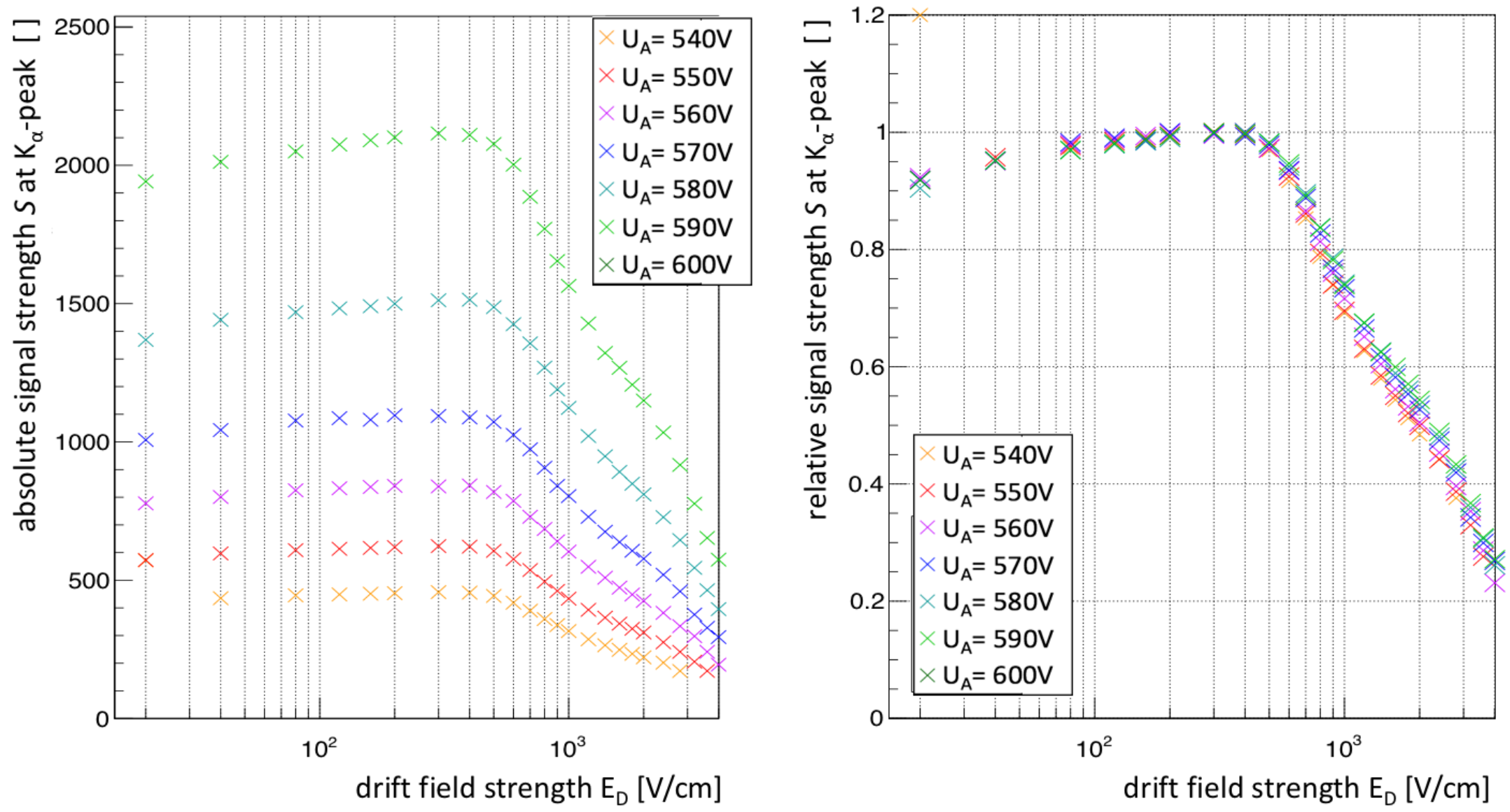

**Figure 3.28.:** Signal strength $S$ at the $K_\alpha$ peak as a function of the drift field strength $E_D$ for different amplification voltages, (left) the absolute values for S are plotted, (right) the normalized curves.

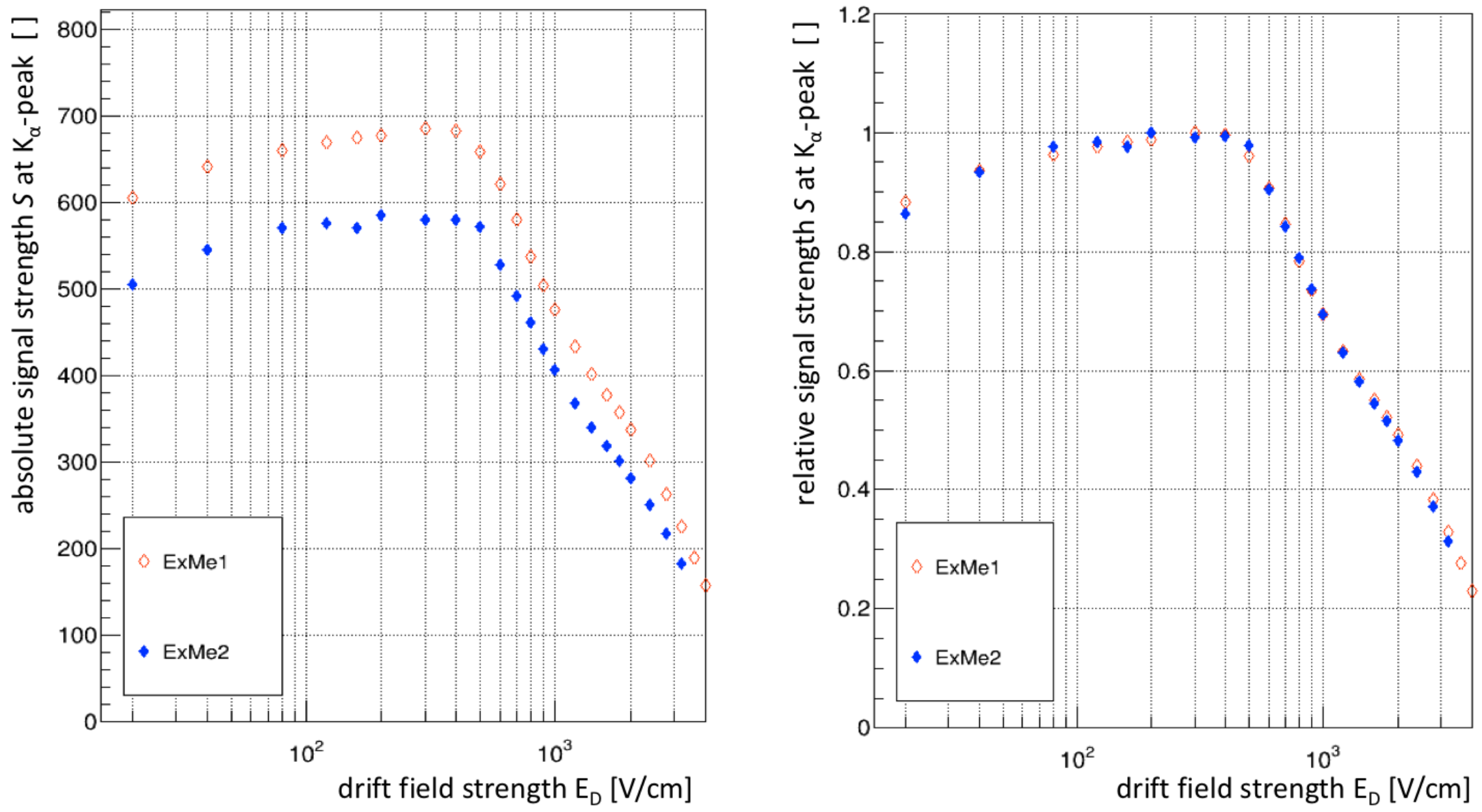

**Figure 3.29.:** Signal strength $S$ at the $K_\alpha$ peak as a function of the drift field strength $E_D$ in the two ExMe detectors for $U_A = 580\,\mathrm{V}$, (left) the absolute values for S are plotted, (right) the normalized curves.

Combining the electron loss simulation for the mesh absorption and gas attachment processes, the latter according to (3.14), yields an improved estimator for the fraction of non-lost electrons. The comparison of the $(1 - A) \cdot T$ curve with experimental data is shown for different meshes in the figures 3.30 to **??** - right. Therein the experimental data is again scaled to the maximum of the simulated curve, which can be well below 100 % due to the combination of attachment losses to Oxygen at low $E_D$ and the decreasing transparency for $E_D \geq \max(E_D)|_{T\geq 95\,\%}$. This overlap is crucial for the interpretation of the experimental data, especially when extracting transparency curves. A direct comparison of those can only be accurate if the gas mixture is sufficiently clean to justify the 100 % scaling. Previous comparison of simulation and experimental electron transparency measurement as presented in [66] lack this accounting for attachment losses, introducing a systematic disagreement.

In the computation of the attachment losses the level of Oxygen contamination $c_{O_2}$ is estimated by fitting the experimental data with the simulated $(1 - A) \cdot T$ curve in the range of low electric fields $E_D \leq 100\,\mathrm{V/cm}$. With even the lowest drift field value being above the recombination threshold and no other mechanism yielding a significant electron loss contribution for low scattering- or mesh approaching energies, $c_{O_2}$ can be extracted as the sole fit parameter. Without a similarly distinct region for the determination of $c_{H_2O}$, the known lower limit for water contamination resulting from humid atmospheric air, discussed in chapter 3.2.1, has been assumed. While treating $c_{H_2O}$ as an additional fit parameter might improve the description of electron losses on the high drift field end, it contributes little to the overall (dis-)agreement around the transparency breakdown threshold.

It is noticeable that even with $c_{O_2} = c_{H_2O} \simeq 0$ the attachment losses contribute significantly to the predicted fraction of non-lost electrons, due to the attachment to $CO_2$ in strong drift fields. Considering this contribution the kink visible in the comparison of the simulated transparency curve, which leads to a crossing with the experimental data, is straightened and the agreement between the shape of the curves improves significantly (figure 3.30 - left).

Still, a systematic discrepancy between experimental data and simulation results remains: the predicted transparency breakdown threshold $\max(E_D)|_{T\geq 95\,\%}$ extracted from simulation is systematically lower than in experimental data. With increasing field strength the simulation results remain below the measured values for a given field strength. A shift along the $E_D$ axis could easily compensate for the discrepancy and yield an excellent overlap. However, no justification for such a shift could be found. While in simulation $E_D$ can be well controlled and easily double-checked by extracting values from the field maps, the experimental drift field in the detector can not be measured directly. While the supplied voltages being precisely controlled, a systematic shift could be introduced due to a current induced voltage drop at the cathode. Since the majority of the ions being absorbed at the grounded mesh, the current through the cathode is limited to the pA range, therefore, even the $< \mathrm{M}\Omega$ resistor in the RC filter applied at the HV connector could not yield a significant voltage drop. The second determinant for the drift field is the coarse geometry of the detector, which could be altered by deformation caused by over-pressure in the gas gap. Such a blow up is, however, counteracted by the stiffening panels and no change in the detector's thickness larger than 50 µm at the panel center has been observed for an overpressure of up to 10 mbar, limiting a geometrical effect to < %-level.

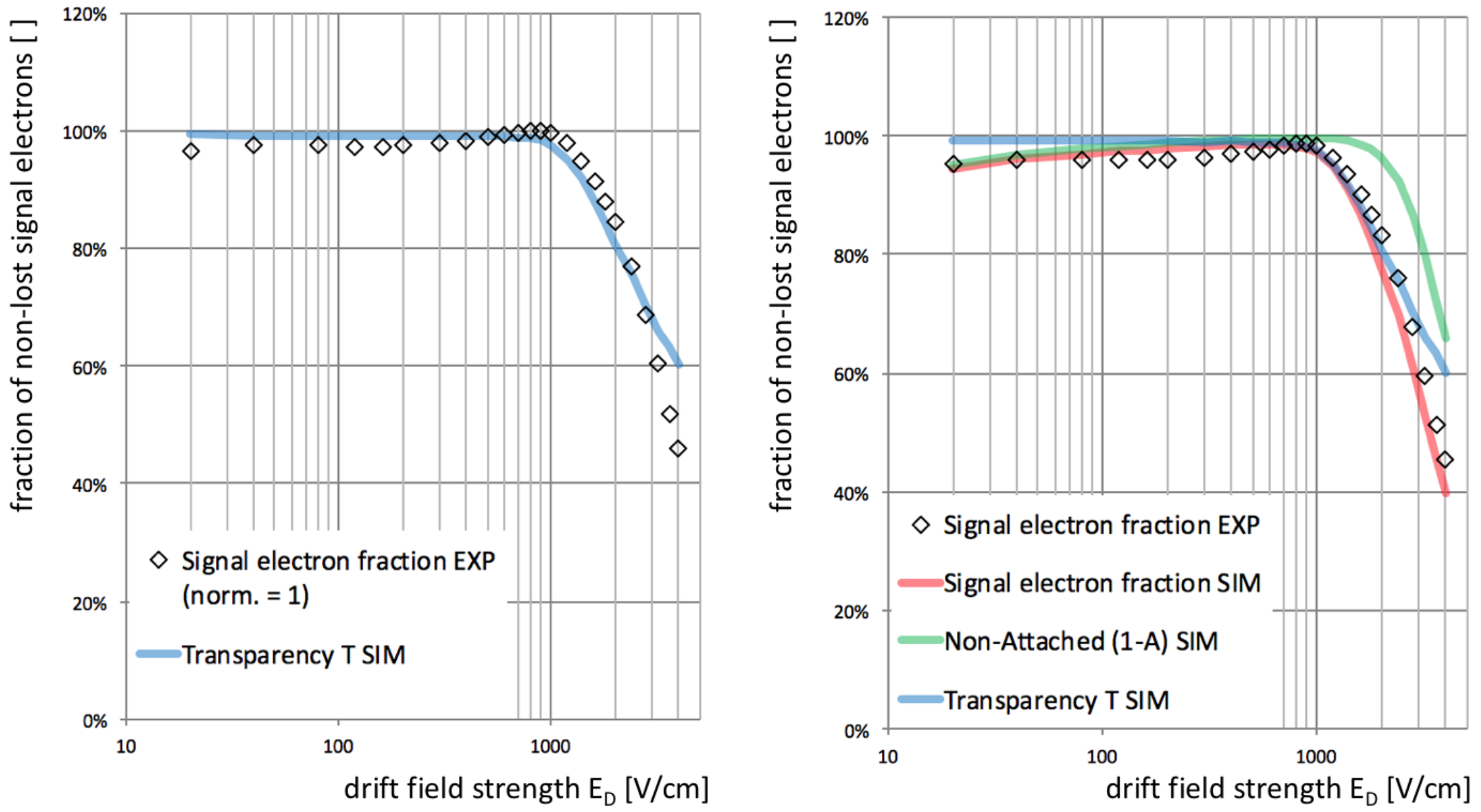

**Figure 3.30.:** Measured fraction of non-lost electrons as a function of the drift field strength $E_D$ for a 60-18: 59.2 % mesh in ExMe1. Values are scaled to equal the maximum of the simulation it is compared with: the electron transparency (blue) (left) and its combination (red) with non-attachment electrons (green) (right).

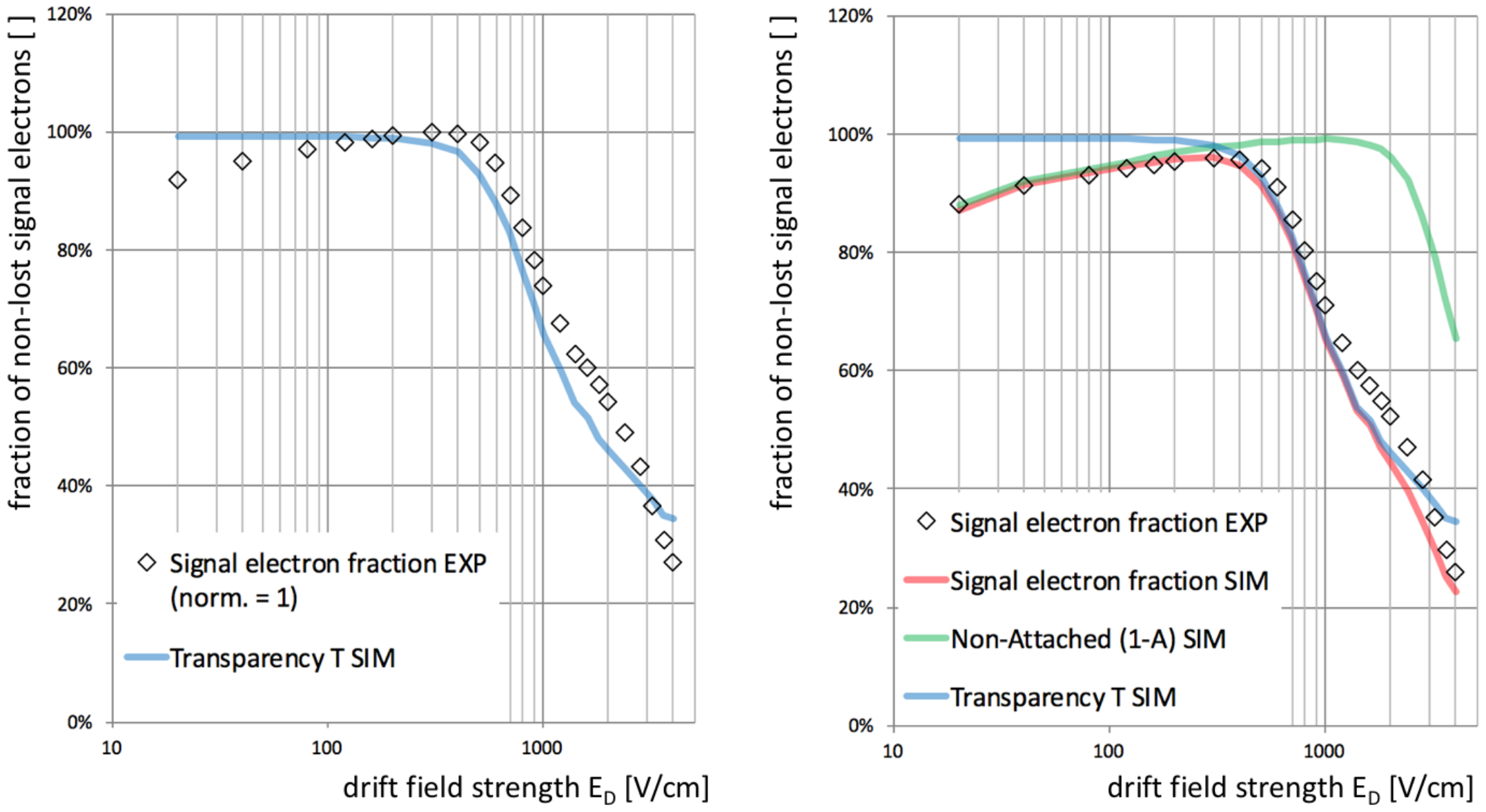

**Figure 3.31.:** Measured fraction of non-lost electrons as a function of the drift field strength $E_D$ for a 45-18: 51.0 % mesh in ExMe1. Values are scaled to equal the maximum of the simulation it is compared with: the electron transparency (blue) (left) and its combination (red) with non-attachment electrons (green) (right).

This discrepancy is noticeably more pronounced for meshes with lower optical transparency. Additionally, the effect is larger for meshes with 30 µm wires compared to the 18 µm wire meshes with similar open area. This dependency on the mesh geometry indicates towards the simulation as source of a systematic effect: an inherent difference between real electron behavior and the simulation model is the finite step size for electron drift in the Garfield algorithm. While real electrons gain energy continuously along their path, the path of their simulated counterparts is composed of small but finite steps. Accordingly, the change in electron energy is only evaluated after each of those steps. This leads to a non-physical crossing of field lines where they are strongly bent, which is the case above the mesh wires. Similar to the physical electron inertia effect, explained in section 3.2.2 figure 3.18 - (B), this simulation owed displacement has a preferred direction towards the wires and, hence, introduces a systematic decrease in electron transparency. For low drift velocities, the physical inertia effect is small and, thus, the simulation step introduced electron displacement across field lines is dominant. With increasing electron momentum the real displacement from the starting field line due to electron inertia is increasing and becomes dominant over the, in first order velocity independent, simulation bias. Therefore, the non-physical bias in the simulation is reduced with larger $E_D$ and a better agreement between simulation results and experimental data is reached for meshes with a high $\max(E_D)|_{T\geq 95\,\%}$, as consistently observed in all measurement to simulation comparisons. This bias effect could be reduced by increasing the fraction of null-collisions in the Magboltz source code and hence decrease the electron drift step size, requiring significantly increased computing resources. In the given configuration the simulation yields a definite lower limit for the real electron transparency of a mesh. With increasing open area and thus large $\max(E_D)|_{T\geq 95\,\%}$, the quality of the prediction increases significantly.

Besides the remaining bias, this study yielded the up-to-date most precise comparison of experimental electron loss data with simulation prediction for an extended set of Micromeshes. The results are graphically summarized in the figures 3.34 and 3.35. The dependency of electron loss mechanisms to the drift field, the geometry parameters and gas contamination of the $ArCO_2$ mixture have been systematically assessed, increasing our understanding of the involved processes. The simulation tools and detectors for experimental measurement of electron losses, developed in the scope of this study, can easily be adapted for an extension of the parameter space, to other gas mixtures, temperatures or pressures.

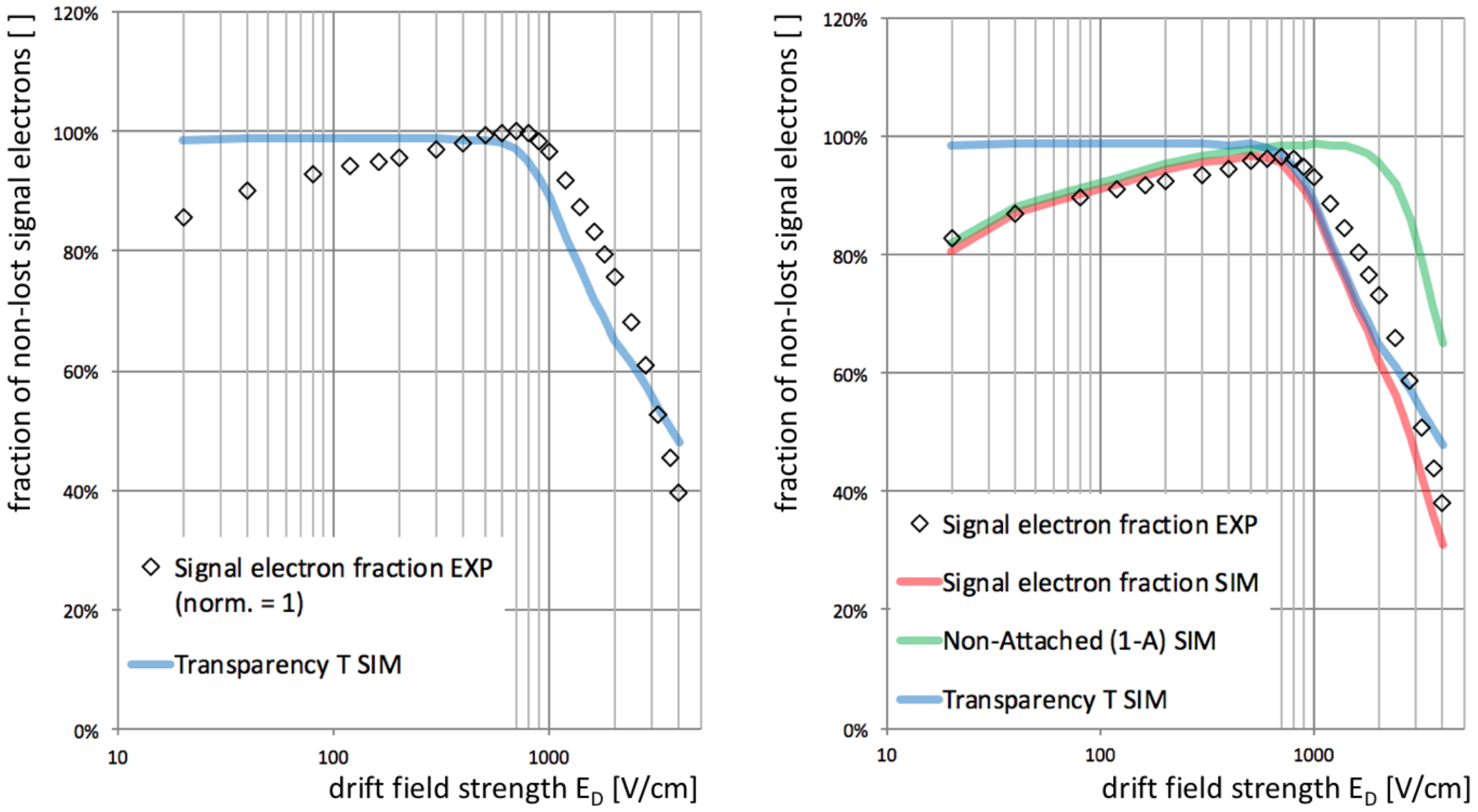

**Figure 3.32.:** Measured fraction of non-lost electrons as a function of the drift field strength $E_D$ for a 80-30: 52.9 % mesh in ExMe1. Values are scaled to equal the maximum of the simulation it is compared with: the electron transparency (blue) (left) and its combination (red) with non-attachment electrons (green) (right).

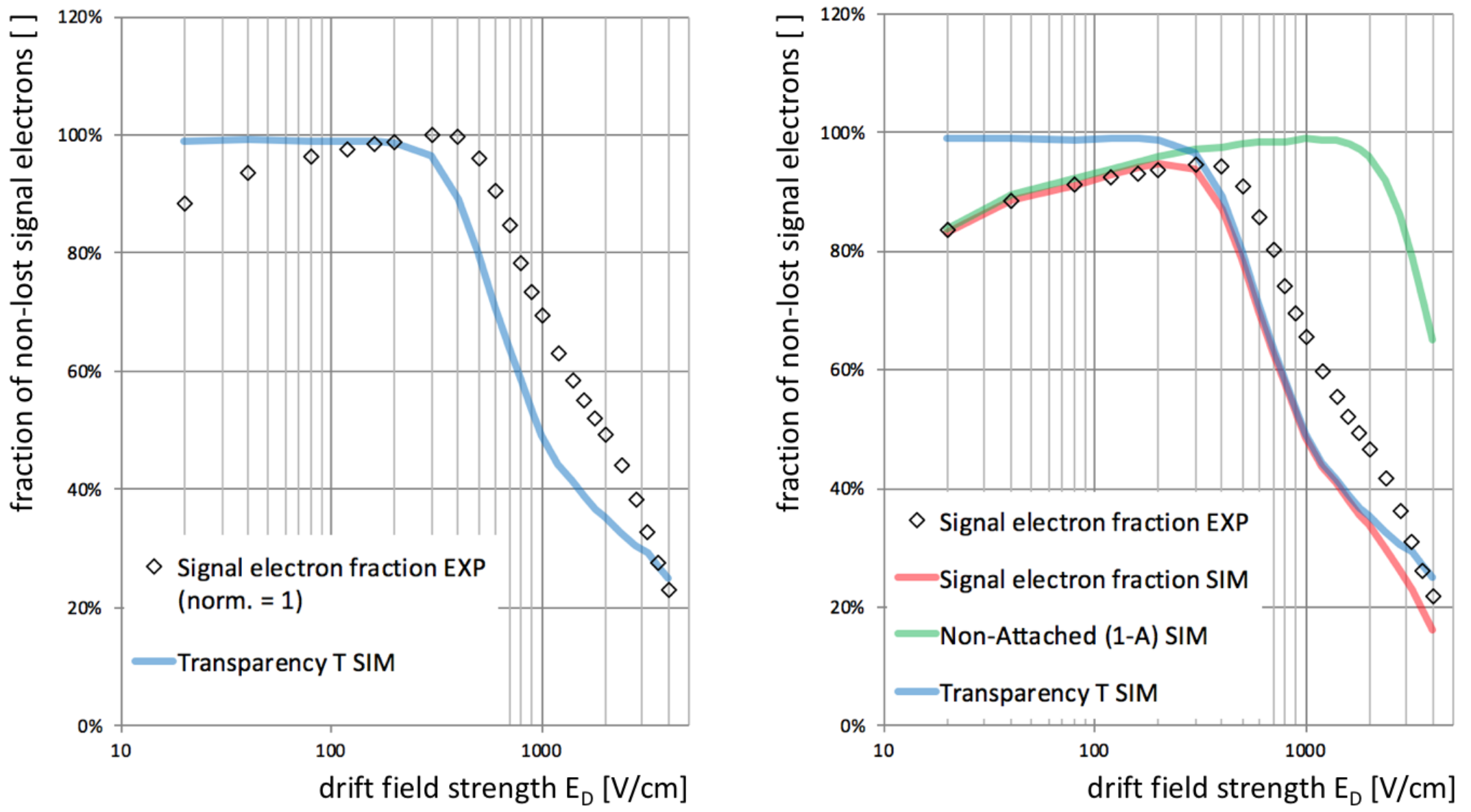

**Figure 3.33.:** Measured fraction of non-lost electrons as a function of the drift field strength $E_D$ for a 60-30: 44.4 % mesh in ExMe1. Values are scaled to equal the maximum of the simulation it is compared with: the electron transparency (blue) (left) and its combination (red) with non-attachment electrons (green) (right).

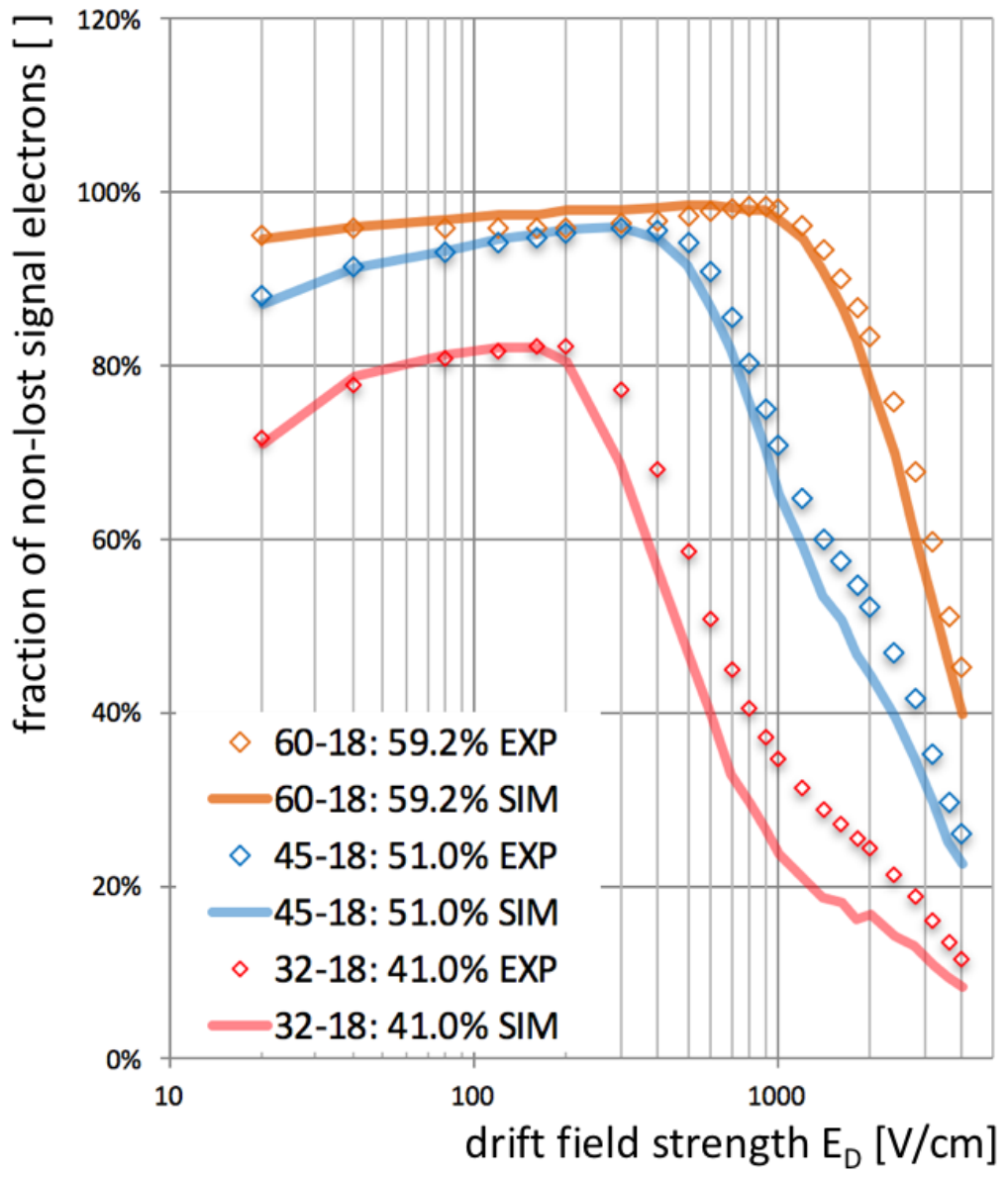

**Figure 3.34.:** Comparison of experimental data and simulation prediction of non-lost signal electrons as a function of the drift field strength $E_D$ for different 18 µm meshes.

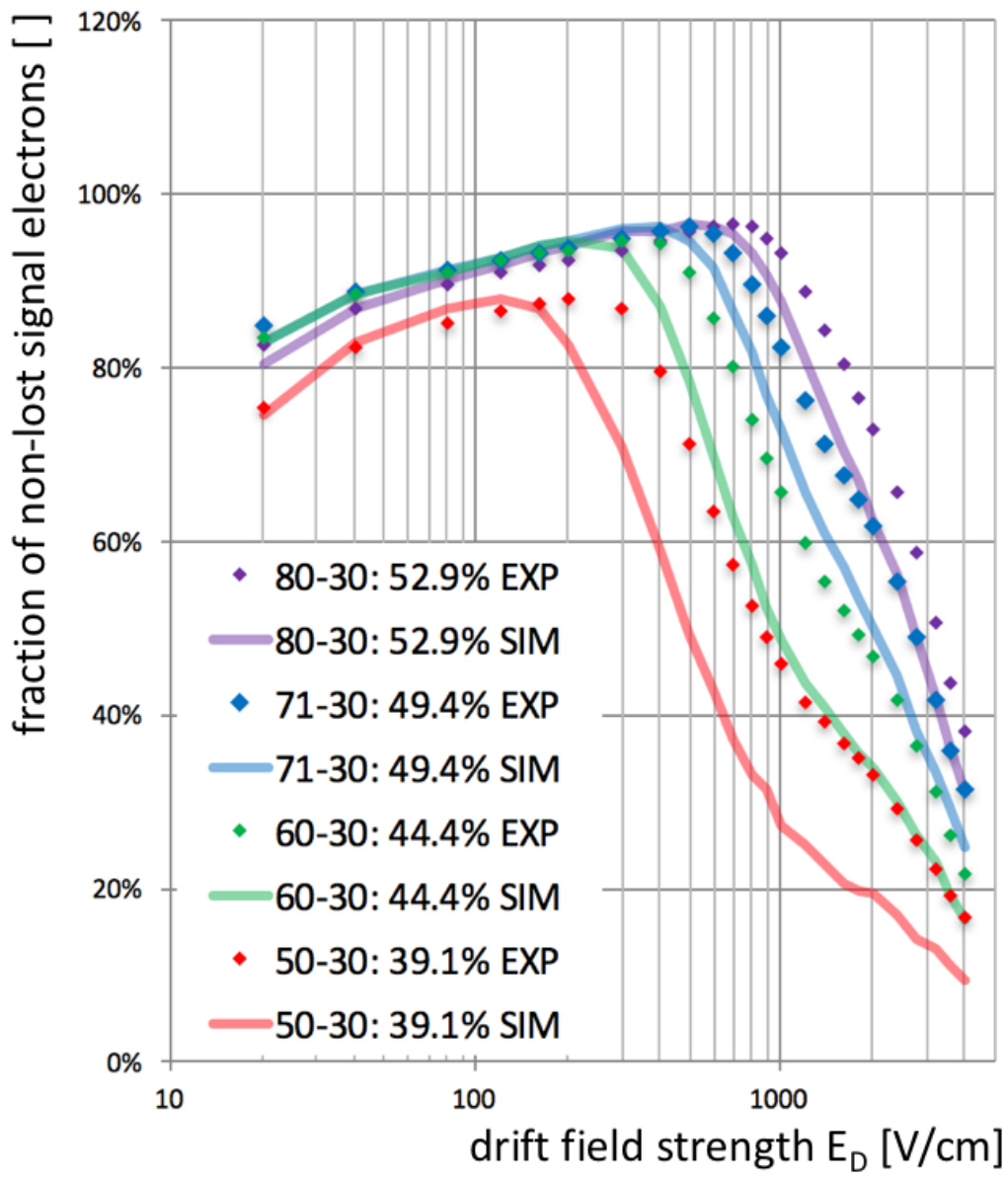

**Figure 3.35.:** Comparison of experimental data and simulation prediction of non-lost signal electrons as a function of the drift field strength $E_D$ for different 30 µm meshes.

# 4. Electron Amplification and Avalanche Formation

Drifting into a gas region with a strong electric field electrons can accumulate sufficient energy in-between collisions to cause ionization, thus yielding an additional free electron. Repetition of this charge amplification process by the initial as well as the newly freed electrons, causes a cascade of electron multiplication, typically referred to as an *electron avalanche*. In a Micromegas the avalanche formation process is constraint to the thin gap between the micromesh and the anode structure, yielding a well controlled multiplication process, low gain variance and fast ion evacuation.

To understand the effects of the different dependency parameters, depicted in figure 1.7 the avalanche formation process is first reviewed in section 4.1 on an analytic level. Thereafter, the algorithms and limits of the avalanche growth simulation is discussed alongside with preparatory simulation studies in section 4.2. The results from experimental studies of the gain dependence on the fine geometry parameters defining the ExMe Micromegas amplification gap are discusses in section 4.3. To reduce computing time in the simulation of amplification processes, we introduce a method of electron avalanche extrapolation and present validation studies to prove the method in section 4.4. Finally, we report on the very first direct measurement of the relative gain variance with a single electron response (SER) experiment, conducted in a collaboration with the University Paris-Sud (Orsay) in section 4.5 and compare these results for different gas mixtures with the prediction obtained by simulation.

## 4.1. Analytic Description of Electron Avalanches

While the formalism for describing microscopic scattering processes between electrons and the gas as described in chapter 3.1 remains, the electrons' energy gain in-between collisions is significantly increased. This allows for collisions yielding the excitation of an atom's electron shell ($A + e^- \rightarrow A^* + e^-$) or its ionization ($A + e^- \rightarrow A^+ + e^- + e^-$), and opens up the corresponding cross sections shown in figures 3.1 to 3.5.

### 4.1.1. Direct and Indirect Ionization Processes - Penning Transfer

In a pure gas the only electron amplification process is direct ionization of the atom or molecule. The mean free ionization path $\lambda_{ion}$ defines the average distance an electron travels before having an ionizing collision. Accordingly, its inverse $\alpha$ represents the number of ionizing collisions per unit length and is called the first Townsend coefficient. Both parameters can be expressed in terms of the ionization cross section $\sigma_{ion}(\epsilon)$ by

$$\alpha = \frac{1}{\lambda_{ion}} = N\sigma_{ion}(\epsilon). \tag{4.1}$$

Herein $N$ is the number of molecules per unit volume, which scales linear with the gas density. Furthermore, the electron scattering energy $\epsilon$ and thus the ionization cross section depends on the mean energy uptake between collisions and, therefore, the total mean free path $\lambda_{tot} \leq \lambda_{ion}$. Depending on the electrical field strength either the increase in electron gas interaction $(\lambda_{tot})^{-1}$, or the decrease in ionization probability per collision are dominant. Thus, an increased gas density causes a higher (stronger field) or lower (weaker field) ionization yield per unit length.

The successive ionization of the gas by the initial electron and any electron freed in subsequent ionization leads to an exponential growth of the electron avalanche along its spatial extent $x$. The total number of electrons in an avalanche triggered by a single initial electron corresponds to the gas gain $G$ introduced in chapter 1.3.1 and utilized in (1.7) With the cross sections strongly depending on the collision energy $\epsilon$, the Townsend coefficient becomes a function of the potentially position dependent field strength $E(x)$. In this purely ionization based description, the gas gain can be estimated as [13]

$$G = \exp\left(\int_{x_1}^{x_2} \alpha(E(x))dx\right) \equiv |_{E(x)=E} \exp\left(\alpha(E)(x_2 - x_1)\right) \tag{4.2}$$

where $x_1$ and $x_2$ are the initial and final coordinates of the multiplication path and the simplification requires a uniform electric field with constant field strength $E$.

In gas mixtures the total cross section for direct ionization can be composed according to (3.7). Besides the direct ionization of each gas species additional energy exchange mechanisms may occur, dependent on the energy levels for excitation and ionization in the involved gas species $A$ and $B$. These processes include the energy transfer between excited stated ($A^* + B \to A + B*$, collisional de-excitation) or ionization ($A^+ + B \to A + B^+$, charge exchange) and the ionization of the second species by an excited atom of the first species ($A^* + B \to A + B^+ + e^-$, Penning transfer). This Penning transfer [73] can occur if the excited state $A^*$ has an energy level above the ionization threshold of $B$ and a lifetime sufficiently long to cause the energy transfer before de-exciting via other channels, like for instance photo emission ($A^* \to A + \gamma$). Yielding an additional free electron the penning transfer can contribute significantly to the total gain if the above conditions are met, as studied in depth by Ö. Sahin [74]. In $ArCO_2$ mixtures for example Argon features four excitation level above the $CO_2$ ionization threshold. During avalanche formation each of them occurs with a frequency $\nu_i^{exc}$ and once excited the Argon has a probability $r_i$ to cause a Penning transfer. With the first Townsend coefficient being defined only by the probability for direct ionization, (4.2) can be altered to account for penning transfer occurrence relative to the ionization rate $\nu_j^{ion}$.

$$G = \exp\left(\int_{x_1}^{x_2} \alpha(E(x)) \frac{\sum\limits_j \nu_j^{ion}(E(x)) + \sum\limits_i r_i \nu_i^{exc}(E(x))}{\sum\limits_j \nu_j^{ion}(E(x))} dx\right) \tag{4.3}$$

While all the frequencies can be determined by Magboltz based on the cross sections of the involved processes, the transfer rates $r_i$ remain to be determined. Depending on the excited state's lifetime and possibly on the energy difference to the ionization level of the second gas species the transfer rates are assumed to be different for each state. They further depend on the gas density and the concentration $c_B$ of the second gas constituent. Lacking a method to experimentally measure the transfer rates independently, typically a total transfer rate $r(c_B)|_{NTP}$ is determined, as shown for the example of $ArCO_2$ mixtures in figure 4.1.

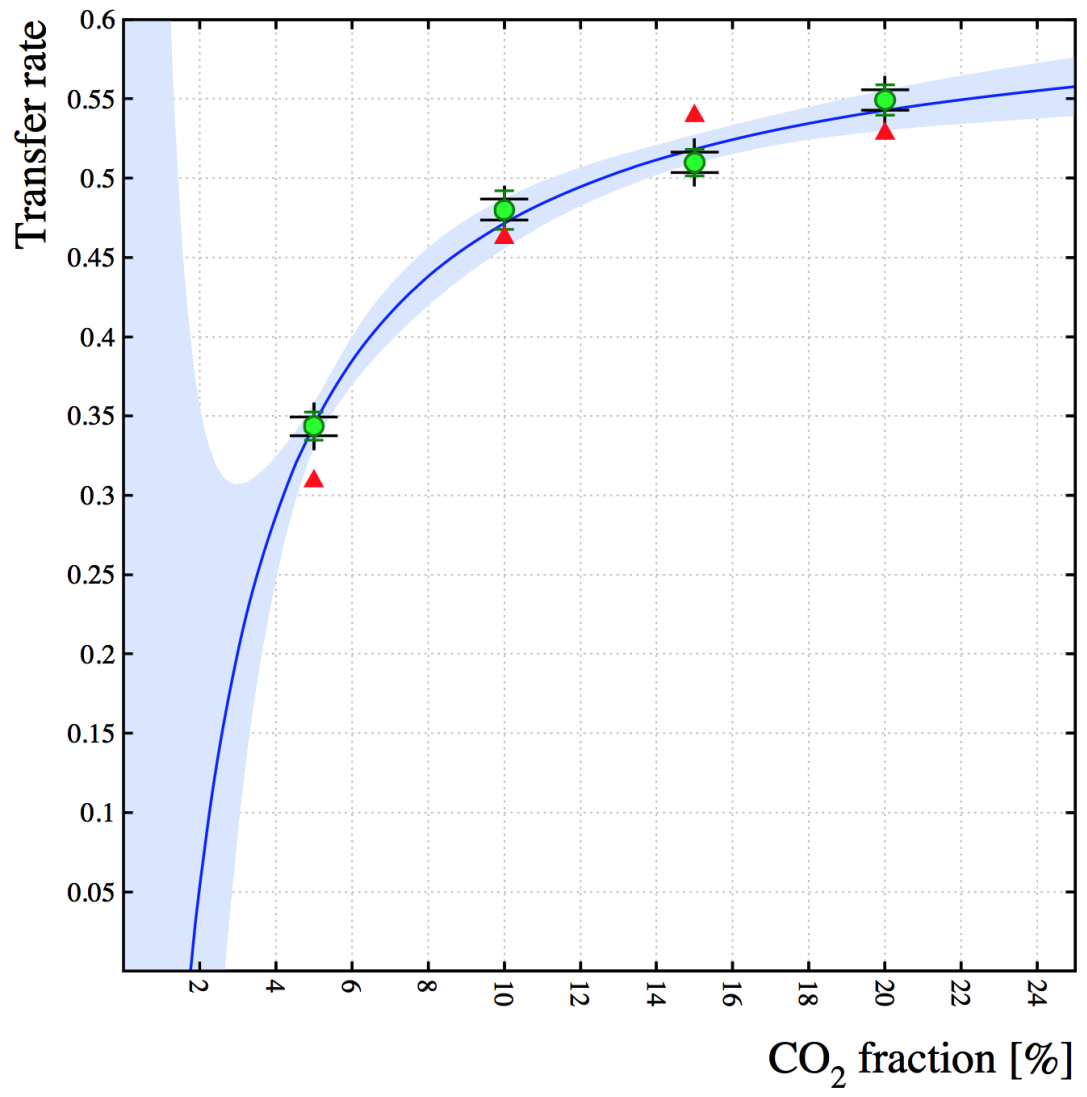

**Figure 4.1.:** Penning transfer rates in NTP $ArCO_2$ as a function of the $CO_2$ concentration $c$, fitted with $(a_1c + a_3)/(c + a_2)$ model (blue curve) and the uncertainty on this parametrisation (blue error band). The four measurement points are obtained with different constraints on the gain curve parameters. Taken from [74], where a detailed discussion of the model can be found.

### 4.1.2. Photon Induced Secondary Avalanches and Photon Quenching

With the increased occurrence rate of excited states at higher scattering energies, the number of photons emitted in radiant de-excitation ($A^* \to A + \gamma$) increases as well. Those photons travel through the gas unhindered by the electric field and can, therefore, interact in a position with a higher electrical potential compared to their origin. In a pure gas the energy of these photons is not sufficient to cause ionization of the gas, but it is well above the ionization threshold of metallic solids utilized as the detector's electrode. A photo effect at the cathode of a parallel plate detector or the mesh of a Micromegas detector frees an electron which in turn causes a secondary avalanche. The latter again causing a photon yield probably triggering another avalanche and so forth.

While these secondary avalanches occur potentially delayed and displaced with respect to the initial avalanche, they are often experimentally not distinguishable. Therefore, a measured gas gain $G'$ is biased by the statistical occurrence of secondary avalanches:

$$G' = \frac{G}{1 - \beta G} \tag{4.4}$$

where $\beta$ is the second Townsend coefficient. It represents the mean probability to trigger a secondary avalanche per electron in the initial avalanche.

While the rate of photon emission from an excited noble gas atom is difficult to reduce, the probability of this photon to reach the electrode and free a photo electron can be significantly suppressed by adding a poly-atomic quenching gas to the mixture. Provided the second species features suitable excitation levels energy can be transferred in gas-gas-scattering ($A^* + B \to A + B^*$, collisional de-excitation) or via a photon mitigated process ($A^* \to A+\gamma$ followed by $B+\gamma \to B^*$, photo-emission and -excitation). While de-excitation is dominated by the radiative channel in mono-atomic noble gases, molecular gases de-excite dominantly via non-radiative, rotational or vibrational transfers [13]. Adding a several %-fraction of a suitable quenching gas to the noble gas, therefore, allows for the dissipation of a sizable energy fraction, significantly decreasing the photon range, and therefore $\beta$, cooling the gas and increasing the detector's energy resolution.

Similar to the photon induced secondary avalanches, electrons can be freed by ions scattering into the electrode. In an $ArCO_2$ mixture at the typical ion drift velocities in a Micromegas this process is rare and its contribution to $\beta$ is negligible.

Although setting a limit to the detector's breakdown threshold at $G \simeq \beta^{-1}$, secondary avalanches are not the primary mechanism for the photon-triggered streamer formation discussed in chapter 1.1.2. The dominant contribution is due to photons from radiative recombination ($A^+ + e^- \to A+\gamma$). Their energy matches the ionization threshold of the gas and, therefore, resonant ionization inside the gas is strongly favored, not requiring any other solid or gaseous constituent to be involved. Requiring a sufficiently high ion density, this process is suppressed in detectors operated in the proportional mode.

### 4.1.3. Fluctuation in Avalanche Formation and Polya Statistic

Together with the mean gain $G$ the relative gain variance $f$, quantifying the fluctuation in the avalanche formation process, are the two central macroscopic observables of the electron amplification. While tuning of the gain is a prerequisite for successful detector operation, the gain variance is the determining factor for the detector's energy resolution and, therefore, of great importance for the design and optimization of a detector. As discussed, the gas gain can be accurately described by the enhanced analytic models by taking into account microscopic processes like penning transfer (section 4.1.1) and photon feedback (section 4.1.2). This analytic description is, however, focused on the average electron yield $\overline{N_e} = G$ per avalanche and does not provide any insight to its statistical distribution $P(N_e)$. Therefore, the relative gain variance

$$f = \left( \frac{\sigma_{N_e}}{\overline{N_e}} \right)^2 \tag{4.5}$$

remains undetermined unless the underlying statistic processes can be correctly described.

Originally derived to describe the growth of epidemic processes, like the death caused by a wave of influenza, the Polya distribution [75] proved to be an accurate model for avalanche growth processes. It is determined by the mean number of electrons $\overline{N_e}$ and a shape parameter $\Theta$, called Polya parameter:

$$P(N_e) = \frac{1}{\overline{N_e}} \frac{(1+\Theta)^{1+\Theta}}{\Gamma(1+\Theta)} \left(\frac{N_e}{\overline{N_e}}\right)^{\Theta} \exp\left[-(1+\Theta)\frac{N_e}{\overline{N_e}}\right] \tag{4.6}$$

where $\Gamma(x)$ is the gamma function. Determining the width of the distribution the Polya parameter $\Theta$ can be used to extract the relative gain variance:

$$f = \frac{1}{1+\Theta} \tag{4.7}$$

The Polya statistic has been successfully applied to describe avalanche fluctuations in the measurements by Cookson and Lewis in the 1960s [76, 77] and the historically important measurement by Schlumbohm [78]. The latter represented an experimental benchmark for the statistical models by Byrne [79, 80], Legler [81] and Alkhazov [82].

Byrne derived a model for avalanche formation, which yielded a distribution in good agreement with the Polya statistic. Meanwhile Legler, following a different school of thought, developed an independent statistical model and argued that *'The Polya distributions [...] cannot claim any physical meaning [and] Their agreement with experimental distributions over a certain range of electron numbers has to be taken as accidental'* [81]. The comparative works by Alkhazov in 1970 showed that *'avalanche size distributions in uniform fields are close to those computed by Legler, but [...] the distributions in the non-uniform fields of cylindrical counters are close to those of Polya-type.'* [82].

With the improvements in computation methods and the development of Monte Carlo methods, the subject of avalanche statistics has been revived after almost four decades. Allowing for a precise modeling of the processes on microscopic level, the requirement of analytically describable distributions has been overcome. The in-depth studies by Schindler [83, 84] verified the Polya distribution as suitable tool to describe avalanche statistics in uniform fields and to extract the relative gain variance from simulated as well as experimental data.

## 4.2. Simulation of Electron Amplification

The overall method of simulating electron gas interaction with the Garfield++ program has been discussed in section 3.2 of the previous chapter. Therefore, we focus only on the differences in simulation of the amplification process compared to the drift and thereafter present two simulation studies on the determination of penning transfer values via gain comparison and the gain dependency on the variation of the amplification gap size.

### 4.2.1. Microscopic Avalanche Simulation in Garfield++

While during the drift simulation in chapter 3.2 only a single electron is iterated, meaning it is displaced by one step, its properties are reevaluated before it is possibly scattered, this treatment is extended to every additional electron freed in electron gas interaction. The order of electron propagation is thereby dictated by the electron's continuous identifier: the initial electron is treated and each electron freed along its path receives a consecutive identifier. Once the electron path terminates due to being absorbed at the mesh, the anode or attaching to the gas, the next electron is selected and evaluated and so forth. While being a computing resources efficient method, this yields a temporally non-consistent way of simulating the amplification process and forbids for interaction of, for instance, ions formed during the same avalanche with the electrons. The program is, therefore, intrinsically unsuited to simulate the formation of a streamer, relying on the local field distortion due to high space charge density. Furthermore, the electron numbering follows a systematic and it is required to use an electron ID independent method whenever a 'randomized' fraction of those electrons is evaluated. The set of information yielded for each electron is the same as for single electron drift simulation and comprises endpoint, -time, -energy and termination status. Avoiding huge data sets typically only derived values like position and spread of the endpoint distribution, statistical parameters for the anode arrival energy, or the fraction of electrons attached during avalanche formation are stored per avalanche.

With the increased scattering energy the dominant processes in electron gas scattering become ionization and excitation. As discussed in section 4.1.1, this opens up energy transfer processes yielding additional free electrons. This penning transfer is implemented in Garfield++ in a simple model, represented in (4.3): whenever a state in the gas species $A$, which is capable to cause a penning transfer to $B$, is excited, it causes a penning effect with a probability $r_P$, the penning rate. This penning rate summarizes over all transfer states by taking into account their relative occurrence rate. It furthermore already reflects the concentration of both species and, therefore, the probability of having a recipient gas molecule of species $B$ in sufficient vicinity of the excited atom $A^*$. This procedure does not allow for a multiplicative combination of transfer mechanisms between more than two species:

$$A^* + B \rightarrow A + B^+ + e^- \ , \ A^* + C \rightarrow A + C^+ + e^- \ , \text{ and } \ B^* + C \rightarrow B + C^+ + e^-.$$

Furthermore, the production and propagation of photons is not implemented in Garfield++. An excited state not yielding an ionizing energy transfer is simply 'forgotten' and the energy dissipates, as if a non-radiative de-excitation process occurred. It should be noticed that a possible ionizing energy transfer via photo-emission ($A^* \rightarrow A + \gamma$) followed by photo-ionization $B + \gamma \rightarrow B^+ + e^-$) is, although not explicitly simulated in Garfield, implicitly included in the energy transfer rates for the Penning effect. As a consequence of neglecting photon treatment, the program is not able to simulate the yield of secondary avalanches on a microscopic level.

With the required processing of a multitude of electrons and the increased computational effort for the enabled Penning transfer, the simulation of large avalanches becomes increasingly expensive in terms of computing resources. A common simplification is the approximation of the complex electrical field structure in a Micromegas with a parallel plate setup, where the uniform field can be described analytically, omitting the computing time intense evaluation of the electric field from a field map. The level of agreement between reality and this approximation depends to a large extent on the fine geometry parameters of the Micromegas: while woven wire meshes with comparatively coarse wire structures and strip patterned anodes yield less homogeneous fields, an almost uniform field can be obtained, for instance, with very fine electro-formed meshes in combination with a continuous anode. Such a configuration has been used for our experiments presented in section 4.5. A discussion of the field uniformity dependence on the parameters of the micromesh geometry is given in chapter 8.

### 4.2.2. Penning Rate Determination from Simulation

The penning transfer rate $r$ introduced in section 4.1.1 is typically determined as a fit parameter in (4.3), as described in [74]. It is, however, as well a microscopic parameter utilized as an input to Garfield++ simulations representing a microscopic energy transfer probability. Therefore, it can be determined by direct comparison of an experimental observable like the gas gain $G$ and an iterative simulation, following the scheme presented in figure 1.8.

Since the penning rate is assumed to be in first order independent of the electric field it can be determined from a single gain measurement and thereafter assumed for all measurements with the same gas mixture. For the studies in noble gas + Isobutane mixtures, which will be discussed in detail in section 4.5, we determined the penning rate by equalizing the experimentally measured with the simulated gain under variation of $r$. Covering a wide range of r-values (shown for the example of Ne:Isobutane (95:5) in figure 4.2), a second run was simulated in a more narrow window (figure 4.3), increasing the precision of the simulation to a level where the penning rate uncertainty was dominated by error in the experimental data.

Utilizing this method the Penning transfer rates $r$ for the three mixtures utilized in the SER experiments described in section 4.5 have been determined (table 4.1) and utilized during subsequent simulation in these gases.

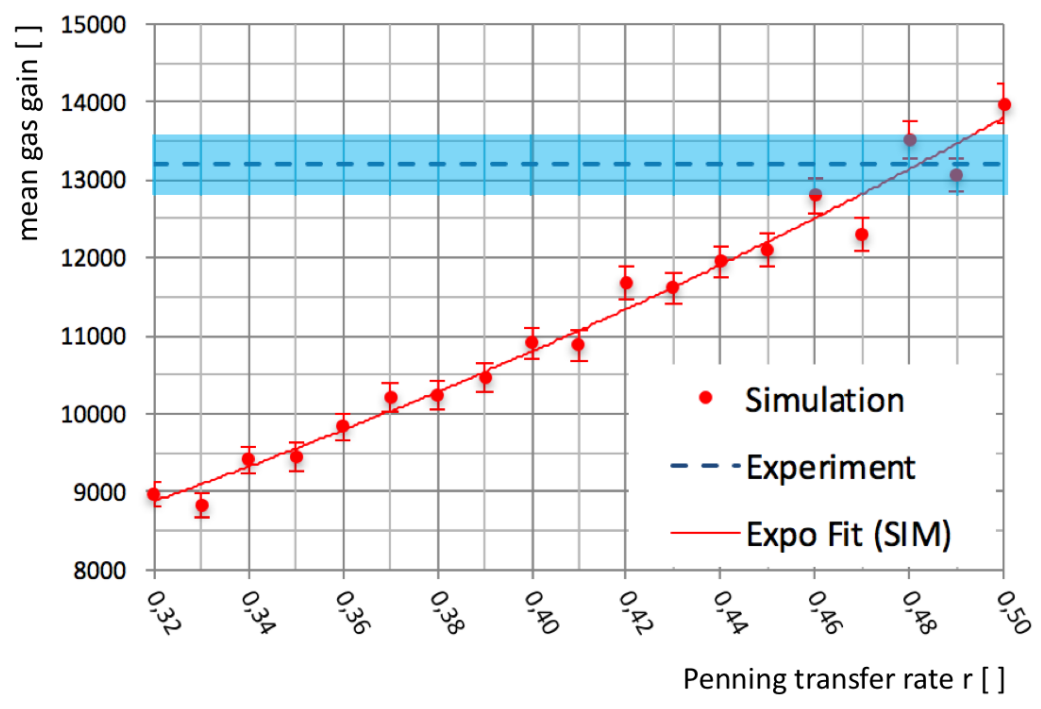

**Figure 4.2.:** Mean gas gain in Neon Isobutane (95:5) NTP at $E_A = 26.25\,\mathrm{kV/cm}$. The Experimental value (blue line + error band) is compared to the simulated gain as a function of the penning transfer rate $r$ (red marker + error bars), fitted with an exponential curve.

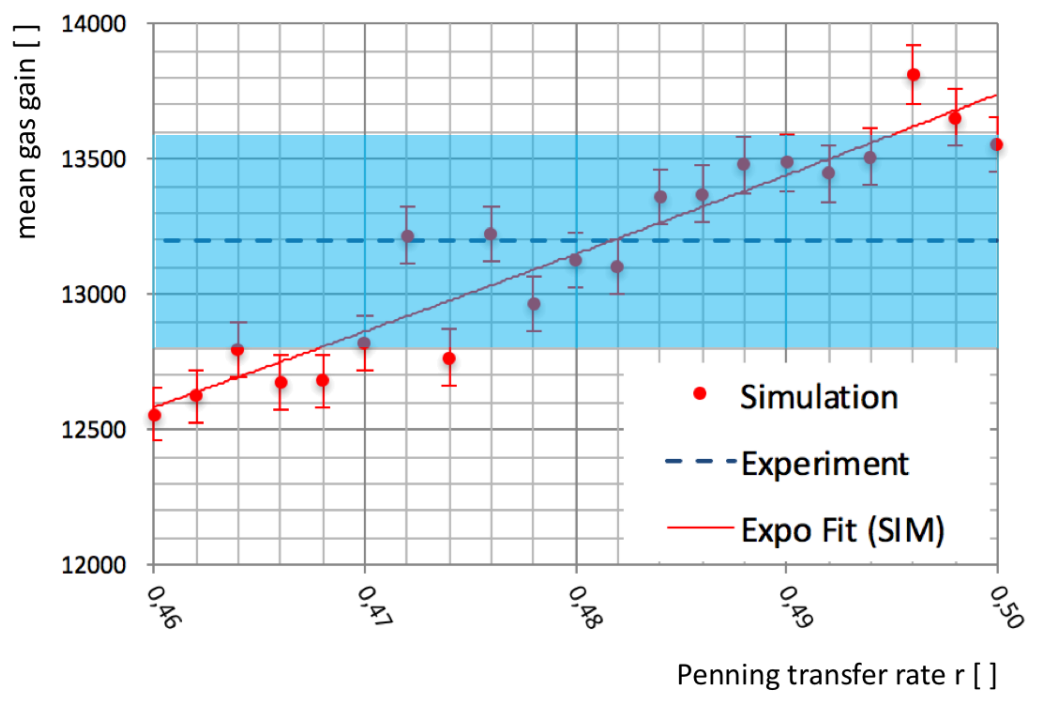

**Figure 4.3.:** Mean gas gain in Neon Isobutane (95:5) NTP at $E_A = 26.25\,\mathrm{kV/cm}$. The Experimental value (blue line + error band) is compared to the simulated gain as a function of the penning transfer rate $r$ (red marker + error bars) in a reduced range with increased simulation data.

| Gas | $E_A$ [kV/cm] | $G$ [ ] | $r$ [ ] |
|---|---|---|---|
| He + Isobutane (95:5) | 26.25 | $(5.70 \pm 0.17) \cdot 10^3$ | $0.1757 \pm 0.025$ |
| Ne + Isobutane (95:5) | 26.25 | $(13.20 \pm 0.40) \cdot 10^3$ | $0.4827 \pm 0.017$ |
| Ar + Isobutane (95:5) | 28.125 | $(5.42 \pm 0.16) \cdot 10^3$ | $0.3217 \pm 0.003$ |

**Table 4.1.:** Penning transfer rates $r$ and the experimental values used for their determination in simulation studies for Argon, Neon and Helium mixed with 5 % Isobutane. Values as published in [85].

### 4.2.3. Simulation of Amplification Gap Size Variation

In avalanche formation in a uniform field the gas gain $G$, as described by the first Townsend coefficient in (4.2), depends on the avalanche development length $x_2 - x_1$ and $\alpha(E)$ which is in turn a function of the gas composition, gas conditions and of the electric field. In a parallel plate (PP) setup with fixed voltage difference $\Delta U$ the electric field becomes a function of the electrodes distance $x$ with

$$E = \frac{U}{x}. \tag{4.8}$$

An electron crossing the distance between the electrodes will in total acquire an energy $\Delta U \cdot e$, where $e$ is the elementary charge. Varying $x$ will not affect this total value, but change the field strength and, therefore, the electron's energy uptake in-between two scatterings. Thus, the effective cross sections and the first Townsend coefficient $\alpha$ are affected. Keeping the gas and its condition constant and, therefore, limiting the changes in the mean free path $\lambda$, one expects two opposing effects on the gain: For larger $x$ the electron experiences more scattering along its path with a total of $x/\lambda$. If the electron's energy is sufficient to cause an ionization in each scattering with high probability, the gain is expected to rise exponentially with $2^{x/\lambda}$. On the other hand, the mean energy gained by the electron inbetween two collisions, corresponding to the mean of the electron's scattering energy $\overline{\epsilon} = \lambda \cdot \Delta U/x$ is decreasing with larger $x$, assuming a constant $\lambda$. Therefore, the probability of yielding an ionization per scattering process is decreasing. The first effect should be dominant for small $x$ where $\lambda \cdot \Delta U/x \gg E_{ion}$ and the second becomes more pronounced for large $x$ where the majority of scattering processes occure at an energy below ionization threshold $E_{ion}$. An empiric formula to describe this dependence has been formulated by Rose-Korff [20, 86] in the early 1940's. It requires, however, the experimental input in form of two parameters which can be hardly interpreted in terms of physical parameters.

Being based on theoretic considerations and validated at different levels of agreement during the last half century, an attempt was taken to scrutinize this behavior on the basis of microscopic processes. Therefore, we simulated a Micromegas in the approximation of a parallel plate setup with the amplification gap thickness as variable parameter. Limiting our study to $ArCO_2$ (93:7) NTP and a typical voltage difference of $\Delta U = 540\,\mathrm{V}$ we simulated the full avalanche growth in a wide range of gap sizes yielding a gain range from $1 - 2.5 \cdot 10^5$. Thereby penning transfer rates according to [74] have been assumed. Besides the mean gain $G$, the obtained distributions yielded the behavior of the relative variance $f$, which could not be obtained from the Rose-Korff description.

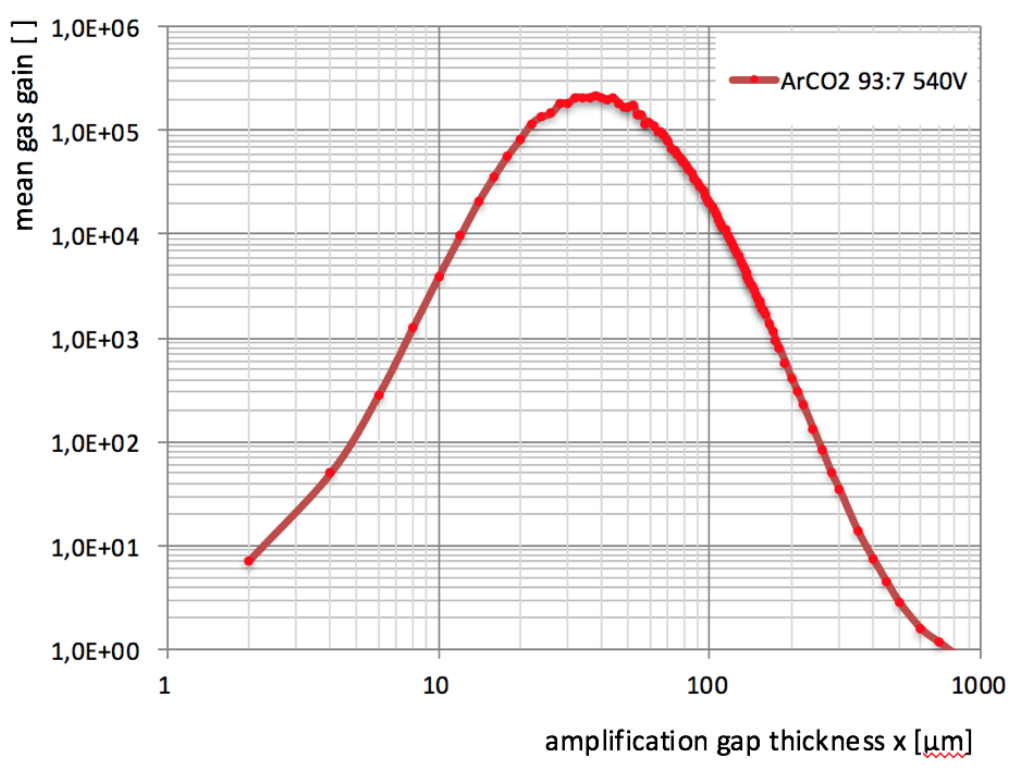

**Figure 4.4.:** Mean gain of a single electron avalanche in a parallel plate setup filled with $ArCO_2$ (93:7) NTP as a function of the amplification gap size. The increased fluctuation for large gain are due to the required reduction of statistics owed to limited computing resources.

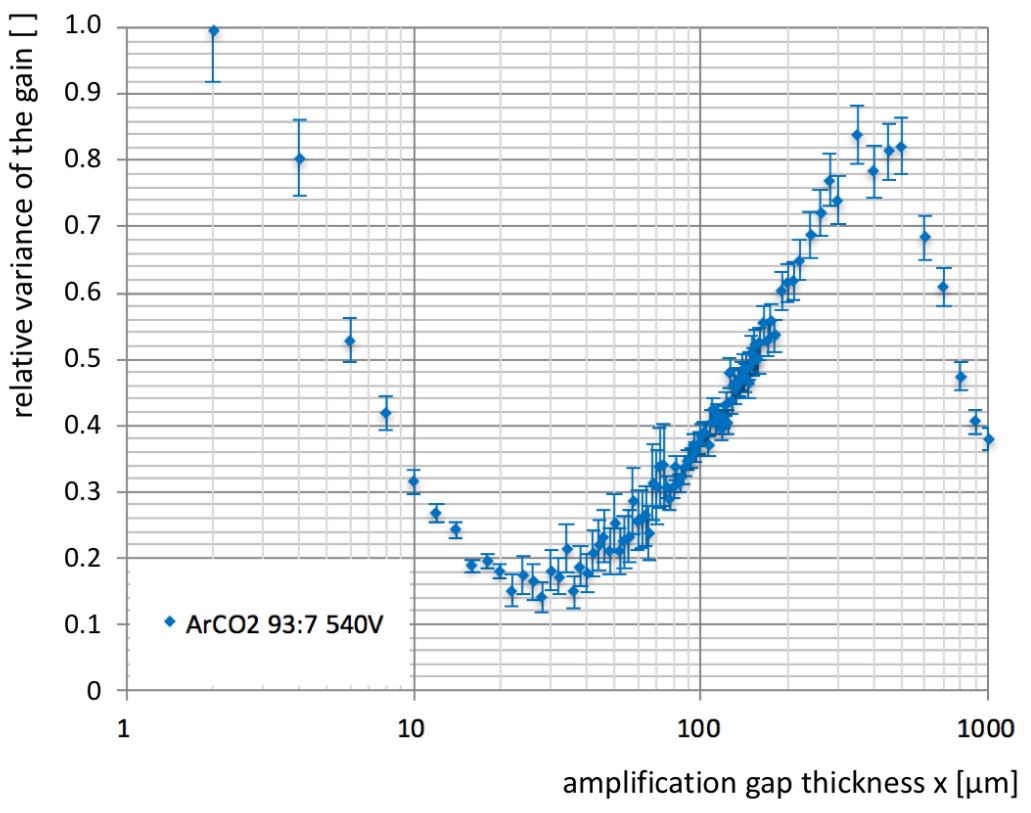

**Figure 4.5.:** Relative variance $f = \mathrm{RMS}^2/G^2$ of the electron gain as a function of the amplification gap thickness in a parallel plate setup filled with $ArCO_2$ (93:7) NTP. A distinct minimum is visible in the region comparable to the maximal gain range in figure 4.4.

Figure 4.4 clearly shows the regions where each of the two effects discussed above are dominant as well as the predicted exponential rise for small $x$. For the 540 V setting the maximum is found at $x_{G_{max}} = 38\,\mu m$. In this region, the reduction of gain due to decreased ionization probability equals with the increase of gain owed to more scattering processes. For the operation of a detector, this region would be ideal, since it provides gain stability under variation of the geometrical parameter and a similar stability to small voltage fluctuations. This region furthermore provides a natural minimum in the relative gain variance $f$ and, therefore, an intrinsically good energy resolution (figure 4.7).

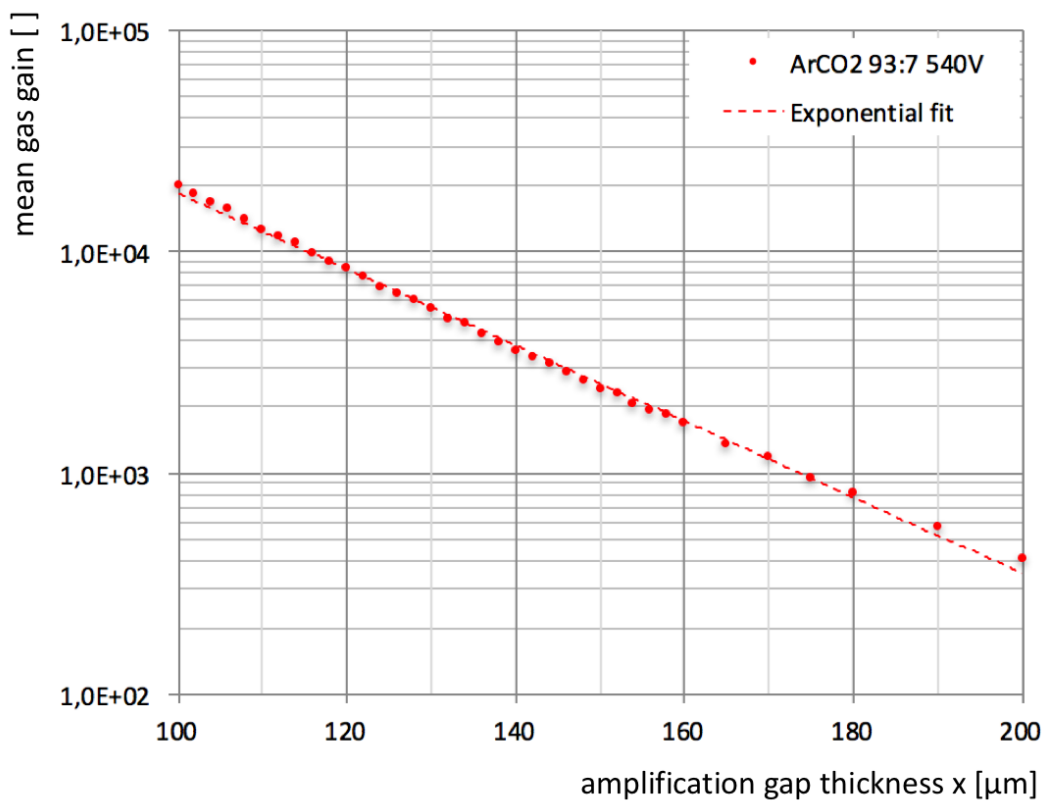

**Figure 4.6.:** Mean gain of a single electron avalanche in a parallel plate setups filled with $ArCO_2$ (93:7) NTP as a function of the amplification gap size. In this limited range typical for amplification gap sizes in Micromegas, the dependence can be approximated with an exponential function.

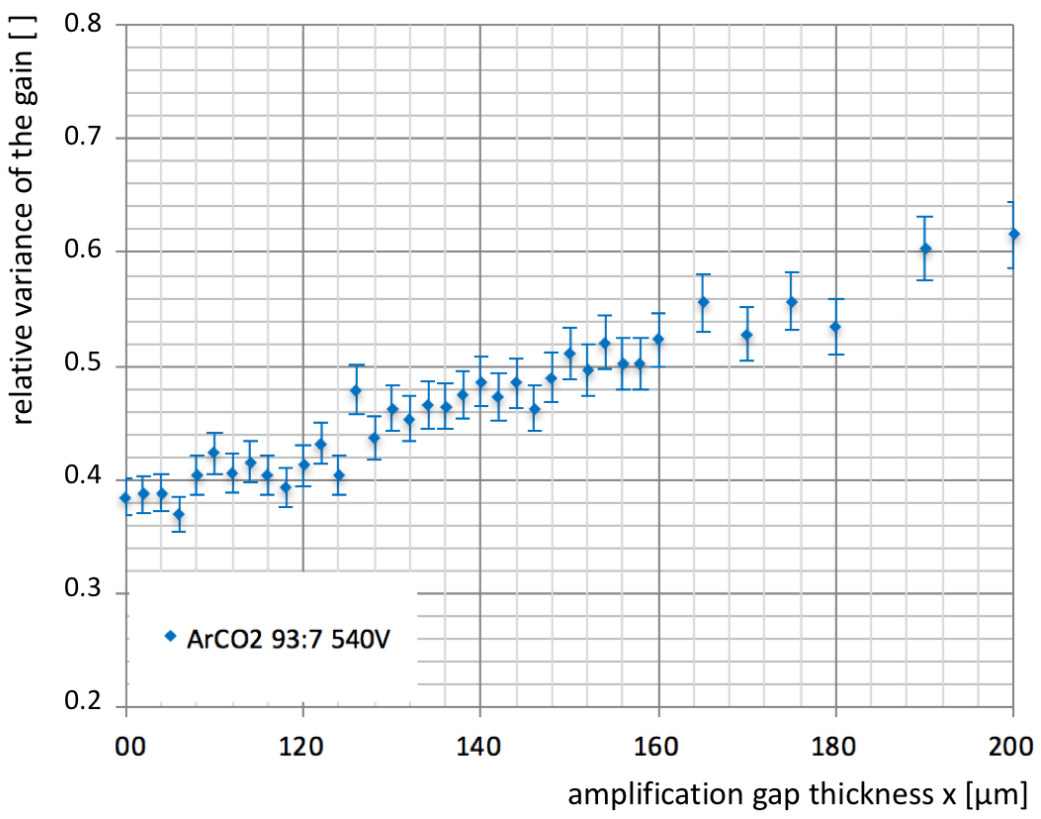

**Figure 4.7.:** Relative variance $f = RMS^2/\overline{G}^2$ of the electron gain as a function of the amplification gap thickness in a parallel plate setup filled with $ArCO_2$ (93:7) NTP. The range of typical values for a Micromegas are shown.

A Micromegas with a $40\,\mu m$ amplification gap is, however, difficult to build and even more difficult to operate stable, given the increased risk of sparks and streamer formation in the high electric field. Furthermore, the expected gain of $2.5 \cdot 10^5$ is by far too high for measuring MIPs or X-Rays causing multiple signal electrons. While the gain could be readjusted by decreasing the potential difference, this would shift the maximum towards an even smaller gap size. Therefore, Micromegas detectors are usually designed to operate on the right flank of this curve. In the range shown in 4.6, the gain dependence on the gap size can in first order be approximated with an exponential decay with $x_{1/2} = 18\,\mu m$ quantifying the high sensitivity of the gain to the amplification gap size. For example, a $\pm 5\,\mu m$ variation in the gap causes a $> 25\,\%$ change in the gain. Therefore, a precise control of the amplification gap thickness is of critical importance for the development and construction of large size Micromegas detectors, as the ATLAS NSW Micromegas.

## 4.3. Gain Dependence on the Micromegas Fine Geometry Parameters

The ExMe Micromegas detectors, introduced in chapter 3.3.1, feature the unique opportunity to perform dependency studies independently for different fine geometry parameters: The mesh geometry by exchanging the micromesh, the inter-pillar distance by probing the different sectors and the anode structure by comparison of the two detectors. Probing the effects of dimension variations on a µm-level the practicability of the strict VOTAT paradigm, in terms of preservation of non-varied parameters, is reaching its limits. Furthermore, the utilized measurement setup, presented in chapter 3.3.2, is not suited nor calibrated for absolute gain measurements. Therefore, the following result review, based on the data sets obtained in the scope of [54], is limited to relative changes in gain and the discsussion focuses on qualitative ordering effects instead of a quantitative analysis.

### 4.3.1. Qualitative Field Comparison with COMSOL Multiphysics®

To understand the impact of a fine geometry parameter variation on the field configuration, eventually defining the amplification process, FEM simulations were performed using COMSOL Multiphysics®. This method is advantageous over a full avalanche simulation in terms of accuracy of the geometric model, required computation time and thus the possibility to cover a larger parameter space. Being the dominant tool for mesh parameter optimization, more details on this simulation will be presented in chapter 8 and only a selection of relevant plots are prepended in this section.

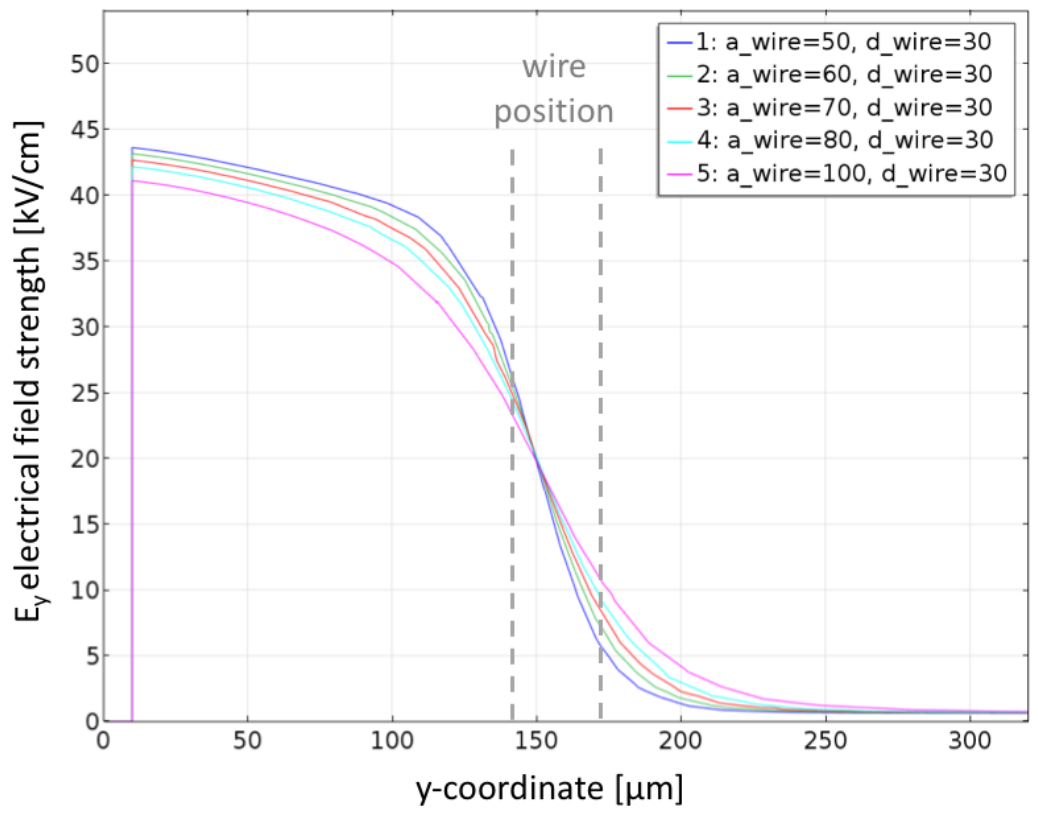

**Figure 4.8.:** Electrical field strength along the electrons' most probable path through the amplification gap with different apertures of a 30 µm-wire mesh. Identical to figure 8.5 in chapter 8.

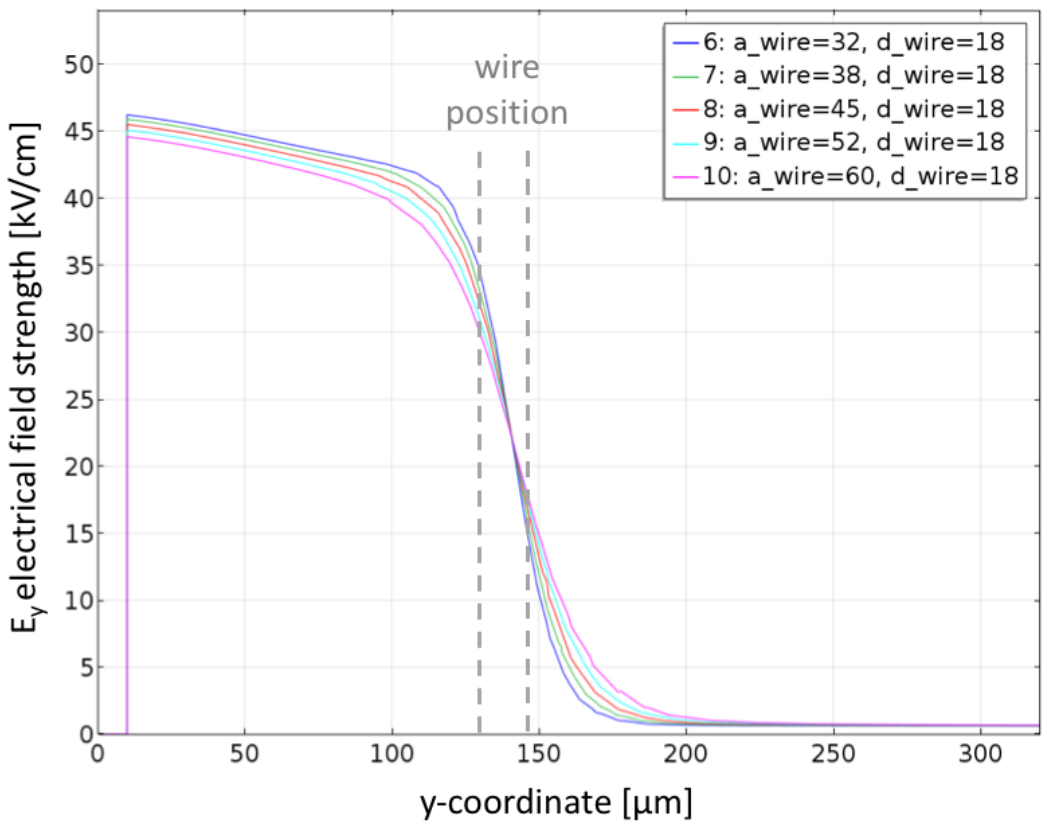

**Figure 4.9.:** Electrical field strength along the electrons' most probable path through the amplification gap with different apertures of a 18 µm-wire mesh. Identical to figure 8.6 in chapter 8.

The electrical field strength along the electrons' most probable path through the Micromegas amplification gap is shown for different mesh geometries in the figures 4.8 and 4.9. A reduction of the distance between the 30 µm- / 18 µm-wires leads to a stronger gradient of the electric field in the transition region between drift and amplification region. It as well causes an increased maximum field strength and subsequently a larger electron energy uptake, which equals the integrated area under the field strength in the boundaries of the amplification gap. Therefore, denser meshes are expected to yield a higher gas gain. Comparing the 30 µm- with the 18 µm-wire meshes shows the steeper slope for the finer structure. Following the same line of argumentation, meshes with finer wires should result in an increased gain.

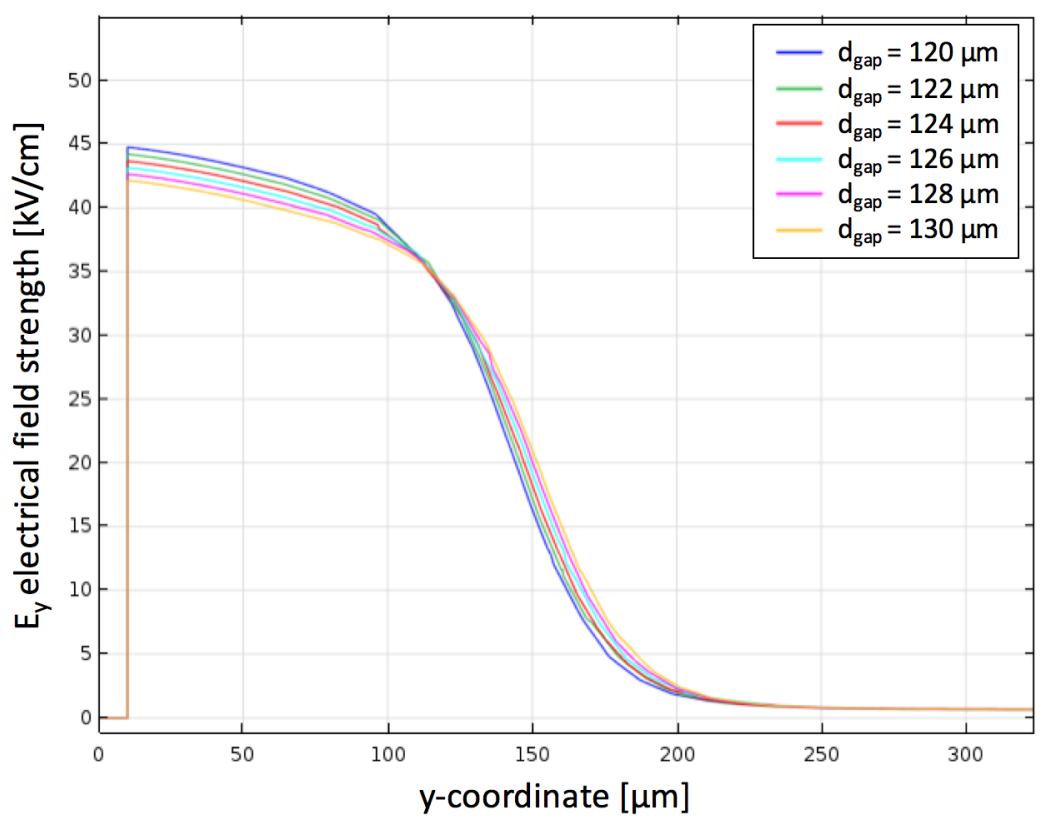

**Figure 4.10.:** Electrical field strength along the electrons' most probable path through the amplification gap with different gap size $d_{gap}$ corresponding to the distance from the flat anode to the bottom of the micromesh.

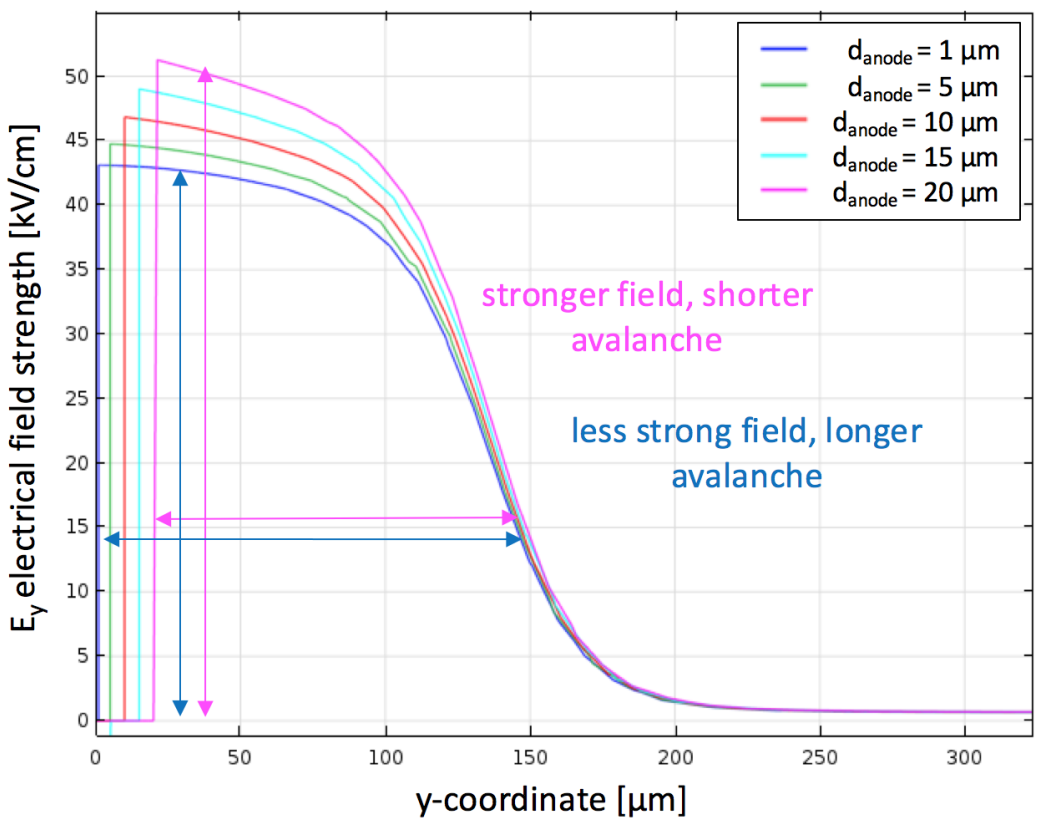

**Figure 4.11.:** Electrical field strength along the electrons' most probable path through the amplification gap with anode structures of different thicknesses $d_{anode}$ under preservation of the micromesh parameters and position.

Figure 4.10 shows the change in the electric field for a small variation of the mesh position above the anode in steps of 2 µm. Experimentally such a discrepancy could be realized by either a change in the pillar height, altering the local mesh reference height, or by the sagging behavior of the mesh in between the pillar support structure, altering the difference between the lower mesh edge and the reference height. The latter effect depends, as quantitatively discussed in chapter 6.3.3, on the mesh tension, its elasticity module, the applied voltage difference and the distance between the pillars. A reduced effective gap distance $d_{gap}$, yields a spatially delayed increase of the field strength, as seen by an electron approaching from the drift field side. On the other hand the electric field reaches stronger maximum values. This resembles the situation of the gap variation in a uniform field which has been simulated and discussed in section 4.2.3.

A similar effect is caused by a variation of the anode structure thickness, shown in figure 4.11. With the mesh position being preserved in this configuration, the rise in field strength deviates only slowly in the region of the mesh, but it reaches significantly increased field strength values closer to the cathode. While figure 4.11 shows the field strength along one probable electron path through the center of the mesh opening to the top of the anode, the varying thickness of the curved anode yields larger discrepancies between the different probable electron paths through the mesh.

### 4.3.2. Influence of the Mesh Geometry

To compare the relative gain between different measurement runs the signal strength maximum in a $U_D$ sweep is extracted. This value corresponds to the normalization factor applied during assessment of the electron losses shown in figure 3.28. Referring to the factorized approach in (1.7) it corresponds to

$$S_{max} = n_e \cdot (1 - R) \cdot (1 - A) \cdot T \cdot G \cdot c_{r/o} \propto G, \tag{4.9}$$

where the proportionality follows under the assumption of equal electron losses $(1-A)\cdot T$ at the drift field corresponding to $S_{max}$.

The relative gain comparison for the full data set obtained in [54] is shown for the two detectors in the figures 4.12 and 4.13. The predicted tendency to an increased gain with finer structures and with a reduced mesh aperture can be observed in both chambers. However, the order of the meshes is not fully compliant as the 60-30: 44.4 % mesh yields values systematically below the less dense 70-30: 49.4 % mesh. While the above mentioned assumption on equal electron losses is reasonable in the case of a amplification voltage variation shown in figure 3.28 it is likely to be broken in the comparison of different meshes as discussed in chapter 3.3.3. The reduced signal electron yield could be accounted for with a correction factor of $([(1 - A) \cdot T]^{-1})$ according to the results presented in 3.3.3. This correction further increases the gain values for less transparent structures and yields the predicted order in the data set of the ExMe2 detector. The larger discrepancy observed in ExMe1 is reduced but not overcome. No satisfying explanation has been found on the basis of the experimental data or conditions [54]. With the unique feature of this mesh being a different stretching and gluing method, a possible impact of the different mesh tension or the utilized glue are the only remaining indications, but non of those yielded a conclusive explanation.

### 4.3.3. Impact of the Inter-Pillar Distance

The influence of the pillar spacing on the signal strength can be observed by comparing measurements between the four sectors of the ExMe Micromegas. Comparing measurements with the same mesh, the identical transparency at a given working point is guaranteed and, therefore, the drift voltage has been fixed at $U_D = 300\,\mathrm{V}$. This sector comparison is shown for the two detectors ExMe1 (figures 4.14 and 4.15) and ExMe2 (figures 4.16 and 4.17) equipped with the same two meshes.

In the ExMe1 data the observed relative order in signal strength is in agreement with a larger inter-pillar distance yielding an increased gain. This order is observed throughout all measurements with the ExMe1 featuring a larger variety of meshes. The effect strength varies slightly between the different runs, but remains in the range of 20 - 30 % referring to the highest values. With the measurements being affected by the atmospheric temperature and pressure as well as small changes in the gas purity, all of which can change within the several hour measurement between the different runs, the variations in the gain on this level are anticipated. This can as well be seen in the the avalanche growth with increasing amplification field strength: The growth exponent is reduced for the long sector B measurements (≈ 12 h per run) compared to the appended faster measurements (≈ 1 h) in the sectors A, C and D. This indicates a small continuous change in the gas parameters, likely due to the continuous flushing during the sector B measurement resulting in a more purified gas volume for the other runs.

Simulating the mesh sagging accordingly to the study presented in 6.3.3 a difference of the effective amplification gap thickness of about 2-3 µm between the 5 mm and 10 mm pillar distance is anticipated[1]. According to the parallel plate simulation in section 4.2.3 and figure 4.6 this corresponds to a gain increase in the order of 8 - 11 %. Although predicting the correct ordering between the sectors, the qualitative estimate is at least a factor of two below the experimentally observed effect. On the experiment side, the effect is reoccurring systematically for all ExMe1 measurements. Neither the gas conditions, altering the relative effect strength as discussed above, nor other considered systematic effects, like for example the mesh position effect on the detector capacitance, could satisfyingly explain this discrepancy. At last a non-homogeneity of the readout structure, in terms of pillar height per irradiation spot, must be considered. Here local fluctuation on µm-level could occur during the production process[2]. Their effect on the gain would be characteristic for the $1 \times 1\,\mathrm{cm}^2$ measurement spot of each sector and thus reproduced in all measurements. If dominated by local differences in the pillar height, the observed order in gain would be purely accidental. A more likely source of the underestimated relative gain change is the approximation of the Micromegas geometry with a PP setup in the simulation in 4.2.3. For the ExMe detectors featuring interrupted anodes and comparatively coarse mesh geometries this assumption is barely justified. While the discussed overall trend remains valid in the non-uniform field, the absolute effect of the gap size on the gain can differ significantly. This effect of a reduced effective avalanche growth length in a Micromegas compared to a PP model will be discussed in section 4.4.2.

While the discussion above focused on the ExMe1 results, another systematic effect can be seen in the comparison with the ExMe2 measurements (figures 4.16 - 4.17): The gain order between Sector B and C is inverted compared to ExMe1 and the prediction. While the relative position of the sector C measurements are in line or only slightly decreased compared to the sectors A and D, the sector B measurements feature a substantial systematical increase. This inversion is reoccurring during all measurements, excluding a gas purity or environmental condition related reason. With the only difference between the two chambers being the anode structure this is likely to be caused by a local geometry deviation, as mentioned above. While during ExMe1 production the flat sputtered anode favors a homogeneous pillar height allover the detector, the varying thickness and width of the screen printed strips of ExMe2 can locally in- or decrease the pillar height (figure 3.20). With changes of ±2 µm causing a sufficient systematic shift in gain such a local deviation from the average at the measurement spot of sector B can easily yield the inverted order.

[1] Assuming a mesh young module of 20 - 50 GPa, a mesh tension of 8 - 10 N/cm and an electrostatic force caused by $\Delta U = 540 - 600\,\mathrm{V}$ yields a minimal / maximal sagging of 0.5 / 1.1 µm in sector A and 2.4 / 3.9 µm in sector D

[2] With the available testing tools such an effect larger than 2 µm could be excluded. Without a nondestructive test procedure to scan the partially optical transparent geometry with sub-µm accuracy smaller local deviations could not be assessed.

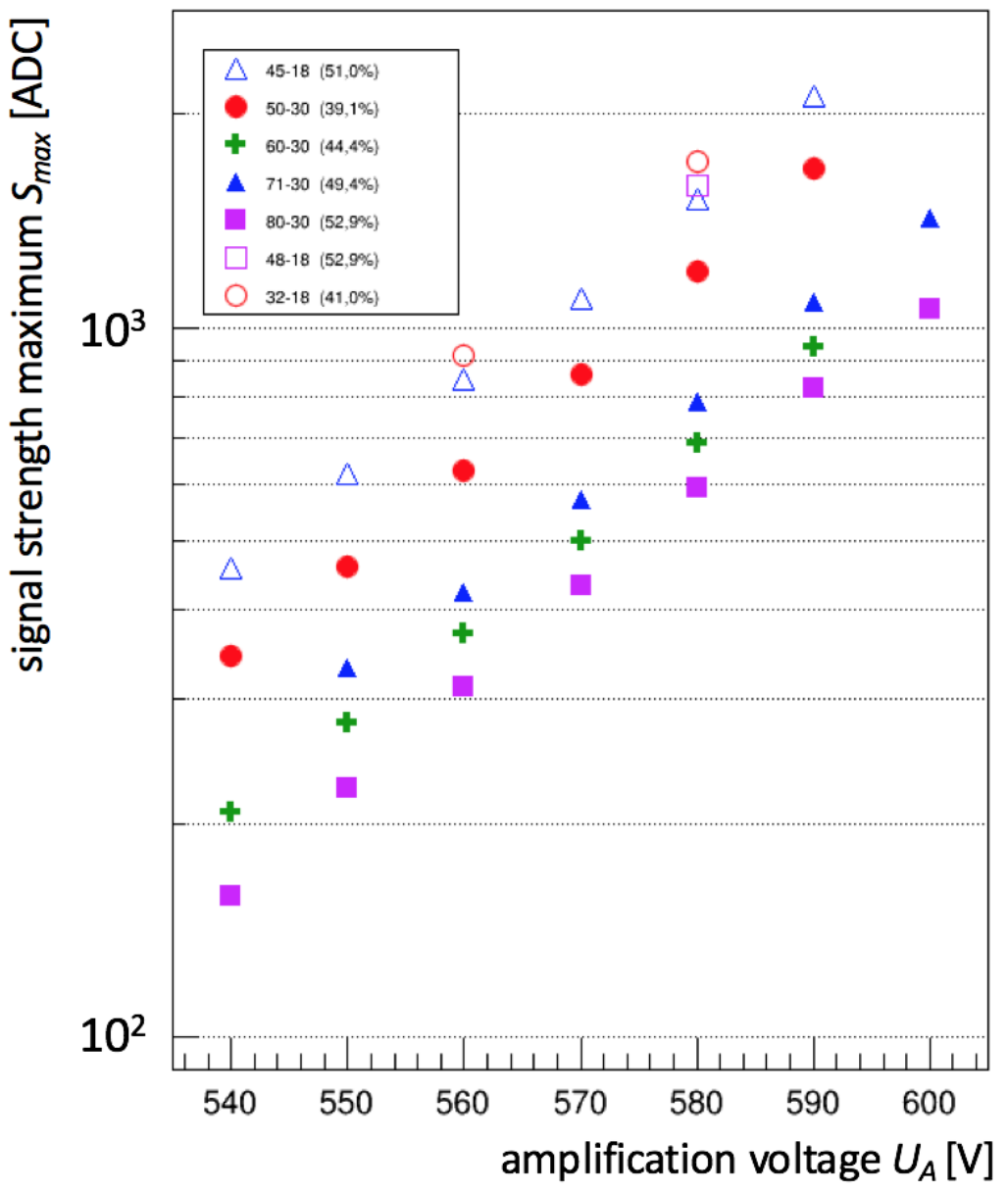

**Figure 4.12.:** Comparison of the maximum signal strength $S_{max}$ over the amplification voltage $U_A$ for different meshes in the ExMe1 detector. [54]

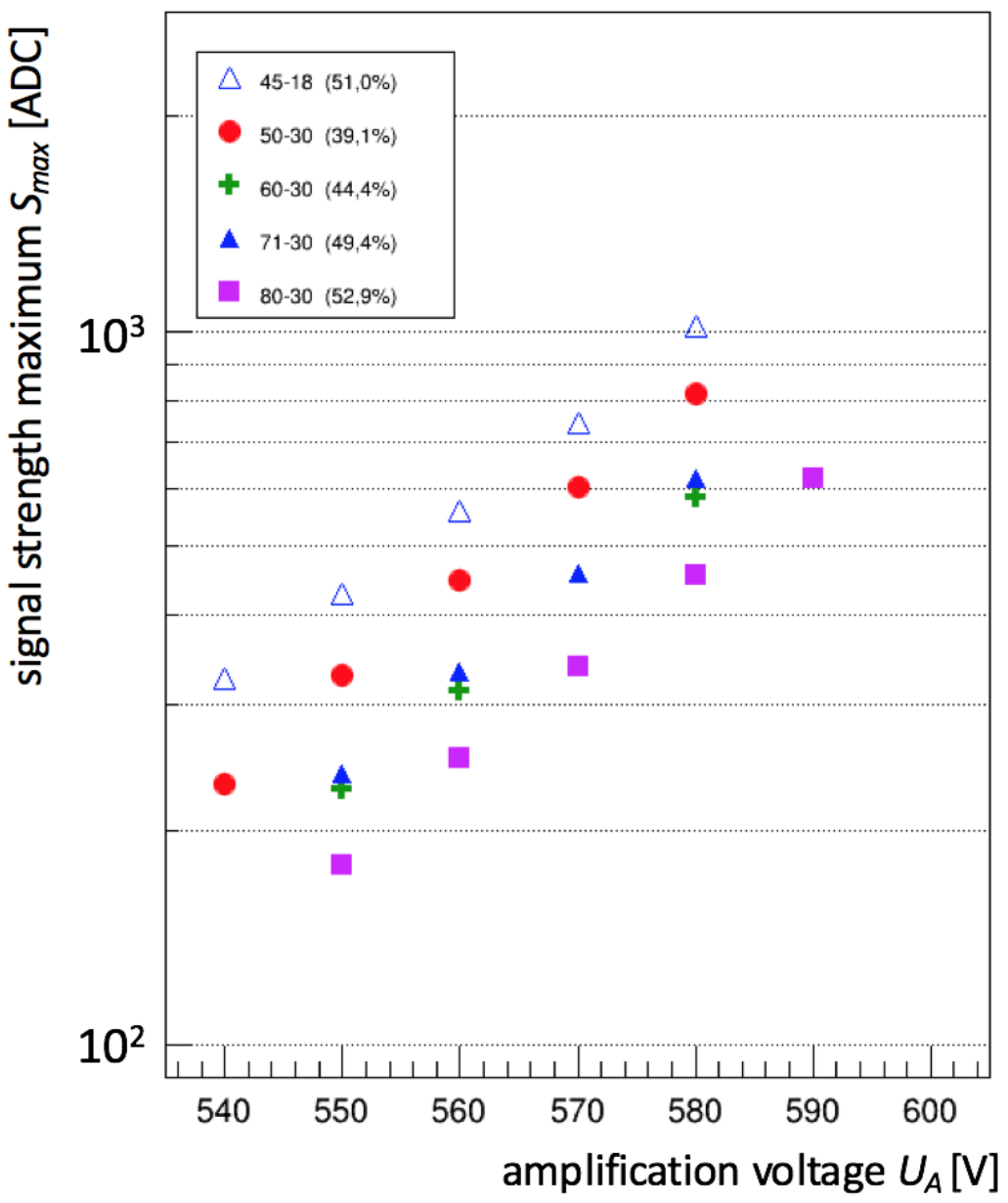

**Figure 4.13.:** Comparison of the maximum signal strength $S_{max}$ over the amplification voltage $U_A$ for different meshes in the ExMe2 detector. [54]

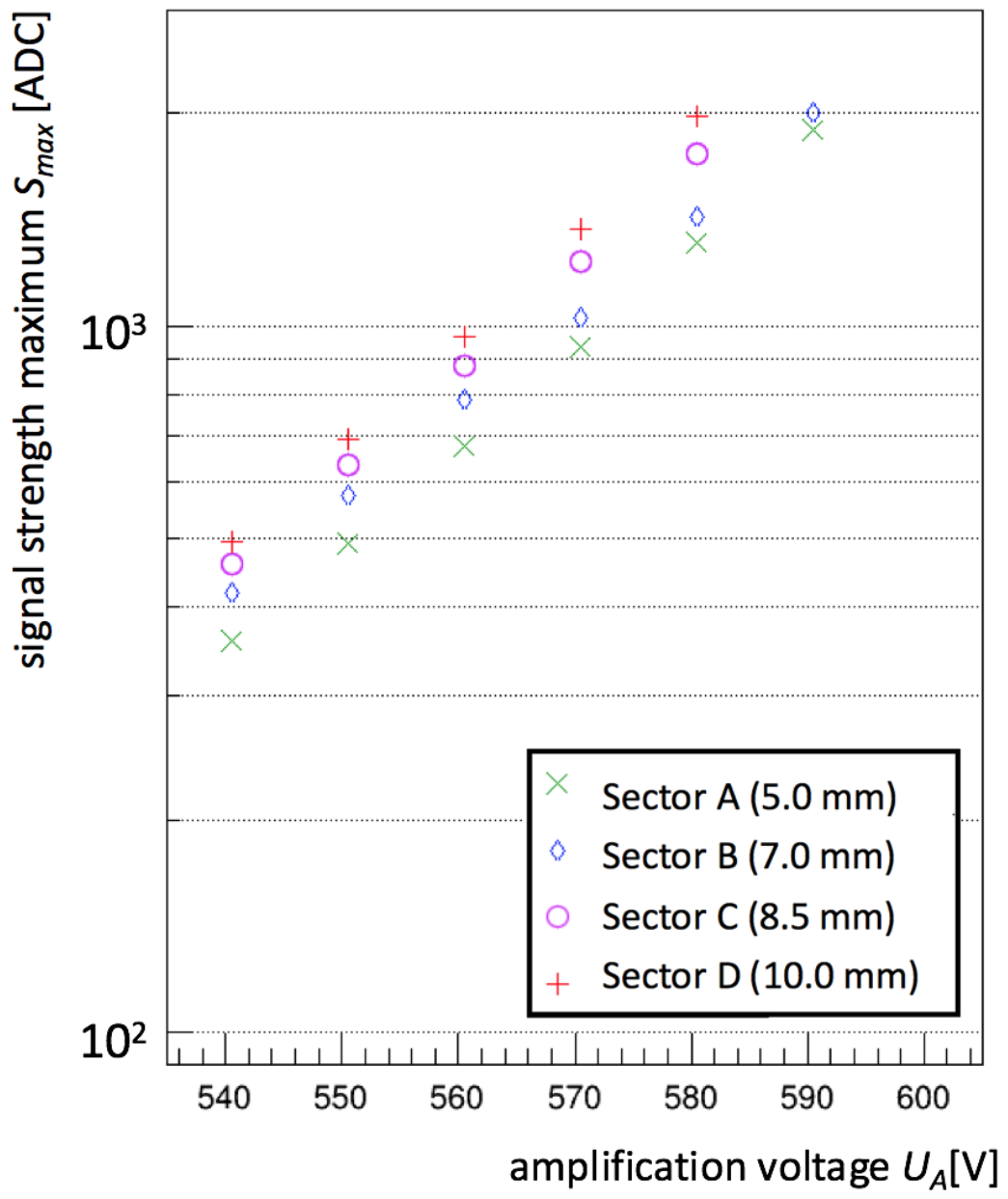

**Figure 4.14.:** Signal strength maximum $S_{max}$ over the amplification voltage $U_A$ for different pillar distances (Sector A-D) in the ExMe1 detector equipped with a 45-18: 51.0 % mesh. [54]

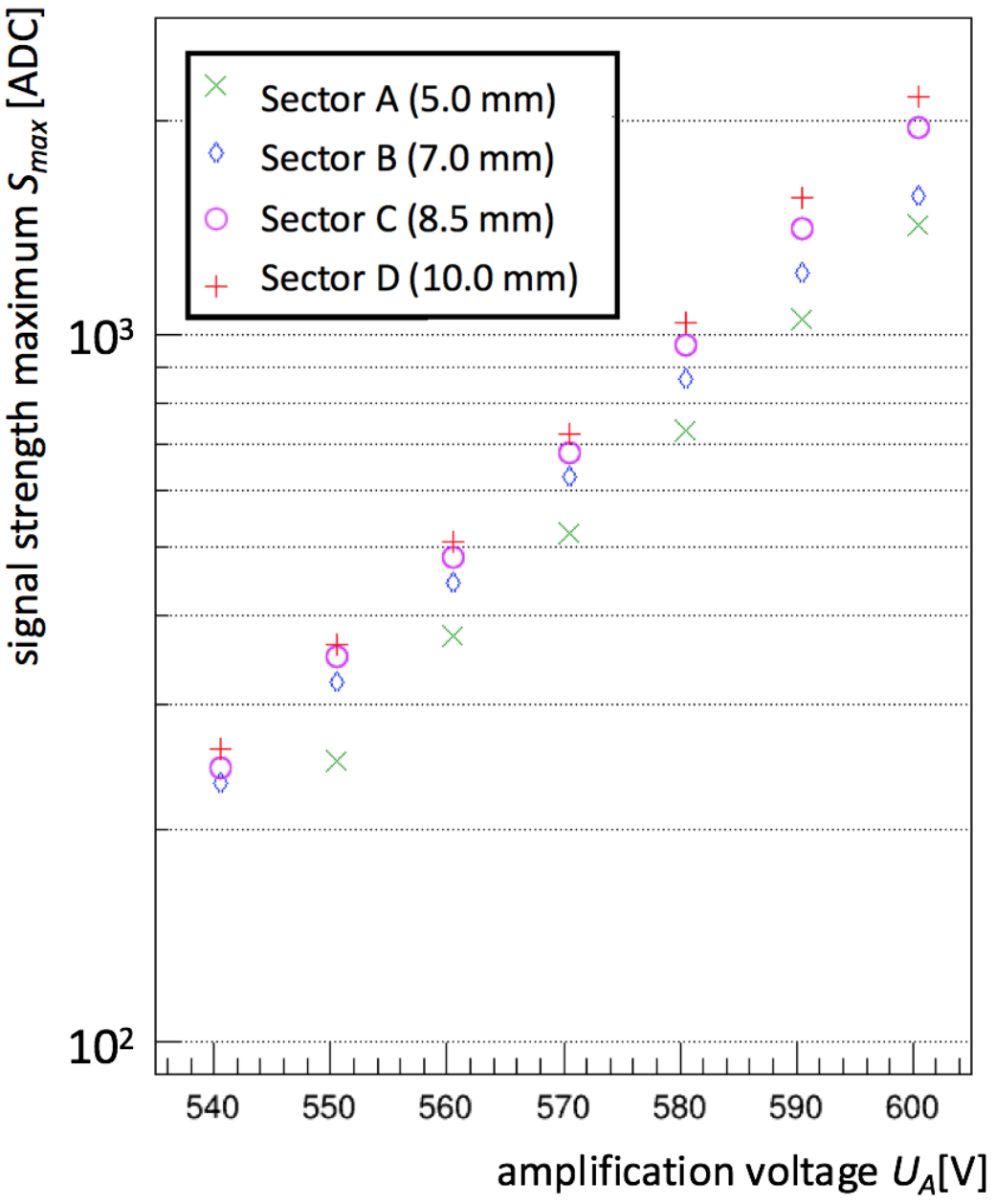

**Figure 4.15.:** Signal strength maximum $S_{max}$ over the amplification voltage $U_A$ for different pillar distances (Sector A-D) in the ExMe1 detector equipped with a 50-30: 39.1 % mesh. [54]

With the order and overall magnitude for relative changes between all other sectors being reproduced in both chambers, the probability for an accidental thickness deviation in line with the pillar distance in those seven sectors is next to negligible. Therefore, we are convinced that, with the restrictions mentioned above and despite lacking a method to correctly quantify the effect, the actual impact of the µm-scale mesh sagging on the gain has been observed.

### 4.3.4. Effect of the Anode Surface

A very clean, repeatable discrepancy was observed throughout all measurements when comparing the two ExMe chambers, featuring two different anode structures. The ExMe2 with the screen printed anode features a consequentially lower gain compared to the ExMe1 with its sputtered anode, as shown in figures 4.18 and 4.19. This observation is in complete contradiction to the expected behavior based on the assumed to be smaller amplification gap in ExMe2 (figure 3.20) which should yield an increased gain (figure 4.6).

With the observation being mesh independent and occurring consequentially during all measurements, basically all not readout PCB related explanations can be ruled. Material related effects based on the different resistivity, like a localized voltage drop resulting in a reduction of the field strength, have been considered. However, the measurements did not yield any rate or current dependence which would be characteristic for voltage drop effects in resistive anodes. Without the capability of a non destructive precision measurement, the dimensions of the anode surface and the chamber inherent pillar structure remain the most likely non excluded explanation for the systematic bias.

This observation again emphasizes that production related fluctuations, despite being thanks to modern production techniques limited to few µm, have a huge impact on the performance of a Micromegas detector. It is, therefore, inevitable to strictly limit parameter variations to the set of parameters under study. While the secondary goal of an inter-sector and inter-chamber comparison with the ExMe detectors yielded only qualitative tendencies, these studies underlined the importance of the chambers design for their primary objective: An assessment of electron losses for different mesh geometries all referring to the literally identical anode and pillar structure.

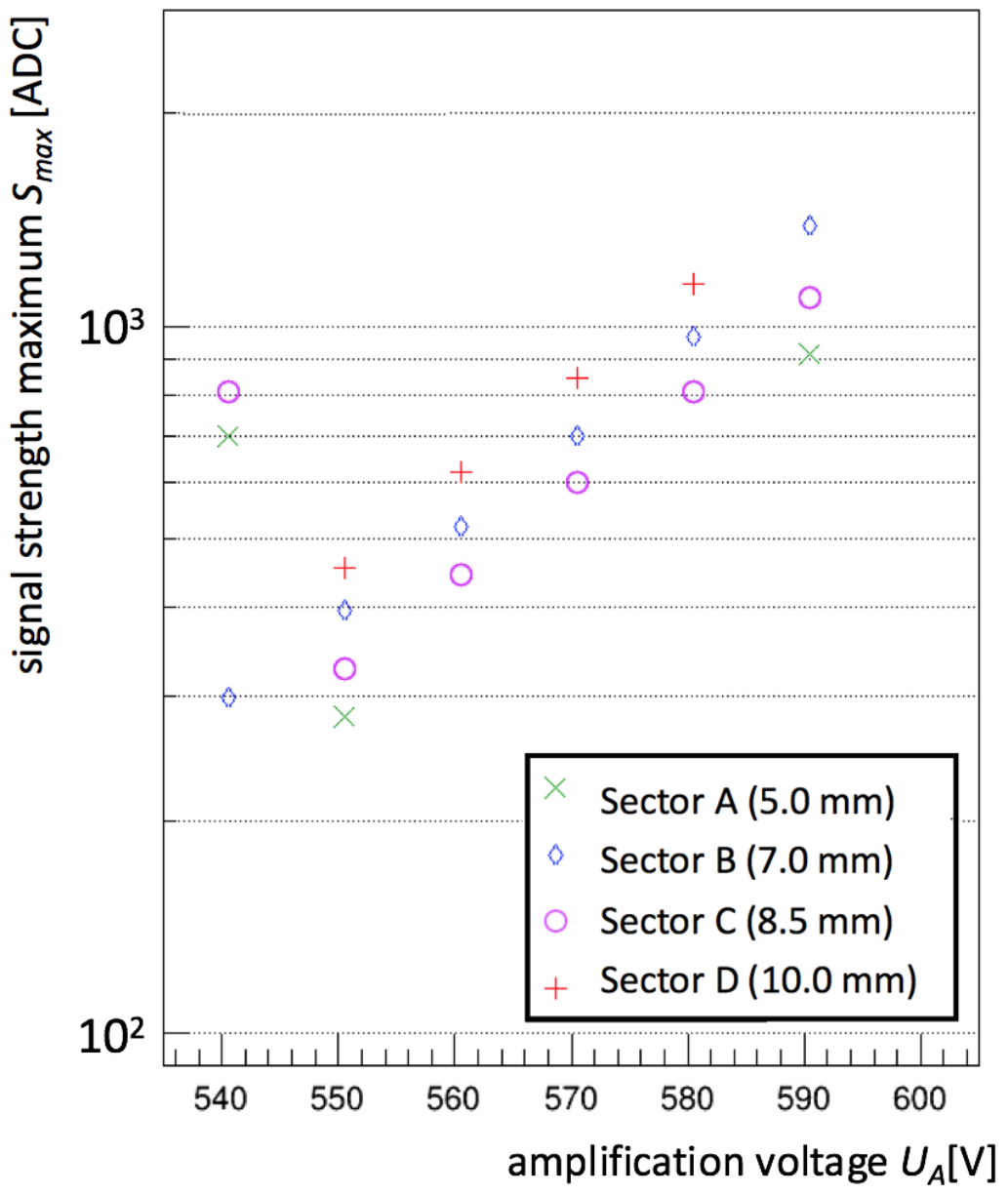

**Figure 4.16.:** Signal strength maximum $S_{max}$ over the amplification voltage $U_A$ for different pillar distances (Sector A-D) in the ExMe2 detector equipped with a 45-18: 51.0 % mesh. [54]

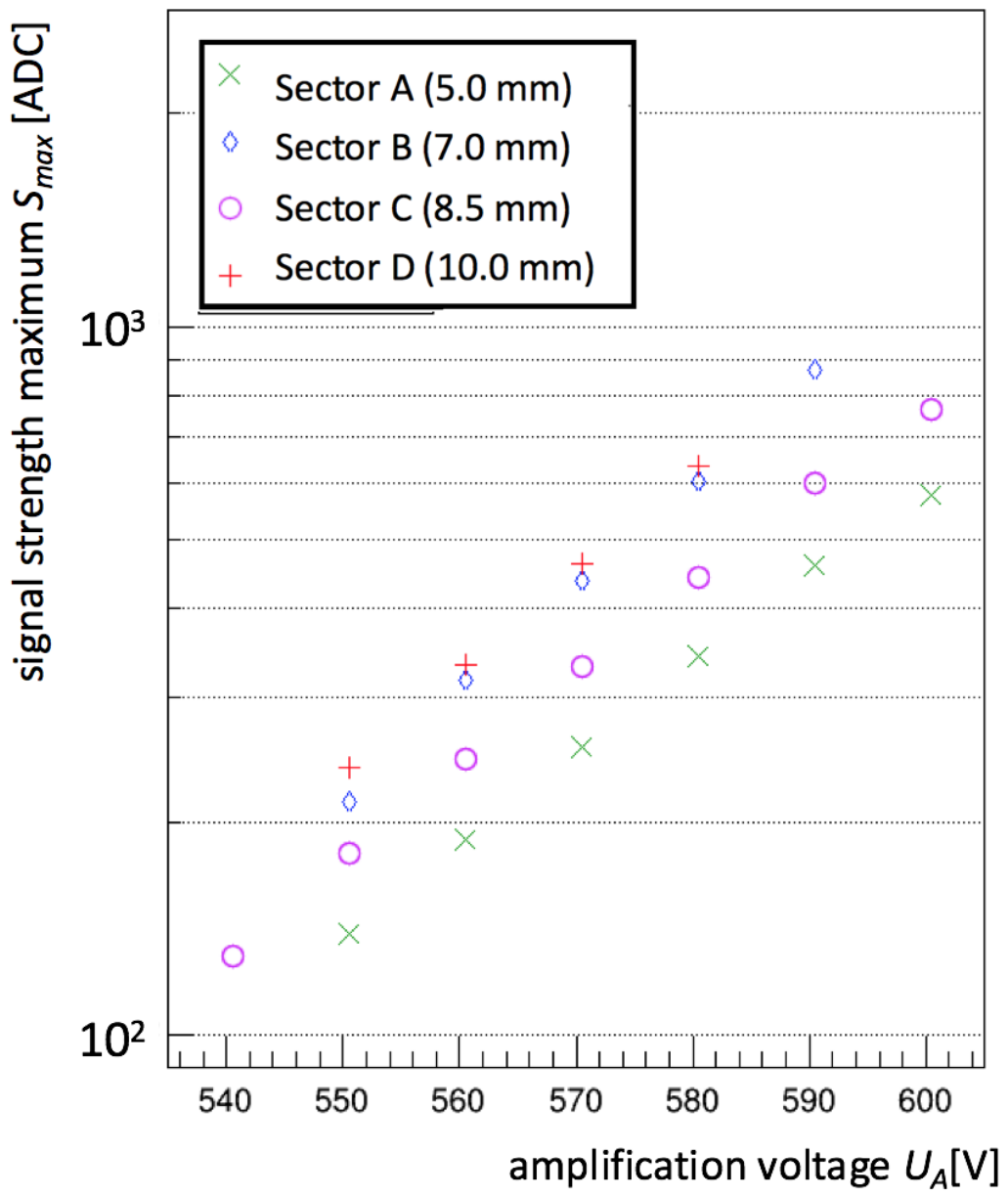

**Figure 4.17.:** Signal strength maximum $S_{max}$ over the amplification voltage $U_A$ for different pillar distances (Sector A-D) in the ExMe2 detector equipped with a 50-30: 39.1 % mesh. [54]

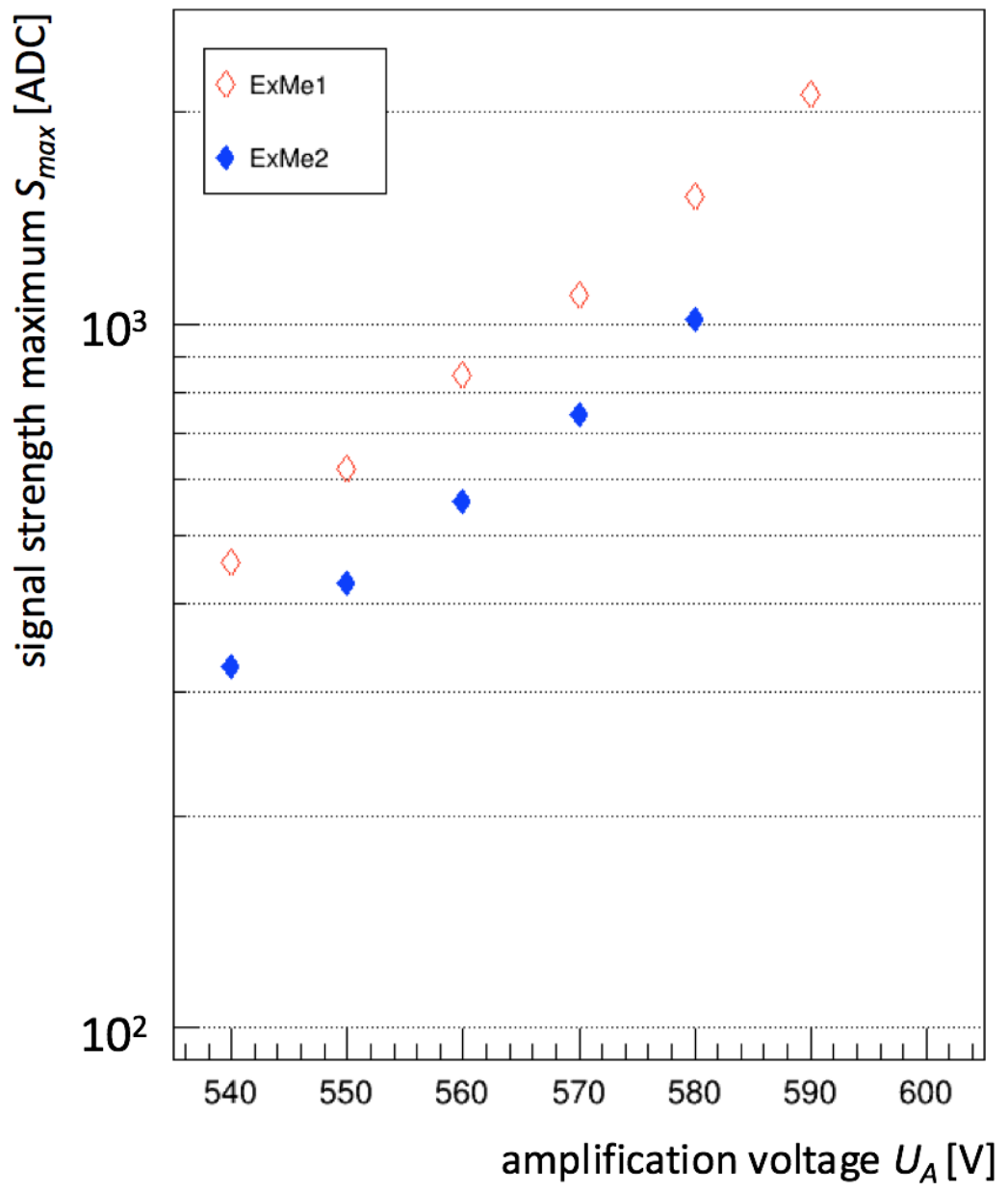

**Figure 4.18.:** Comparison of the maximum signal strength $S_{max}$ over the amplification voltage $U_A$ for the two different detectors ExMe1 and ExMe2, both equipped with a 45-18: 51.0 % mesh. [54]

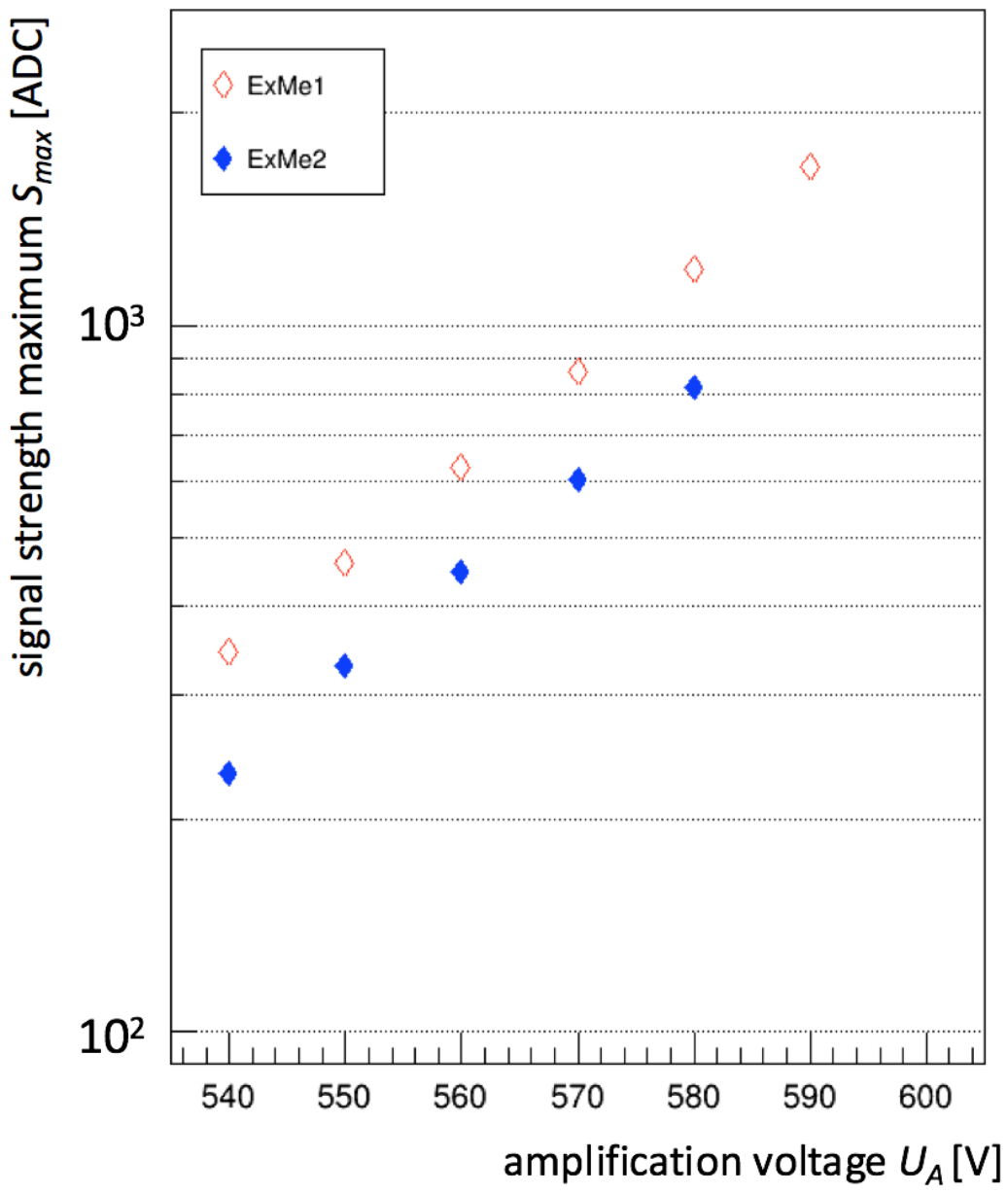

**Figure 4.19.:** Comparison of the maximum signal strength $S_{max}$ over the amplification voltage $U_A$ for the two different detectors ExMe1 and ExMe2, both equipped with a 50-30: 39.1 % mesh. [54]

## 4.4. Statistical Extrapolation towards Large Avalanches

Besides simplification of the geometry and neglection of physical processes, avalanche formation offers an information-loss free method for a significant reduction of computation effort. It is based on the statistical nature of the amplification process and follows a line of thought and formalism proposed by R. Veenhof and implicitly utilized in [83].

In an electron avalanche the majority of the electrons are produced in the last amplification step and, accordingly, the majority of computing resources are consumed in the simulation of these last steps. On the other hand, their contribution to the overall avalanche static fluctuation is small, since describing the parallel amplification of many electrons, their statistic is a convolution of the amplification probability for each electron. This convolution is following the central limit theorem and thus its relative width decreases with increasing number. The statistical fluctuations of the avalanche formation are, therefore, dominantly defined in the first amplification processes, where the small number of electrons involved, one in the very beginning, yield a large impact of the ionization length fluctuations on the full avalanche statistic.

Utilizing a suitable formalism to propagate the statistics of an in detail simulated initial part of the avalanche to the full avalanche statistics could yield precisely the distributions moments like the mean and its variance while reducing the simulation effort substantially.

### 4.4.1. Statistic of (sub-divided) Avalanche Growth

Considering avalanche development along a growth length $x$ as being composed of shorter amplification steps $x_1, x_2..., x_n$, with $n$ being an arbitrary selected number of steps and $x_1 + x_2 + ... + x_n = x$. A single electron triggers an avalanche with a number of electrons distributed by $P_{0\to x_1}(N, E, x_1)$ during step 1. Each of the electrons produced in this first step triggers an avalanche during the second step yielding electrons according to $P_{x_1\to x_2}(N, E, x_2 - x_1)$. This process is repeated in each subsequent step until avalanche growth is stopped.

Maintaining the electrical field and the gas conditions throughout the avalanche development volume and choosing equal step length $x_i = x/n$, the independent development of these sub-avalanches follows the same kind and amount of statistical processes and can, therefore, be described by the same distribution $P_1(N)$ for the electron yield $N$:

$$\begin{aligned} P_1(N) &= P_{0\to x_1}(N, E, x_1) = P_{x_1\to x_2}(N, E, x_2 - x_1) \\ &= P_{x_i\to x_{i+1}}(N, E, x_{i+1} - x_i) \forall i \in \{1, 2, ..., n-1\} \end{aligned} \tag{4.10}$$

The probability distribution $P_2$ for the total number of electrons $N$ after the second step $x_1 + x_2$ can then be calculated by adding up the probabilities for $i = 1, 2, ...N$ electrons being produced in the first step, each multiplied with the convoluted probability of the $i$ electrons being amplified to exactly $N$ electrons in the second step.

$$
\begin{aligned}
P_2(N) &= P_1(1) \cdot P_1(N) + P_1(2) \cdot \left( \sum_{j=1}^{N-1} P_1(j) \cdot P_1(N-j) \right) + \\
&+ P_1(3) \cdot \left( \sum_{k=1}^{N-2} \sum_{j=1}^{k} P_1(j) \cdot P_1(k) \cdot P_1(N-k-j) \right) + ... \\
&= \sum_{i=1}^{N} P_1(i) \cdot P_1^{*i}(N),
\end{aligned} \tag{4.11}
$$

where $P_1^{*i}$ is the i-fold convolution of $P_1$. Induction yields for a subsequent 3$^{\mathrm{rd}}$ and l$^{\mathrm{th}}$ step:

$$
P_3(N) = \sum_{i=1}^{N} P_2(i) \cdot P_1^{*i}(N) \quad \text{and} \quad P_l(N) = \sum_{i=1}^{N} P_{l-1}(i) \cdot P_1^{*i}(N) \tag{4.12}
$$

Based on the distribution $P$ its first moment $S_1(P) = \overline{P}$, corresponding to its mean, can be derived. For the electron distribution after the second step we obtain:

$$
\begin{aligned}
\overline{P_2} &= \sum_{i=1}^{\infty} iP_2(i) = P_1(1) \sum_{i=1}^{\infty} iP_1(i) + P_1(2) \sum_{i=1}^{\infty} iP_1^{*2}(i) + P_1(3) \sum_{i=1}^{\infty} iP_1^{*3}(i) + ... \\
&= P_1(1)\overline{P_1} + P_1(2)\overline{P_1^{*2}} + P_1(3)\overline{P_1^{*3}} + ... \\
&= P_1(1)\overline{P_1} + P_1(2) \cdot 2 \cdot \overline{P_1} + P_1(3) \cdot 3 \cdot \overline{P_1} + ... \\
&= \overline{P_1} \sum_{i=1}^{\infty} iP_1(i) = \overline{P_1}^2
\end{aligned} \tag{4.13}
$$

using in (4.13) that the mean is a cumulant under convolution and, therefore, $\overline{P_1^{*i}} = i\overline{P_1}$. The mean of the distribution after subsequent steps can be calculated similarly and induction yields

$$
\overline{P_3} = \overline{P_1}^3 \quad \text{and} \quad \overline{P_l} = \overline{P_1}^l, \tag{4.14}
$$

confirming the expected exponential growth of the distributions mean. The more interesting parameter is the distributions second moment $S_2$

$$
S_2(P_2) = \sum_{i=1}^{\infty} i^2 P_2(i) = \sum_{i=1}^{\infty} P_1(i) S_2(P_1^{*i}) \tag{4.15}
$$

since it relates to the distributions variance with $S_2(P) - (S_1(P))^2 = RMS^2(P)$. For the variance of the distribution after the second avalanche step one obtains

$$\begin{aligned} RMS^2(P_2) &= S_2(P_2) - (S_1(P_2))^2 \\ &= \sum_{i=1}^{\infty} P_1(i)\left((S_1(p_1^{*i})^2 + i\left(S_2(P_1) - (S_1(P_1))^2\right)\right) - (S_1(P_2))^2 \\ &= \sum_{i=1}^{\infty} P_1(i)\left(i^2(S_1(p_1)^2 + i\left(S_2(P_1) - (S_1(P_1))^2\right)\right) - (S_1(P_1))^4 \\ &= S_1(P_1)\left(1 + S_1(P_1)\right)\left(S_2(P_1) - S_1^2(P_1)\right) = \overline{P_1}(1+\overline{P_1})RMS^2(P_1) \end{aligned} \tag{4.16}$$

which can be generalized to

$$RMS^2(P_3) = \overline{P_1}^2(1 + \overline{P_1} + \overline{P_1}^2)RMS^2(P_1) \tag{4.17}$$

and

$$RMS^2(P_l) = \overline{P_1}^{l-1}\sum_{i=0}^{l-1}\overline{P_1}^i \cdot RMS^2(P_1) \tag{4.18}$$

The relevant parameter effecting the detector's energy resolution is the relative width $\sqrt{f}$ or rather the relative variance $f$ of the electron-number distribution:

$$\begin{aligned} \sqrt{f} = \frac{RMS(P_l)}{\overline{P_l}} &= \frac{RMS(P_1)\sqrt{\overline{P_1}^{l-1}\sum_{i=0}^{l-1}\overline{P_1}^i}}{\overline{P_1}^l} \\ &= \frac{RMS(P_1)}{\overline{P_1}}\sqrt{\sum_{i=0}^{l-1}\frac{1}{\overline{P_1}^{l-i-1}}} = \frac{RMS(P_1)}{\overline{P_1}}\sqrt{\frac{1-\frac{1}{\overline{P_1}^l}}{1-\frac{1}{\overline{P_1}}}} \end{aligned} \tag{4.19}$$

Derived for natural numbers $l$, the formalism can without constraint be applied to rational numbers, as can be seen by increasing the number of steps $n$ and re-group the steps arbitrarily. From a physics perspective, however, a lower limit is set for the step size, since an avalanche requires a minimal distance to grow and its statistic for $x_1 \lesssim \lambda_{ion}$ does not yield physically meaningful results.

### 4.4.2. Simulation Studies based on Avalanche Extrapolation

In our application, we simulated the growth of an avalanche along the 160 µm long amplification gap of the Micromegas used in the Single Electron Response (SER) measurements reported on in section 4.5 and published in [85]. With the requirement to cover a broad range of electric fields or rather gas gain of 50 to $5{\cdot}10^5$, we applied the avalanche extrapolation method described in the previous section 4.4.1.

The avalanche formation has been simulated over the full growth length $z$ for comparatively weak amplification fields ($E_A \leq 28\,\text{kV/cm}$), yielding a gain $G_{full}(E_A)$ of up to $10^4$. In parallel the amplification process has been simulated for a reduced step length $z_1$ within the full electric field range up to 35 kV/cm, yielding $G_{step}(E_A)$. Both gain values are extracted via a Polya fit, according to (4.6) with an additional normalization parameter, on the electron number distributions obtained from the corresponding step- or full simulation. Relying on the well established exponential growth of the avalanche derived in (4.13) and (4.14), we extract the avalanche growth exponent $\xi$ between the full simulation with respect to the shorter step.

$$\xi(E_A) = \frac{\ln G_{full}(E_A)}{\ln G_{step}(E_A)} \tag{4.20}$$

This factor $\xi$ can be used to predict the mean (4.14), $RMS$ (4.18) and form factor $\sqrt{f}$ (4.19) of the full avalanche distribution $P_\xi$ based on the results from the step simulation $P_1$. The errors in $G_{full}$ and $G_{step}$ were propagated to estimate the statistical uncertainty of $\xi$. Similarly the Polya parameter $\Theta$ including its uncertainties could be determined from the fit to the step- and full simulation data. Their uncertainties are propagated to the relative variance $f_{step}$ calculated by (4.7) and combined with the uncertainty of $\xi$ in the extrapolation of $\sqrt{f_{full}}$ according to (4.19).

Two validation studies were performed to understand the robustness of the model and test for its limits. To probe for effects induced by the gas the simulation was carried out for three mixtures of Helium, Neon and Argon, each with an admixture of 5 % Isobutane, according to the corresponding experimental setup described in section 4.5.

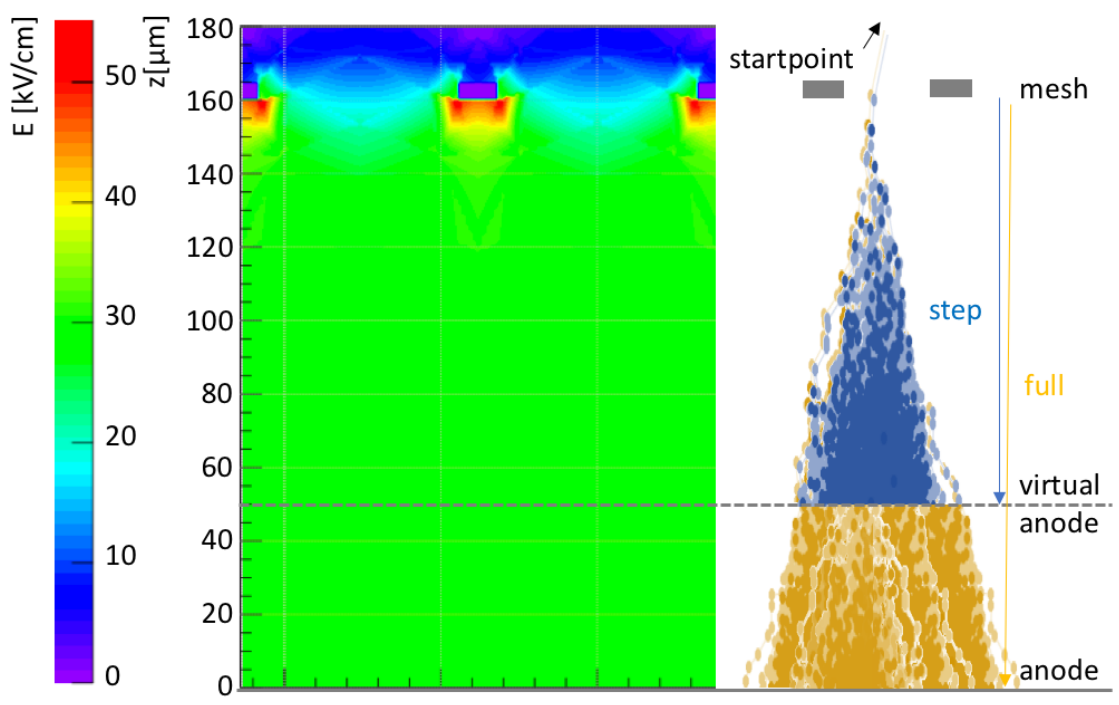

**Figure 4.20.:** Scheme of the first avalanche extrapolation study based on a FEM model of the Micromegas utilized in the experiment in 4.5. The step length of 110 µm is selected and electrons are started in the drift Volume.

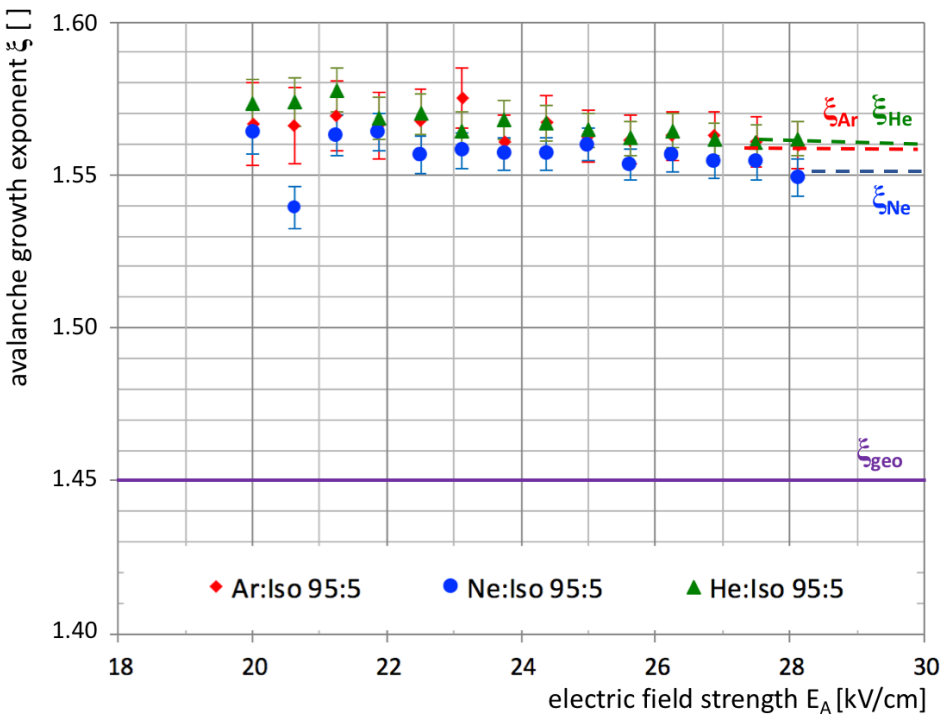

**Figure 4.21.:** Avalanche growth exponent $\xi$ derived from step- and full avalanche simulation in a FEM model Micromegas as a function of the electrical field strength in the center of the amplification gap. Colors represent the three gases.

In the first study we simulated the Micromegas in full geometric detail (figure 4.20) using FEM methods and thus including the non-uniformity of the amplification field. The initial electrons were seeded in the drift region with a small randomized momentum to ensure they reached energy equilibrium according to the field before entering the amplification volume. Their growth into an avalanche was simulated along the 160 µm long amplification gap (full simulation) or up to a 'virtual anode' at $z = 50\,\mu m$, resulting in an 110 µm distance from the bottom of the micromesh to the anode (step simulation). The resulting growth exponents $\xi$ calculated by (4.20) are shown as a function of the field strength in figure 4.21.

While for low electrical field strength the values fluctuate stronger, they converge for large $E_A$ in a region of $1.55 < \xi_{eff} < 1.56$. A discrepancy between the convergence limits for the different gas mixtures can be observed, but remains within the statistical uncertainty. It is noticeable that all values differ significantly from the geometrically expected growth exponent $\xi_{geo} = z/z_1 = 1.45$, with $\xi_{eff} > \xi_{geo}$. This indicates a reduced effective avalanche growth length, which calculates to $z_1 \approx 103\,\mu m$ assuming the $\xi_{eff}$ extracted from figure 4.21.

Following the electrical field strength along the electrons' most probable path through the center of the mesh aperture, the non-uniform region of the electric field reaching into the amplification gap (figure 4.20 - left) is clearly visible. Along these first $\leq 10\,\mu m$ the field strength increases continuously and so does the ionization probability of the electron. With the increased probability of non-ionizing collisions due to the comparatively weak electric field, the avalanche's effective starting point is shifted into the amplification gap and, therefore, the effective growth length is reduced with respect to the geometrically expected position.

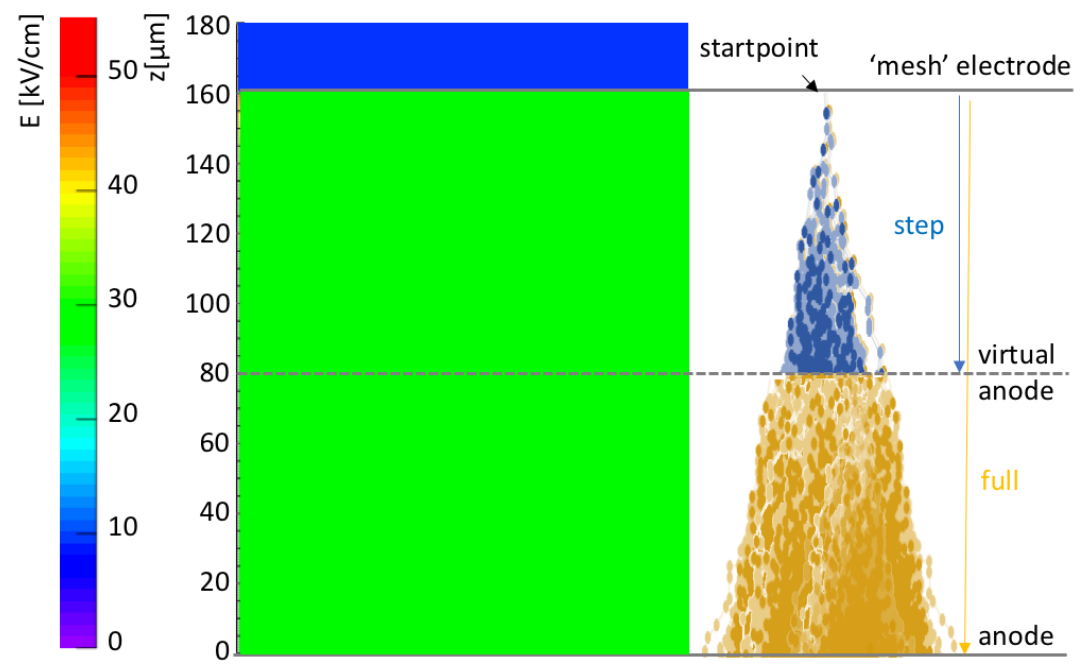

**Figure 4.22.:** Scheme of the second avalanche extrapolation study based on a parallel plate approximation of the Micromegas. The step length of 80 µm is selected and electrons are started at the edge of the amplification volume.

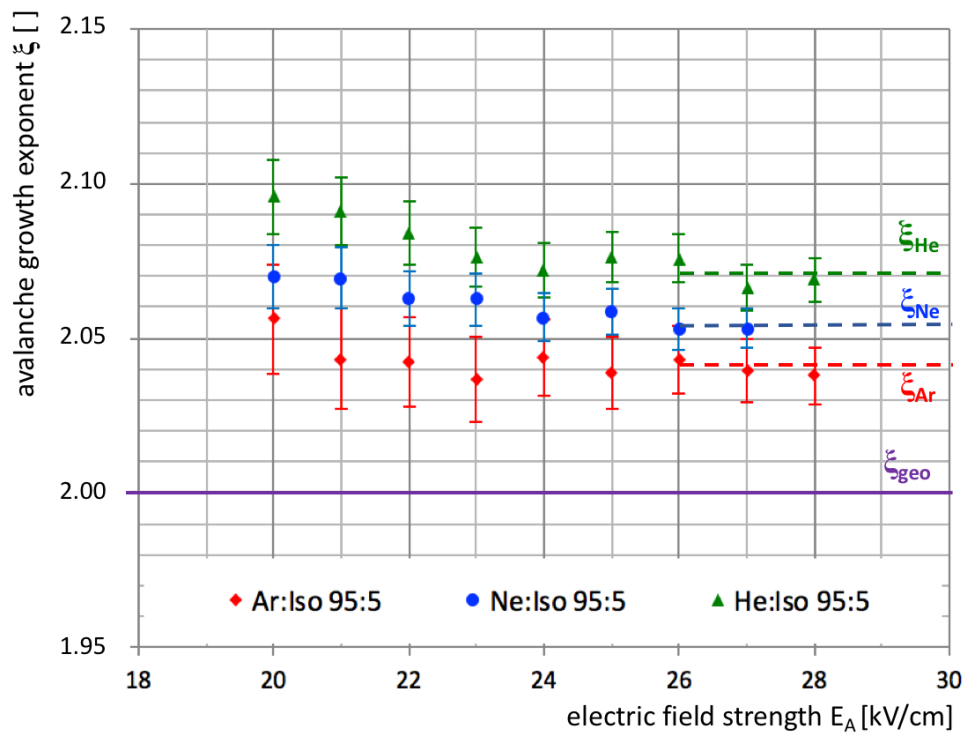

**Figure 4.23.:** Avalanche growth exponent $\xi$ derived from step- and full avalanche simulation in a uniform field as a function of the electrical field strength in the center of the amplification gap. Colors represent the three gases.

Since this effect is a result of the non-uniformity of the amplification field in a real Micromegas, it should vanish in the simulation of avalanche growth in a uniform field. This parallel plate (PP) approximation was used for the second validation study (figure 4.22), where the step simulation has been limited to the first half of the 160 µm long amplification gap by a virtual anode at $z = 80\,\mu m$. Both measures, the simplification of the field geometry and the reduction of the step size, contributed significantly to a reduction of computation time. The initial electrons were started at the edge of the amplification region with a directed momentum towards the amplification gap, corresponding to a starting energy $\epsilon_0$. The latter has been extracted from preceding simulations of the drift process by measuring the energy distribution of electrons at the moment of their transition from the drift- into the amplification region ($z = 160\,\mu$m). Depending on the drift field strength and the gas mixture the mean values vary in the range of $3\,\mathrm{eV} < \epsilon_0 < 6\,\mathrm{eV}$.

The resulting growth exponents according to (4.20) are shown in figure 4.23. Similar to the previous case, the exponent fluctuates for low amplification fields and stabilizes in the range of $2.04 < \xi_{eff} < 2.07$, with a systematic difference for the three gas mixtures. All values differ again from $x_{geo} = 2$ by 2.0 - 3.5 %. The corresponding effective avalanche growth length $z_1 \approx 78\,\mu$m is reduced by $2\,\mu$m with respect to the geometrical step length. Since a non-uniformity of the field can be excluded as an explanation, the differences between the full simulation of the avalanche and the step extrapolation was again scrutinized. While the starting energy $\epsilon_0$ is the same in full and step simulation, the mean electron energy at the virtual anode at $z = z_1 = 80\,\mu$m differs in both scenarios: in a free full development of the avalanche the electrons' energy $\epsilon_{z_1}$ while crossing the $z_1$-plane follows a distribution according to the field strength $E_A$ with a mean typically in O(10 eV) for the Argon based mixture and above for Neon and Helium, due to their increased ionization thresholds. Utilizing the extrapolation method, the electron's energy is artificially reset to $\epsilon_0 < \epsilon_{z_1}$, causing a net loss of energy. The energy uptake of an electron in a $E = 25\,\mathrm{kV/cm}$ field corresponds to $2.5\,\mathrm{eV}/\mu$m and, hence, an average initial electron travels $\approx 2\,\mu$m in Argon and again slightly further in Neon and Helium before reaching a condition comparable to those trespassing the $z_1$-plane. Similarly to the effect of a non-uniform field, this leads to a spatial delay of the first ionization and thus a reduced effective avalanche growth length. The discrepancy in the growth exponent between the gas mixtures reflects their different ionization threshold. In the first study the offset in the growth exponent is dominated by the non-uniformity in the field, allowing for a continuous energy uptake of the electrons, the effect of starting energy difference is suppressed and hence this ordering effect does not appear.

Since the starting energy $\epsilon_0$ is not an arbitrary parameter but transfers information from the end condition of the drift process to the amplification process, it has been kept according to the physical meaningful values, instead of enforcing a $\xi = 2$ by increasing $\epsilon_0$. As the extrapolation method does not constrain the growth exponent to natural numbers, as discussed in 4.4.1, it automatically mitigates the effect of reduced growth in the first step, either caused by field non-uniformity, or reduced electron starting energy.

The stability of the extrapolation exponent and a good understanding for its gas dependent behavior alone are without greater use, unless the extrapolation method for the distributions second momentum, derived in (4.18) and (4.19), can be validated. Figure 4.24 shows the comparison of the directly extracted $f_{full}$ for $E_A \leq 28\,\mathrm{kV/cm}$ with the relative width obtained by extrapolating $f_{step}$ with $\xi_{eff}$ of the corresponding gas. Within their $\pm 1\,\sigma$ uncertainty all but one (He:Iso $|_{27\,\mathrm{kV/cm}}$) values are in agreement with the extrapolated values.

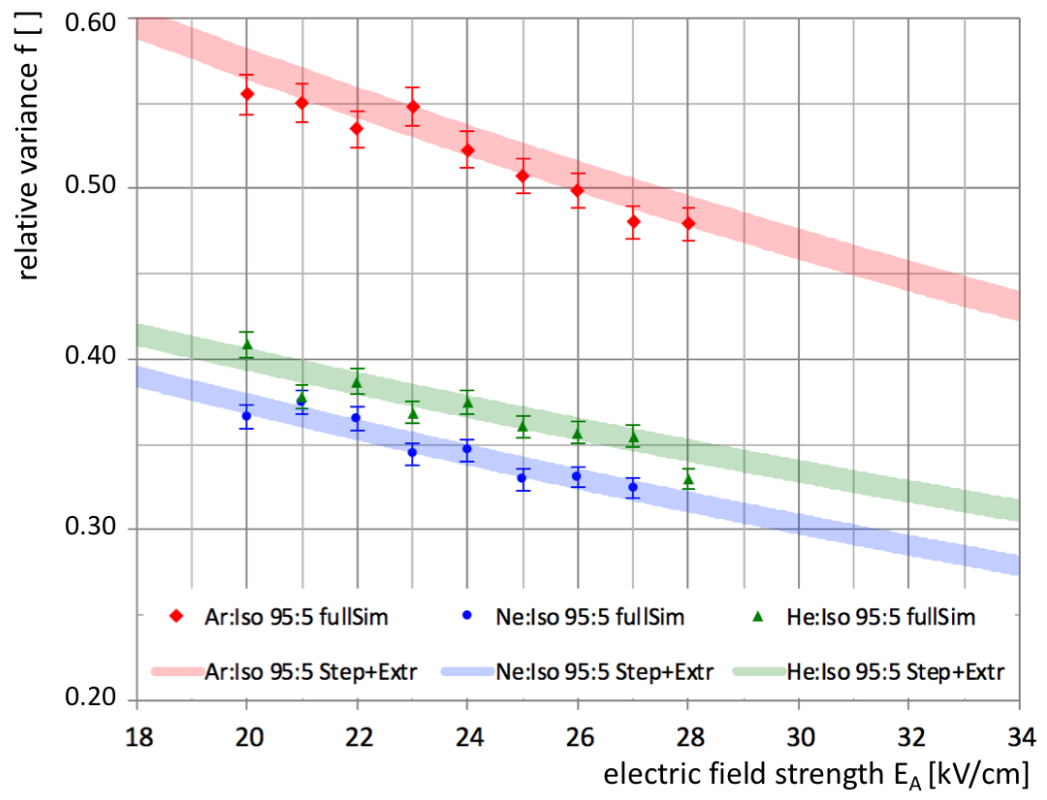

**Figure 4.24.:** Comparison of the relative variance $f = RMS^2(P)/\overline{P}$ extracted directly from the full simulation electron distribution (colored markers with error bars) to the prediction obtained by step simulation and statistic extrapolation, fitted with an exponential curve (color shaded error band).

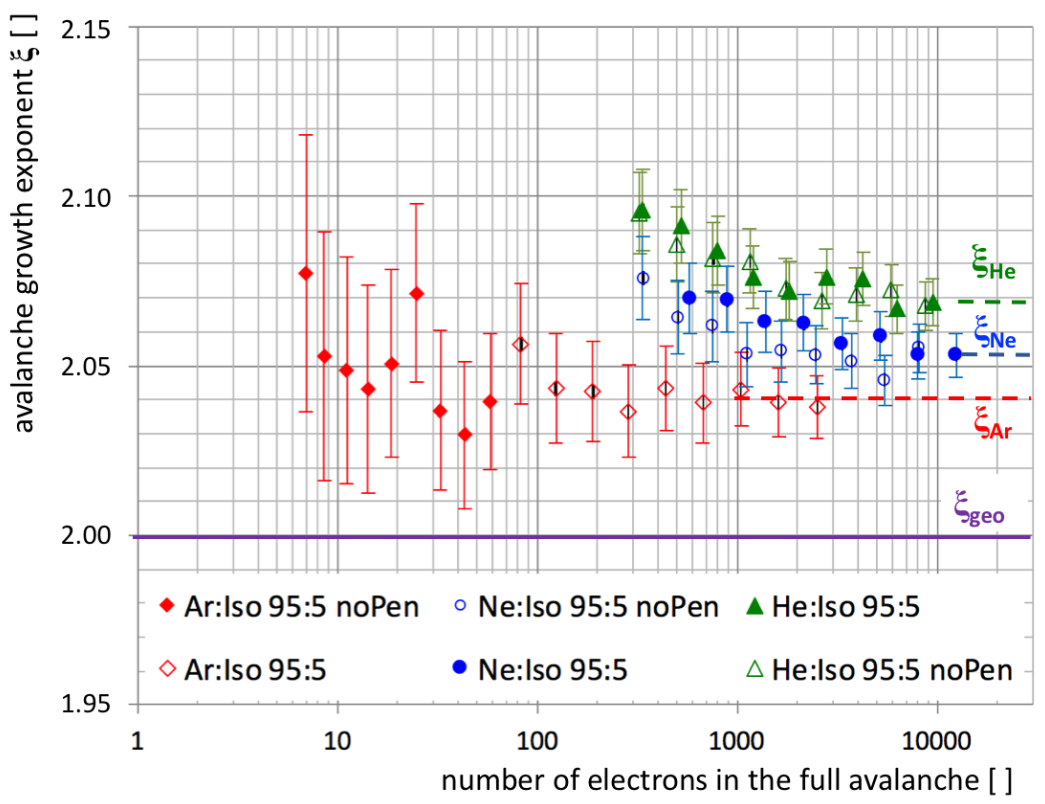

**Figure 4.25.:** Avalanche growth exponent $\xi$ derived from step- and full avalanche simulation in a uniform field as a function of the mean electron yield of the full avalanche. Different colors represent the three gas mixtures, empty / full markers correspond to simulation including / neglecting Penning transfer.

As an additional cross-check, the second validation study has been widened to a comparison of avalanche formation with and without penning transfer. While differences are comparably small for Neon and Helium, the increase in gain is substantial in the Argon mixture. Figure 4.25 shows the high fluctuation of $\xi$ for small avalanche sizes calculated without penning transfer. As the underlying condition of the extrapolation method is an avalanche development length larger than several mean ionization length $\lambda_{ion}$, and this condition is violated for very small avalanches yielding only a few electrons, this increase is expected. In both studies it can be seen that the statistical uncertainty decreases with increasing field strength. In a comparison between the gases in figures 4.21 and 4.23 it appears that the Argon values have an increased statistical error. This is due to the reduced ionization yield in argon compared to the other two mixtures. The errors are mainly determined by the size of the avalanches. Plotting the growth exponent $\xi$ as a function of the mean gain of the full avalanche, as shown in figure 4.25, allows for a direct comparison.

The second model has been selected for the comparison of simulated avalanche growth with experimental results from single electron response mainly due to the reduced consumption of computational resources allowing to span a wider range in electric fields and an increased statistic required for direct comparison of the electron yield distributions. The results of this comparison will be discussed in the next section 4.5 and have been published in [85].

## 4.5. Avalanche Statistics in Single Electron Response (SER)

Experimentally all signal formation processes start with the primary ionization. Therefore, it is impossible to observe and measure subsequent processes directly and independently of the primary ionization. For the measurement of mean quantities like the gas gain $G$ or the drift velocity $v_D$ a higher number of signal electrons can be beneficial to mitigate the impact of other statistically distributed processes like electron losses. In contrast the primary ionization statistic is strongly perturbing the assessment of statistical properties, like the gain fluctuation. A work around is provided by a **S**ingle **E**lectron **R**esponse (SER) measurement, where the signal caused by a single primary electron can be directly observed. The statistic of such a primary ionization process has been scrutinized in chapter 2.2.3. It has been concluded, that despite aiming for a single electron, the occurrence of multiple electron events can not be completely neglected, but efficiently suppressed depending on the experimentally adjustable non-zero event probability.

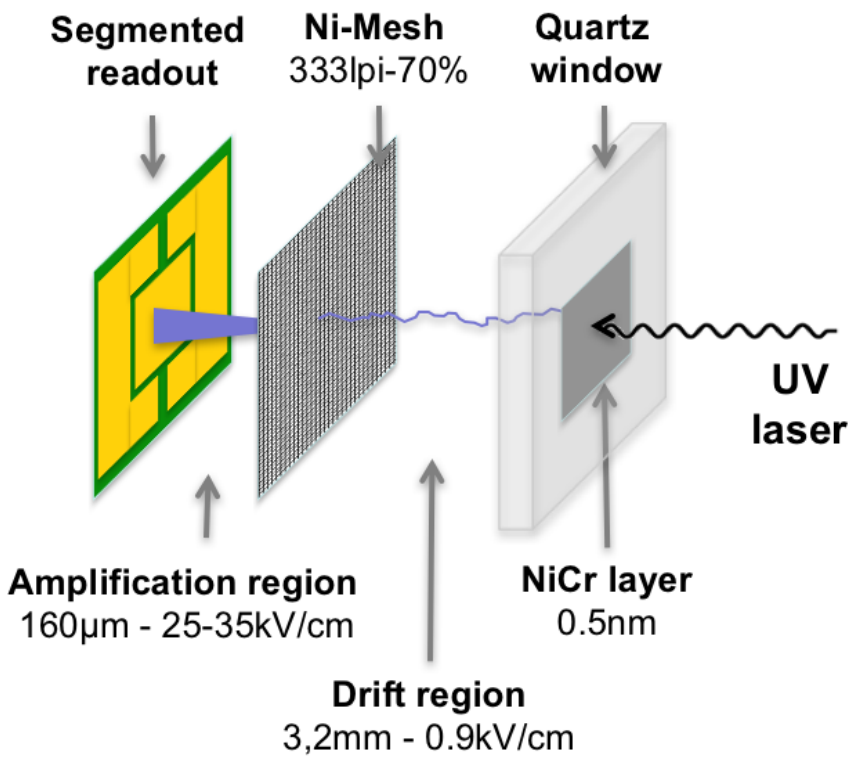

**Figure 4.26.:** Schematic of the SER setup: An UV laser illuminates the nickel-chromium cathode, frees a single photo electron which drifts to the mesh and triggers the formation of an avalanche.

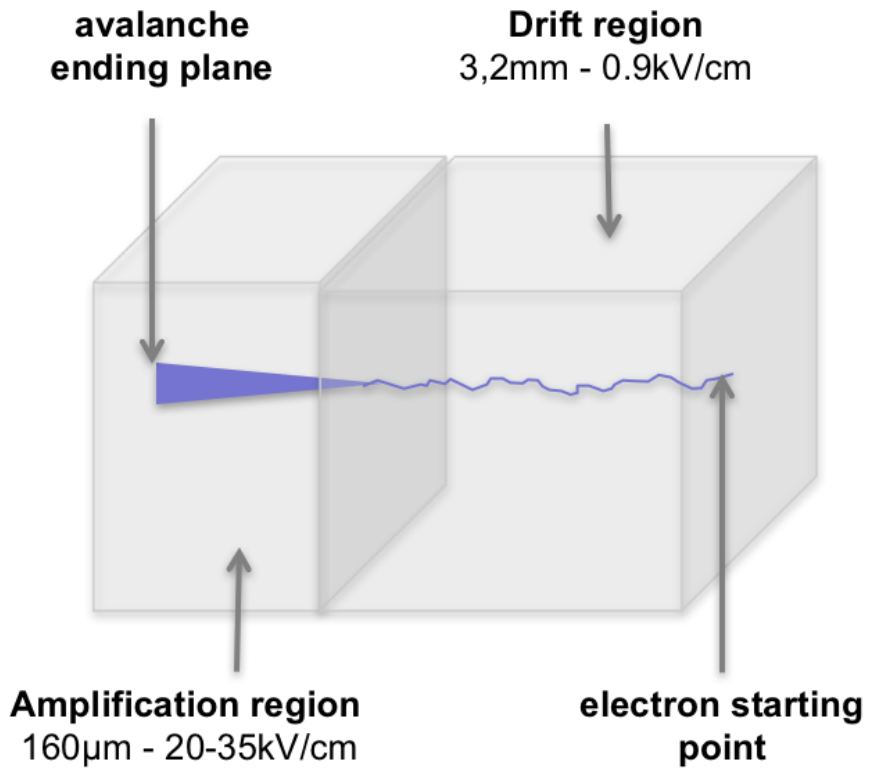

**Figure 4.27.:** Schematic of the MC simulation method for SER: An electron is started at the cathode and drifts to the mesh. Its amplification is a separate simulation step.

An experimental realization of the SER approach, schematically shown in figure 4.26, has been built by the group of T. Zerguerras (University Paris-Sud, Orsay) [87]. In the scope of this more recent study the setup has been refurnished and enhanced with a more precise gas system and electronics with reduced noise level, down to 200 $e^-$ / 380 $e^-$ RMS in stand alone mode or connected to the Micromegas detector. This Micromegas features a flat copper anode pad, enclosed by two veto brackets and a very precise electroformed nickel mesh with 333 lpi and an optical transparency of 70 % held at a distance of 160 µm by nylon spacers. Therefore, it is a very close approximation to a parallel plate geometry. The SER is realized by a 337 nm pulsed laser focused through a quartz window onto the 0.5 nm thin nickel-chromium layer on the detector's cathode. The non-zero event rate has be adjusted to $p = 0.05$ by attenuation of the intensity of the laser beam, which is used before ahead to cause the trigger signal. For reference measurements a $^{55}$Fe source is included in the setup and can be moved for calibration measurements in front of the detector. A full description as well as a technical drawing can be found in [85].

Complementary to the experimental measurement the author conducted simulation studies, to predict the single electron spectra, showing the (probability-) distribution of the measured gas gain in SER. The simulation followed a two stage scheme visualized in figure 4.27.

The drift process was simulated separately and the required electron transition parameters like the energy distribution $\epsilon$ were determined. Due to high gas purity and mesh transparency the electron losses are almost zero. Furthermore, any electron loss mechanism would in first order only lead to an increase in the zero-event rate and, therefore, not spoil the measurements. In higher order the statistic on multiple electron events derived in chapter 2.2.3 are perturbed, requiring a differentiation between the rate of initially freed electrons and the rate of amplified electrons. This effect negligibly small for the expected tiny fraction of lost electron.

The results from the drift process were used to determine a representative start condition for the electrons in the amplification stage. The latter was simulated in a PP approximation utilizing the statistical extrapolation method discussed in section 4.4.2 with the half step scheme. The penning transfer rates were determined in a preceding simulation run discussed in section 4.2.2.

Following the established threefold scheme, described in chapter 1.3.3, the experimental data was additionally compared to the analytic description developed by Ö. Sahin [74, 88] and discussed in section 4.1. Here a three parameter fit (penning transfer rate $r$, second Townsend coefficient $\beta$ and a gain scaling factor) was used to reach agreement with the experimental data. Experimental measurement, simulation and analytic analysis were performed for three gas mixtures of 95 % Helium, Neon or Argon with 5 % Isobutane, experimentally limited to the operation voltage range of the detector.

### 4.5.1. Gain Comparison between Experiment, Simulation and Analytic Model

The experimentally measured gain as a function of the amplification field is compared to the fit based on the analytic model and the simulation, as shown in figure 4.28 for the three gas mixtures.

In all three gases the experimental data is non-surprisingly best described by the analytic fit which includes a gain scaling and the second Townsend coefficient $\beta$ as additional fit parameters (black line). This shows the non negligible impact of secondary avalanches on the measurement. While the fit parameter is meant to describe secondary avalanches caused by photon- and to a lesser extent ion-feedback, the experimental data contains a feedback contribution as well as the occurrence of multi-electron events. Experimentally those are not distinguishable within one signal.

On the level of $\beta = 0$ (blue lines) the analytic model and the simulation yield similar values with a slight systematic towards higher gain values in the MC simulation. This can be understood by considering the reference measurement for the penning trans-fer rate determination (see table 4.1). While the simulation assumes zero secondary avalanches in this calibration, the small but not negligible impact of these at the low reference voltages are already reflected in the contribution of $\beta G$ in the analytic model. Therefore, the analytic model under suppression of the feedback contribution yields slightly lower values. The relative development of the avalanche growth with increasing voltages is, however, described similarly in both approaches.

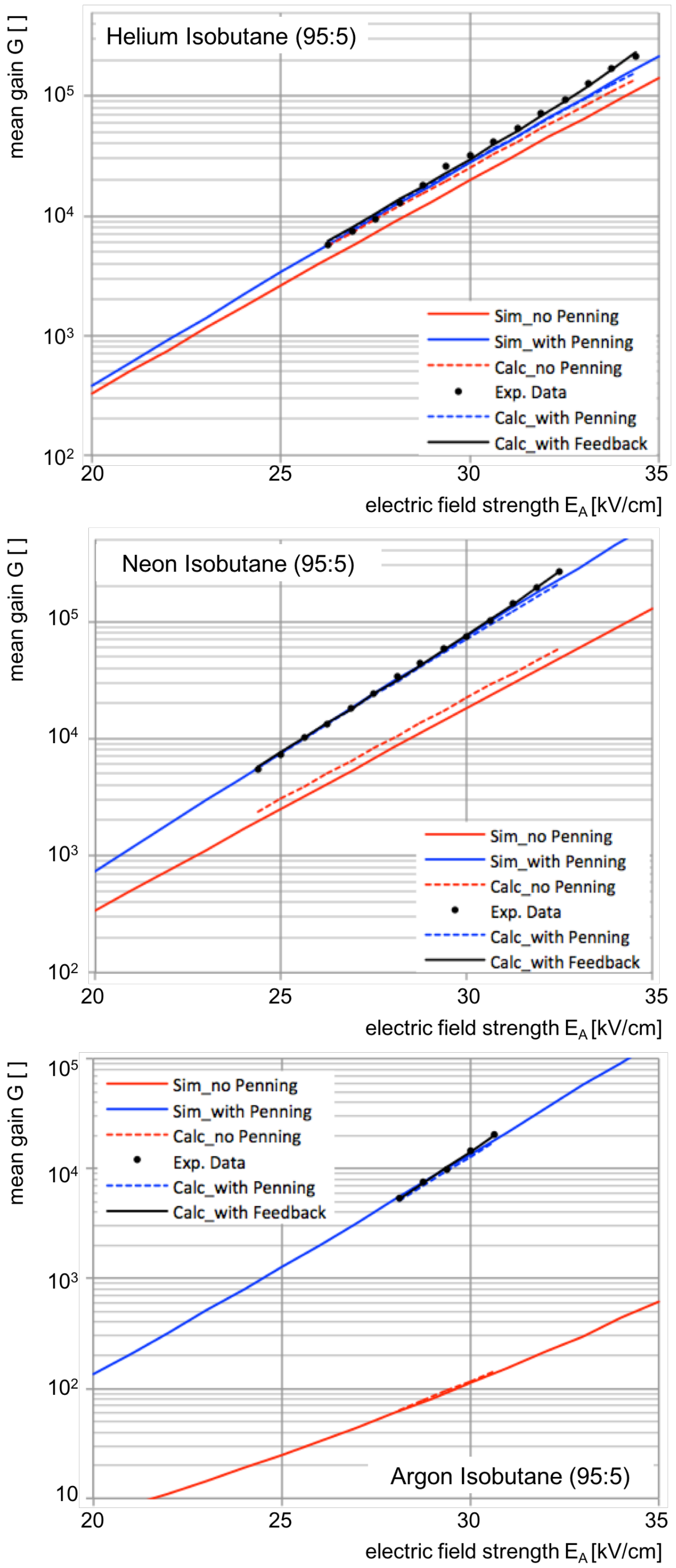

**Figure 4.28.:** Mean gain $G$ comparison between experimental data (dots) with the analytic fit (black line), the reduced analytic model (dashed lines) and the simulation results (full lines), under consideration (blue) or suppression (red) of the penning effect. Shown for 95 % Helium (top), Neon (center) and Argon (bottom) mixture with 5 % Isobutane.

The systematic discrepancy with inverted order is observed for the values under suppression of penning transfer mechanisms (red lines). It is due to the addition of a third gain scaling parameter $\gtrsim 1$ in the analytic model. This re-scaling reduces the estimated contribution of penning transfer to the avalanche size and, therefore, the relative increase of the gain by allowing penning transfer (difference red to blue) is reduced. The discrepancy becomes less pronounced for a less probable penning transfer, as can be seen in a comparison between the gases.

For the Helium mixture, the increase in gain due to the penning transfer is small compared to Neon and Argon. The latter is showing an immense contribution to the total gain by almost 99 %, in other words: Only each hundredths electron is freed via direct ionization while 99 % origin from penning transfer between the excited Argon to an Isobutane molecule. The relative difference in the effect strength between the three gases can be understood by considering their energy schemata. Helium only allows for a few excited states, all of which have a comparably energy threshold, much above the ionization energy of Isobutane ($E_{ion,\mathrm{Isob}} = 10.67\,\mathrm{eV}$). Therefore, large scattering energies are required to excite a Helium atom and the corresponding rate is rather small, compared to the direct ionization rate in the 5 % Isobutane admixture. This energy gap is significantly smaller for Neon, yielding a higher probability of Neon excitement and consequently penning transfer. For Argon the lowest excitation threshold is only slightly above the Isobutane ionization energy. Given the almost twenty fold abundance of Argon atoms compared to Isobutane molecules, excitation of Argon become much more probable than direct ionization. Although only a fraction of the excited states yields a free electron via Penning transfer, this becomes the dominant contribution to the avalanche formation.

### 4.5.2. Assessment of Gain Fluctuation in Experiment and Simulation

The relative gain variance $f$, defined in (4.5) and expressed in terms of the Polya parameter $\Theta$ in (4.7), is a measure of the gain fluctuation during avalanche formation. As discussed in chapter 1.3.1 it is one of the main determinants for the energy resolution of a detector and in turn depends on the gas, the electrical field strength and the avalanche development length, as shown in figure 1.7.

In the SER experiment, the relative fluctuation can be directly measured by extracting the Polya parameter from the fit to the measured spectrum. While up to now no analytic theory properly describes the gain fluctuation, it can be predicted by MC simulation. Figure 4.29 shows the comparison between experimental data and the results of the simulation study.

Under suppression of penning transfer mechanisms (red lines), the simulation yields a systematically increased variance, compared to the simulations where penning-effect is enabled (blue lines). This is a consequence of the energy dissipation occurring when penning transfer is forbidden. If all energy stored in the electric field would be directly transfered to ionization, the gain fluctuation would converge to zero. The more statistical processes of energy dissipation occur, the more increases the fluctuation. Some of these processes are inevitable and may be even welcome to cool the gas, like the energy dissipation caused by quenching gases with vibration and rotation states. However, adding another energy dissipation mechanism by forbidding energy transfer from excited states to ionization artificially increases the obtained gain fluctuation. This effect is much more pronounced for gas mixtures with a larger fraction of the electrons being freed due to Penning transfer. In figure 4.29 the relative variance changes most pronounced in the comparison with / without penning transfer for the Argon mixture, followed by Neon and Helium. This is in agreement with the penning effect strength order observed and discussed in the previous section 4.5.1.

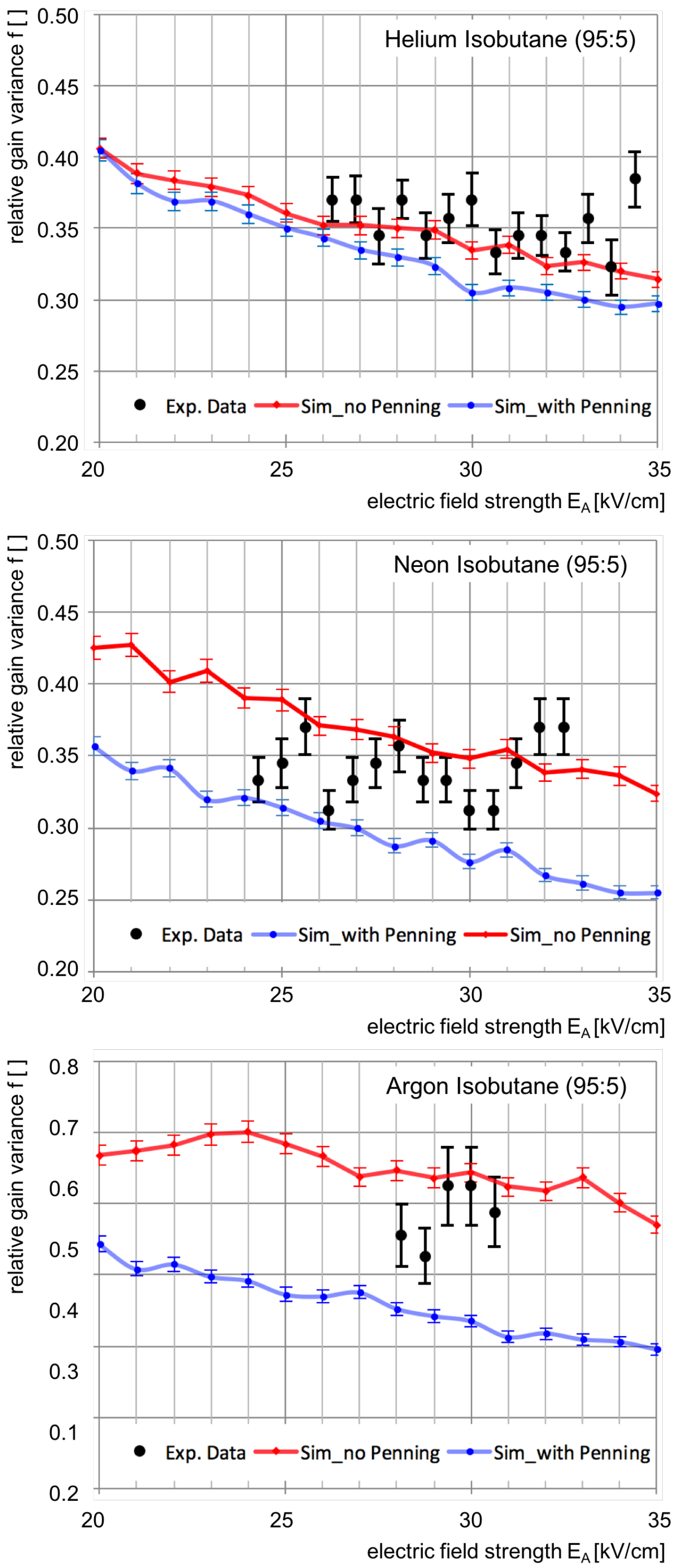

**Figure 4.29.:** Relative gain variance $f$ as a function of the amplification field. Experimental data (black dots + uncertainty) is compared to the simulated values (color dots + uncertainties and lines for optical guidance) under consideration (blue) or suppression (red) of the penning effect. Shown for 95 % Helium (top), Neon (center) and Argon (bottom) mixture with 5 % Isobutane.

Furthermore, the relative gain variance is systematically lower in the simulation with penning transfer compared to the experimental data. This disagreement becomes more pronounced with increasing electric field, or rather gain. While the simulation yields the relative variance in an ideal single electron response, the experimental data always includes the occurrence of multi-electron events and secondary avalanches, the latter increasing in rate with higher gain. Both effects increase the gain variance and, therefore, the experimental values must be interpreted as an upper limit for the real relative gain variance in the gas.

### 4.5.3. SER Spectra Comparison and Conclusive Results

A deeper insight to the discrepancy can be obtained by direct comparison of the experimentally measured and the simulated gain spectra. Such a comparison for each of the gas mixtures is shown in figure 4.30. The experimental data (black) is fitted (blue) with a Gaussian contribution for the noise introduced by the zero-electron events and a Polya distribution according to (4.6). The simulation results (red) are normalized to the experimental data but contain no other scaling or adjustment parameter. The agreement for all three gases is on an unprecedented level.

Slight systematic discrepancies are visible on the large avalanche tail for high $N_e$, where the simulation prediction is decreasing faster than the experimental data. This is again consistent with the above discussion on the bias caused by secondary avalanches. Accordingly, the Polya fit is biased by these high tail events and, therefore, the width of the Polya distribution $\Theta$ and ultimately $f$ is slightly increased. This leads to the systematic discrepancy in the above discussed relative gain variance.

For Neon the increased gain ($G_{Ne} = 1.3 \cdot 10^4$) allows for an easy separation of the noise contribution and the avalanche statistic fitted with the Polya curve. This separation is more pronounced in Helium compared to Argon, although both spectra correspond to the approximately same gain ($G_{He} = 5.7 \cdot 10^3$, $G_{Ar} = 5.5 \cdot 10^3$). This indicates another gas inherent property, as well visible in the comparison of the sub-figures of 4.29. Among the three gas mixtures Argon features a significantly higher relative variance. Accordingly, the width of the Polya curve is wider and the distinction between noise and signal is reduced.

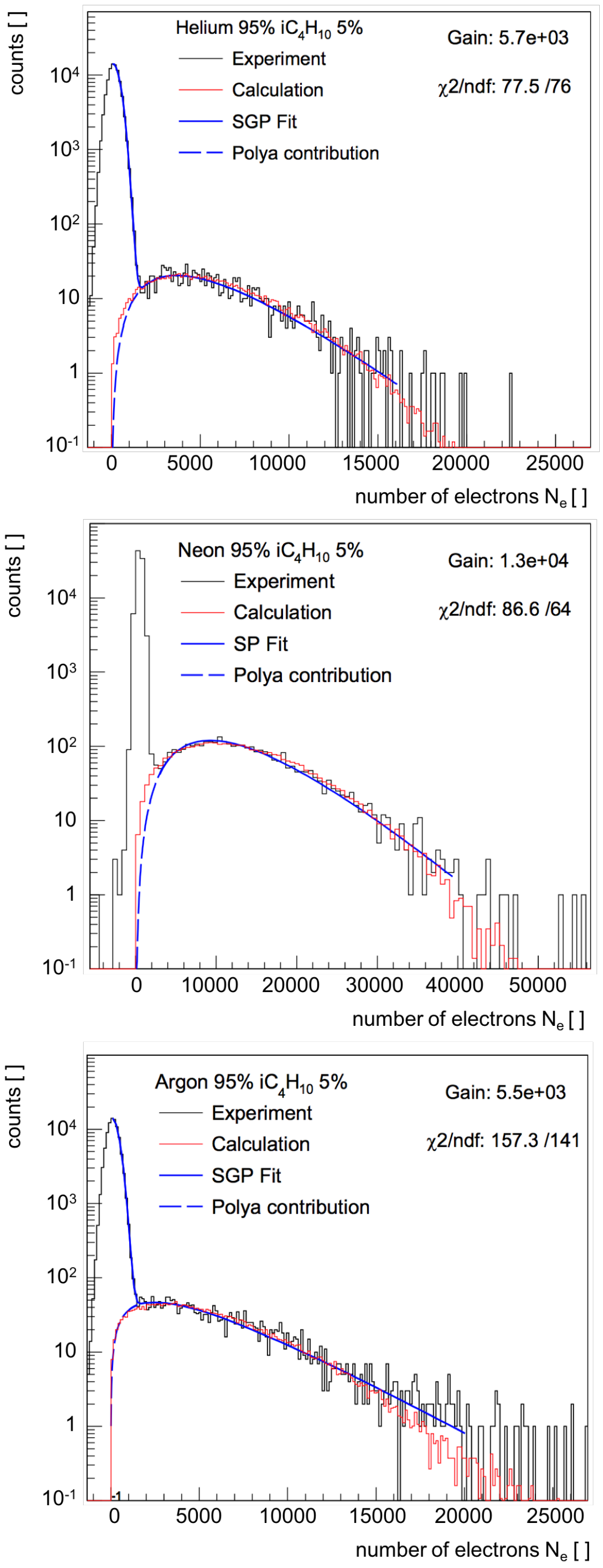

**Figure 4.30.:** Experimental SER spectra together with the best-fit curves (blue contineous lines), the contribution of the Polya distribution (blue dashed lines), and the simulated SER distribution (red lines). The $\chi^2$/ndf ratio refers to the fit likelihood. Shown for 95 % Helium (top), Neon (center) and Argon (bottom) mixture with 5 % Isobutane. [85]

The conclusive comparison of gain and relative gain variance between the three gas mixtures is shown in figures 4.31 and 4.32. The Argon mixture yields the lowest gain and the highest relative variance. This is in agreement with early predictions by Alkhazov [82] attributing the increased variance to the lower ionization threshold. The reduced gain similarly can be explained by the higher number of energy dissipation mechanisms close to the ionization threshold of the second gas species. For the other two mixtures the gain hierarchy is inverted compared to their ionization energy. With the energy gap between excitation states of the noble gas and ionization threshold of Isobutane molecules being sufficiently large, energy dissipation via the excited states of the noble gas becomes less dominant in both cases. Neon mixtures, however, feature an increased drift velocity due to a larger mean free path. Accordingly, the electrons scatter with a higher mean energy, yielding an increased ionization probability and a higher gas gain, compared to Helium. While it seems that the variance hierarchy is inverted as well, this can be found to be an artifact of the comparison scale. With the relative gain variance decreasing with increasing avalanche size, as can be seen for example in figure 4.5 and implicitly in figure 4.29, a more representative comparison should refer the corresponding gain, not the electric field strength. Correcting for this effect, the relative gain fluctuation at a given gain follows the expected hierarchy and is largest for Argon followed by Neon and Helium.

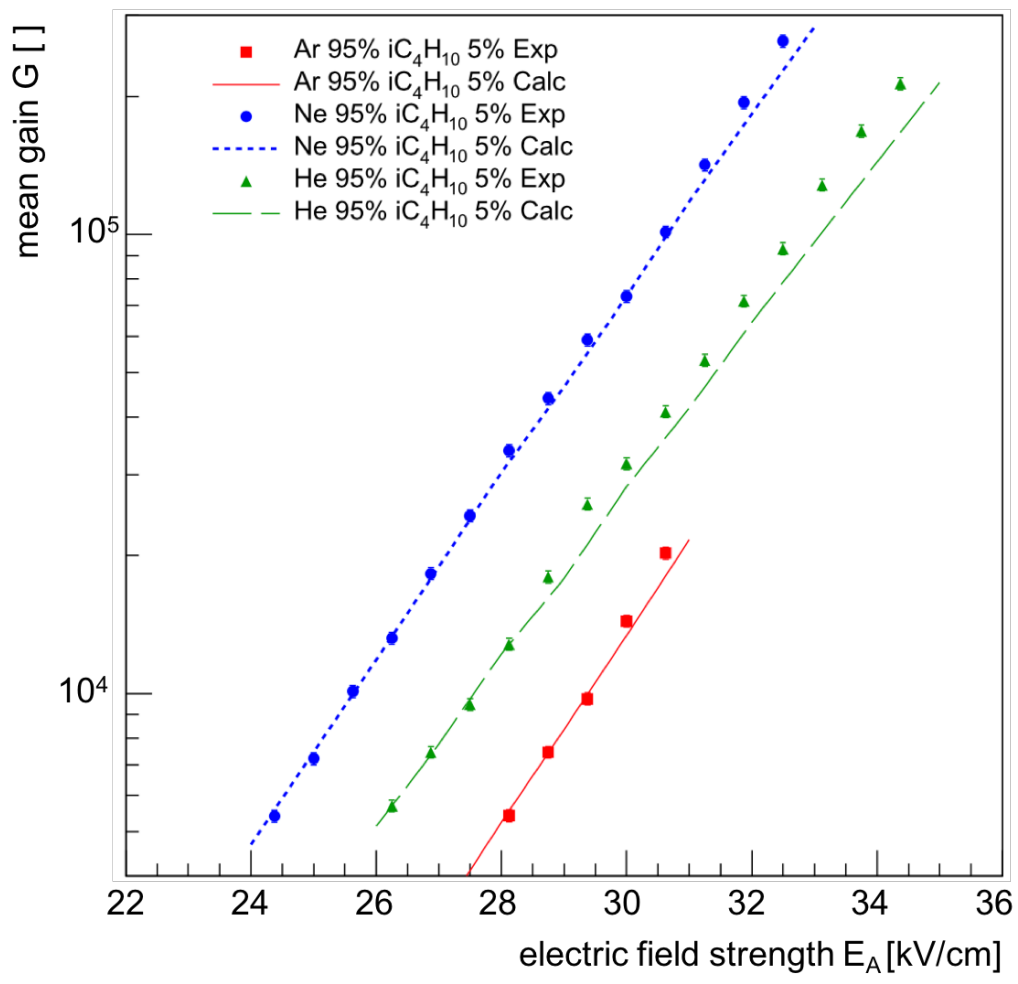

**Figure 4.31.:** Experimental mean gain $G$ as a function of the amplification field $E_A$ compared with the Monte Carlo simulations for the Argon, Neon and Helium gas mixtures. [85]

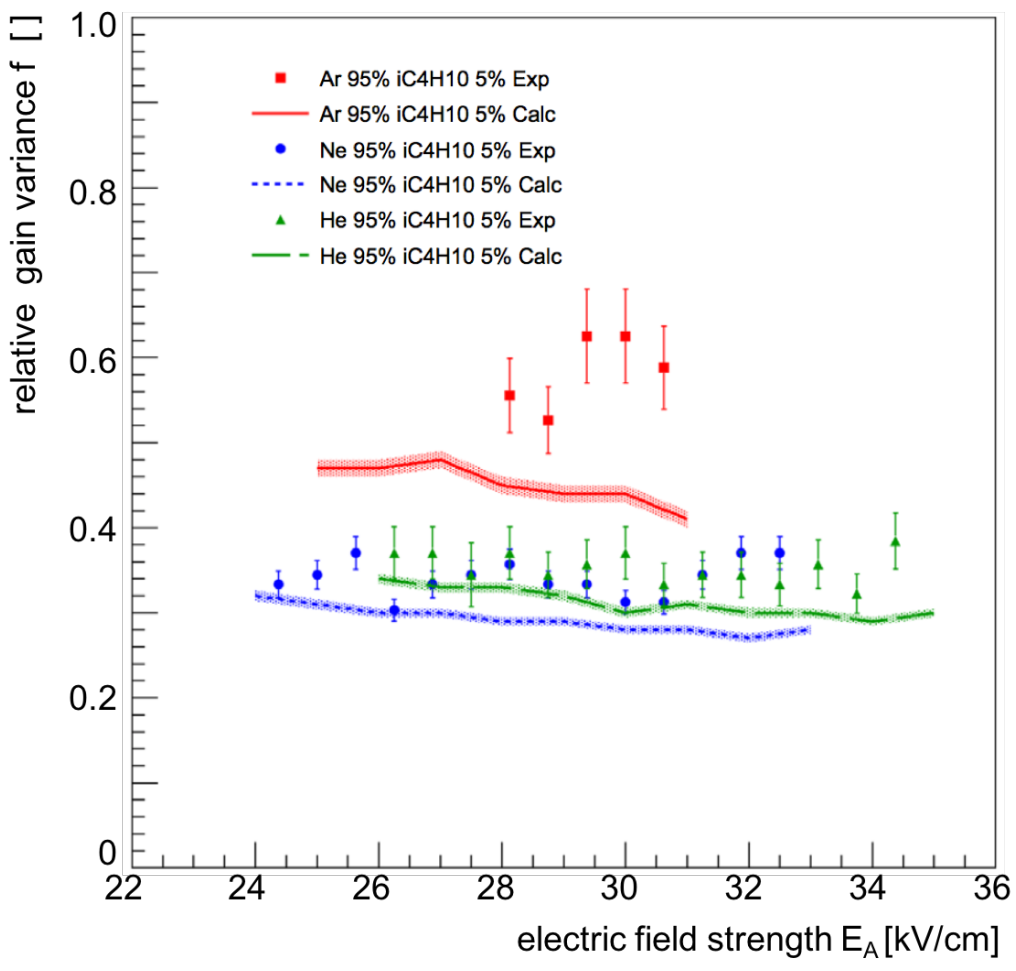

**Figure 4.32.:** Experimental relative gain variances $f$ as a function of the amplification field $E_A$ compared with the Monte Carlo simulations for the Argon, Neon and Helium gas mixtures. The shaded areas represent the statistical errors. [85]

# Part II.

# Micromegas Detectors for the ATLAS New Small Wheel

# 5. The ATLAS Detector and its Upgrade Program

The ATLAS (**A** **T**oroidal **LHC** **A**pparatu**S** [6]) is a general purpose detector at the LHC (**L**arge **H**adron **C**ollider [89]). Together with the CMS (**C**ompact **M**uon **S**olenoid [31]) experiment, the ATLAS collaboration searches for new particles, evidence of new physics and performs precision measurements in the full range of Standard Model physics. Both detectors are designed for operation with an LHC luminosity of up to $10^{34}\,\mathrm{cm}^{-2}\mathrm{s}^{-1}$ in p-p-collisions at a center of mass energy $\sqrt{s} \leq 14\,\mathrm{TeV}$. In addition to these two complementary general purpose detectors the other two major LHC experiments focus on more specialized topics of particle physics: The ALICE (**A** **L**arge **I**on **C**ollider **E**xperiment [30]) detector is optimized for the research on quark gluon plasma. Its full potential is exploited during the LHCs heavy ion operation, where $\mathrm{Pb}^{82+}$ ions are collided with an energy of up to 2.76 TeV/nucleon yielding a total center of mass energy of 1.15 PeV and a nominal luminosity of $1.0 \times 10^{27}\mathrm{cm}^{-2}\mathrm{s}^{-1}$ [89]. The LHCb ( [90]) collaboration primarily investigates matter-/antimatter asymmetries and CP-violation in rare B-Meson decays. Therefore, LHCb is designed as a forward detector with emphasis on precise vertex reconstruction and measurement of the B-decay end-states.

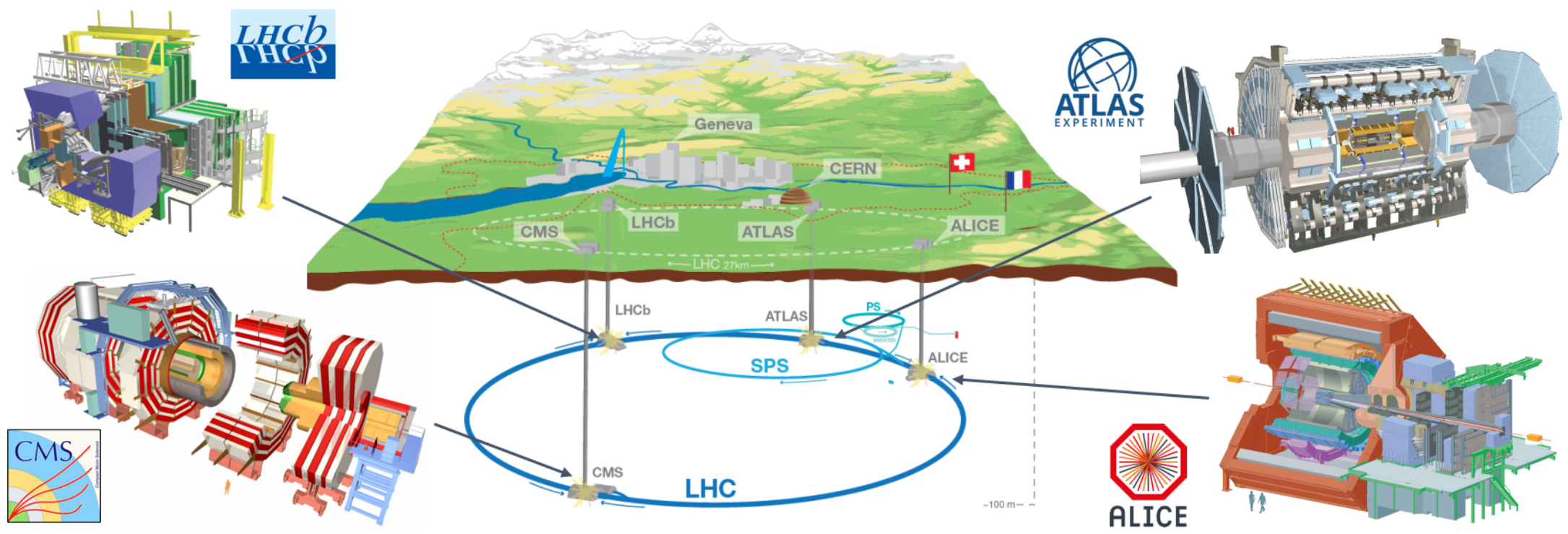

**Figure 5.1.:** The Large Hadron Collider (LHC) and its major experiments, located under the Swiss-French countryside close to Geneva. © 2016 CERN

This chapter shall provide an overview of the ATLAS experiment and its detector systems in section 5.1, with an emphasis on the ATLAS Muon Spectrometer, which utilizes several gaseous detector technologies. Thereafter the implications of the **H**igh **L**uminosity upgrade of the **LHC** (HL-LHC [91]) on the ATLAS detector and the extensive upgrade program to maintain its excellent performance are discussed in section 5.2. In section 5.3 the **N**ew **S**mall **W**heel (NSW) upgrade, the first major intervention on the ATLAS Muon System will be presented in detail.

## 5.1. The ATLAS Detector

The ATLAS detector is composed of several subsystems, arranged in a cylindrical configuration around the LHC beam pipe with the interaction point in the geometrical center (figure 5.2). The cylindrical layers are commonly referred to as *barrel* section, while the disk shaped detector assemblies, closing the barrel in both forward directions perpendicular to the beam pipe, are called *end-cap* sections.

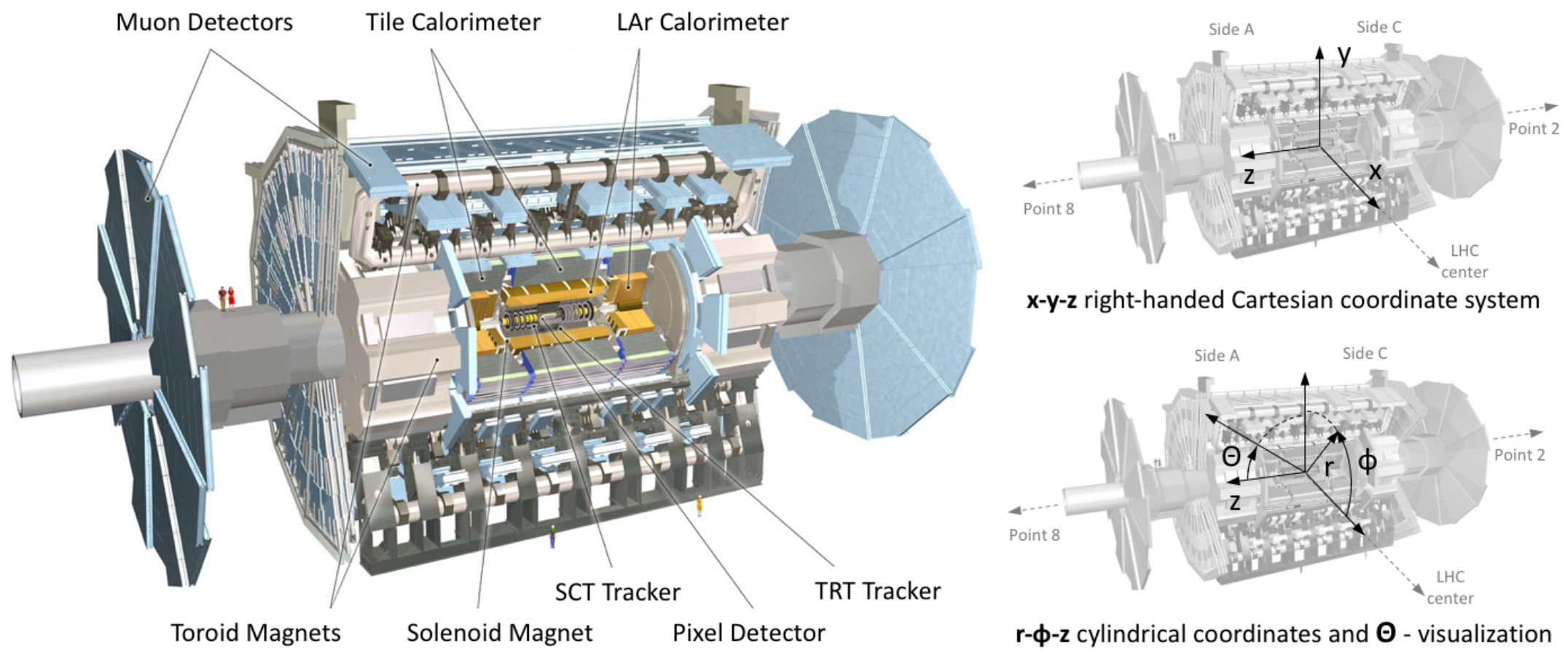

**Figure 5.2.:** The ATLAS detector and its subsystems. The two figures on the right visualize two representations of the ATLAS coordinate system. ATLAS Experiment © 2016 CERN, modified according to [6].

### 5.1.1. The ATLAS Coordinate System

The position of the detector components are commonly expressed in the right-handed Cartesian coordinate system of ATLAS (figure 5.2 - right top), which is aligned with the z-axis along the beam and the positive x-direction pointing towards the LHC center-point. An equivalent description expresses the x-y-plane position in r-$\phi$-coordinates in conjunction with the z-coordinate (figure 5.2 - right bottom). Trajectories of particles originating from the interaction point are often described by the azimuthal angle $\phi$ in the x-y-plane in conjunction with a polar angle $\Theta$ measured in the r-z-plane as the angle from the positive direction of the z-axis. Exploiting the ATLAS forward/backward symmetry, $\Theta$ is frequently substituted by the pseudorapidity $\eta$:

$$\eta = -\ln\left(\tan\left(\frac{\Theta}{2}\right)\right). \tag{5.1}$$

While (5.1) gives a geometrical definition of $\eta$, it can as well be interpreted as the zero-mass limit of a relativistic particle's rapidity $y$ ($\lim_{M\to 0} y = \eta$) with

$$y = \frac{1}{2}\ln\left[\frac{E+p_z}{E-p_z}\right], \tag{5.2}$$

where $E = \sqrt{|\overrightarrow{p}|^2 + M^2}$ is the energy of a particle of mass $M$ traveling with momentum $\overrightarrow{p}$ and the momentum z-component $p_z$.

### 5.1.2. The Inner Detector (ID)

The **I**nner **D**etector of ATLAS (ATLAS ID) is an ensemble of three precision tracking detectors: The **pixel** detector, the **S**emi**C**onductor **T**racker (SCT) and the **T**ransistion **R**adiation **T**racker (TRT) (figure 5.3). The ATLAS ID provides high precision track information of charged particles, allows for reconstruction of primary and secondary vertices and discriminates electrons, muons and heavier charged mesons utilizing their specific transition radiation in the TRT. It is enclosed by a superconducting solenoidal magnet providing a roughly uniform axial magnetic field of 2 T.

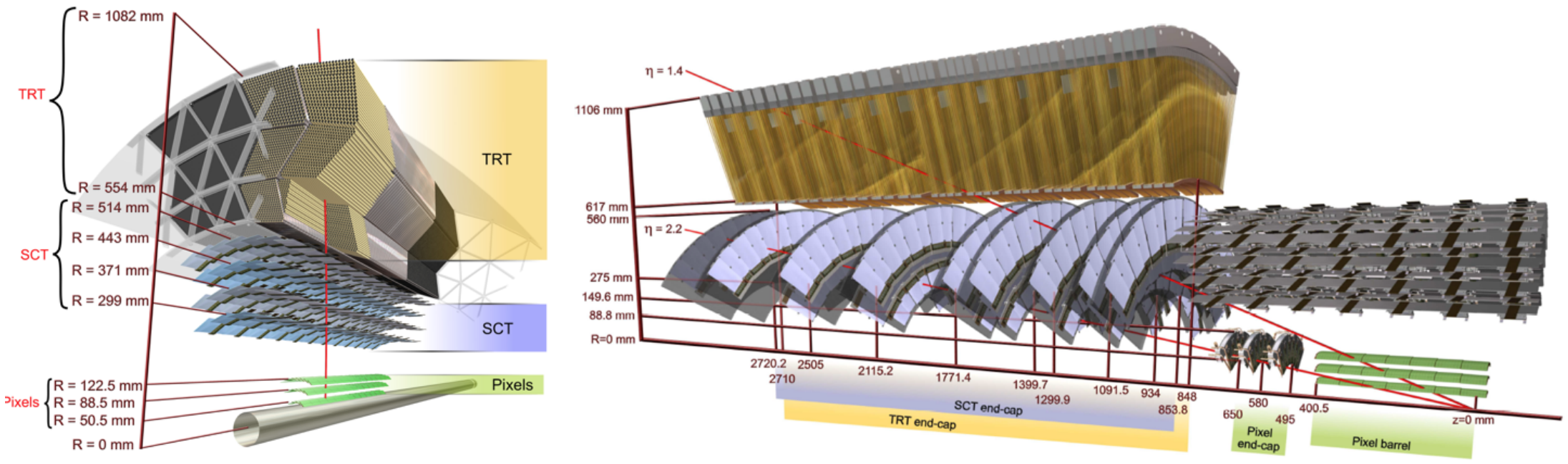

**Figure 5.3.:** Drawing of the sensors and structural elements of the ATLAS ID in the barrel (left) and the end-cap (right) region. Red lines indicate the tracks of charged particles originating from the IP. [6]

The pixel detector is composed of 1744 modules comprising more than $4.7\times10^4$ silicon pixels of 250 µm thickness and $50\times400$ µm[1] surface distributed over an area of $16.4\times60.8\,\text{mm}^2$. The charge deployed in a silicon pixel by a trespassing charged particle is measured in one of the 16 radiation-hard front-end chips on each module. The pixel detectors provide a space point measurement with a resolution of 10 µm in the $\phi$-coordinate and 115 µm in z-direction for the barrel or r-direction for the end-cap modules. Requiring a penetration of three pixel layers by a particle emerging from the interaction point, the pixel detector covers an $\eta$-range up to $|\eta| \leq 2.5$.

The rectangular silicon strips of 6 cm length of the SemiConductor Tracker in the barrel section are daisy-chained in pairs and arranged with a pitch of 80 µm. In the end-cap modules the radial strips vary slightly in dimensions depending on their position. They maintain in average the same pitch to provide an intrinsic spatial accuracy of 17 µm in the $\phi$-coordinate. On the second strip layer of each module, the strip orientation is rotated by a stereo angle of 40 mrad to allow for reconstruction of the z-coordinate (barrel) or r-coordinate (end-cap) with a spatial resolution of 580 µm. In total, the SCT comprises of over $1.6\times10^4$ sensor modules with $2\times768$ daisy-chained strips per sensor. Like the pixel detector, the SCT covers the $|\eta| \leq 2.5$ range, but with a four layer hit requirement.

[1] $50\times400$ µm is the nominal size of about 90% of the pixel on a module. The remaining 10 % have slightly increased dimensions of $50\times600$ µm.

The Transition Radiation Tracker is based on 4 mm diameter straw tubes made of two 35 µm thick multi-layer films[2] and stabilized by carbon fibers. A 31 µm gold-plated tungsten wire is stretched along the axis of the tube with a positioning precision of ≤ 400 µm and serves as the anode in this drift tube geometry (see chapter 1.2.1). The straws operate with a 70:27:3 Xe:$CO_2$:$O_2$ gas mixture at a potential difference of ≈1530 V, yielding a mean gas gain of $2.5 \times 10^4$. While the minimum ionizing particle's trajectory can be measured with a spatial resolution of 130 µm along the tubes orientation, the low energetic transition radiation photons absorbed in the Xenon gas yield much larger signal amplitudes and can be discriminated on a straw-by-straw basis using separate high and low thresholds. The almost $3{\times}10^5$ straw tubes, orientated parallel to the z-axis (barrel) or in radial direction (end-cap), contribute to the tracking and provide particle identification for $|\eta| \leq 2$.

### 5.1.3. The Calorimeter System (Calo)

The sampling **Calo**rimeter system (Calo) is designed to measure the energy, position and direction of electrons, photons, $\tau$-leptons and hadronic jets. It is composed of several **L**iquid **Ar**gon (LAr) detectors for electromagnetic calorimetry and hadronic calorimeters based on scintillating tiles in the barrel and LAr in the end-cap and forward region, as shown in figure 5.4.

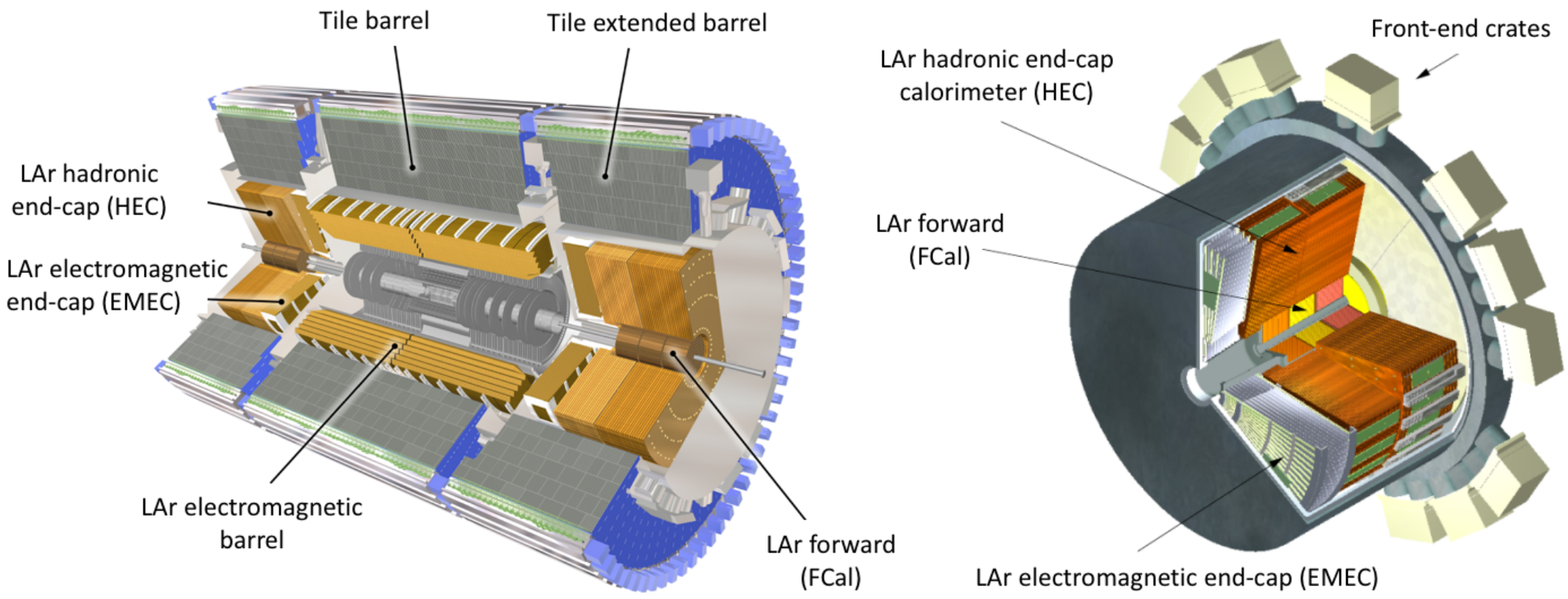

**Figure 5.4.:** Drawing showing the subsystems of the ATLAS calorimeter (left) and an enlarged view of the end-cap calorimeters (right). [6]

The **E**lectro**M**agnetic (EM) calorimeter is composed of accordion-shaped lead-LAr stacks, with Kapton® supported electrodes sensing the charge deposited by ionization of the liquid Argon. The linear response behavior combined with its intrinsic radiation hardness predestines liquid Argon as active medium in large volume calorimetry. The thickness of the lead absorber plates is optimized towards the detector's performance and energy resolution according to the covered $\eta$-range. For the precision physics required in $|\eta| < 2.5$ the EM calorimeter is segmented in three sections in-depth, while the **E**lectro**M**agnetic **E**nd-$C$ap (EMEC) modules, covering $2.5 < |\eta| < 3.2$, are more coarse in lateral granularity and depth-segmented in two sections only. In the region of $|\eta| < 1.8$ the EM calorimeters are complemented by presamplers, an instrumented argon layer, which provides a measurement of the energy lost in front of the detector.

[2] 25 µm polyimide film coated with 0.2 µm Aluminum with a 5−6 µm graphite-polyimite protection layer on one and a 5 µm polyurethane layer on the other side. The first is utilized as a cathode with a resistance of <300 Ω/m, the latter heat seals the two films back-to-back during straw production.

The hadronic sampling calorimeter in the barrel region is based on steel absorbers and scintillating tiles, read-out by wavelength shifting fibers into two separate photomultiplier tubes. The Tile calorimeter is split in a central barrel- ($|\eta| < 1.0$) and two extended barrel- ($0.8 < |\eta| < 1.7$) sections each composed of 64 azimuthally segmented modules. The **H**adronic **E**nd-**C**ap detectors (HEC) are composed of copper absorbers interleaved with LAr active layers. Each end-cap comprises of two HEC wheels of four meter diameter build from 32 wedge-shaped modules with two depth segments each. The wheels are located directly behind the EMEC and extend out to $|\eta| > 1.5$ along their circumference and $|\eta| < 3.2$ in their center, overlapping with the coverage of the Tile extended barrels and the LAr forward calorimeter (FCal).

The FCal covers a range of $3.1 < |\eta| < 4.9$ and consists of three modules in each end-cap: the innermost utilizes copper absorbers and is optimized towards EM calorimetry while the tungsten absorbers in the other two modules measure predominantly the energy of hadronic interactions. The active LAr medium is flushed through small tubes with concentric anode rods running parallel to the beam axis through the absorber. This allows for a tuning of the LAr gaps down to 0.25 mm close to the beam pipe to avoid charge recombination in ion buildup under high irradiation.

The LAr EM barrel calorimeter is hosted in a separate cryostat, sharing the vacuum vessel with the ATLAS solenoid, reducing the material budget towards the IP. In both forward directions the EMEC, HEC and LAr share a common end-cap cryostat. In total the calorimeters summarize to a thickness of $> 22$ / $> 24$ radiation lengths ($X_0$) in the barrel/end-cap EM calorimeter and $\approx 9.7$ / 10 interaction lengths ($\lambda$) in the barrel/end-cap region of the hadronic calorimeters. This provides sufficient containment of electromagnetic and hadronic showers and minimizes punch-through into the muon system to a level below the irreducible background by prompt or muon decay. Combined with the wide $\eta$-coverage this allows for a good missing energy $E_T^{miss}$ measurement, which is essential for many signatures of new physics like, SUSY particle searches.

### 5.1.4. The Muon Spectrometer (MS)

The **M**uon **S**pectrometer (MS) detects muons and other charged particles penetrating the calorimeter system, measures their momenta and allows for triggering on high-$p_T$ muons. It provides a momentum resolution of 10 % for 1 TeV muon tracks, which translates into a sagitta measurement accuracy of ±50 µm for an expected 500 µm sagitta along the beam axis. The magnetic field of 0.5 - 1.0 T is generated by eight large air-core superconducting toroid magnets embedded in the barrel muon system and two sets of eight smaller magnet coils located between the first and second wheel of the muon end-cap system (figure 5.5).

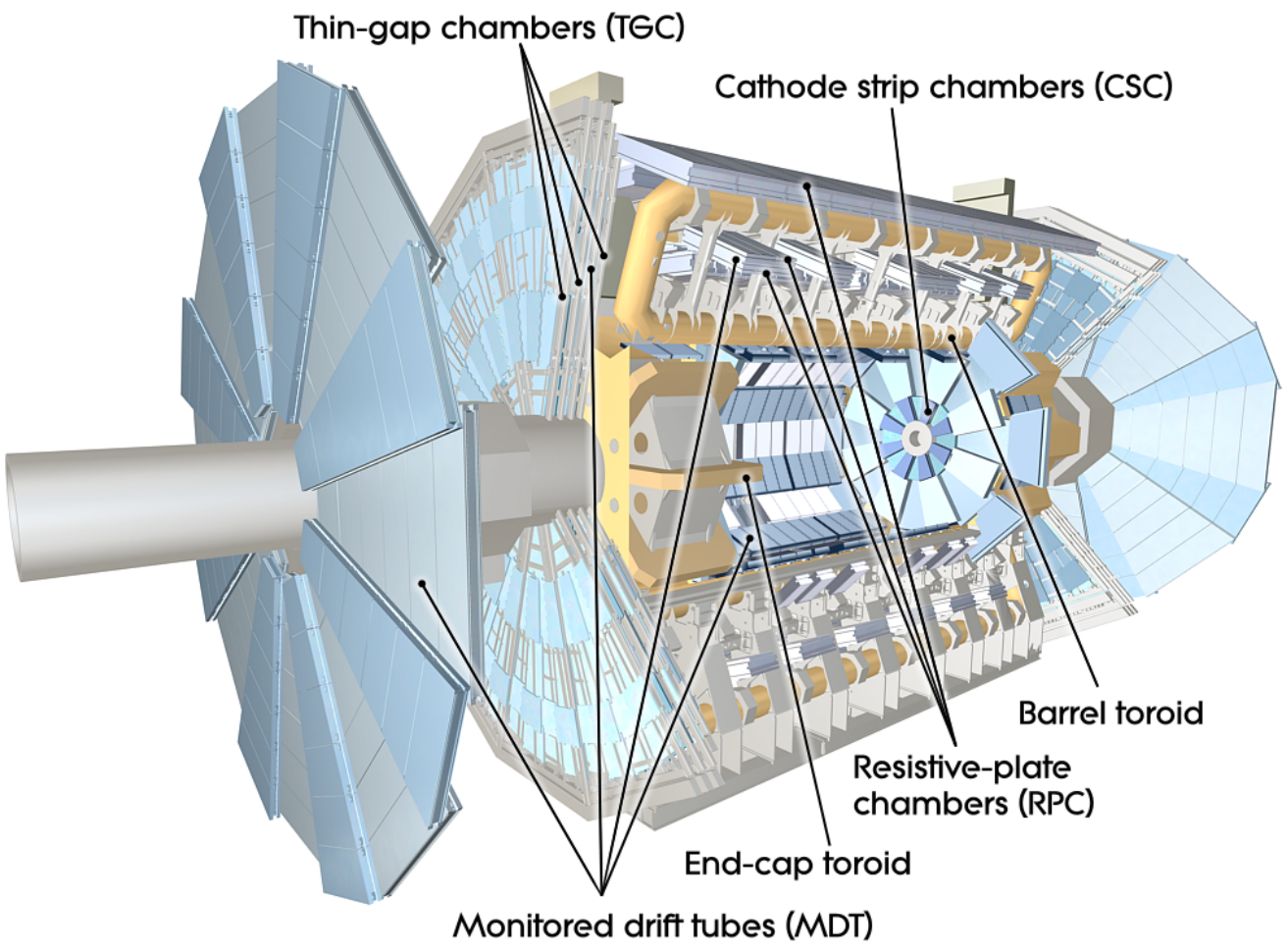

**Figure 5.5.:** Layout of the ATLAS Muon Spectrometer comprising MDT, CSC, RPC and TGC detectors, embedded in the ATLAS toroidal magnet systems in barrel and end-cap region. [6]

Similar to the other ATLAS subsystems the Muon Spectrometer is arranged in a barrel region (B) and two end-caps (E), each composed of an inner (I), middle (M) and outer (O) layer of muon detectors. The octant structure of the magnets is reflected in the $\phi$ symmetry of the MS which consists of eight radial segments of 45° each comprising a small (S) sector (even numbers: 2,4,6,...16) and a large (L) sector (odd numbers: 1,3,5,...15), as shown in figure 5.6. The large and small sectors are slightly displaced in r (barrel) or z (end-cap) to allow for an overlap in $\phi$. Besides increasing detector coverage, this overlap allows for a relative alignment of adjacent sectors using straight muon tracks passing through both modules.

The precision momentum measurement is primarily performed by **M**onitored **D**rift **T**ube (MDT) chambers (see chapter 1.2.1), combining high measurement accuracy with a simple design. Each chamber consists of two multi-layers[3] of close-packed 29.97 mm diameter drift tubes operated with Ar:$CO_2$ (93:7) at 3 bar pressure. The better than 20 µm accurately positioned wires yield a spatial hit resolution, perpendicular to the wire, of 80 µm per tube, or about 35 µm per chamber. To maintain this precision under thermal and gravitational deformation of the tubes and the wires, a sag-adjustment system is implemented in all barrel chambers. Therefore, each MDT chamber comprises an in-plane alignment system, measuring deformations on µm-level, as well as temperature and magnetic field sensors. Similarly, an optical alignment system is used to monitor the relative position of each MS chamber, referring to their neighbors [92]. This yields a stable and dense grid of accurate positions and allows for an online reconstruction of the coordinates of each chamber with a precision $\leq$30 µm. The more than $3.5\times10^5$ monitored drift tubes cover a pseudorapidity range of $|\eta| < 2.7$ on the outer two layers

[3]with four layers drift tubes in the inner layer and three in the middle and outer MS layer.

and $|\eta| < 2.0$ in the innermost layer of the MS summing up to a chamber surface of more than 5500 m$^2$.

On the MS first end-cap layer, the Small Wheels, the MDTs have been partially replaced with **C**athode **S**trip **C**hambers (CSC), providing higher rate capability with safe operation up to 1000 Hz/cm$^2$, compared to 150 Hz/cm$^2$ for the MDT. The CSCs combine the geometry of a multi-wire-proportional chamber (see chapter 1.2.3) with a segmentation of the cathode planes into strips. The charge from ionization of a muon transpassing the Ar:$CO_2$ (80:20) gas mixture is amplified in the vicinity of the anode wires and inductively measured on the strips. While the 30 µm anode wires are orientated in radial direction[4], the O(1.5 mm) wide strips on one cathode plane are running orthogonal to the wires, segmented by 0.25 mm gaps. To reduce the amount of read out channels, only each third strip is connected to the readout, inductively measuring the signal on their neighboring floating strips. This results in an effective readout-pitch of O(5.3 − 5.6 mm). The CSCs reach a hit position resolution in the bending direction of 60 µm per plane and additionally provide a O(5 mm)-resolution in the transverse plane utilizing the more coarse segmented second array of cathode strips, running in the wire direction. The 32 CSC modules cover an area of ≈65 m$^2$ and consists of four active layers each. They comprise in total more than $4{\times}10^4$ wires and almost $8{\times}10^4$ strips.

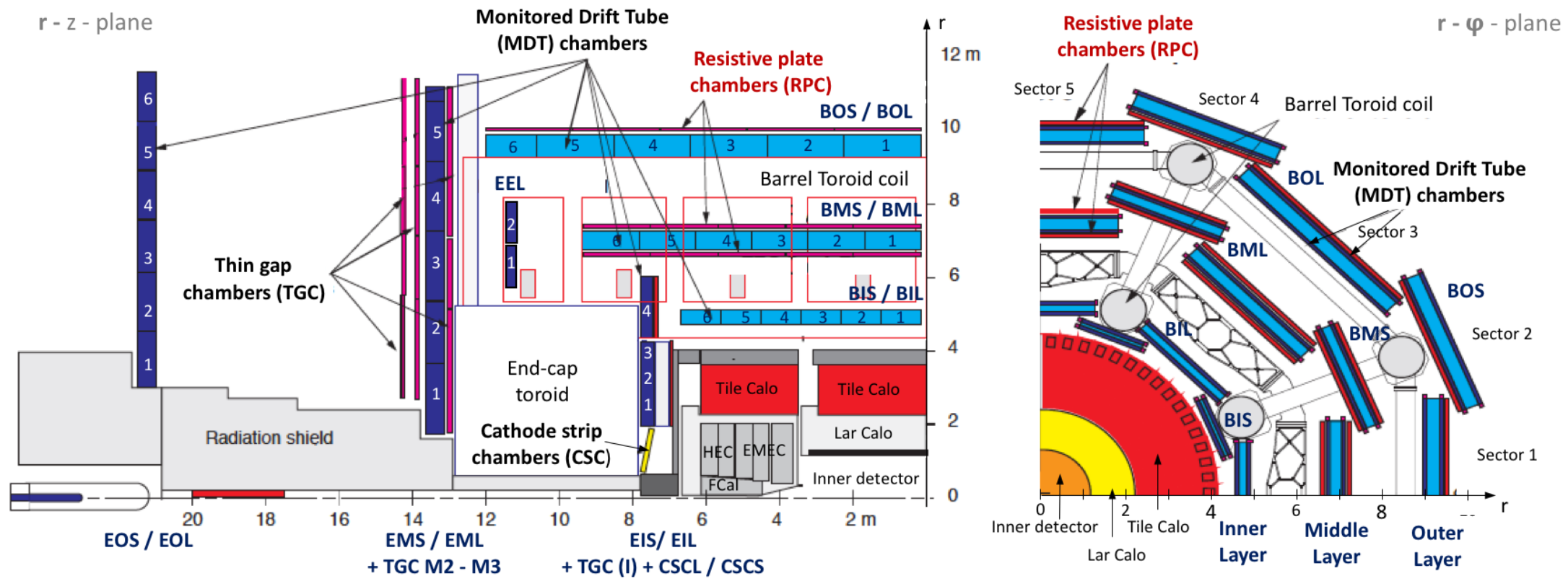

**Figure 5.6.:** Cross-section of a quadrant of the ATLAS Muon Spectrometer in the r-z plane (left) and the r-$\phi$ plane (right) comprising all detector modules. The naming of MDT chambers is based on their location in the barrel or end-cap (B,E), in the inner, middle, or outer layer (I, M, O) and in either the a large or a small sector (L,S). Drawings from [93] modified according to [6].

The MS precision detectors are complemented with a trigger system, providing fast identification of low transverse momentum ($p_T$: 6 - 9 GeV) and high $p_T$ (9 - 35 GeV) muons and their bunch crossing correlation and adding measurement of the non-bending coordinate to the MDT output. For optimized $p_T$ measurement the spatial resolution of the trigger detectors must be adapted to their $|\eta|$-region, since the $p_T/p$-ratio de-

[4]The central wire of each chamber is orientated radially with all other wires being parallel.

creases with increasing $|\eta|$ and accordingly does the bending power of the magnetic field. Therefore, two different technologies have been selected:

In the barrel region ($|\eta| < 1.05$) **R**esisitive **P**late **C**hambers (RPCs) provide a good spatial and time resolution at moderate rate capability with a simple to construct setup (see chapter 1.2.2). The ATLAS RPCs are parallel plate chambers with a gap of 2 mm filled with a $C_2H_2F_4$:Iso-$C_4H_{10}$:$SF_6$ (94.7:5:0.3) gas mixture. The ionizing track of a passing muon immediately triggers an avalanche, avoiding time delay due to charge drift. Thus a fast signal with a width of 5 ns is generated. Position measurement is performed by two sets of orthogonal orientated strips inductively coupled to the resistive plates. Each of the three RPC trigger stations comprises two overlapping units of two RPC gas gaps. Two stations are mounted on both sides of the middle layer MDT chambers and a third station is located on the respective inner (small sector) and outer face (large sector) of the outer layer MDTs (figure 5.6), providing full $\phi$-coverage with three stations.

The end-cap region is equipped with **T**hin **G**ap **C**hambers (TGCs), a multi wire proportional chamber (chapter 1.2.3) with a wider wire spacing (1.8 mm) than the wire-cathode distance (1.4 mm). The highly quenching gas mixture of $CO_2$ and n-$C_5H_{12}$ (n-pentane) (55:45) allows for operation in a quasi saturated mode, suppressing streamers effectively. While tracks passing between two wires perpendicular to the wire plane would suffer from long drift times, the orientation of the TGCs in the ATLAS end-cap forbids muon penetration angles $\Theta \leq 10°$. Thus, signals arrive with a 99 % probability within the 25 ns window, required for bunch crossing identification. The hit position in radial (bending) direction is measured via the wires, grouped in 6 - 31 wires per channel to adjust the granularity according to the $|\eta|$-position. The azimuthal coordinate is determined with a granularity of 2 -3 mrad using radial strips inductively coupled to the cathode. One triplet (M1) and two doublets (M2, M3) of TGCs are mounted on the middle big wheel MDTs, outside of the magnetic field. The inner layer chambers (I), mounted on the small wheels, comprise another TGC doublet, yielding a coverage of $|\eta| < 2.4$ divided in an outer ($1.05 < |\eta| < 1.92$) and an inner ring ($1.92 < |\eta| < 2.4$).

### 5.1.5. The Trigger and Data Acquisition Scheme (TDAQ)

With a collision rate of 40 MHz (20 MHz during the initial run period) the LHC provides several magnitudes more proton-proton-event data than can be handled by the **D**ata **A**cquisition **S**ystem (DAQ) and than can be transferred to the long-term storage in the CERN datacenter. To reduce the amount of data, the ATLAS **T**rigger and **DAQ** (TDAQ) system identifies events potentially containing interesting physics on the basis of the high-$p_T$-paradigm[5] and discards the remaining majority within a short latency to minimize buffer storage. The save-or-discard-decision is based on subsequent instances: The **L**evel **1** (L1) Trigger and the **H**igh-**L**evel **T**rigger (HLT) composed of **L**evel **2** (L2) and event filter. Each step selects events with successively tighter requirements and stronger cuts to step wise reduce the rate (figure 5.7).

[5] With the Standard Model particle spectra being well understood within the eV-GeV range, new physics signatures like the occurance of SUSY particles are expected in a mass scale >100 GeV and, therefore, their detectable decay products are likely to carry a high momentum. With the longitudinal momentum of the scattering system being undefined in p-p-collisions, the transverse momentum remains as conserved quantity. Tagging events containing high-$p_T$ signals offers a convenient way to identify events potentially including new physics and is, therefore, applied as first trigger paradigm in many collider experiments conducting universal searches.

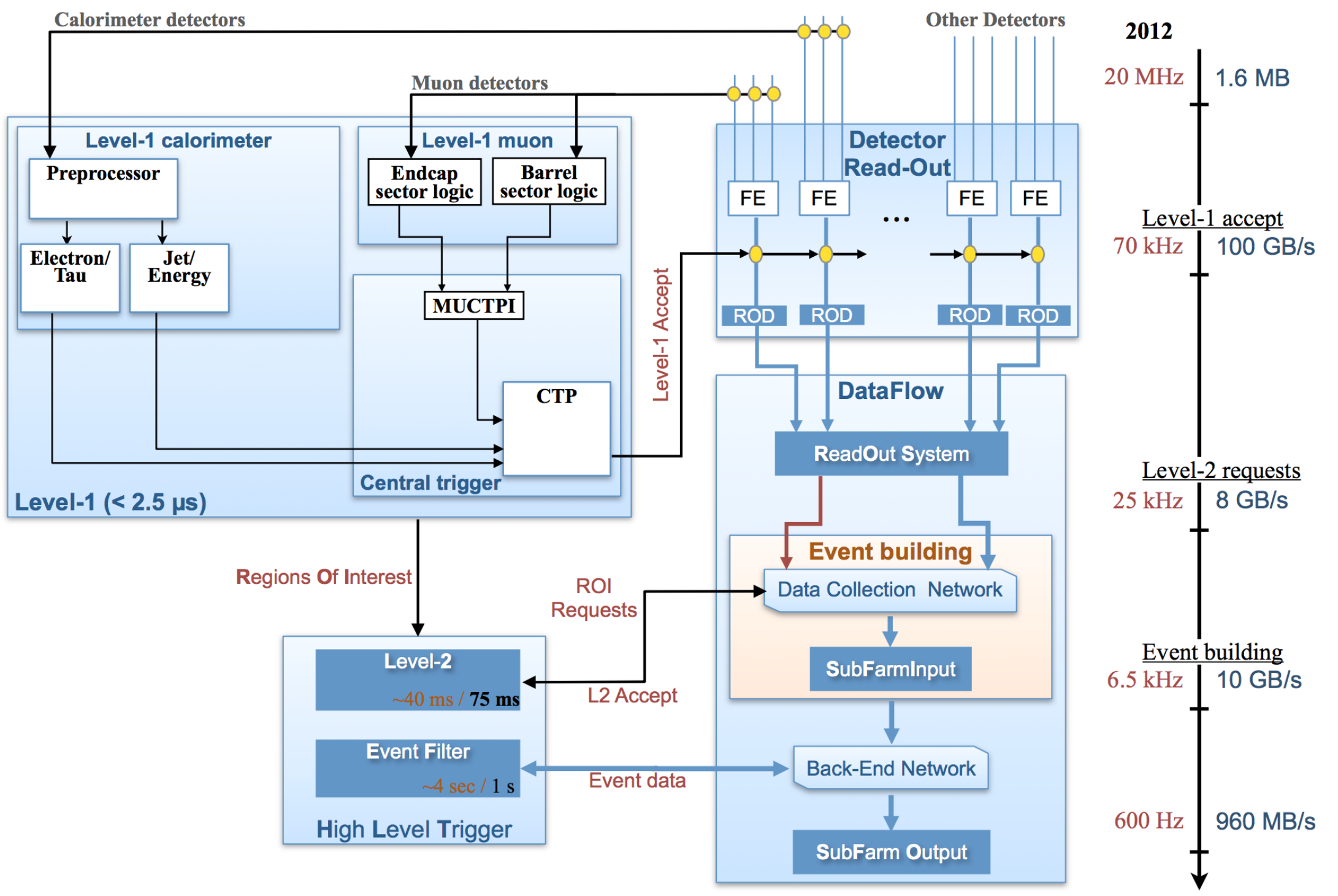

**Figure 5.7.:** Schematic of the ATLAS Trigger and DAQ system as implemented during run 1, with average data flow numbers from the 2012 run. [94]

The L1 trigger uses hit signals with reduced detector granularity from either the RPCs and TGCs to identify high $p_T$ muons as well as from the calorimeter sub-systems to flag events comprising energetic electromagnetic clusters, jets, $\tau$-leptons or events with large $E_T^{miss}$. The L1 is hardware based and detector specific and must provide a decision to all front-end electronics within 2.5 µs, reducing the detector output rate to 70 kHz.

On L2 the trigger combines information of the different detectors in the **R**egions **o**f **I**nterest (RoI) determined on L1 and applies additional constraints on the quality of a trigger object for example by combining the trigger information passed on by different sub-systems. With an average processing time of 40 ms, the L2 reduces the data rate to below 25 kHz.

The L2 is complemented with the **event filter**, forming the **H**igh **L**evel **T**rigger (HLT). This full-event-building and -selection process requires about four seconds and only ≈600 Hz event rate is passed on. During the HLT the full granularity and precision of the Calo, the MS and data from the ATLAS ID as well is taken into account to improve threshold cuts, particle identification and reduce the fake trigger rate.

## 5.2. ATLAS Upgrades towards High Luminosity LHC

To fully exploit the LHC discovery potential a series of interventions and upgrades is foreseen. While the first modifications have been applied to ensure secure operation at the LHC design collision energy, instantaneous luminosity and bunch spacing, the upcoming upgrades aim at an increase of the luminosity beyond design parameters and shall conclude in the **H**igh **L**uminosity **LHC** (HL-LHC).

In parallel to the accelerator complex the experiments require upgrades to cope with the increased challenges of higher event pile-up, detector occupancy and increased background rates. The major modifications to the ATLAS experiment undertaken during LS1 (Phase-0) and scheduled for the upcoming LS2 (Phase-I) and LS3 (Phase-II) are presented in this chapter. Among those the New Small Wheel (NSW) upgrade, discussed in more detail in section 5.3, scheduled for LS2 is of major importance to ensure reliable and accurate muon reconstruction in the ATLAS end-cap region under an increased muon rate. A survey of the ATLAS upgrade program can be found in [7, 8].

### 5.2.1. LHC Performance and High Luminosity LHC Schedule

With the initial commissioning of the LHC being overshadowed by the September 2008 incident [95], the world largest particle accelerator restarted operation in late 2009. After being fully commissioned, the first 3.5 TeV protons collided in March 2010, initiating a physics run period lasting three years. During this Run 1 the LHC delivered about 30 $\mathrm{fb}^{-1}$ of p-p-collisions with $\sqrt{s}$ = 7 TeV (2011) and $\sqrt{s}$ = 8 TeV (2012). The exceptional performance of the LHC and its experiments lead among other findings to the discovery of the Higgs-Boson by the ATLAS [3] and the CMS collaboration [4].

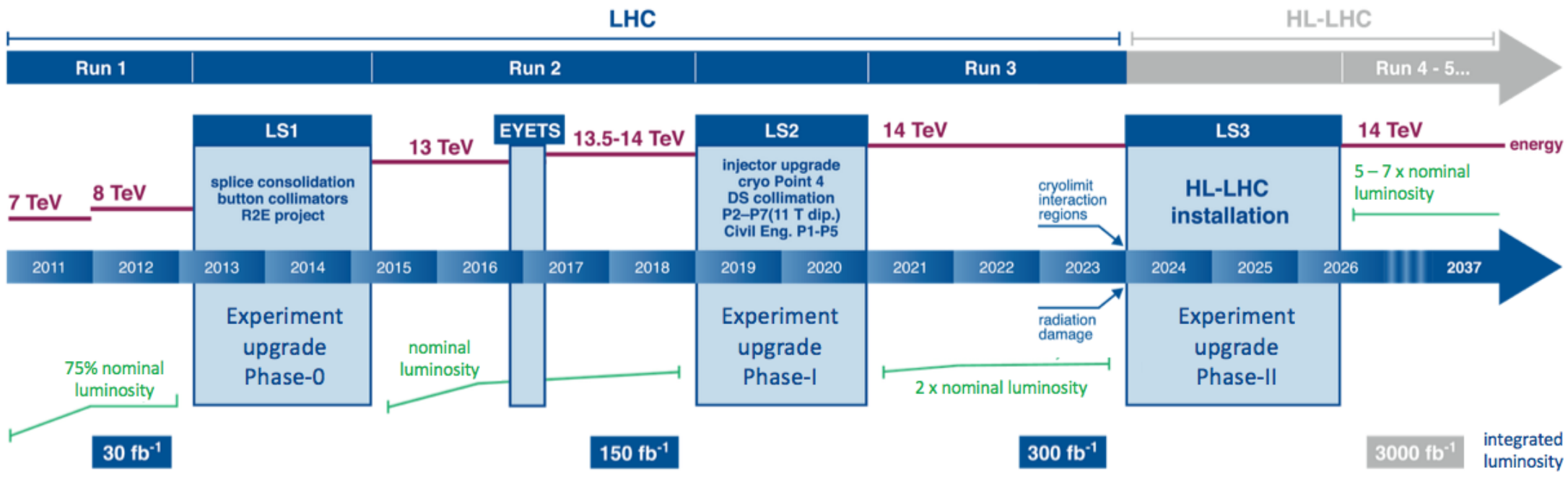

**Figure 5.8.:** Timeline for the LHC / HL-LHC data taking (run) and upgrade (long shutdown, LS) periods, including the energy and luminosity goals within each run-period. The NSW upgrade is scheduled in LS2. Taken from [91], modified for legibility.

During the two years of the **L**ong **S**hutdown **1** (LS1) between 2013-14, the conductive splices between the LHC s magnet modules have been repaired to allow the LHC to reach out for its design operation parameters in Run 2. Restarting data taking in 2015, the LHC has continued its successful journey towards 14 TeV collision energy and surpassed all performance expectations. In 2016 it delivered an integrated luminosity of 40 $\mathrm{fb}^{-1}$ at $\sqrt{s}$ = 13 TeV, a surplus of 60 % compared to the envisaged 25 $\mathrm{fb}^{-1}$ [96]. During the currently ongoing **E**xtended **Y**ear **E**nd **T**echnical **S**top (EYETS), lasting until the second quarter of 2017, the accelerator will be prepared for the first 14 TeV run, foreseen within the upcoming two years.

The LS2 shall be primarily used to upgrade the injector complex including the connection of the LINAC4 [97]. Furthermore, the commissioning of the newly built cryogenic plant at point 4 and the installation of the first 11 T dipole magnets and dispersion suppression collimators around point 2 are foreseen [98].

The final conversion to HL-LHC is scheduled for LS3 and will require about 30 months. The interventions comprise among others the exchange of the insertion magnets, new superconducting RF crab cavities and radiation-hard absorber with increased aperture [91]. These interventions will allow the LHC to surpass its design instantaneous luminosity by a factor of up to 5, leading to a high luminosity operation from 2024 onward and an expected total integrated luminosity of $3000\,\text{fb}^{-1}$ by 2035.

### 5.2.2. ATLAS Phase-0 Interventions and the IBL

In parallel to the LHC consolidation during LS1 (2013-2014) the ATLAS collaboration applied several modifications to the initial detector.

- The installation of a new pixel layer in the ATLAS ID complemented the original three layers of pixel modules in the barrel region with a forth **I**nsertable **B**-**L**ayer (IBL) [99] around a new beam pipe with reduced diameter. Adding this fourth layer closer to the IP improved the vertex reconstruction capability significantly. Furthermore, it recovered for losses of efficiency in the previous innermost pixel layer and shall ensure stable detector performance despite accumulating radiation damage until the ATLAS ID exchange in LS3.

- Several detector chambers have been added to the Muon Spectrometer in the sector 12 and 14 of the barrel middle layer to improve coverage of the MS in the region of the ATLAS support structure. These **BMF** (barrel middle layer feet) chambers utilize the sMDT technology for precise muon tracking. An additional set of RPC trigger detectors has been installed in this 'feet'-region.

- Besides the modifications on the detector systems several interventions have taken place on the ATLAS services and infrastructure. Most noticeable among them is the merging of the TDAQ level 2, the event building and the event filter into a single processing unit running in a homogeneous HLT farm [94].

### 5.2.3. ATLAS Phase-I Upgrades and the NSW

The ATLAS Phase-I upgrade [100] foreseen for LS2 (2019-2020) is driven by the need to maintain an optimal trigger rate despite the increased event rate, while avoiding an adjustment of the current cut thresholds, for example on lepton $p_T$. Therefore, additional techniques for background suppression are required to reduce the rate of foremost jets mimicking electrons in the calorimeters and fake muons in the forward spectrometer. The upgrade program comprises several major interventions:

- The **F**ast **T**rac**k**er (FTk) will provide track reconstruction in the SCT, the pixel detector and the IBL on level 1 with nearly offline resolution. The FTk utilizes a highly parallel hardware architecture to first match the hits to one of $10^9$ pre-stored patterns, using associative memories, and second perform a fast FPGA based linear fit, all within a 100 µs latency. [101]

- An exchange of the **LAr EM Calo**rimeter trigger readout electronics is foreseen in order to access the electromagnetic shower information with higher-granularity, higher-resolution and add shower depth information at L1. Therefore, the current concept of *'trigger towers'*, summing up the energy deposition within a $\Delta\eta \times \Delta\phi = 0.1, \times 0.1$ cone, is replaced with the scheme of so-called *'super cells'* with a finer granularity of $\Delta\eta \times \Delta\phi = 0.025, \times 0.1$ and an additional longitudinal segmentation. An improved rejection of low $p_T$ jets faking electrons by a factor of $3-5$ is expected. [102]

- The **N**ew **S**mall **W**heels (NSW) will replace the current inner station of the forward muon spectrometer (EI). The deployed Micromegas and sTGC detectors will be able to cope with the increased event and background rate, provide tracking and trigger information and significantly reduce the fake muon rate in $1.3 < |\eta| < 2.5$ [9]. This upgrade is discussed in more detail in chapter 5.3. Furthermore, a replacement of the muon chambers in the **B**arrel **I**nner layer **S**mall sections 7 and 8 (BIS78) is foreseen to extend the coverage of the improved fake muon rejection scheme to the barrel-end-cap transition region $1.05 < |\eta| < 1.3$. A combination of sMDTs (section 1.2.1) and RPCs (section 1.2.2) will be used as precision tracking and trigger detectors.

- To exploit the full potential of these upgrades, the ATLAS central **T**rigger and **D**ata **Aq**uisition (TDAQ) system will be adapted. While retaining its original structure a topological combination of MS and Calo trigger information will be included on level 1 [94].

- The **A**TLAS **F**orward **P**roton (AFP) project will add detectors in the far up- and downstream region, about 210 m from the interaction point. Utilizing a silicon tracking device combined with a very fast quartz-based Cerenkov timing detector, protons originating from diffractive collisions will be measured with high spatial-, temporal- and momentum resolution. AFP adds a physics program complementary to the existing ATLAS forward detectors [103].

### 5.2.4. ATLAS Phase-II Program and TDAQ Revision

A further upgrade (Phase-II) is scheduled for LS3 in 2024-2025, preparing ATLAS for HL-LHC operation. As in Phase-I the challenges remain a stable trigger rate without selection threshold adjustment as well as increasing detector occupancy and pile-up. Furthermore, the accumulated radiation damage at the ATLAS subsystems needs to be taken care of during LS3, demanding the exchange of detector and electronics components. The measures to be taken have been outlined in [104] and discussed in several costing scenarios in [105]. We refer hereafter to the preferred reference scenario:

- A drastic change is foreseen for the ATLAS **T**rigger and **D**ata **AQ**uisition scheme (TDAQ), splitting the current level 1 trigger into two levels of mainly FPGA based hardware triggers: level 0 (L0) and level 1 (L1). In the reference scenario, the L0 and L1 triggers are designed to operate at rates up to 1 MHz and 400 kHz. L0 shall provide a trigger decision, including the RoI, based on muon and calorimeter data and their topological combination with a latency of 6 µs. The L1 *'track trigger'* searches the L0 RoIs for ITk tracks with high transverse momentum $p_T$ >4 GeV. In parallel the L1 *'global'* combines entire detector data to reconstruct global event quantities such as $E_T^{miss}$. Obtaining up to 400 kHz from the 30 µs latency L1, the **H**igh-**L**evel **T**rigger (HLT) will combine an **upgraded** **F**ast **T**rac**k**er (FTk++) and the **e**vent **f**ilter to further decrease the recorded data by a factor of 40. Therefore, tracks corresponding to particle momenta as low as $p_T$ >1 GeV will be identified in the RoIs allowing for tight isolation cuts with a minimal sensitivity to pileup.

- The complete ATLAS ID will be exchanged by the new **I**nner **T**rac**k**er (ITk), completely consisting of silicon detectors. Being exposed to an estimated 300 fb$^{-1}$ of p-p-collisions by the end of Run 2, the current ATLAS ID is expected to degrade in performance due to radiation damage. Furthermore, the current detector will not be able to cope with the increased occupancy (SCT, TRT) and suffer from bandwidth saturation (pixel, SCT) under HL-LHC conditions. In the reference scenario the ITk would be composed of 4 pixel layers and 4 + 2 layers of short-strip and long-strip Si detectors in the barrel as well as 12 pixel and 7 strip layers in the end-caps. It would cover a $|\eta| < 4.0$ range and would comprise more than 600 million pixel and 74 million strips. Featuring a decreased pixel size (25×150 µm$^2$ in the inner two pixel layers and 50×250 µm$^2$ in the outer two layers) compared to the current Pixel detector, the ITk shall allow for reliable reconstruction of primary and secondary vertices in 200-pile-up events and measure the transverse momentum and direction of isolated particles.

- The performance of the calorimeter system is expected to degrade under the increased instantaneous luminosity only in the forward region. Therefore, an exchange of the FCal with a **s**mall-gap **F**orward **Cal**orimeter (sFCal) with higher transverse granularity (100 µm LAr gaps compared to 250 µm in the current FCal) is considered. An alternative scenario features the installation of a MiniFCal in front of the existing calorimeter in order to reduce the flux in the high-$|\eta|$ FCal region to levels where the current system can operate normally. Different options of an either LAr-based cold MiniFCal or a warm MiniFCal comprising Si-detectors alternating with copper and tungsten absorber layers are under study. Besides this detector upgrade, the major intervention on the Calo system will be the exchange of the front- and back-end electronics of the LAr and Tile calorimeters, continuing the efforts started in the Phase-I upgrade. This exchange is required not only due to expected radiation damage on the current electronics, but also to fully access the extended trigger information.

- The upgrade of the **M**uon **S**pectrometer (MS) is dedicated to improve its fake muon reduction and provide trigger information in accordance with the new L0/L1 scheme. While the barrel inner layer small sector chambers (BIS) will be replaced with a combination of sMDT and RPC chambers the large sector MDT chambers (BIL) will be topped up with additional RPCs, providing a robust L0 trigger and closing existing coverage holes. In the $2.0 < |\eta| < 2.4$ forward region a replacement of the current trigger chambers with sTGC detectors is planned to improve trigger selectivity under the expected high event and background rates. While the remaining MDT, RPC and TGC chambers are expected to meet the requirements for operations in HL-LHC conditions, the existing electronics will have to be adapted to the new trigger scheme[6]. The state of the art electronics shall provide fast input to L0 as well as improved $p_T$ measurements satisfying the sharpened threshold definition for L1 trigger at 400 kHz. This sharpening of the high-$p_T$ thresholds is expected to reduce the fake muon rate by a factor of 4.

## 5.3. The New Small Wheel (NSW) Upgrade

The **N**ew **S**mall **W**heel (NSW) upgrade is the first major intervention on the ATLAS muon detectors and is scheduled to take place during LS2 (5.2.3). It features the complete replacement of the innermost station of the MS in forward region, including its support structure, an alignment system and new shielding to cope with increased particle flux (5.3.1), as well as new detectors. These trigger- and tracking-detectors are based on the sTGC and the Micromegas technology (5.3.2) and will be equipped with up-to-date electronics (5.3.3), which comply with the envisaged trigger scheme after Phase-II. The NSW upgrade is mandatory to extend the ATLAS barrel muon spectrometer reconstruction efficiency and precision to the end-cap region and maintain these under increased event rate. With the occurrence of high energetic muons being an indicator for interesting and possibly new physics and muon signatures contributing to physics analysis in the whole range from Standard Model physics to Higgs property measurements and beyond Standard Model searches, the NSW upgrade is of major importance to the ATLAS physics program.

### 5.3.1. NSW Structure, Alignment System and Shielding

The overall layout of the NSW follows the design of the current MS Small Wheel: A staggered detector arrangement of eight small and eight large sectors, providing overlap at their edges to fully cover the disk with active detector area (figure 5.9).

The support structure is built of several spokes attached to a central rim, called Hub, and supported by two feet-spokes in the sectors 11 and 15. The sectors are kinematically mounted [106, chapter 10] on this support structure, with the small sectors being closer to the **I**nteraction **P**oint (IP) (figure 5.10).

[6] With the exception of the NSW electronics, which will be compatible with the L0/L1 trigger scheme.

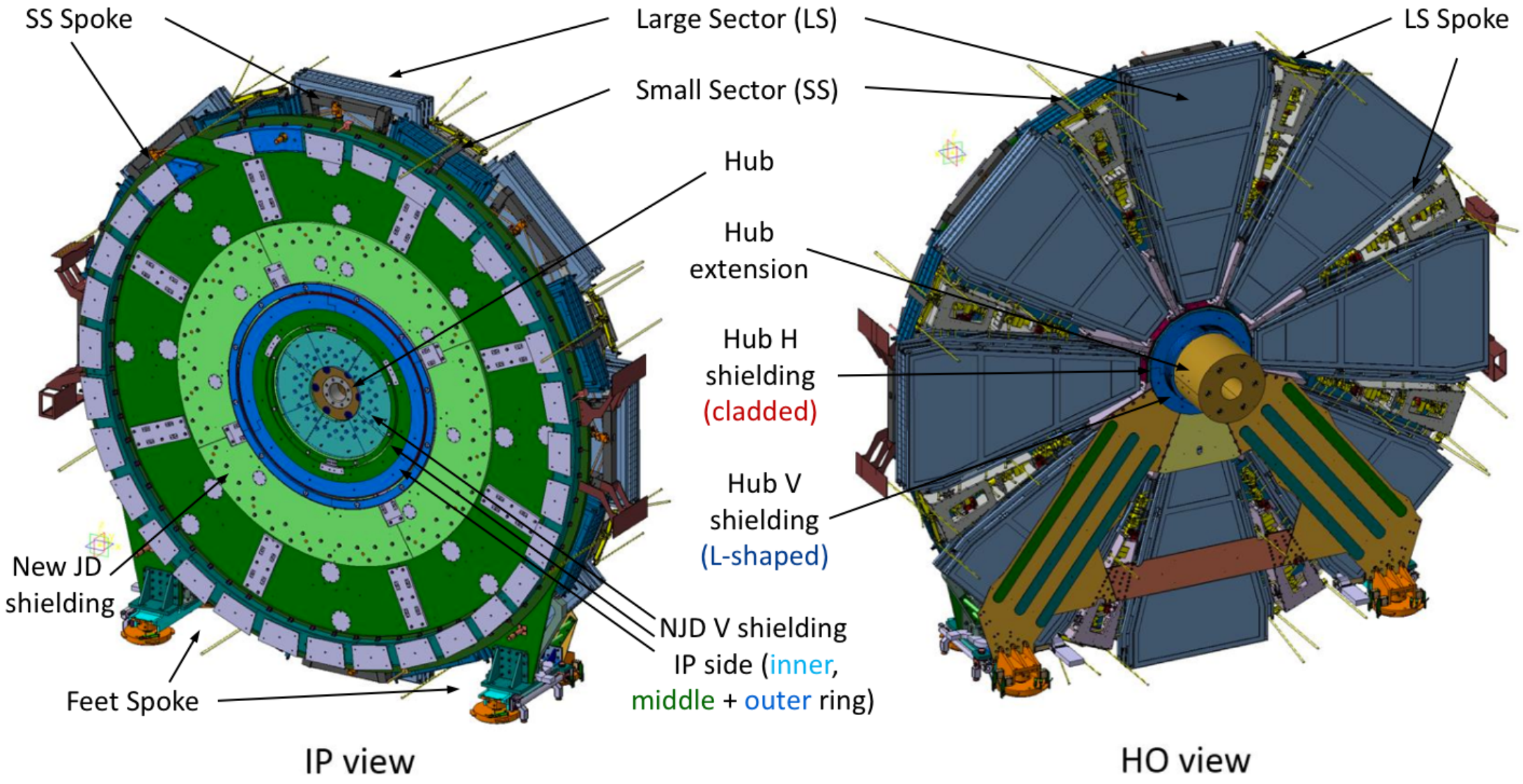

**Figure 5.9.:** Overall layout of the New Small Wheels (NSW) with its main constituent parts seen from the interaction point (left) and the HO side (right). [106]

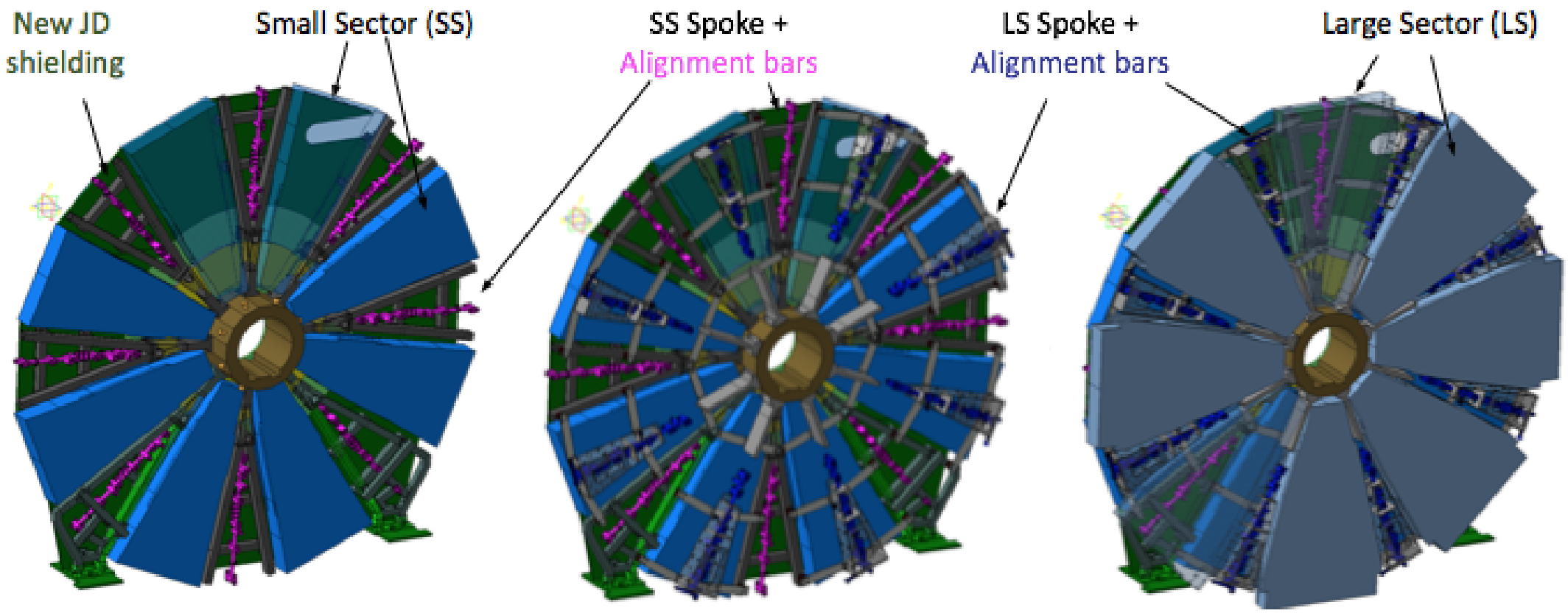

**Figure 5.10.:** Depiction of the NSW assembly steps, starting with the mounting of the small sectors and their spokes on the support structure already equipped with the JD (left), assembly of the large sector spokes (center) and completion with the large sectors (right). The position of the 16 alignment bars inside the spokes is shown.

Each of the sectors is composed of four wedges of detector modules (figure 5.11), with the two tracking detector-wedges (Micromegas) being mounted on both sides of a central spacer frame and the trigger chambers (sTGC) attached to the outside of the sector. This arrangement provides the largest possible lever arm between the trigger detectors, maximizing their angular resolution once signals from both wedges are combined.

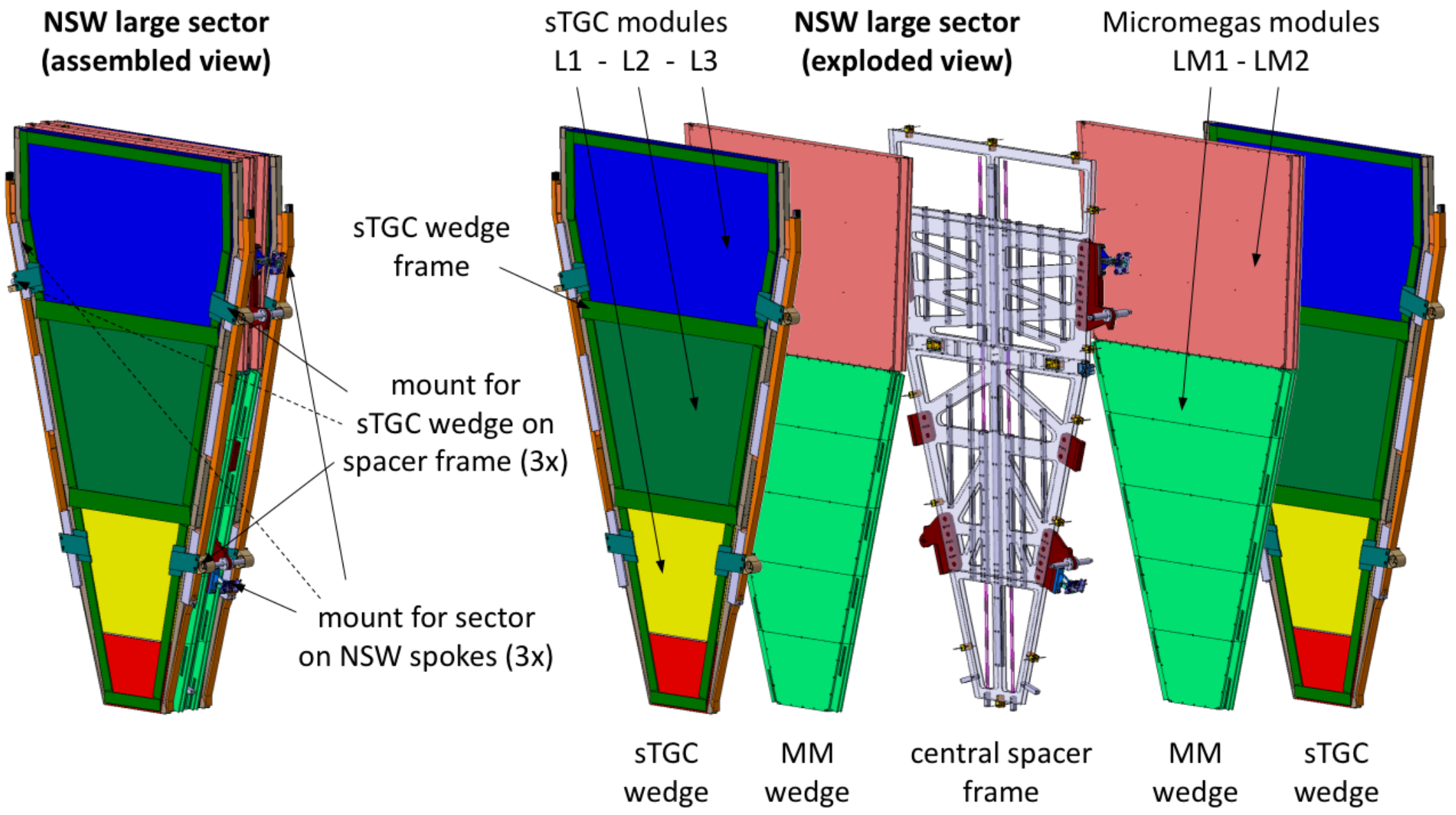

**Figure 5.11.:** Drawing of one NSW large sector in assembled (left) and exploded view (right). The sector's segmentation in wedges and their subdivision into modules is shown as well as the positions of the kinematic mounts. On the Micromegas LM1 the segmentation of the readout plane into 5 read-out boards is indicated, as is the distinction of the highest-rate region on the sTGC modules.

Accurate track reconstruction requires a precise knowledge of the detector position, which is limited by the mounting accuracy of the sectors, their deformation under weight and thermal expansion. The NSW online alignment system is based on alignment bars, mounted inside the spider-web like structure of the spokes (figure 5.10). They comprise cameras to constantly monitor the position of the chambers, defined by light sources which are precisely glued[7] to the detector's outer skin. Additionally, the internal deformation of the alignment bar and its relative position with respect to other bars and to the current MS end-cap alignment system is measured.

To reduce the background particle flux in the detectors, the NSW detectors are protected by a disk of absorbing materials, the so-called **N**ew **JD** (NJD) shielding. In the high-$|\eta|$ region, corresponding to $r \leq 1775$ mm, additional vertical shielding will be added to both sides of the disk (NJD V, IP / HO side) to mitigate the impact of the higher particle flux. Additional material layers are applied around the shielding copper hub as a cladding (horizontal, Hub H) and in-between the NSW as well as in the toroid end-cap magnets (vertical, Hub V) [106]. The shielding material composition of mainly copper, polyboron and lead has been optimized towards the absorption of electrons, neutrons, protons and other charged particles within the envelope and weight allowance [107].

[7]The position of the calibrated alignment platforms on the detector chambers will be verified with 20 µm and 50 µrad accuracy.

### 5.3.2. NSW Detector Modules and -Requirements

The NSW detectors shall provide an offline muon $p_T$ measurement with 10 % resolution at 1 TeV/c, in the $1.3 \leq |\eta| \leq 2.7$-range. This will extend the excellent muon reconstruction capability of the ATLAS barrel to the end-cap region, causing a sizable effect on the ATLAS physics analysis. So far the $p_T$ threshold of a muon candidate in the end-cap region is increased for many ATLAS searches to suppress the higher muon fake rate compared to a barrel muon. Therefore, many signatures including one, or exclusively, muons registered in the end-cap detectors, are discarded, reducing the utilizable statistic significantly. To reach this requirement a track segment reconstruction on 50 µm-level corresponding to a single plane resolution of 100 µm in the bending coordinate ($\eta$) must be reached by the NSW detectors. A track segment efficiency of ≥97 % for $p_T$ >10 GeV is required as well as a rather coarse O(mm) spatial resolution in the second coordinate ($\phi$). Furthermore, an online angular resolution of ≤1 mrad on L1 trigger must be provided for effective fake muon rejection. Similar to the other ATLAS MS stations, two gaseous detector technologies will be utilized to achieve these requisites: **s**mall-strip **T**hin **G**ap **C**hambers (sTGC) as primary trigger detector and **Microme**sh **ga**seous **s**tructure (Micromegas, MM) detectors (figure 5.12) for precise muon track reconstruction. While the distinction between trigger and tracking detectors is comparatively stringent in the current system, comprising of slow but precise MDTs combined with fast but coarsely segmented RPCs, the NSW sTGC detectors are designed to contribute substantially to the tracking. Vice versa the Micromegas hit information will be used for trigger decision, adding additional redundancy and robustness to the system.

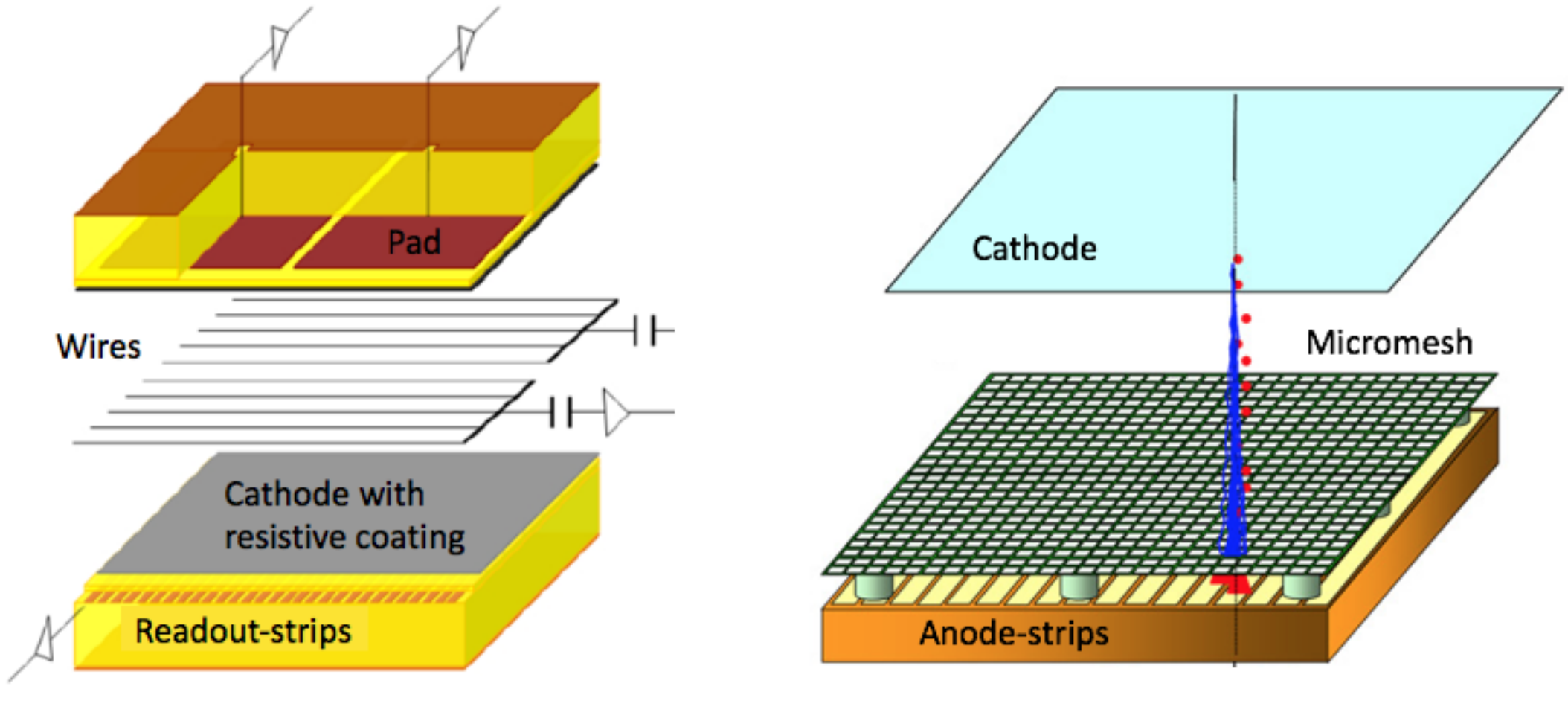

**Figure 5.12.:** Schematic drawing of a sTGC (left) and a Micromegas (right). [9]

The sTGC detectors can be regarded as an evolution of the TGC detectors of the current ATLAS MS (5.1.4). They are multi-wire proportional chambers (1.2.3) with 50 µm thick gold-plated tungsten wires floating in 1.4 mm distance from the chamber's cathodes. Those surfaces are coated with graphite-epoxy to render the detector high rate capable up to 20 kHz/cm$^2$ [8]. The resistive cathodes couple capacitively to an arrangement of pads on one side and strips, running orthogonal to the wires, on the other. While the pad signals provide fast trigger information, the strips can be used to assess the precision coordinate ($\eta$). Additionally, the signal read-out from the wires provides a coarse measurement of the second coordinate ($\phi$) with reduced accuracy, due to wire grouping and channel reduction. The sTGC operation parameters are listed in table 5.1.

The Micromegas technology (1.2.5), implemented for the precision tracking detectors, has been introduced and discussed in detail in part I of this thesis. Several choices concerning the specific detector details for the NSW Micromegas will be scrutinized in the following chapter 6. Furthermore, the challenges related to design, production and quality assurance of two major detector components, the anode PCBs and the micromeshes, are discussed in chapter 7 and 8. Meanwhile an overview of the main Micromegas parameters is given in table 5.1.

| **Item** | **Symbol** | **NSW-sTGC** | **NSW-Micromegas** |
|---|---|---|---|
| Overall geometry | | MWPC | 2-stage PP |
| Gas gap thickness<br>Gas mixture<br>Gas pressure<br>Gas temperature | $d_{gas_gap}$<br><br>$p_{gas}$<br>$T_{gas}$ | 2.8 mm<br>$CO_2$:n-$C_5H_{12}$ (55:45)<br>ambient + O(4 mbar)<br>ambient, ≈ 20°C | 5.2 mm<br>Ar:$CO_2$ (93:7)<br>ambient + O(4 mbar)<br>ambient, ≈ 20°C |
| Readout structure<br>and pitch | | strips, 3.2 mm<br>pad, ≈8 cm<br>wire, 1.8 cm | strips, 425 / 450 µm<br>(large / small sector MM) |
| Surface resistivity | $R_{electrode}$ | 100 or 200 kΩ/□ | 1 MΩ/□ |
| Voltages | $U_{(...)}$ | +2.9 kV (wires)<br>GND (cathodes) | +580 V (anode)<br>-300 V (cathode)<br>GND (mesh) |
| Spatial resolution<br>(single plane) | $\sigma_\eta$<br>$\sigma_\phi$ | O(100 µm)<br>O(≈3 mm) | |

**Table 5.1.:** Main parameters of the NSW sTGC and Micromegas detectors [9].

While the sTGC wedges are segmented radially into three modules, the Micromegas wedges are composed of two chambers (figure 5.11). This allows for a coverage of the inactive areas at the module edges of one detector system by the other and vice versa. Each of the modules comprises of four active detector layers, therefore, they are often referred to as quadruplets. The two Micromegas modules will be mounted directly on the spacer frame, whereas the sTGC quadruplets will be glued to an FR4 frame to form a full wedge before being attached to the sector. With the alignment system monitoring the position and deformation of the modules (Micromegas) or wedges (sTGC), the intrinsic accuracy of the quadruplets (and their relative alignment) are crucial for precise track reconstruction. Therefore, each detection layer is required to be flat with a non-planarity $\leq$50 µm and their stacking must be precise as well as parallel, ensuring a deviation from its nominal z-coordinate of $\leq$80 µm for each layer. Furthermore, the position of each readout channel within the module must be known on $<$ 100 µm with respect to the alignment platforms. This requires high precision of the utilized components, accurate in-plane alignment during the construction of each read-out layer, precise inter-plane alignment during quadruplet assembly and exact positioning of the alignment platforms on the module's outer skin.

[8] A higher surface resistivity is desirable for lower rate operation and, therefore, the cathode resistivity is increased for sTGC detectors located at larger radii.

### 5.3.3. NSW Electronics, DAQ and Trigger

The readout chain of the NSW detectors [9,108], as visualized in figure 5.13, is based on the VMM [109], a customized **A**pplication **S**pecific **I**ntegrated **C**ircuit (ASIC) designed in the radiation tolerant IBM 130 nm process. For the **M**icro**M**egas detectors, the **F**ront-**E**nd boards (MMFE8) comprise of eight VMMs connected to up to 512 readout strips. The sTGC detectors will be read-out by two types of **F**ront-**E**nd **B**oards (FEB): the **s**(trip)FEB reading up to 448 strips via 8 VMMs and the **p**(ad)FEB comprising of two VMM chips to read up to 104 pad channels and one VMM connected to the sTGC wires.

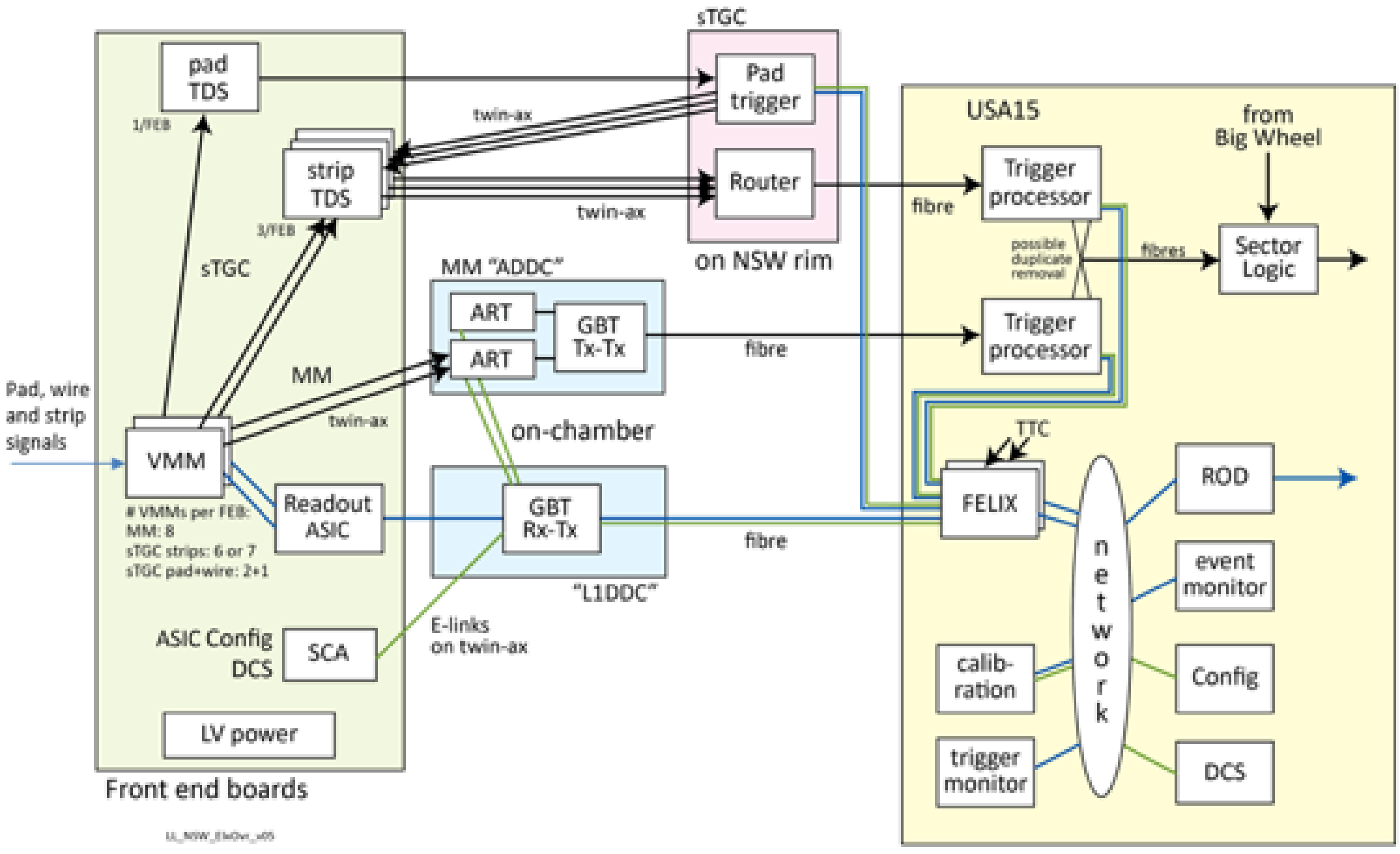

**Figure 5.13.:** Trigger and DAQ scheme for the NSW detectors comprising of front-end boards (left), on-chamber and rim electronics (center) and off cavern processing units (right). [108]

Each VMM provides selectable polarity, low-noise charge amplification with an adjustable gain (0.5, 1.0, 3.0, 9.0 mV/fC), baseline stabilized signal shaping with changeable integration time (25, 50, 100, 200 ns) and discriminator circuits for 64 channels. It measures and digitizes the peak amplitude, the **T**ime-to-**P**eak (TtP) and the **T**ime-over-**T**hreshold (ToT) referring to an external trigger signal, like the ATLAS bunch crossing clock. While the output of all channels above the individually adaptable threshold is digitized and stored in a de-randomizing buffer to be read once a trigger is fired, the address of the first hit per VMM is transmitted with minimal latency to contribute to the trigger decision.

For the Micromegas trigger this **A**ddress-in-**R**eal**T**ime (ART) is serially transmitted to a dedicated companion ASIC, combining the input from 32 VMMs, identifying the strip address of the hit and appending the VMM geographical address before transmitting it to the MM trigger processor via **G**iga**B**it **T**ransceiver (GBT) links, mounted on the same **A**RT **D**ata **D**river **C**ard (ADDC). The sTGC trigger signal is sent to the **p**ad /**s**trip **T**rigger **D**ata **S**erializer (pTDS / sTDS) located directly on the corresponding front-end boards. The sorted trigger data is forwarded to the rim electronics, where a 3-out-of-4 pad coincidence is used for L0 trigger decision. The strips under the fired pad coverage are sequentially used for L1 trigger decision. The trigger data from both detectors will be merged in FPGA based processors and then send to the **F**ront-**E**nd **L**ink **I**nterface e**X**change (FELIX) [110] and the ATLAS MS sector logic to be combined with Big Wheel trigger data for final trigger acceptance.

For the data acquisition, the full information from up to eight VMM chips are aggregated in a **R**ead-**O**ut-**C**ontroller (ROC) ASIC [111], located on each front-end board, once the L1 (or L0 after Phase-II) trigger is fired. The reformatted data is then transmitted unfiltered (before Phase-II) or filtered according to L1 trigger (after Phase-II) via the **L**evel **1** **D**ata **D**river **C**ard (L1DDC) [112] to the FELIX for DAQ and event building.

Apart from trigger and DAQ components the front-end cards contain the **S**low **C**ontrol **A**dapter (SCA) to configure and monitor the electronics and provide input to the **D**etector **C**ontrol **S**ystem (DCS).

The VMM and the companion ASICs are in different stages of prototyping [108], with the main functionality proven, but with several features yet unavailable. Further iterations are ongoing in order to finalize the chip design and start with ASIC and board production. A total of 4096 MMFE8 and 768 each sFEB and pFEB will be installed on the two NSWs, providing readout for more than 2 million Micromegas strips and approximately 354 thousand sTGC channels. Twice the amount of 512 L1DDC cards will be used for DAQ of Micromegas and the sTGC data. An addtional 32 L1DDCs are required for the sTGC trigger path as well as 512 ADDCs for the Micromegas trigger.

# 6. Micromegas Technology Choices for the NSW

Introduced more than 20 years ago [36], the Micromegas concept (chapter 1.2.5) is still one of the most versatile and up-to-date gaseous detector technologies. This is owed to the constant improvements and adjustments in the realization and production techniques of Micromegas detectors. It allowed to overcome the weaknesses of the initial concept, such as vulnerability to discharges, complex and time consuming construction and limitation in active detection area.

The NSW Micromegas follow the original design in terms of a separation of drift and amplification region, but include many of the concepts developed during the last two decades of R & D, like the resistive anode layer for spark protection and a floating mesh design to allow large module sizes and efficient detector construction. These technological choices have been scrutinized, improved and finalized within the last years and several decisions have been influenced by or based on the research performed during this thesis.

## 6.1. Readout Pattern and Hit Reconstruction

The readout pattern is the main determinant for the achievable spatial resolution of a Micromegas detector. These patterns can be conveniently produced in industries using photo-lithographic etching [113] of a thin copper layer supported by a fiber glass epoxy panel (see chapter 7.2 for more details) and are, therefore, suitable for large size detector construction.

For the ATLAS NSW Micromegas the active area will be covered by readout strips, providing hit position information in the coordinate perpendicular to their running direction. The layout of the readout pattern must be compatible with the requirement of a spatial resolution of $\leq$ 100 µm in the precision ($\eta$) coordinate[1]. Furthermore, each detector module shall provide the hit position in the second ($\phi$) coordinate[2] with a resolution of O(2 - 3 mm). The hit position in z-direction can be assessed via the chamber position and the planarity of the readout plane.

[1]It should be noted that the $\eta$-coordinate in a Micromegas module, partially referred to as radial coordinate, is not identical with the ATLAS radial coordinate introduced in 5.1.1. Only the center line (geometrical symmetry axis) of each module is aligned with the ATLAS radial coordinate, and the $\eta$-strips are straight and orthogonal to this line, providing measurements along parallel lines to the modules symmetry axis.

[2]The second coordinate $\phi$ is, accordingly, not identical with $\phi_{\mathrm{ATLAS}}$, but defined as the coordinate along the $\eta$-readout strips.

### 6.1.1. Centroid- and μ-TPC Hit Reconstruction

The most basic hit reconstruction only requires a **s**ingle **s**trip **o**ver **t**hreshold (sSoT) signal. Assuming a flat probability distribution for the hit position above the strip pitch $P_{x-hit}(x) = 1/pitch$ , the mean-root-square $\sigma_{sSoT}$ of the actual hit position to the position of its registration $x_0$, the center of the strip, can be calculated as:

$$\sigma^2_{sSoT} = \int_{x_0-\frac{pitch}{2}}^{x_0+\frac{pitch}{2}} \left(P_{x-hit}(x) \times (x - x_0)\right) dx, \tag{6.1}$$

resulting in a geometrical resolution limit of

$$\sigma_{sSoT} = \frac{pitch}{\sqrt{12}}. \tag{6.2}$$

This limit is applicable for detectors where the signal caused by the hit is limited to one readout strip or channel, like some silicon strip and -pixel detectors without charge sharing or gaseous detectors with coarse readout pads. In a Micromegas with strip readout this case would only be realized in a Single Electron Response like situation (discussed in chapter 4.5) and under the suppression of transverse diffusion, or rather if the strip pitch is much larger than the diffusion. In the NSW like Micromegas a muon traversing the 5 mm thick conversion gap is expected to free a larger number of signal electrons, as described in chapter 2.2.1. Furthermore, these signal electrons undergo scattering during their drift, as discussed in chapter 3, and therefore their position when causing an avalanche in the amplification gap is affected by diffusion. As a result the charge induced on the strips after one muon hits the detector is spread over several readout strips, as depicted in figure 6.1. Given sufficient charge sensitivity and a fine temporal sampling in the readout electronics, the charge-over-position-over-time information of a hit can be used for hit reconstruction, improving the achievable spatial resolution.

In the **centroid** reconstruction method, the charge-over-position distribution of a hit is fitted with a (double-)Gaussian and the center of charge is retrieved. Although the spatial binning within the distribution equals the strip pitch, the reconstructed hit position can be more precise than the more simple single strip over threshold reconstruction and therefore $\sigma_{centroid} \leq \sigma_{sSoT}$. The resolution of a centroid reconstruction depends largely on the initial position of the signal electrons and is best if they are aligned perpendicular to the readout plane, as shown in figures 6.1 and 6.2 (b) and (d). A large number of signal electrons contribute to a better reconstruction, as does a good ratio of transverse diffusion to strip pitch: the method can be effective only if the charge is spread over several strips. On the other hand, it is worsening if the electrons are spread over too many strips, since single electron signals might not be recognized due to statistically occurring small avalanche charge not surpassing the threshold in the readout electronics.

An alternative approach uses a TPC-like reconstruction algorithm, within the Micromegas community referred to as **μ-TPC** due to the thin drift volume. By taking into account the arrival time of each cluster[3] of primary electrons at the mesh using the hit position-over-time distribution. The distance of the primary ionization from the micro-mesh can thus be derived utilizing the known drift velocity of the electrons. Combining the time and position information of signals caused by different clusters of electrons created by the traversing particle, the particle's track can be reconstructed, as shown in figure 6.2 (c). The extracted hit position is the crossing point of this track with a predefined plane, like in the center of the drift region. Successful implementation of these techniques requires a fine timing segmentation in the readout electronics and a sufficient separation of the signal caused by a cluster of primary electrons, where the segmentation can be either in space or in time. Accordingly, this method fails for perpendicular tracks with very small spatial separation and becomes more precise for steeper inclined tracks. It is, therefore, complementary to the centroid method.

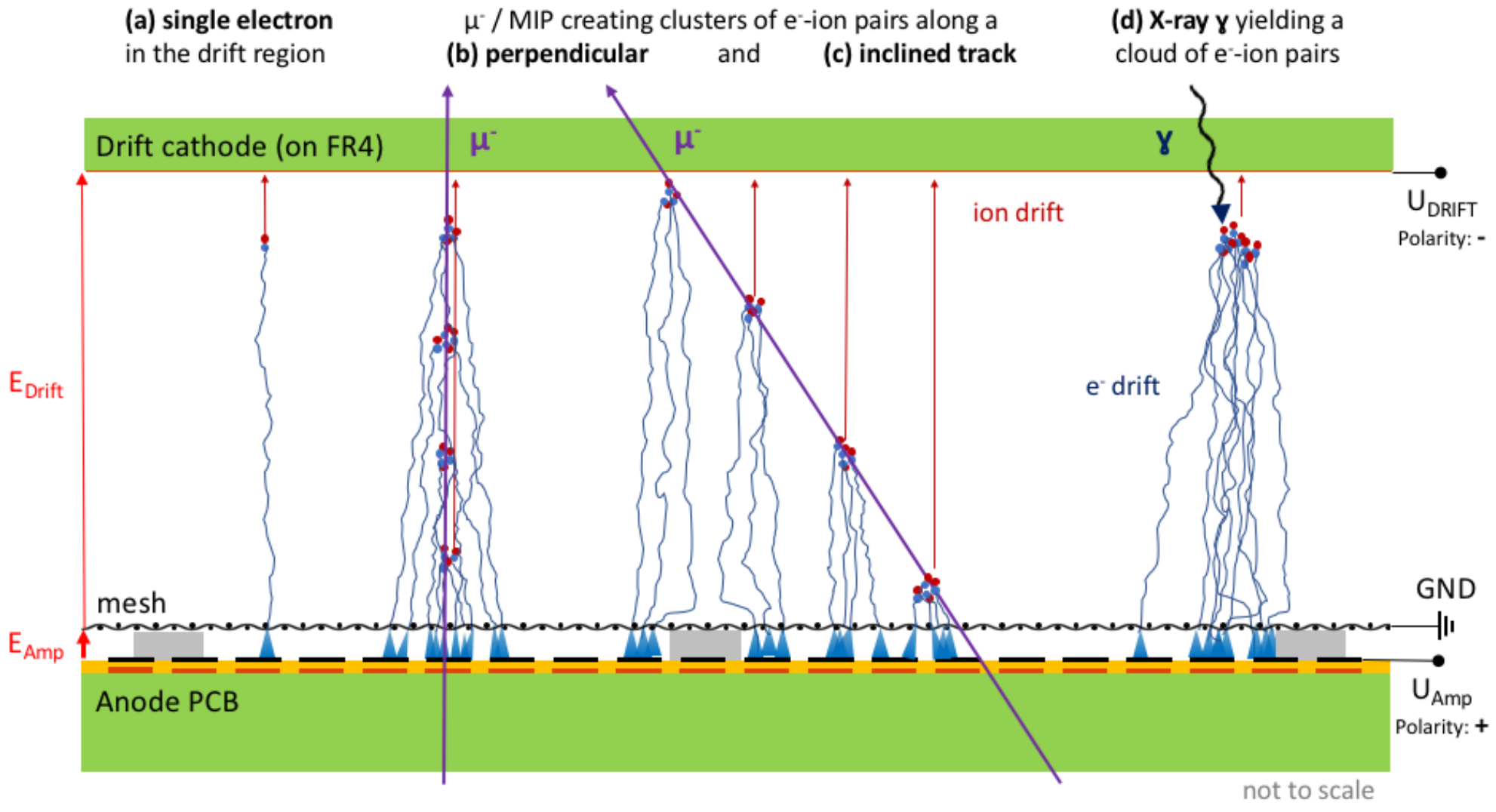

**Figure 6.1.:** Structure of a Micromegas detector and signal formation processes for different event types as shown in figure 2.4. Electron diffusion and redirection around the pillars are indicated. The number of depicted clusters and electrons are smaller compared to a real event.

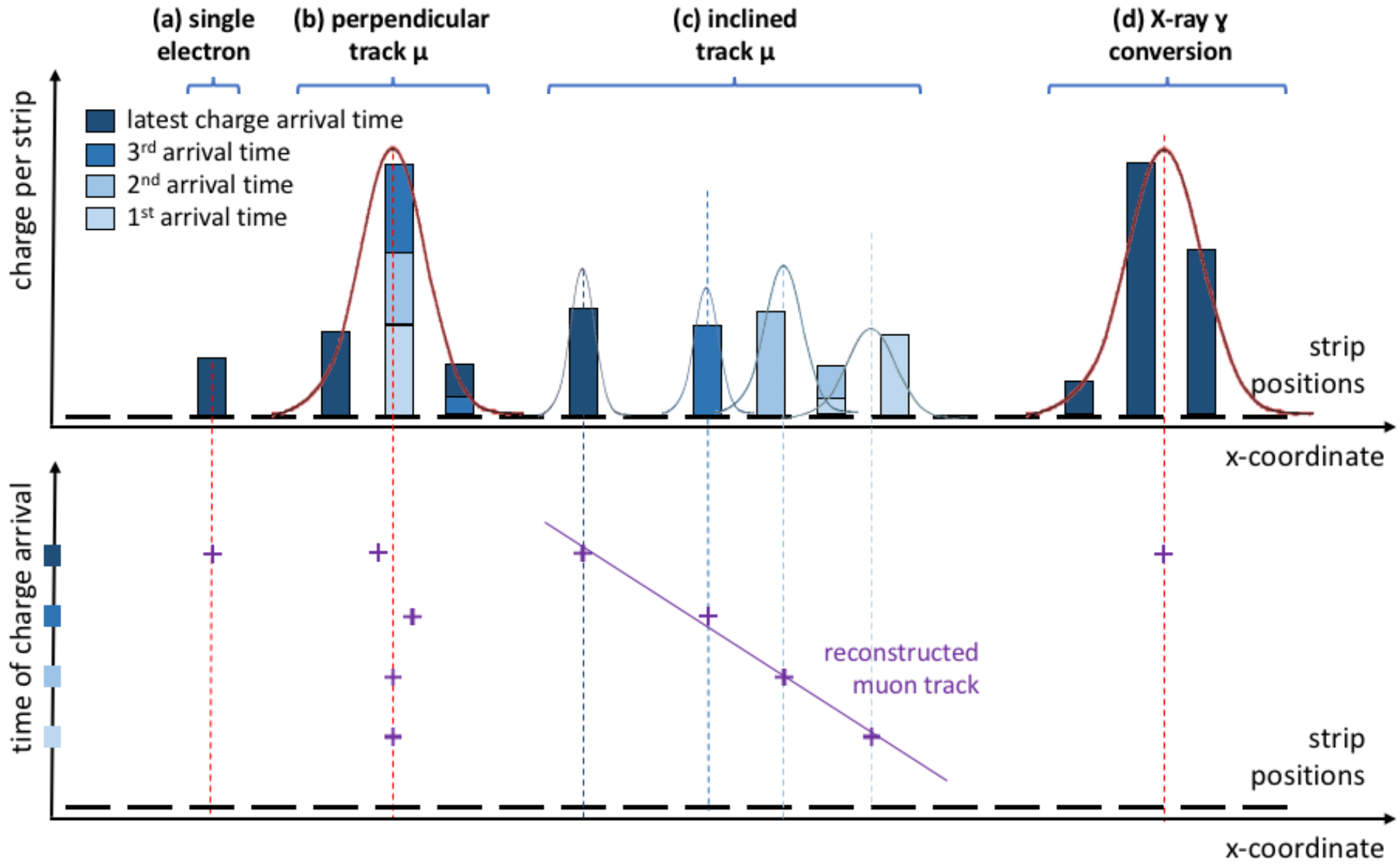

**Figure 6.2.:** Visualization of the charge deposition per strip (top) and the cluster arrival time over the spatial coordinate (bottom) for the events shown in figure 6.1. While for (a) only a single strip above threshold measurement is possible, the spatial accuracy is increased using the centroid method for (b) and (d). Considering the charge arrival time the reconstruction of inclined tracks (c) profits from a TPC-like mode.

We reported on the performance of different test detectors with O(400 µm) readout strip pitch in [114] and [115]. As shown in figures 6.3 and 6.4, the combination of both methods yields the required O(100 µm) spatial resolution for inclination angles up to 40°.

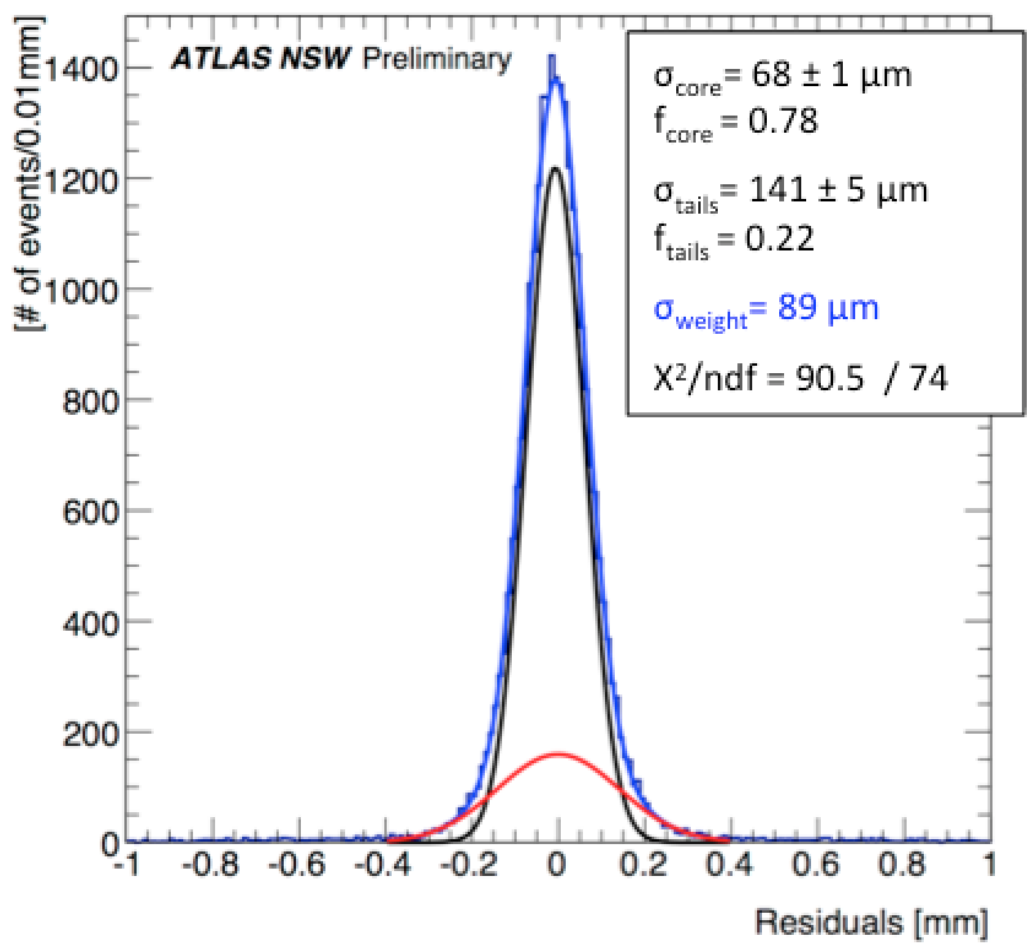

**Figure 6.3.:** Distribution of the residuals between the hit position in two T-type Micromegas chambers, divided by $\sqrt{2}$. The measurements were performed with an amplification voltage $HV_{amp}$ = 550 V utilizing a 150 GeV/c µ / $\pi^+$ beam. [114]

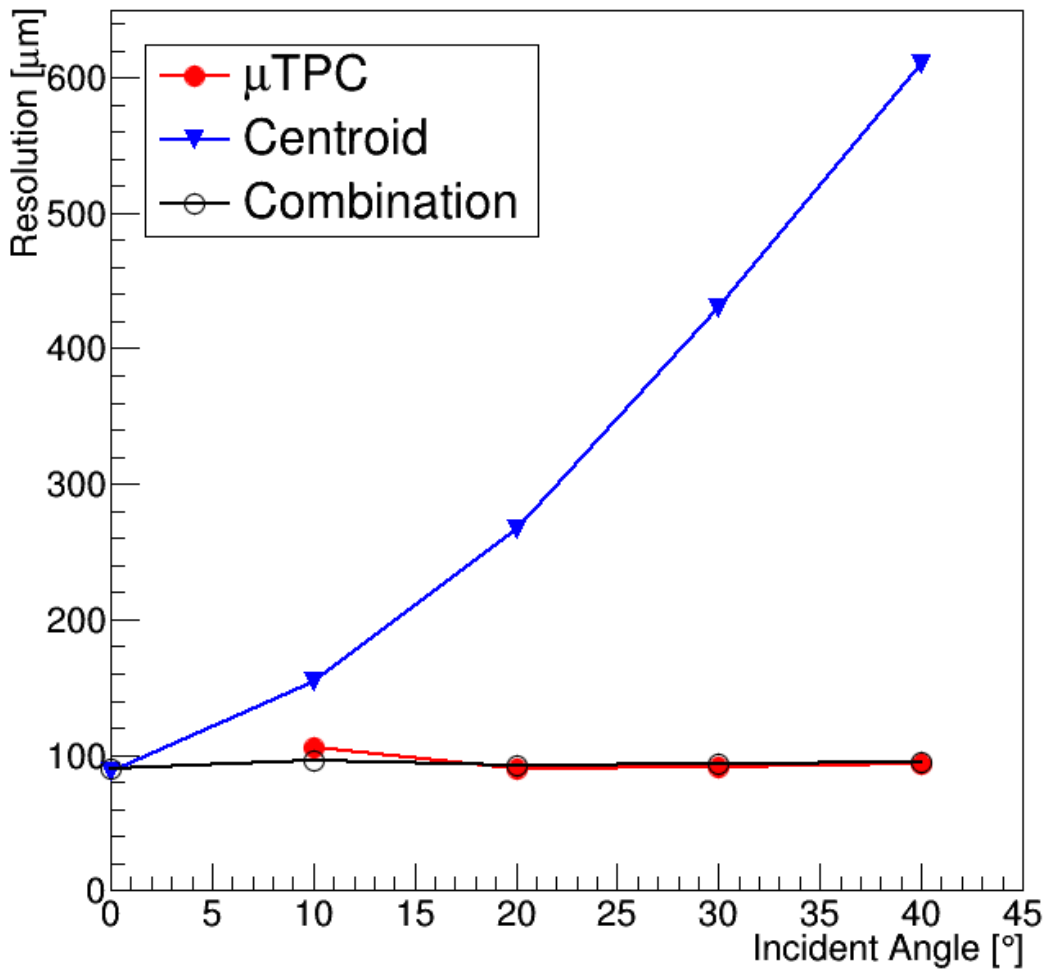

**Figure 6.4.:** Spatial resolution using the charge centroid method (blue triangles), the µ-TPC method (full red circles) and the combination of both (black open circles) as a function of the particle incident angle. Lines are for optical guidance only. [114]

Optimizing the number of readout channels towards the segmentation of the NSW Micromegas modules into three or five boards per readout plane, with the constraint of grouping strips in units of 512, read out by one front end board, a strip pitch of 450 µm and 425 µm has been selected for the large and small sector modules. For both types, the strip width is fixed at 300 µm, causing the variation in the pitch by adjustment of the inter-strip gap.

[3]It should be noted that efficient and robust cluster identification requires complicated algorithms, taking into account various effects like for example dead strips, losses of signal electrons and charge sharing or cross-talk between strips on PCB and readout electronics level.

### 6.1.2. Stereo Strips for Second Coordinate Measurement

Additionally to the excellent resolution in the precision coordinate ($\eta$), the NSW Micromegas detectors shall provide a measurement of the second coordinate ($\phi$) with reduced resolution. Instead of a further segmentation of the readout structure, yielding additional channels and conflicts in channel routing, a stereo measurement scheme [116] is applied: by rotation of the readout strips by an angle $\Theta$, the one dimensional hit position ($\eta'$) measured by the strips refers to a rotated coordinate system ($\phi', \eta'$). Combining this measurement with a second hit ($\eta''$) in a system rotated by $-\Theta$ ($\phi'', \eta''$), both coordinates in ($\phi, \eta$) can be reconstructed (figure 6.5).

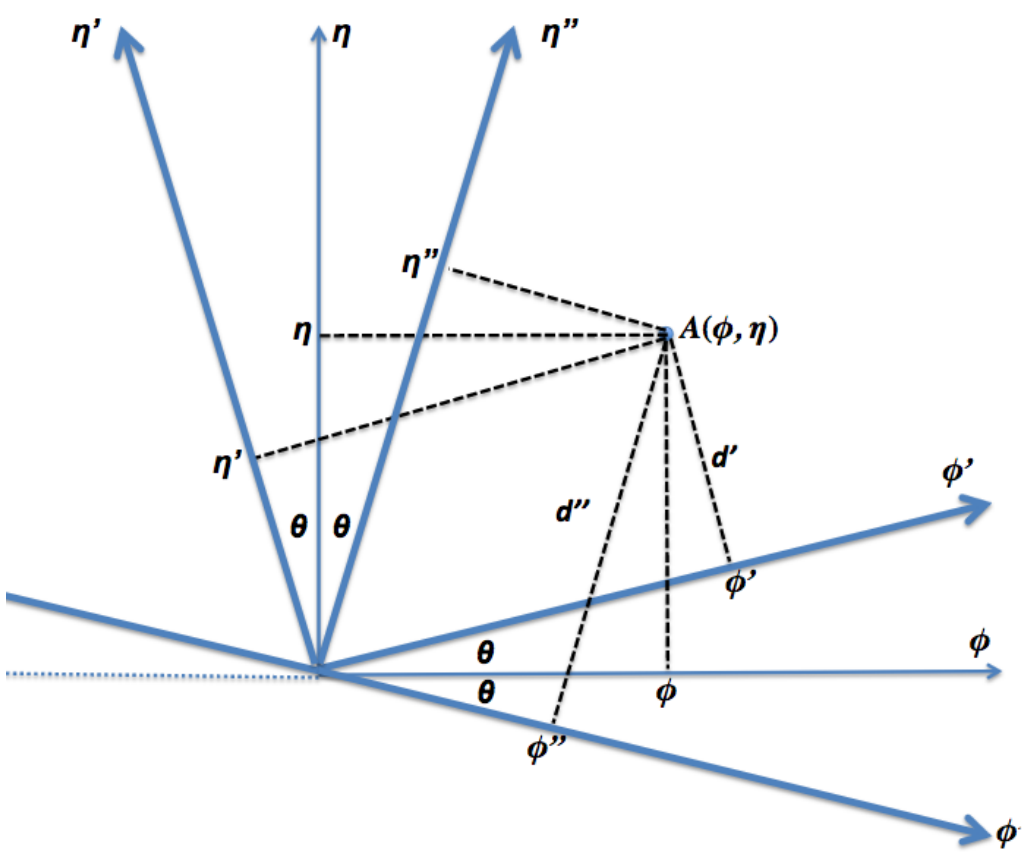

**Figure 6.5.:** Two coordinate systems ($\phi', \eta'$) and ($\phi'', \eta''$) are rotated by an angle $\Theta$ or $-\Theta$ with respect to the system ($\phi, \eta$). An arbitrary point $A$ can be equivalently described by any combination of two coordinates, for instance by ($\eta', \eta''$). [116]

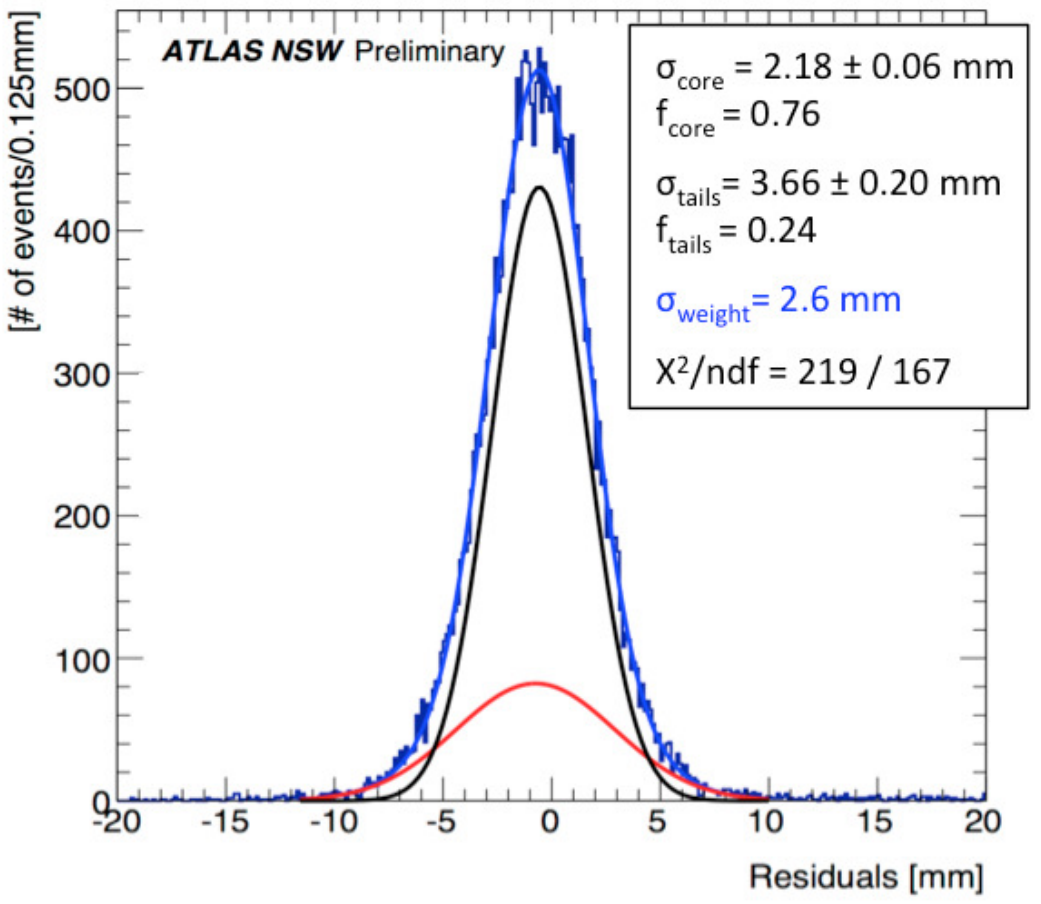

**Figure 6.6.:** Residual distributions from the hit position difference between the second coordinate hit reconstructed with the stereo doublet of the MMSW detector and the hit position extrapolated from the more accurate reference telescope. [114]

For $\Theta = 45°$ the stereo scheme matches a classical x-y-readout with the resolution of the single plane measurement in both directions. With decreasing stereo angle the measurement in one coordinate becomes more precise, while the resolution in the second coordinate decreases. Based on the resolution in the rotated coordinate systems $\sigma_{\eta'} = \sigma_{\eta''}$ the stereo angle dependent resolution in the ($\phi, \eta$) coordinates $\sigma_\phi$ and $\sigma_\eta$ can be calculated [116]:

$$\sigma_\phi = \frac{\sigma_{\eta'}}{\sqrt{2}\sin\Theta} \quad \text{and} \quad \sigma_\eta = \frac{\sigma_{\eta'}}{\sqrt{2}\cos\Theta} \tag{6.3}$$

Assuming a measurement resolution of $\sigma_{\eta'} = 100\,\mu m$, a small stereo angle of $\Theta = \pm 1.5°$ allows for a $\phi$-coordinate reconstruction with an accuracy of $\sigma_\phi = 2.7\,mm$, causing a negligible sub-%-effect on $\sigma_\eta$. This theoretically predicted resolution ratio has been confirmed in measurements (see figure 6.6) with the MMSW detector as published in detail in [115] and [114].

### 6.1.3. Quadruplet Arrangement of the Readout Planes

The 80 mm envelope of the NSW Micromegas modules can accommodate four active detector layers, each comprising of a 5.2 mm gas gap and independent drift- and readout planes, as depicted in figure 6.7.

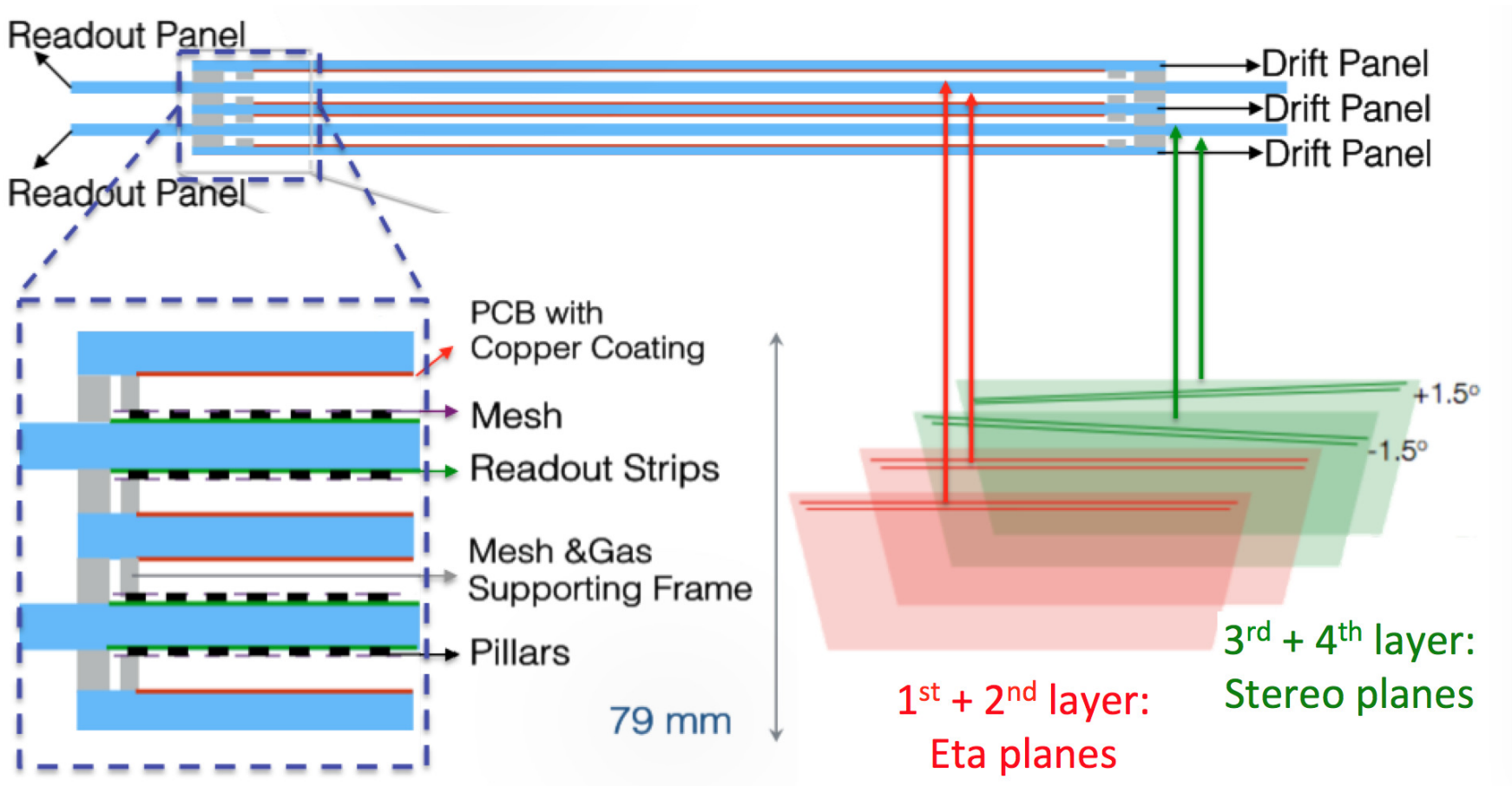

**Figure 6.7.:** Layout of a NSW quadruplet consisting of three drift and two double sided readout panels, forming four 5.2 mm thick gas gaps. In the 1st and 2nd layer the readout strips are parallel to each other and perpendicular to the radial direction (eta), in the 3rd and 4th layer the strips are inclined by $\pm 1.5°$ (stereo). [114]

The four readout planes are arranged in a back-to-back configuration forming two double sided readout panels. On one of them (eta-panel), the strips on both sides are orientated perpendicular to the $\eta$-coordinate and provide independent hit position measurements with $\sigma_\eta \leq 100\,\mu m$, and $\sigma_{\eta,doublet} \approx 70\,\mu m$ when hit information from both layers is combined. On the second (stereo-) panel the strips are inclined by $\Theta = \pm 1.5°$, forming a stereo pattern as previously described (section 6.1.2). Their combined hit information is used to reconstruct the $\eta$ hit position ($\sigma_{\eta,doublet} \approx 70\,\mu m$) and the $\phi$ hit position ($\sigma_{\phi,doublet} \approx 2.7\,mm$).

An inefficiency in one of the eta-layers results in a reduced spatial accuracy, denying the combination of the hit measurements. If the traversing particle is not recognized in one of the stereo layers (inefficiency), the second coordinate can still be reconstructed using the $\eta$-hit position obtained from the eta-doublet. Therefore, a 3-out-of-4-hits reconstruction scheme can be applied, allowing the detector to cope with inefficiencies as long as their positioning is randomized between the layers.

The back-to-back arrangement of the readout planes provides several advantages over a staggering of layers with the same orientation:

- The construction of the readout panels requires high accuracy in PCB positioning. Therefore, the combination to two readout panels minimizes the number of panels requiring high accuracy and allows to build the three drift panels with lower precision.
- Mounting the stereo layers back-to-back allows the use of the same PCB layout on both planes, introducing the change in strip rotation direction ($\pm 1.5°$) by flipping over the PCB. This minimizes the number of different PCB patterns required to build the detector and, therefore, facilitates the anode board production.
- Combining two readout planes into a doublet reduces the lever arm between the reconstructed hits and thus allows for an improved combination of the hit information, which is crucial for the stereo-scheme.

In cooperation with the CERN DT group and the University of Mainz (Germany) we constructed, tested and validated a set of two Micromegas quadruplet prototypes. These two MMSW detectors are the first ever built Micromegas quadruplets and follow closely the NSW Micromegas design envisaged at the time of their construction. While some of the design details and construction methods had to be altered or refined before being applied to the larger size NSW modules, those two detectors have been regarded as milestones in the NSW project. Especially the proof of an inter-plane positioning accuracy better than 30 µm between two readout layers in back-to-back configuration as well as the high planarity and parallelism of the panels were of tremendous importance for the further development of the construction method. The performance of the MMSW detectors and their commissioning in comprehensive laboratory and test beam measurements was a proof of principle for the quadruplet layout, the stereo reconstruction scheme and the selected anode structure.

Besides a personal contribution to this project it is clearly a collaborative effort and success, especially owed to the expertise of our colleagues in the CERN DT group. In the interest of conciseness in this thesis, no detailed report on the construction method, the laboratory tests and the MMSW performance is given and we instead refer to our publication in [115].

## 6.2. Resistive Micromegas with Screen-Printed Layers

The vulnerability to discharges has always been one of the main drawbacks of the Micromegas technology. These violent discharges, often referred to as sparks, cause a drop of voltage, rendering the detector less efficient until the potential difference is restored. The comparable high currents can damage the readout electronics and potentially the anode structure as well. Aiming for an application of Micromegas detectors in ATLAS for almost 20 years, it is essential to overcome this vulnerability.

### 6.2.1. HV Breakdown and Sparks in Micromegas

In a Micromegas, a high voltage breakdown can be caused by imperfections of the detector, such as a contamination of the amplification gap, current bridges between the mesh and the anode or regions of increased field strength due to locally reduced gap size, or pointy structures on the electrodes. Given the small spatial extent of the amplification structure Micromegas are especially prone to these type of imperfections. Accordingly, high quality standards need to be applied during the anode board production (chapter 7), mesh electrode selection (chapter 8) and the detector module construction and assembly. However, the occurrence of sparks is not limited to detector imperfections, but can be triggered by unavoidable fundamental processes, as discussed in chapter 1. If the total electron charge in an amplification process exceeds the Raether limit [14, 117] of $\approx 10^6 - 10^7 e^-$ in an O(100 µm) amplification gap of a Micromegas [118], a self sustained streamer develops and causes a discharge. Even with a properly set gain this charge threshold can be occasionally surpassed by gain fluctuations (chapter 4.1.3), events including highly $\delta$-electrons (chapter 2.2.1) or a rate dependent spatial and temporal coincidence of two events. Furthermore, ion- and photon feedback, as discussed in chapter 4.1.2, contribute to the sparking rate. Unable to exclude the occurrence of sparks in the Micromegas, the occurrence rate and their effect on the detector must be minimized.

### 6.2.2. Resistive Pattern and Inter-Layer Alignment

To achieve efficient spark protection, the NSW community opted for a pattern of resistive strips covering the readout, similar to those first proposed in [119]. This scheme provides a protection for the readout electronics, reduction in spark intensity due to the high impedance limiting the current flow and localization of the voltage drop based on the surface resistivity, the latter similar to an RPC discussed in chapter 1.2.2.

Aiming for a homogeneous impedance between the high voltage supply point and each position on the anode, the resistive strips are connected every 10 mm in an alternating pattern with their top or bottom neighbor (figure 6.9 and figure 6.10 in 6.2.3). Thus, the deposited charge flows via several paths instead of following only one strip and, therefore, the measured impedance is much more position independent, compared to the linear increase along a single strip. Additionally, these interconnection bridges yield the pattern less prone to local defects like interrupted or damaged strips (figure 6.13 in 6.2.3). The charge spread, however, remains limited to the initially hit strip and if applicable one neighboring strip, if the hit occurred in the vicinity to an interconnection.

Ideally each resistive strip would couple capacitively only to its readout counterpart. Considering capacities between neighboring strips, a charge sharing is inevitable, but can be minimized by an accurate alignment of the strips on top of each other. Since a $< 50\,\mu m$ alignment of $> 2\,m$ wide patterns is difficult to achieve during full NSW scale production, the effect of a displacement between the two layers has been studied with a dedicated Micromegas prototype (TQF), comprising of four sectors with different resistive layer to readout strip (mis-)alignment, as shown in figure 6.8 - left.

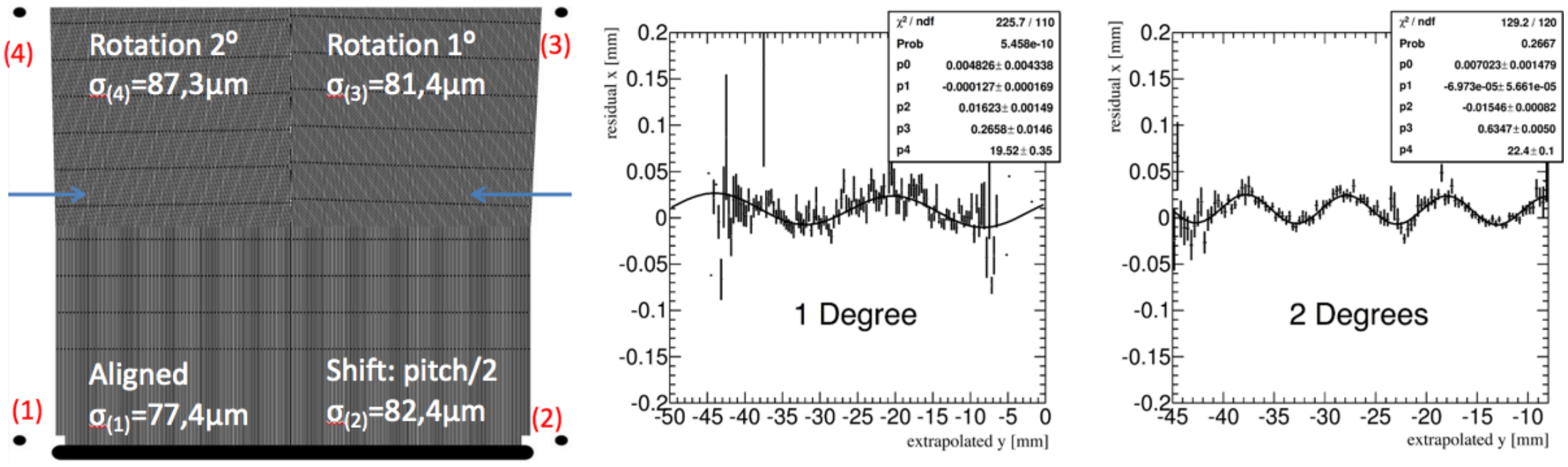

**Figure 6.8.:** Left: Relative (mis-) alignment between readout and resistive strips in the TQF sectors and position measurement resolution. Center and right: Correlation between the residuals in the x-axis and the extrapolated position in the y for the 1° (center) and 2° (right) inclination-regions of the TQF chamber. [120]

A small degradation of spatial resolution of $\Delta\sigma_{(2)-(1)} \approx 5\,\mu m$ results from the constant shift, proving the hereby utilized centroid method quite robust against the charge sharing between neighboring strips. For the rotated patterns, an average degradation of $\Delta\sigma_{(4)-(1)} < 10\,\mu m$ is observed. In sector (3) and (4) a systematic shift between real and reconstructed hit position of $\pm 15\,\mu m$, modulating with a periodicity of 25 mm at 1° and 10 mm at 2°, has been observed. The periodicity reflects the strip crossing distance at the given rotation angle. Proving that these misalignment effects on the resolution are comparatively small, no highly accurate alignment between the readout and resistive layer of the NSW Micromegas is required.

### 6.2.3. Material Choice and Production Method

Two production options have been investigated, comprising different techniques, different materials and, hence, different properties of the resistive layer.

The **carbon sputtering** is a comparatively novel technique to grow sub-μm thin layers of carbon on a substrate. The strips are created by covering a negative of the pattern before the sputtering, using conventional photolithography, and removing this protection thereafter. Thus, only the exposed areas remain coated, as visualized in figure 6.11.

The thickness of the layer is determined by the exposure time to the carbon atoms during the sputtering and the carbon density during the process. It can be well controlled on O(10 nm)-level, and thus the surface resistivity of the layer can be well adjusted. The resulting strips are thin, flat, free of surface irregularities and accurate on $< 10\,\mu$m-level in their lateral dimension (figure 6.9), all desirable attributes for an application as a Micromegas anode. Requiring a photo-lithographic processing of each foil and a long O(1 h) carbon exposure during sputtering, the production of sputtered resistive patterns is quite complex and expensive, yielding the application of this technology questionable for large size detector systems.

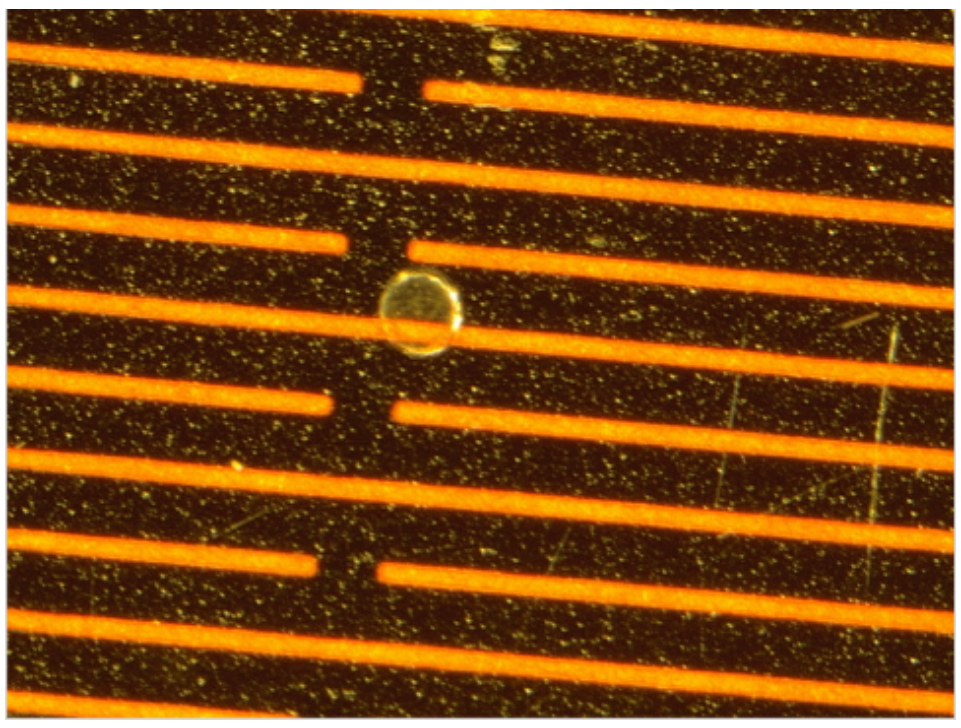

**Figure 6.9.:** Microscope picture of a sputtered resistive layer in the MMSW detector. The pattern is very accurate comprising sharp boundaries.

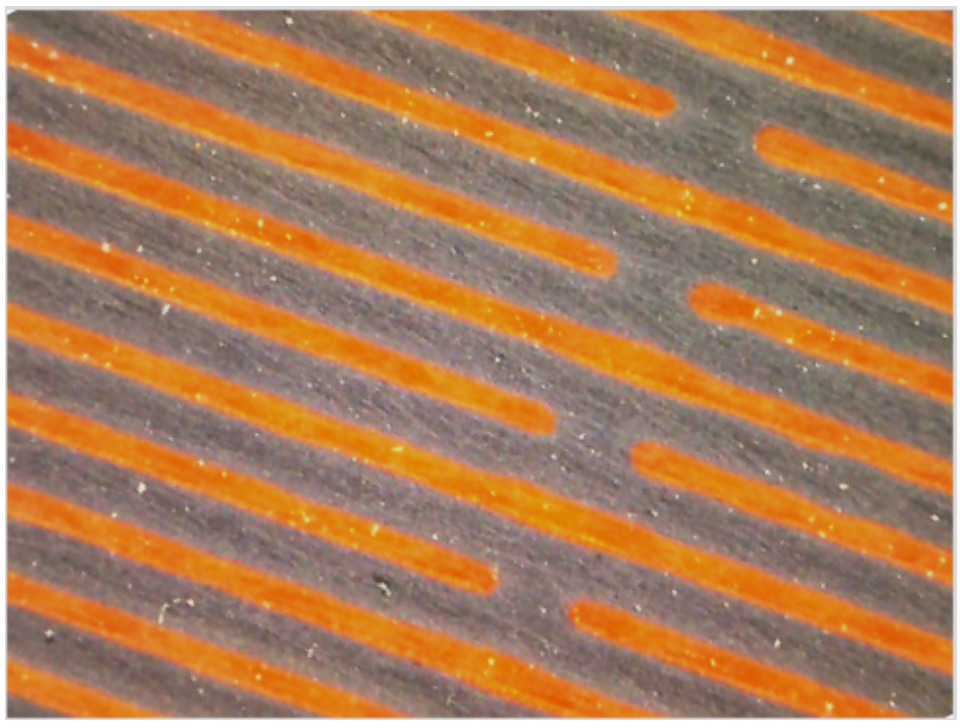

**Figure 6.10.:** Microscope picture of a screen printed resistive layer after surface polishing. Slight fluctuations in the line-/gap width and pattern accuracy are visible.

**Screen printing** is a well-established method to produce pattern of sub-100 µm accuracy over large surfaces. Squeezing resistive, carbon-doped ink through a screen with a photo-lithographically deployed negative of the pattern, the ink is deposited on the Kapton® foil, as shown in figure 6.12.

The composition and viscosity of the ink, the permeability of the mesh used as screen and the pressure and speed during the ink application influence the layer thickness and consequentially the lines' resistivity. Accordingly, tuning of the resistivity is possible but with a lower reliability compared to the sputtering process. Given the liquid nature of the ink, the resulting strips have a bump-like cross section with a height of typically $10 - 15\,\mu$m and their boundaries are not sharp, but prone to ink bleeding (figure 6.13 and 6.14). Once the pattern is transferred from the mask to the screen (figure 6.12, step 1-4), the same screen can be used for printing a large number of foils (figure 6.12, step 5-7), reducing production time and -cost significantly if many foils carrying the same pattern are required. The degradation in pattern accuracy during repeated use of the same screen can be avoided by monitoring the printing result and exchanging the screen if necessary.

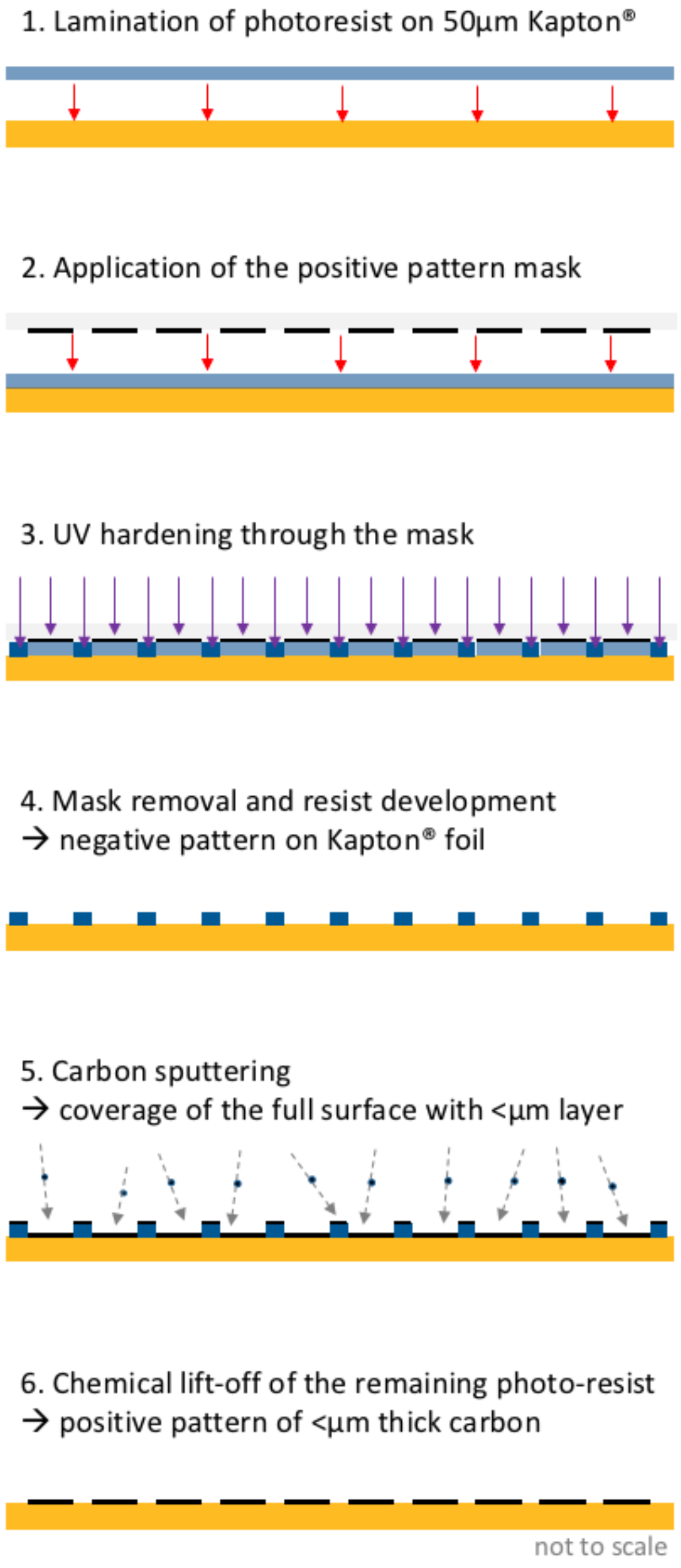

**Figure 6.11.:** Process-flow of sputtering production of a resistive foil. A photo lithographic process (1-4) is required before the foil is exposed to the carbon sputtering (5), yielding a very thin and accurate pattern after cover lift-off (6).

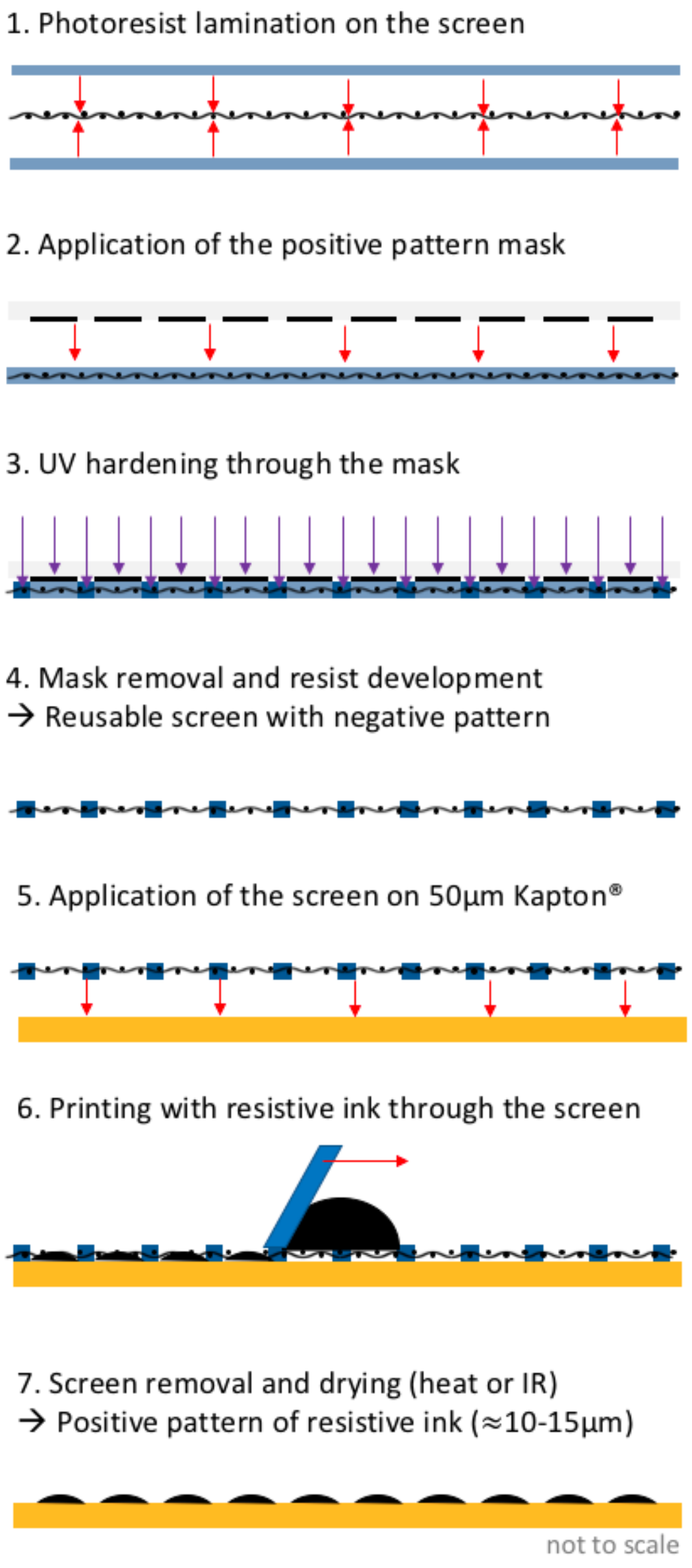

**Figure 6.12.:** Process-flow of screen preparation (1-4) and resistive foil printing by squeezing ink through the screen (6). Utilizing a reusable screen, only steps (5-7) need to be repeated to print a new pattern.

Forming the top surface of the Micromegas anode, defects in the resistive layer can have a huge impact on the detector's properties. Besides being intrinsically less accurate, screen printing of resistive layers is prone to production blemishes caused by, for example, dust or filaments on the screen, leading to non-printed areas and possibly causing ink bleeding around the blemish (figure 6.13). Irregularities in the printed pattern can as well be caused by non-uniform pressuring of the ink through the screen, causing reduced ink application (figure 6.14). A bumpy surface of the resistive Ink and filaments trapped in the layer which are pointing into the amplification gap, will yield increased electric fields and enhance the risk of discharges. To remove this bumpiness, all foils are polished using fine sandpaper.

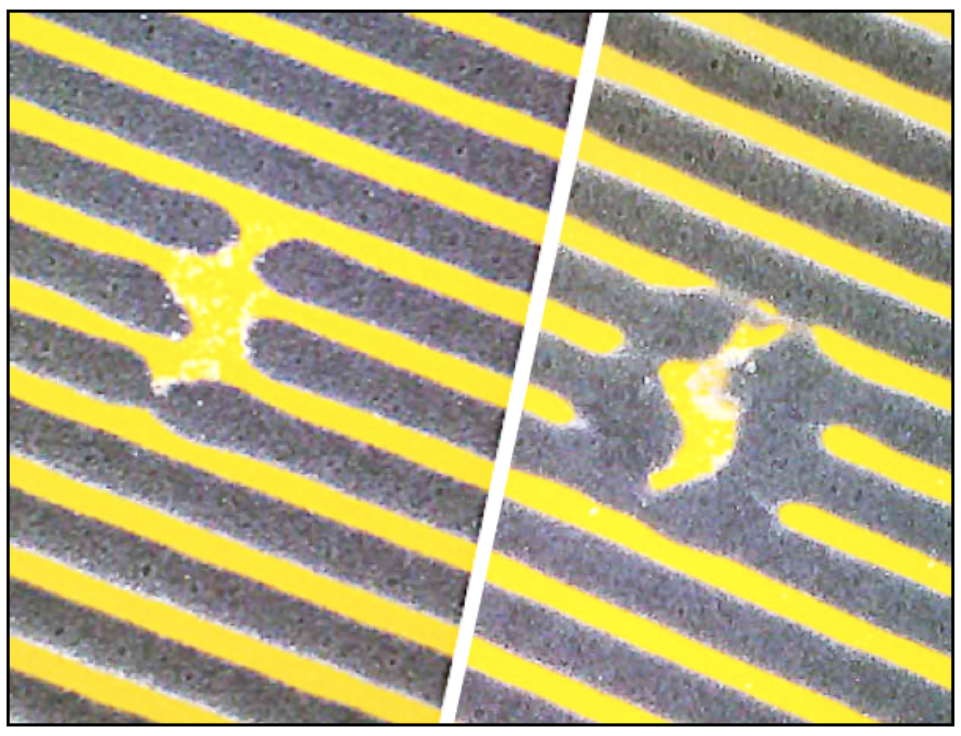

**Figure 6.13.:** Microscope picture of two screen printed resistive layers with a printing defect due to a filament on the screen, without (left) and with (right) ink-bleeding around the defect.

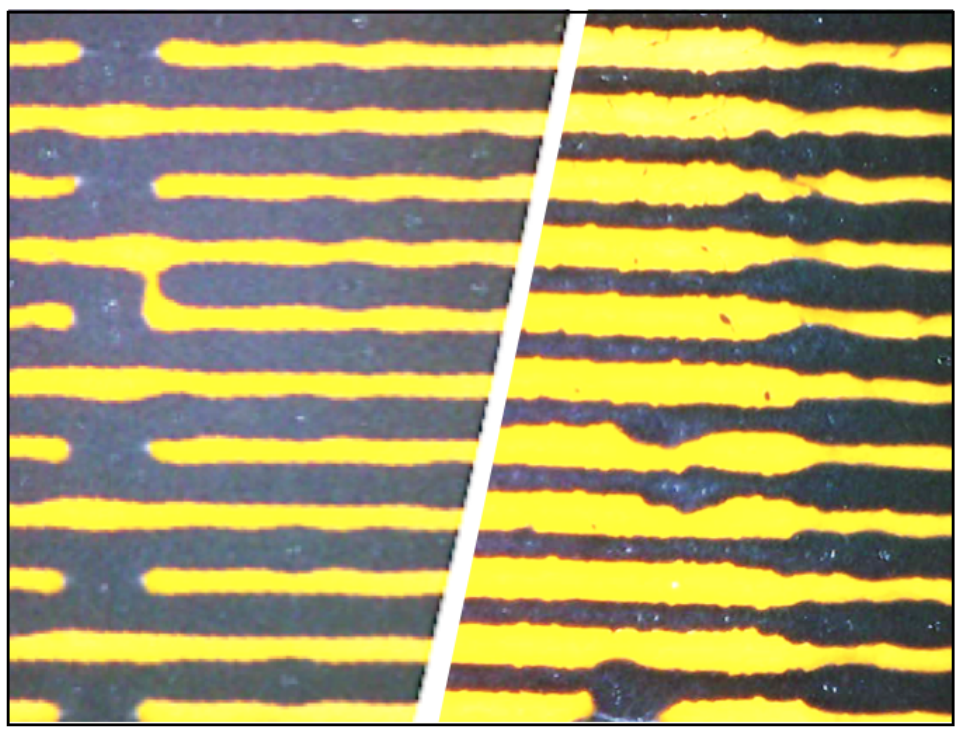

**Figure 6.14.:** Microscope picture of two screen printed resistive layers in bad printing quality, yielding single unconnected lines (left) and an area with very thin strips (right).

While being insusceptible to these type of production defects, sputtered resistive layers lack the mechanical robustness of their screen printed counterparts. The high pressure involved in PCB processing (see chapter 7) can cause mirco-cracks in the very thin carbon structure. The cracked layers are prone to delamination of the carbon, either by adhesive materials used during PCB production, removing very small point like regions of the layer (figure 6.15), or during the subsequent cleaning of the boards or panels where larger flakes of cracked carbon can be removed (figure 6.16).

Although desirable in terms of pattern accuracy and anode flatness, the insufficient robustness and high production costs disqualified sputtered resistive foils for a use in the NSW Micromegas and a screen printing production scheme has been chosen instead.

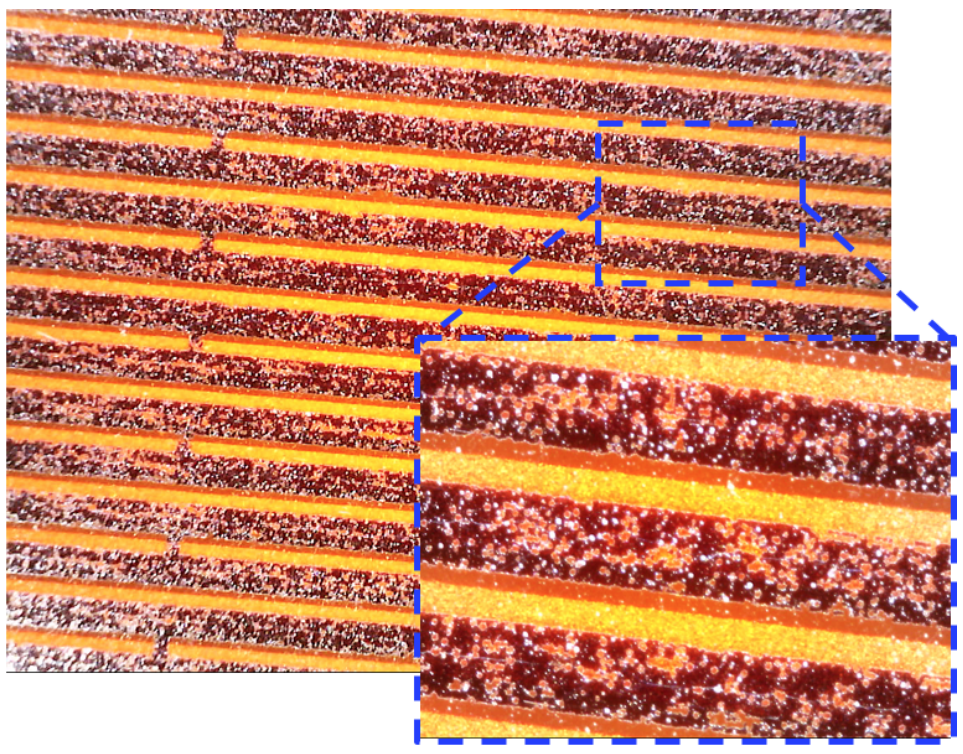

**Figure 6.15.:** Microscope picture of a sputtered resistive layer with significant damage after the PCB processing. A large number of ≈ 20 µm diameter spots is missing in the carbon pattern.

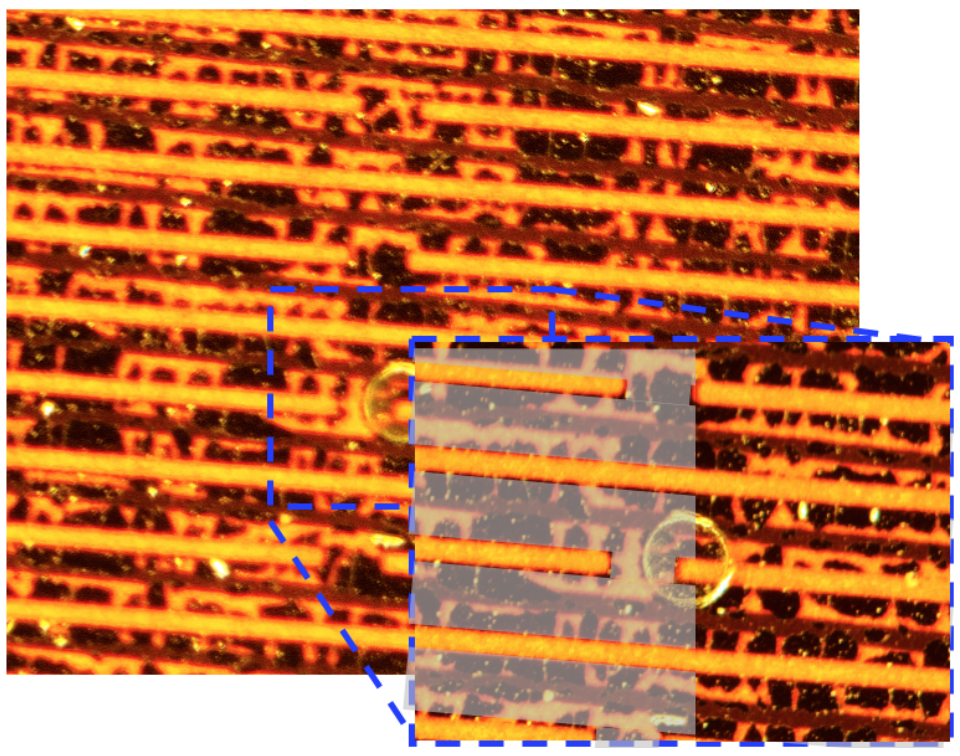

**Figure 6.16.:** Microscope picture of a sputtered resistive layer with significant damage after wet cleaning. Larger flakes of the carbon layer have been removed. The shaded area indicates the nominal position of the resistive lines.

## 6.3. Mechanically Floating Meshes on Pyralux® Pillars

The positioning of the mesh above the Micromegas anode is crucial to obtain a homogeneous gain on the full detector surface and consequentially an optimized energy resolution, as discussed in chapter 4. In a Micromegas the distance between the anode and the mesh is typically defined by precise insulating spacers. In the first Micromegas construction [36] quartz fibers where glued on the anode in regular distances and the mesh thereafter was stretched over these fibers and glued onto their top. Some modern small size Micromegas still use nylon fishing lines as precise spacers [87]. However, the manual positioning and gluing of the wires is tedious mechanical work and the required time scales linearly with the detector's surface, disfavoring this concept for large detectors like the NSW. Furthermore, the continuity of the spacing lines yields a non-negligible non-sensitive area of up to several percent, by blocking the amplification volume. Two approaches have been developed to overcome these constraints and allow for Micromegas mass-production: the *bulk* Micromegas and the mechanically *floating mesh* technique.

### 6.3.1. Floating Mesh vs. Bulk Micromegas

Only five years after the first Micromegas has been build using quartz lines, Derré et al. replaced these wire spacers with cylindrical pillars [121, 122] developed with lithography on a photo-resistive polyamide film, as visualized in figure 6.17. These structures can be as small as 200 µm in diameter and ensure a precise height according to the thickness of the (multiple) layer(s) of photo-resistive coverlay like DuPond Varcel® or Pyralux®. If sufficiently stretched within an external frame, the meshes do not need to be glued on this pillar pattern, but can be mechanically pressed on top. The electrostatic force, resulting from the voltage difference between mesh and anode, once polarized, strengthens the attachment of the mesh on the pillars and holds it in a well-defined position. Due to the lack of mechanical fixation within the active detector area this technique is in the following referred to as (mechanically-) floating mesh.

An alternative use of photo-lithography is the encapsulation of the mesh in the photo-resistive layer, resulting in a so called 'bulk' Micromegas [123]. Therefore, the mesh is laminated on top of one (or several pre-laminated) layer(s) of photo-resist and covered with an additional layer. The lamination at high pressure and temperatures of O(90°C) causes the photo-resist to melt, fill the holes of the mesh and fuse with the layer below. The full stack is thereafter hardened at the pillars' positions using UV light and developed to remove all non-hardened material. The mesh remains unaffected by the development process and is finally held in position above the anode by the pillars, see figure 6.18.

Both methods significantly reduce the anode surface occupied by the spacer structure and can reach sensitive areas above 99 %. The maximum unit size is for the bulk process limited by the size of the printed circuit board (PCB) and the equipment for lamination and the photo-lithography process. The limiting industrial standard width is 60 cm and boards of more than 2 m length can be processed by only a few companies worldwide. Electrical integrity of the detector forbids free mesh wire ends and thus demands a coverlay frame to embed the wires around the circumference of the active area, reducing the efficient detector surface when joined to larger modules. Following the floating mesh scheme, several PCBs can be joined to one readout plane before applying the mesh, requiring only the continuity of the support structure at the PCB joints. Therefore, the size of an uninterrupted detection unit is only limited by the size of the mesh. Woven wire meshes, discussed in chapter 8, are available on 2 m width, basically unlimited in the second dimension, as industrial standard and with larger width on demand from specialized companies.

As Micromegas are prone to contamination in the amplification gap, causing discharges once mesh and anode are polarized, this narrow gap must be completely free of dust particles, filaments and enclosures. High cleanliness standards during stacking and lamination reduce the occurrence of such a contamination, but a sufficiently high yield in terms of perfect cleanliness in industrial production poses a very difficult challenge. While small dust particles trapped between the anode and the mesh in a bulk Micromegas can be flushed out through the mesh openings, larger filaments or enclosures are almost impossible to remove, disqualifying the full board. With a removable and re-attachable floating mesh, the amplification gap can be thoroughly cleaned right before the closure of the detector, thus mitigating the impact of unavoidable contamination during the production process and reducing the cleanliness requirements during panel construction.

For both of these arguments, clearly disfavoring a bulk Micromegas design for the several $m^2$ size detectors of the ATLAS NSW, the floating mesh technique has been chosen. The pillars will consist of two 64 µm layers of DuPond Pyralux®, resulting in a comparable thick amplification volume, which yields the Micromegas less sensitive to small dimensional variations, like the structure of the anode (section 6.2) or the micromesh (chapter 8).

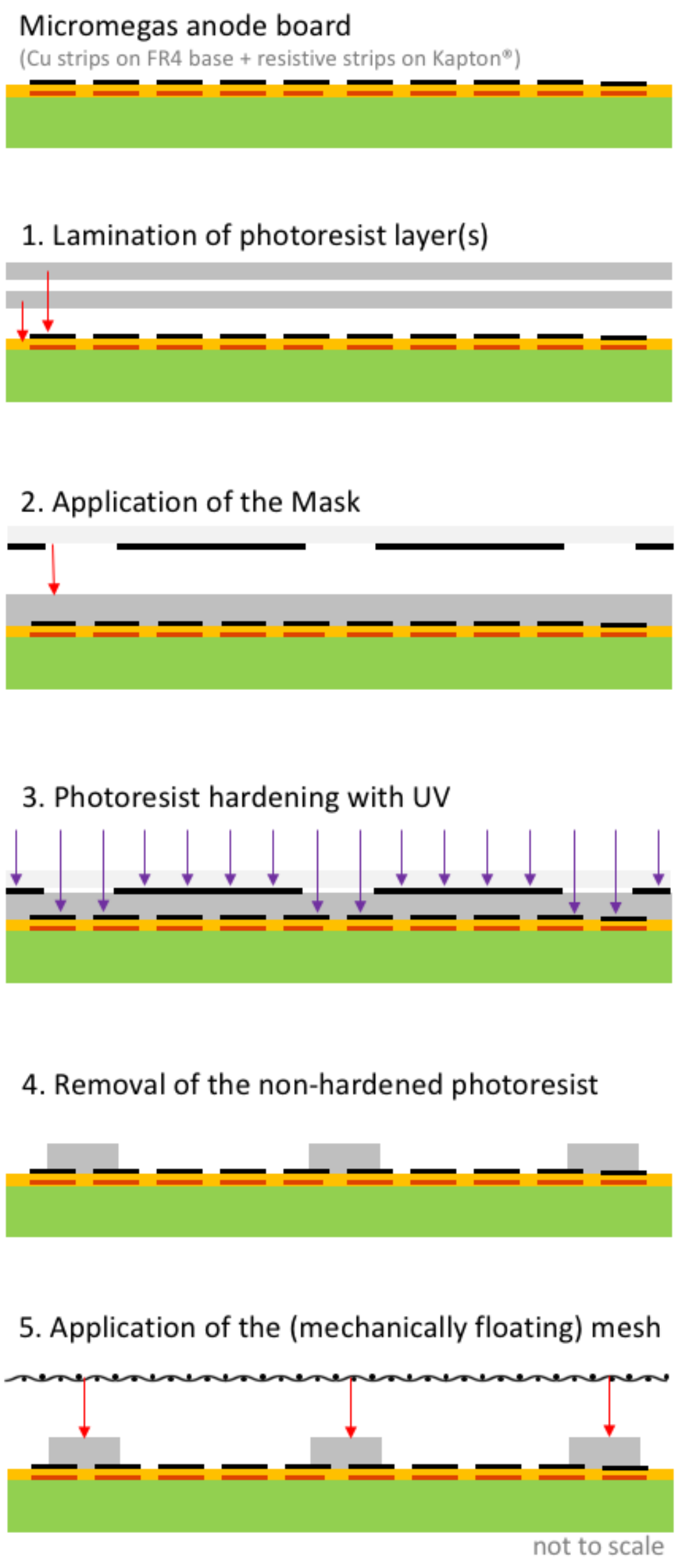

**Figure 6.17.:** Process-flow of photo-lithographical deposition of pillars on-top of a Micromegas anode board and subsequent mechanical positioning of the micromesh.

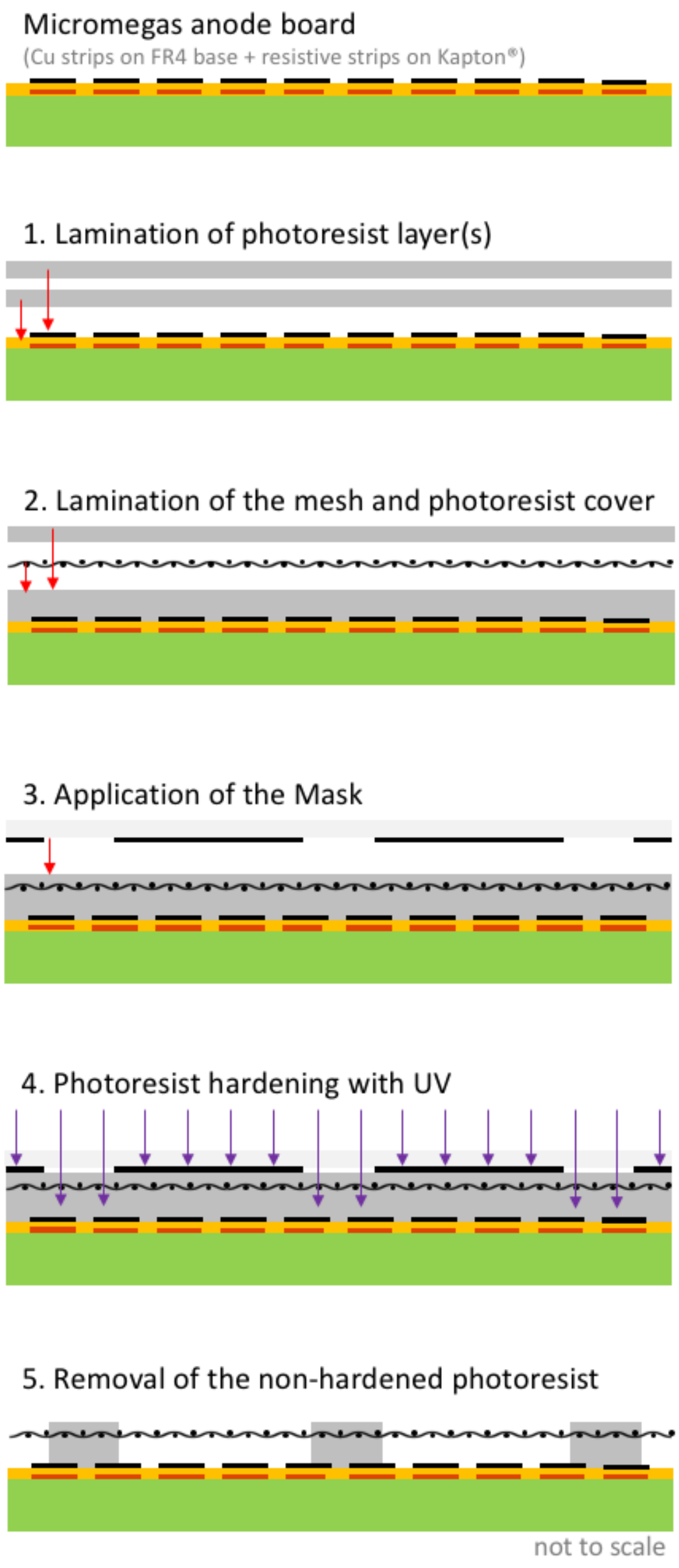

**Figure 6.18.:** Process-flow of a bulk Micromegas production [123] using photo-lithography on photo-resistive coverlay laminates with an embedded micromesh.

### 6.3.2. Pillar Shape, Dimensions and Arrangement

Following the established approach of cylindrical pillars [121, 122] the first test chambers (Tmm-type, bulk Micromegas) supported the mesh with a 2.5 mm spaced square arrangement of 500 µm diameter pillars. The effect of this comparable dense and large pillar structure on the detector's efficiency has been studied and the two representative plots in figure 6.19 clearly show the local reduction of detection efficiency in the area occupied by the pillars.

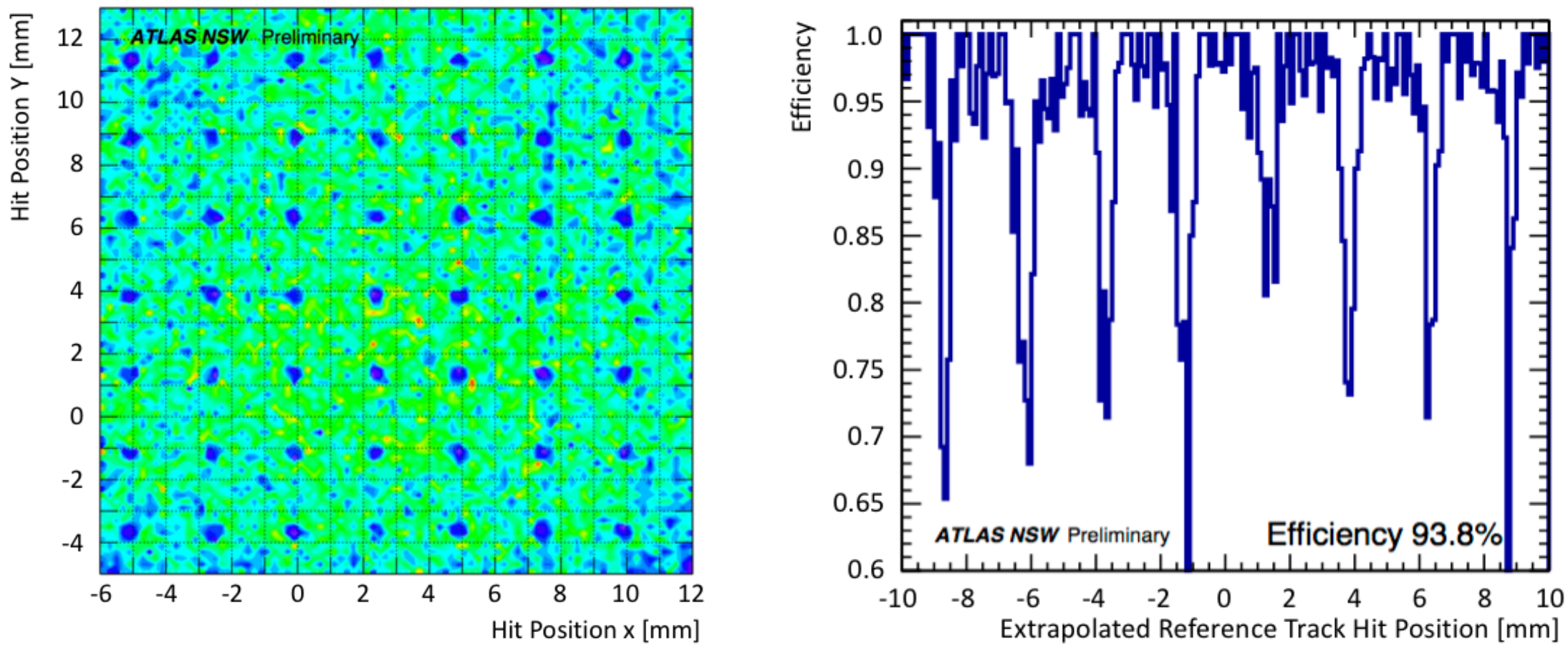

**Figure 6.19.:** Efficiency of the hit reconstruction in a Tmm-type Bulk Micromegas on the full detector surface (left) and along a 500 µm wide line crossing the pillars (right). The inefficient spots appearing every 2.5 mm, corresponding to the pillar structure supporting the mesh of the chamber, are visible. The efficiency reducing effect of the pillars is locally much more pronounced, reaching efficiency dips of the order of 40%. For this study the chamber was kept perpendicular to the beam axis. The hit position in both X and Y readouts is calculated using the centroid method and only events with a single cluster per readout (perpendicular tracks) are used. The measurements were performed with a Tmm type MM bulk resistive chamber operated with an amplification voltage of $U_{amp}$ = 540 V. The data was acquired during PS/T9 with a 10 GeV/c $\pi+$/p beam.

It is, however, noticeable that the reconstruction efficiency for a particle traversing perpendicular through the detector in a pillar occupied region is still well above 50 %. This is caused by the combination of two effects: the transverse diffusion of the signal electrons in the drift gap allows (a part of them) to approach the amplification gap in positions deviating from the position of the primary ionization. Furthermore, the electrical field configuration in the vicinity of the pillars causes a redirection of the signal electrons into the amplification gap around the pillar occupied volume, as can be seen in figure 6.20 [124]. The fraction of successfully redirected electrons, as well as the direction of this position shift, depends on the electrical field configuration, the gas composition and most of all on the size of the pillar.

Therefore, the pillar diameter and the number of pillars per area should be reduced to minimize efficiency losses. For the NSW Micromegas a triangular pillar lattice has been selected, yielding maximal mechanical support per occupied area. The pillar dia-meter has been progressively reduced in different test detectors reaching 230 µm, without observing an negative impact on the mesh positioning or voltage integrity. This optimization decreased the occupied area to sub-1 %-level. Taking into account the deviation of the signal electrons around the pillar structure, as described in [124], these structures yield a negligible efficiency loss.

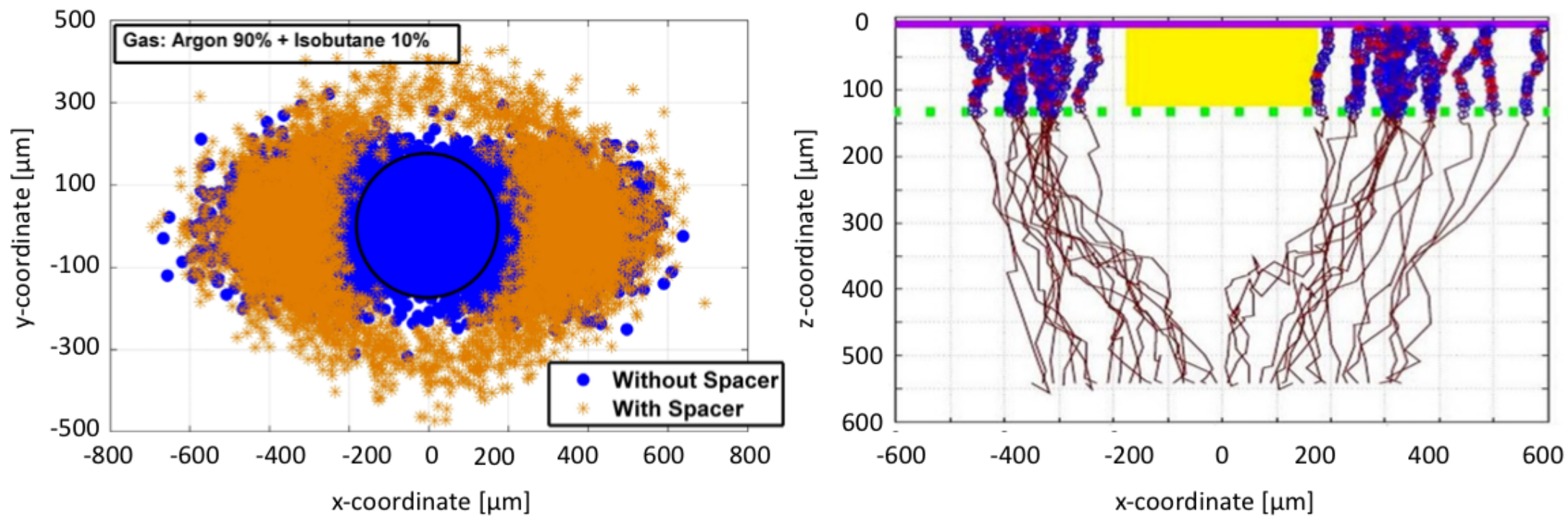

**Figure 6.20.:** Left: Endpoint (x-y-plane) of electrons in argon-isobutane mixture (90:10) approaching the empty (blue) amplification gap (thickness= 128 µm, mesh pitch= 63 µm, $E_{amp} \approx 34\,\mathrm{kV/cm}$, $E_{drift} = 200\,\mathrm{V/cm}$) compared to a region occupied with a pillar (orange). Right: Paths and endpoints of the electron drift towards a pillar. The starting locations randomized along the x-axis ($y = 0$) with $-300 < x < 300$. Simulation conducted with neBEM and Garfield [124].

During industrial production of the first large size NSW Micromegas PCBs the small adherent surface of the pillars on the anode structure turned out to cause substantial loss of pillars during PCB processing. Especially in cases where the pillars where completely deposited on top of the resistive strips, the adhesion of the uncured Pyralux® turned out to be insufficient to reliably sustain the development process. Thus, the pillars were partially washed off on a significant number of the boards. If occurring consecutively these missing pillars cause a substantial increase of the mesh sagging between pillars (see next section 6.3.3) and therefore the risk of high voltage discharges and instability, disqualifying the board for NSW Micromegas production. To circumvent this conflict between sufficient adherent surface and minimal inefficient detector area, the NSW community opted for line shaped support structures, instead of the well-established cylindrical pillars. These lines of 200 µm width and 1000 µm length are orientated perpendicular to the readout strips, thus spanning over several strips and contacting the Kapton® on at least two positions. This significantly increased the adherence of the coverlay to the anode and completely solved the missing pillar problem. The pillars' small elongation along the readout strips allows for a comparable small covered area in the sub-% range. Furthermore, it facilitates signal electron deflection along the strip direction, causing no bias in the reconstructed position, while disfavoring deflection towards neighboring strips.

Joining several PCBs to one readout plane, as foreseen for the NSW Micromegas modules, carries the risk of creating areas of lower pillar density at the joint of two boards, like in case of slight displacement of the pillar pattern and / or an inaccurate milling of the board edges. These possibly increased inter-pillar distances could cause, similarly to missing consecutive pillars, discharges or high voltage breakdowns. Although adding a full coverlay frame in this region could easily solve this issue, the created non-sensitive area would be far from negligible and significantly reduce the detector's coverage systematically on all layers. To account for this, a rim of increased pillar density is foreseen along the anodes edges mitigating the effect of slightly shifted pillars or inaccuracies in edge milling.

### 6.3.3. Mesh Tension and Inter Pillar Distance

Besides a precise pillar height, the flatness of the micromesh contributes to variations of the local amplification gap thickness. The structure of the mesh (chapter 8), is independent of the pillar structure and causes typically periodic deviations on the scale of the mesh pitch, which is comparable to the detector's spatial resolution. The sagging of the mesh in-between the supporting pillars, however, yields a non-homogeneity in the amplification gap thickness and thus in the gas gain, as discussed in chapter 4.3, with a periodicity of the chosen pillar distance of $2 - 10\,\text{mm}$. The sagging is primarily caused by the electrostatic force between the mesh and the anode and is $O(20\,\text{mN/cm}^2)$ for the 540 V - 600 V working range of the NSW Micromegas. Gravitational forces are much smaller and therefore the impact of the detector's orientation in space is negligible.

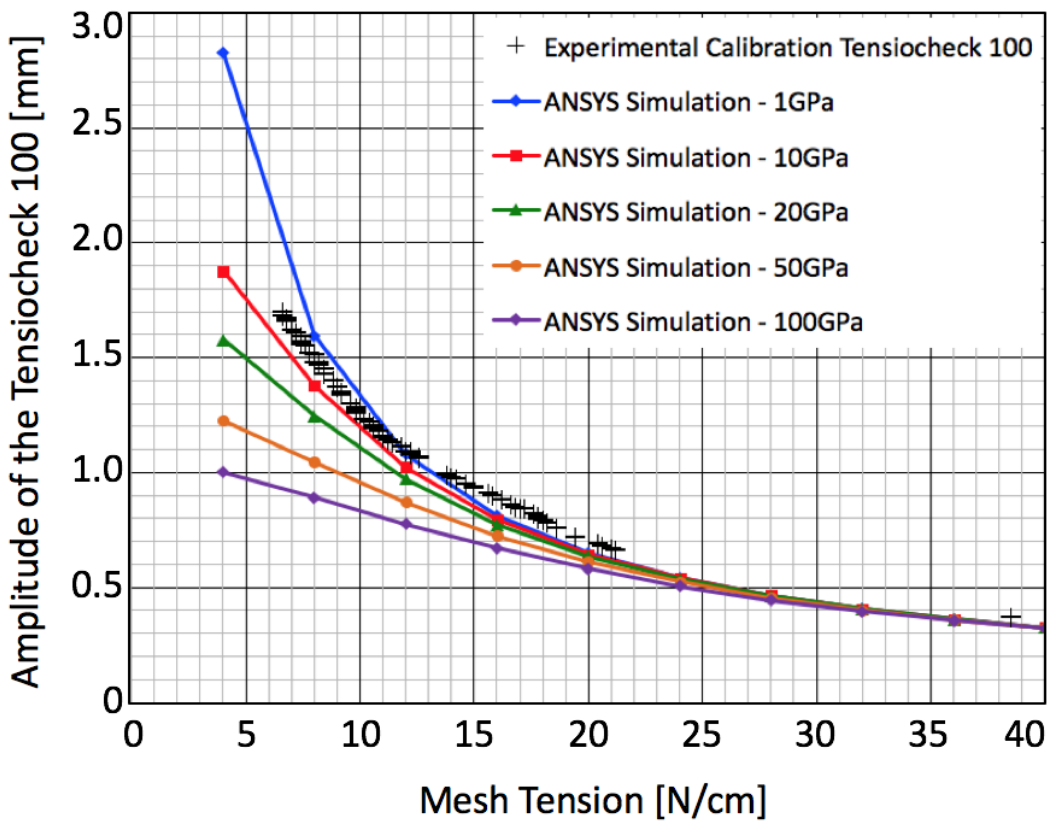

**Figure 6.21.:** Calibration curve of the Tensiocheck 100R probe amplitude over the mesh tension compared to ANSYS® simuation for different mesh elasticity.

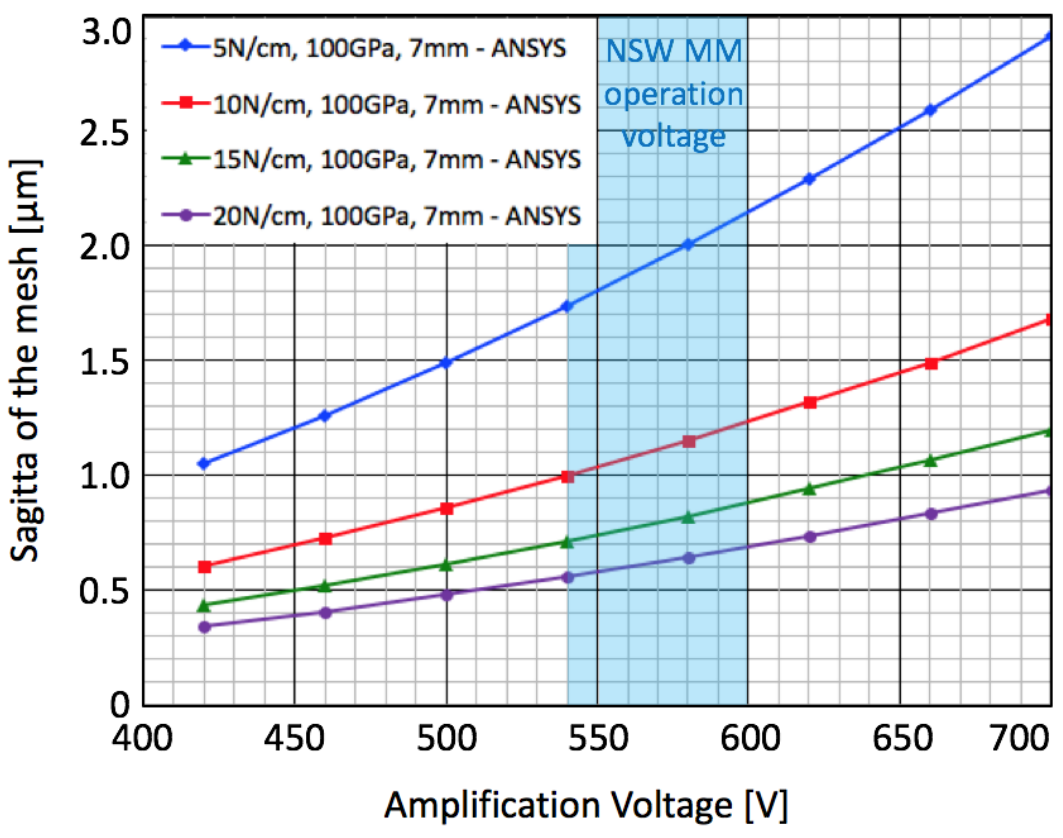

**Figure 6.22.:** Mesh sagitta over the amplification voltage pulling the mesh for four different mesh tensions with fixed 7 mm pillar distance and $E = 100\,\text{GPa}$.

The optimized choice of the inter-pillar distance can reduce the non-uniformity to an acceptable level, while keeping support structure and, hence, occupied area to a minimum. Since the effect of mesh sagging on the Micromegas is difficult to quantify experimentally (as discussed in chapter 4.3), numerical simulations based on finite element calculations have been conducted in cooperation with S. Lauciani (ANSYS®) and S. Karentzos (COMSOL®) to determine the mesh-tension to mesh maximum sagging (sagitta) relation and its dependencies.

The ANSYS® [62] simulation approach has been validated by modeling, besides the mesh sagging, the experimental quantifiable behavior of a commercial tensiometer (Sefar Tensiocheck 100R) and comparing its mesh-tension over probe amplitude relation with the measured calibration curve of the tool. Figure 6.21 shows the agreement of simulation and experimental results, taking into account that the tool calibration has been performed with a rather elastic poly-amid mesh with a young module of $\approx 8 - 10\,\text{GPa}$. As a further validation, a second independent simulation has been performed using COMSOL Multiphysics® [63]. The comparison of the simulated sagitta of the mesh under the electrostatic force caused by a 540 V potential difference are shown in figure 6.24. The good agreement between the independently computed results establishes trust in both simulation approaches.

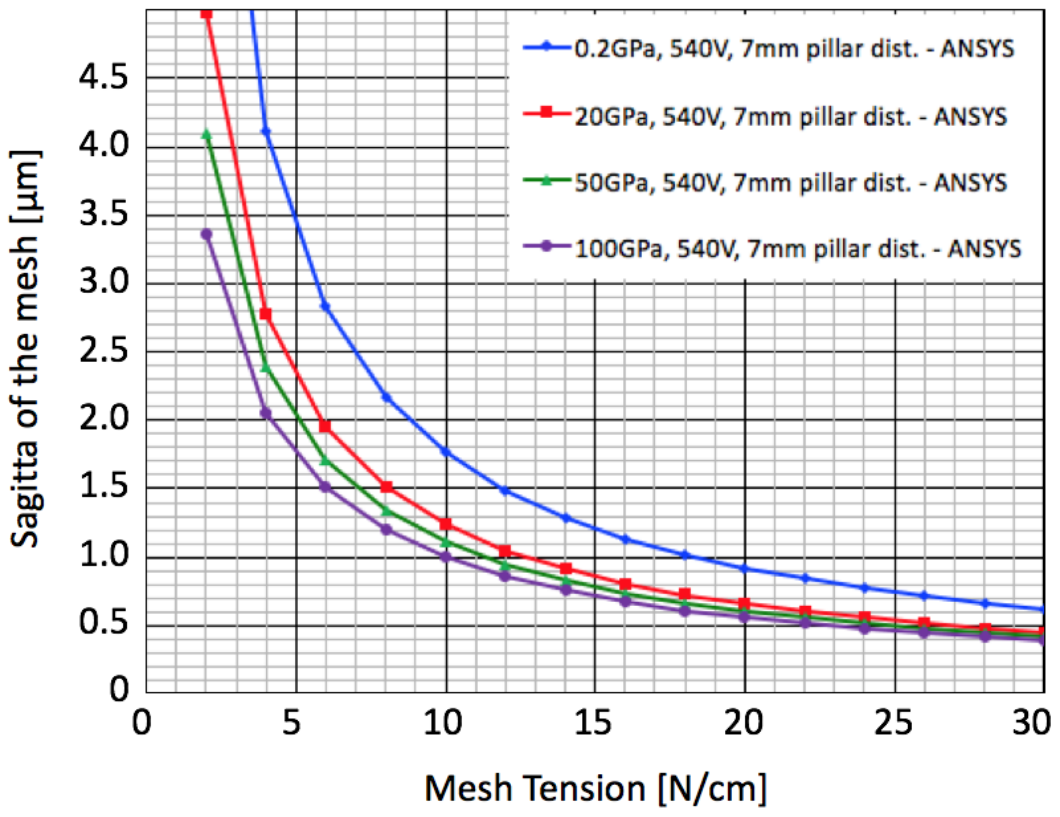

**Figure 6.23.:** Mesh sagitta over mesh tension for four different Young Modules at $U_{amp} = 540\,\text{V}$ and 7 mm pillar distance.

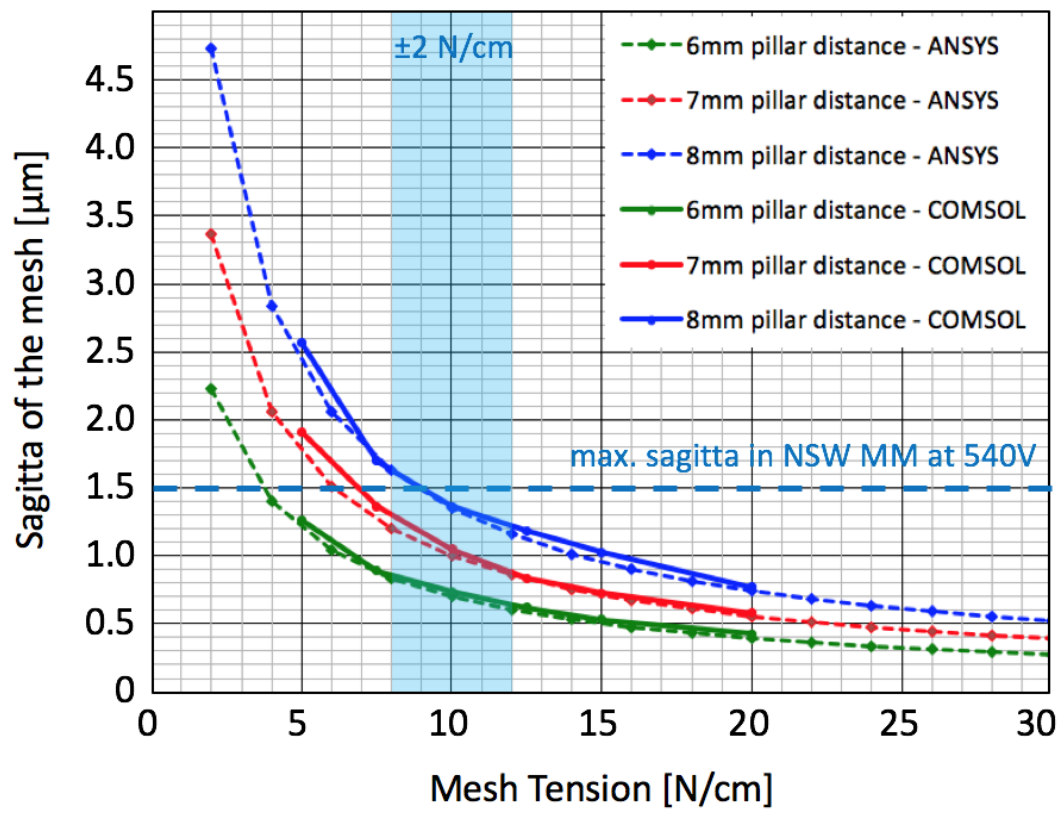

**Figure 6.24.:** Mesh sagitta over mesh tension for three different pillar distances at $U_{amp} = 540\,\text{V}$ and $E = 100\,\text{GPa}$.

The sagitta scales quadraticly with the electrostatic force, or rather the applied voltage, causing a sagging increase of $\approx 25\,\%$ from the lower 540 V to the 600 V upper boundary of the NSW Micromegas operation voltage (figure 6.22). A variation of the mesh's elasticity (Young Module $E$) within the typical range for stainless steel woven wire meshes of $50 - 100\,\text{GPa}$ only has a small impact of $O(0.1\,\mu\text{m})$ on the mesh sagging at nominal mesh tension of $10\,\text{N/cm}$ (figure 6.23). An increased distance of the pillars causes a quadratic increase on the sagitta and an increased mesh tension reduces the sagging hyperbolical, as can be seen in figure 6.24.

The main constraint for the mesh tension in the NSW Micromegas is the mechanical stress caused by the mesh stretched on the drift panels. The nominal mesh tension has been therefore limited to $(10 \pm 2)\,\text{N/cm}$, still causing a mechanical stress of more than 2 kN on the panels. By choosing a pillar distance of 7 mm, the mesh sagging is expected to be $\leq 1.5\,\mu\text{m}$ at 540 V and $\leq 2\,\mu\text{m}$ at 600 V amplification voltage. This tight constraint accounts for the occurrence of missing pillars (see previous section 6.3.2), doubling the inter-pillar distance and thus quadrupling the sagitta. According to figure 4.5 in chapter 4.2.3, this causes a local gain increase of close to 50 %, the envisaged maximum allowed deviation from nominal gain in the ATLAS NSW Micromegas.

# 7. Production and Quality Control on Readout Anode PCBs

Comprising the structures for signal readout, the spark protection layer and the pillar pattern to support the mechanically floating mesh, all discussed in detail in the previous chapter, the Micromegas anode boards represent the heart of the detector. The large board size of up to 2.2 m and the high quantity of more than 2000 boards forbid an in-house processing of these components in the CERN DT workshops. The inevitable production in industries requires a comprehensive technology transfer, an intense follow-up during the prototyping and production phase and a dense and reliable quality control (QC) and quality assurance (QA) scheme. All this has been prepared and established as part of this thesis, in very close collaboration with the CERN PCB workshop, in particular Rui de Oliveira, and the two supplying companies ELTOS (Italy) and ELVIA (France). Both companies have been involved in the PCB production for the first NSW modules and are currently producing the full quantity of NSW Micromegas anode PCBs which will last until the end of 2017.

The hereafter presented design of the anode boards (section 7.1), production processes and quality requirements (section 7.2) and established QC/QA scheme (7.3) follow to a large extent our publications in [125] and [126]. It shall be noted that several changes in design details and process description have been applied for the final version presented here with respect to [125]. The chapter is closed with an introduction to the developed quality control markers (section7.4) as well as the tooling developed for the acceptance QC procedure at CERN (section 7.5).

## 7.1. Anode Board Design

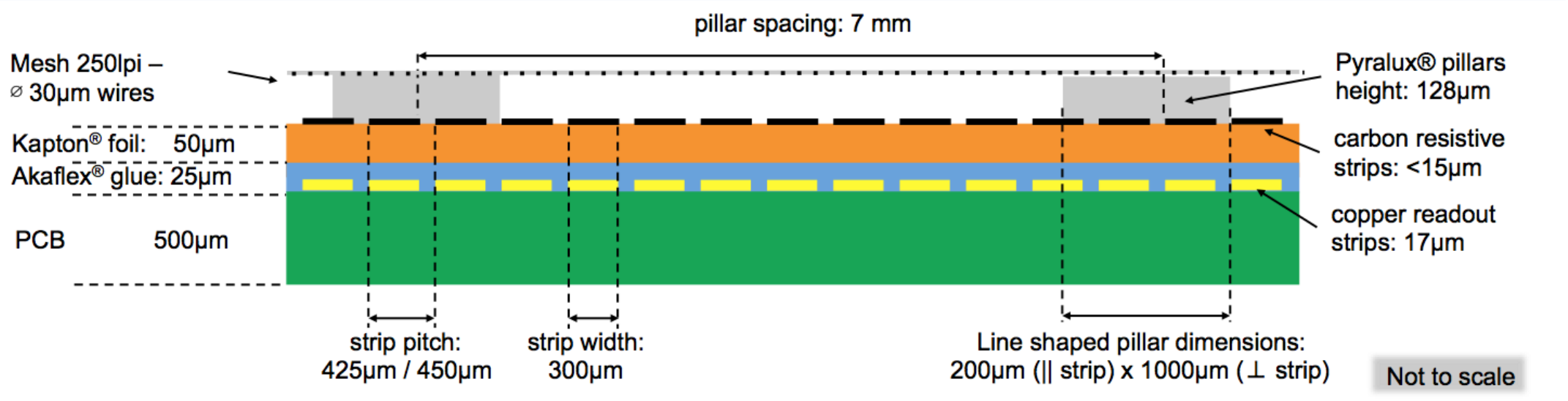

**Figure 7.1.:** Schematic of the NSW Micromegas anode board layers and dimensions of its structures (different strip pitch corresponds to small / large NSW sector PCBs). [126]

The NSW Micromegas anode boards are stacked structures of several layers, as shown in figure 7.1. They are composed of fiber glass epoxy (FR4) panels, serving as a base for the copper readout structure. Kapton® foils carrying the screen printed resistive pattern are glued on top utilizing a glue film to fill the gaps in the copper pattern and equalize the anode board's thickness, before the pillar structure is applied. The different layers are produced or applied sub-sequentially, as described in detail in section 7.2.

In the lateral extent each anode board can be divided into an active area, suitable for particle detection, and rim areas on the left and right of each board (figure 7.2) and additionally on the top / bottom on the largest / smallest board of each module.

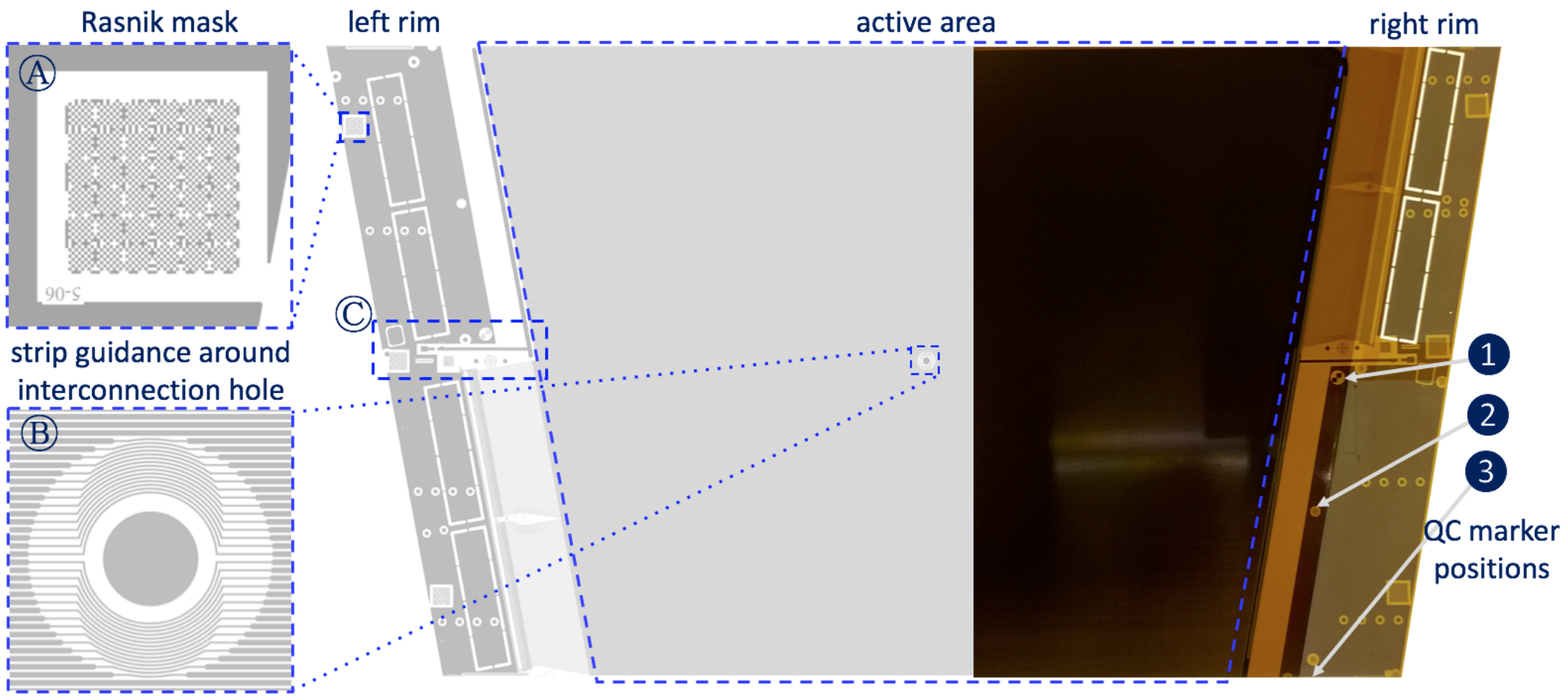

**Figure 7.2.:** Drawing of a Micromegas anode PCB copper pattern (left) and picture of the finalized board (right). The location and structure of Rasnik masks (A), strip routing around holes (B) and the center of the rim area (C) (see figure 7.3) are shown, as well as the location of three quality control markers (see figures 7.5, 7.6 and 7.7).

Following the technological decisions discussed in chapter 6, the active area is covered with 1022 readout strips of 300 µm width and a pitch of 425 / 450 µm for small / large sector boards. Although each PCB could accommodate for 1024 strips according to its width, the first and the last strips have been removed to allow a larger mechanical clearance for edge cutting and joining of the PCBs to a readout plane. The large extent of the modules calls for central stiffening screws, to counteract mechanical deformation, occurring due to gas over-pressure. Interconnection holes are foreseen in the active area to allow for a penetration of the screw. The readout strips are not interrupted by these interconnection holes, but guided around as shown in figure 7.2 (B).

The strips are routing half-and-half to the right and left rim where the front-end electronics will be connected by elastomeric connectors, avoiding soldering. Next to these connection regions, cut-outs are foreseen to establish direct contact of the electronics with the cooling channel, embedded in the panel. The board's position references are located in the center of the rim, surrounded by different targets for precise board-to-board alignment. This area, furthermore, contains soldering pads for ground, high voltage supply and electronic components, as displayed in figure 7.3. In total, each side rim carries four Rasnik masks (figure 7.2 (A)) for precise measurement of the board's dimensions and position. Additionally, each PCB carries a number of Quality Control markers which will be described in 7.3.

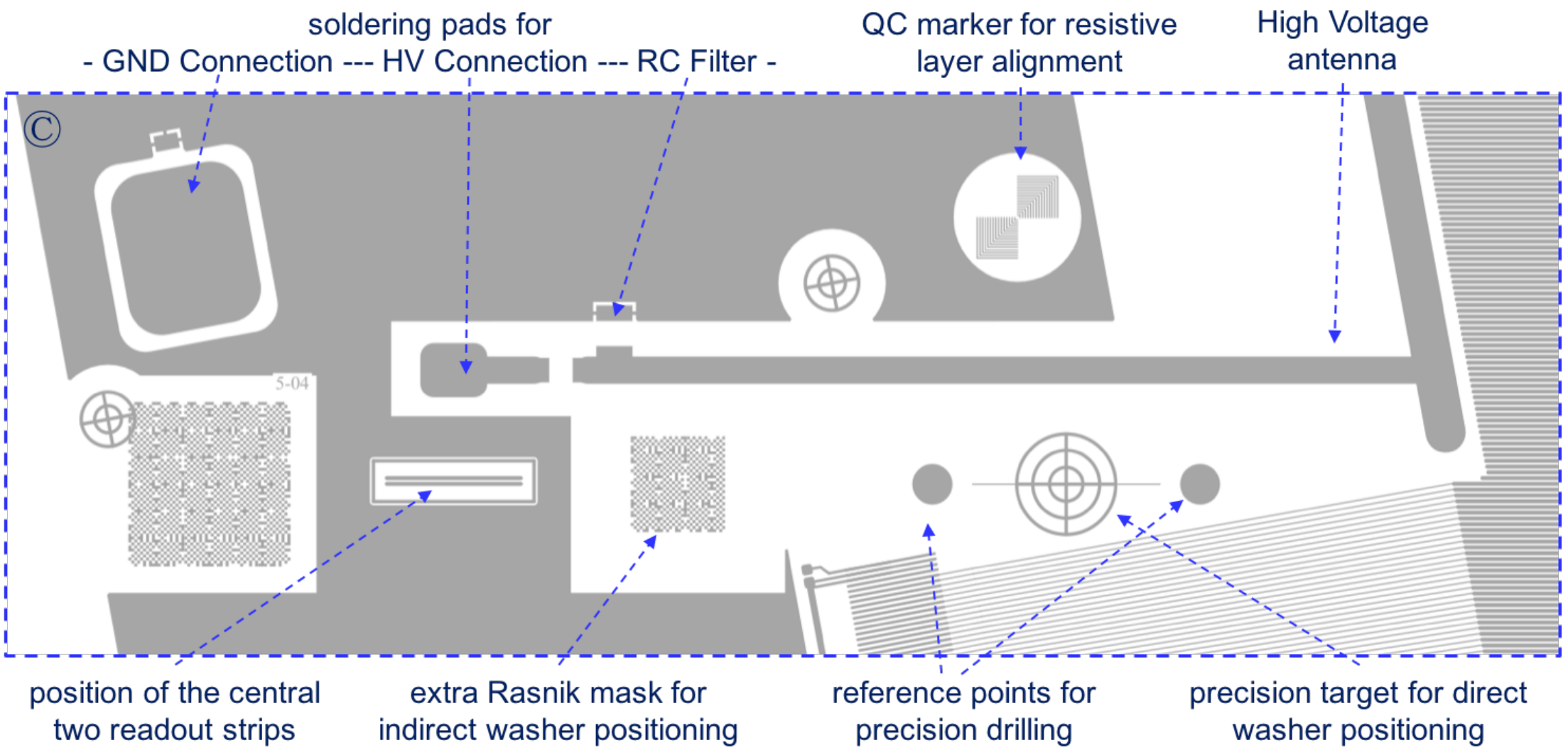

**Figure 7.3.:** Copper pattern at the rim center of the NSW Micromegas anode PCBs with the positions of several references (bottom), soldering pads and a quality control marker (top).

The resistive lines are congruent to the readout strips, but with an array of bridges connecting each strip alternating with its top or bottom neighbor every 10 mm. This yields a more homogeneous surface resistivity which is less affected by damages of single lines. The strips are interrupted in their center to divide the surface into two high voltage sectors per PCB, each of them supplied through a broad distribution line along the rim, interconnecting all resistive lines. The high voltage is applied to the resistive layer from the copper pattern antenna (figure 7.3) below via a hole in the Kapton® foil filled with conductive silver polymer ink. A less than 40 µm high line of this silver polymer is applied on top of the voltage distribution line, and ensures a low impedance connection of all strips. This is mandatory to remove a systematic bias in the resistivity distribution, as observed in the MMSW detectors [115, figure 10].

The line shaped $1000 \times 200\,\mu\text{m}^2$ Pyralux® coverlay pillars are arranged in a triangular lattice with 7 mm spacing. Additionally to the pillars covering the detection surface, a frame of coverlay is surrounding the active area of the joint detection plane to ensure the correct height of the mesh along its circumference. A higher density of pillars is foreseen at the joining edges of the PCBs, as discussed in 6.3.

## 7.2. Production Process and Quality Requirements

The production process follows a multi-step work-flow, which has been introduced in [125] and is graphically summarized in figure 7.4 - right. While all individual processes are standard in PCB industries their uncommon combination, the large size of the boards and the stringent requirements for accuracy and quality, described in detail in the subsequent paragraphs, result in a serious challenge. These acceptance criteria have been derived from physics performance considerations as well as mechanical necessities linked to the detector construction process.

### 7.2.1. Copper Pattern Creation

The readout pattern is etched into the 17 µm thin copper layer on temperature stabilized glass fiber epoxy (FR4, 0.5 mm ± 50 µm) via classical photo-lithographic processes [127] in a temperature and humidity controlled environment. Taking into account the expansion of the base material due to temperature and humidity changes during the subsequent production steps, transport, storing and detector plane assembly, the mask for production of the copper pattern is slightly re-scaled to reach its final dimensions after the board's predicted expansion. Intense testing has been performed at CERN and in both aforementioned companies to properly understand this expansion behavior. The expansion factor has been determined to 430 µm/m respectively 480 µm/m in the short direction and 380 µm/m respectively 430 µm/m in the long direction for boards produced by ELTOS and ELVIA. Deviating values for both companies results from small differences in the processing parameters. The decreased expansion along the strip's direction is due to the stabilizing effect of the copper lines.

The required absolute accuracy, referring to the design files, is ±30 µm for the precision coordinate direction (short direction) to guarantee accurate positioning of the readout strips in the detector, which is a precondition to achieve the required spatial resolution. On the long side ±100 µm/m accuracy is required. Due to the board's expansion the nominal reference values need to be adopted accordingly to the environmental parameters: measurements of the development mask on the dry PCB before etching refer to the re-scaled pattern, while measurements on boards, fully acclimatized to ambient temperature and humidity in the CERN QC laboratory, refer to the envisaged dimensions.

The local copper pattern inaccuracy, like the deviation in line thickness, is tolerated to ±20 µm. This level of accuracy is required to ensure a correct one-to-one strip connection with the elastomeric connectors to the front-end electronics and guarantee the readability of the Rasnik masks.

Shorts between two lines can be repaired and are only tolerated on $< 0.1\,\%$ level, while up to 1 % of non-repairable strip interruptions are accepted as long as they are not located on neighboring strips.

### 7.2.2. Selective Plating on the Connector Pads

To ensure a perfect conductivity between the connector pads and the elastomeric connectors, the copper has to be covered with a noble metal like Au, Ag or Pd using either electroless-/immersion-plating with a mediating Ni layer [128] or electroplating processes [127]. A complete and homogeneous coverage of the pads is required, where the layer thickness depends on the choice of the plating but ranges for all in O(0.1 µm). Alternatively to the scheme presented in figure 7.4, the plating can be performed between any other subsequent production steps.

### 7.2.3. Kapton® Foil Preparation and Cutting

The resistive pattern is applied on a Kapton® foil (50 µm) via screen printing using a carbon loaded ink (ESL D-RS 12115) [129], as explained in detail in section 6.2.2 and figure 6.12. They are industrially produced in Japan and delivered to the PCB contractors. A polishing of the resistive foils is required to remove enclosures in the resistive ink (as discussed in 6.2). This is performed during the foil QC in Kobe (Japan).

The resulting pattern yields an initial surface resistivity of $0.3-0.4\,\mathrm{M\Omega/\square}$, resulting in the target resistivity of $\approx 0.8\,\mathrm{M\Omega/\square}$, after an increase during the subsequent production steps. Since the surface resistivity is sensitive to changes in temperature, pressure and humidity during the production, a frequent surveillance of this parameter is well suited to control for changes in these parameters.

Before further processing the foils have to be cut to size with moderate accuracy of $\pm 1$ mm and cleaned thoroughly.

### 7.2.4. Gluing of Kapton® Foil on the PCB

The Kapton® foil is thereafter glued on the readout side of the PCB with a 25 µm thick Akaflex® glue layer under high pressure of $5-7\,\mathrm{kg/cm^2}$ at 170°C. An alignment accuracy of $< 0.5$ mm is required between the two layers.

Good cleanliness during the stacking of the layers is essential to avoid enclosures in the glue and dents on the anode surface. To reduce non-uniformity in pressure, leading to air inclusion between the copper strips, additional layers of conformity material (Pacoflex® (0.27 mm) or Pacotherm® (0.89 mm)) are added to the stack before pressing and removed afterwards. A homogeneous, enclosure-free gluing with a flat surface is demanded, since small bumps on the surface of the anode could favor the formation of discharges in the detector. The change in resistivity during the pressing has to be monitored.

### 7.2.5. Application of Polymer Silver Conduct

To apply voltage on the resistive pattern, a hole is punched through or cut into the Kapton® above the end of the copper high voltage antenna and filled with conductive ESL 1901 SD silver conduct. Thereafter, a conductive line is manually screen printed over the length of the resistive high voltage distribution line, connecting all resistive strips at their ends. This low impedance line contributes significantly to a more homogeneous resistivity across the anode surface.

A coarse positioning ($\pm 1$ mm) and the flatness of the conduct line ($\leq 40\,\mu$m) has to be guaranteed as well as a low impedance between high voltage connection pad, the silver line ends and the resistive pattern.

### 7.2.6. Coverlay and Pillar Pattern Creation

After lamination with a double layer of 64 µm Pyralux® PC1025, the pillar pattern is transferred via a photo-lithographic process, as described in chapter 6.3.1 and visualized in figure 6.17, with a positioning accuracy of $\pm 1$ mm referring to the copper pattern.

The essential requirement is an (almost) complete pillar pattern, where missing pillars are only allowed if non-consecutive and less than 0.1 % in total. A maximum of ten extra structures and other defects on the coverlay are allowed, if minor in size ($< 1 \times 1\,\mathrm{mm}^2$). To obtain a homogeneous gain in the detector, a constant height of the pillars is required, varying less than 5 µm in between different regions of the board.

### 7.2.7. Board Cutting and Drilling of Assembly Holes

Having undergone several wet etching-, development and cleaning steps, the PCB must expand to their final dimensions before being cut to final size. Therefore, a more than one week acclimatization period is foreseen before the boards are milled with a CNC machine. Along the edge running parallel to the readout strips a precision better than 100 µm is required. The tight margin is necessary to avoid overlapping anode boards when joining them to a full detector plane.

Assembly holes are drilled with the same accuracy requirement along the outside borders of the detector plane and at the positions foreseen for the panel interconnection screws. The precision reference holes for board alignment, located in the center of the rim (figure 7.3), and those required for front-end-electronics positioning, located in the center of each connector pad, are drilled later in the CERN DT workshop with a X-Ray optic assisted precision CNC. This procedure reaches a position accuracy better than 30 µm, referring directly to the copper pattern.

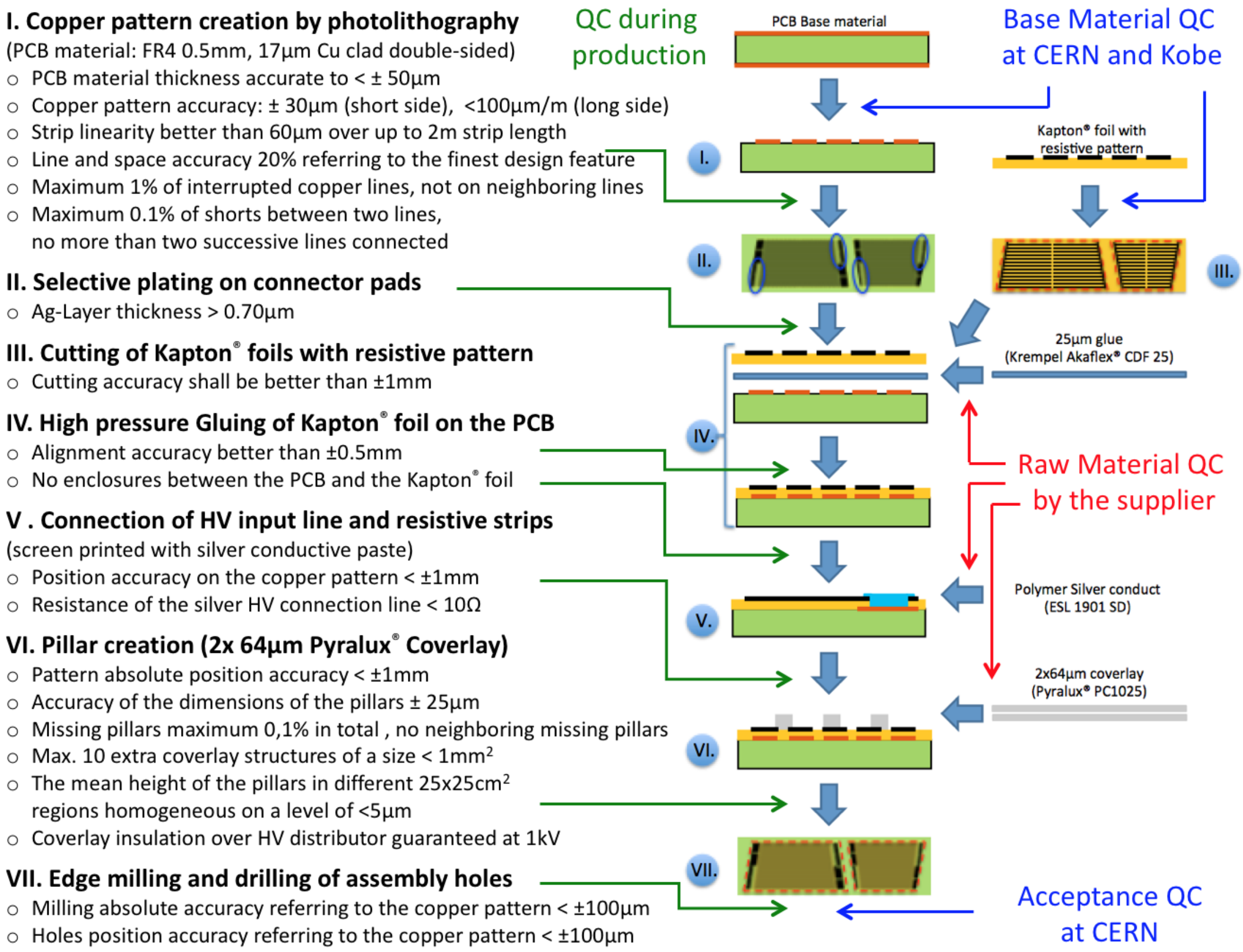

**Figure 7.4.:** Multi-step workflow during NSW Micromegas anode PCB production (right) and quality requirements (left) for each process: I. Copper pattern creation; II. Connector pad plating; III. Kapton® foil cutting; IV. Kapton®-PCB gluing; V. Silver conduct application; VI. Pillar pattern creation; VII. Cutting and drilling. Quality control steps performed by the PCB producer (green), other industrial suppliers (red) and the NSW collaboration (blue) are indicated. Updated figure based on [125, 126].

## 7.3. Quality Control and Quality Assurance Scheme

The multistep QA/QC scheme applied during the NSW anode board production is visualized in figure 7.4. It has been developed based on general QA/QC considerations and guiding thoughts, each of them and their consequences being addressed in one of the subsequent paragraphs. Although the QC plan has undergone multiple adjustments and fine-tuning during the last years of prototyping and pre-series productions, these principles remain reflected in the final scheme.

### 7.3.1. Shifting QC to Industries

To identify flawed items early and minimize material losses, it is inevitable to either establish a QC team permanently following the production directly at the suppliers premises, as partially opted for by the NSW sTGC community, or transfer QA responsibility to the industrial partner. In both cases QA/QC shall be performed in industries without a delay of production by additional logistic steps and involving the suppliers' staff whenever possible to reduce the work load on the NSW community.

This requires an adaptation of our QC scheme to the tooling available in the companies and / or the preparation of tooling and methods to be provided to the supplier. The first case might require slight modification of testing methods dependent on the suppliers equipment. In the latter case the provided QC methods and QC tools must be very reliable, simple and time efficient to use and provide accurate and repeatable QC results independent of the operator.

### 7.3.2. Multistep QC to Optimize Production Yield

Given the complexity of the process and lacking experience on large-area PCB production in industries, a single quality control after finalization of the production will be insufficient to achieve a satisfying yield, limit material losses and trace systematic production-related faults. Therefore, QC testing must be done as closely and small stepped as possible, qualifying or disqualifying each item if possible after each production step.

Following this guideline, all components used during the anode board manufacturing are tested prior to their usage. While some base materials, the copper clad FR4 sheets and the Kapton® foils with screen printed resistive pattern, are under the responsibility of the NSW community and, therefore, qualified either at CERN or at Kobe University, other commercially available materials, like the Akaflex® glue layer, the Pyralux® coverlay, the resistive carbon loaded ink and the conductive silver ink, rely on the quality tests and warranty issued by the producer.

Within the PCB processing, the principle lead to a consequent board testing after each production step, which is well reflected in figure 7.4. While typically only the requirements connected to a processing step are tested thereafter, some properties, like the impedance of the resistive layer, are monitored throughout the production to detect unexpected behavior and identify changes in the production parameters immediately.

### 7.3.3. Tracking and Data Management of QC Results

Shifting QC measurements to industries and splitting the QC into several testing steps enhances the need of proper bookkeeping. A database (DB) structure developed centrally by the NSW collaboration will store all the information and QC results for each single component to be included into the NSW detectors. It will allow for a reliable tracing of the constituents of each detector module.

Therefore, each anode PCB as well as its components are identified either with a unique ID for the PCB and the resistive foil or a batch ID for non-unique components, like raw material FR4, glue layers, resistive ink, Pyralux® coverlay and so forth. Raw material suppliers and PCB producers are obliged to provide QC reports with every delivery, stating the results of the agreed upon QA/QC procedure. In case of a batch information, for example resistivity tests of the carbon loaded ink, this result is passed on to each unique item produced under utilization of this batch of ink.

### 7.3.4. Maintaining High Quality during Industrial Testing

Minimizing the risk of negligent performed tests in industries, the production must be supervised by members of the NSW collaboration, frequently visiting the producers' premises and re-testing individual items sample wise.

Additionally, a full certification of each PCB in a CERN based QC is mandatory before accepting the boards for detector construction. The large quantity of boards and requirements to be cross-checked at CERN requires largely automatized measurements or testing methods which are quick, reliable and operator-independent. The conceptual design and preparation of methods and tooling for this final acceptance QC has been one of the main objectives during this task and is explicated in the following sections.

## 7.4. Layout Implemented Quality Control Markers

To achieve fast and reliable QC we designed and implemented quality control markers into the layout of the copper and the resistive pattern. They allow a quick and accurate visual QC of different characteristics of the board geometry and the inter-layer alignment. Three examples of these QC markers are explained and examples are shown in the following paragraphs.

### 7.4.1. Copper Pattern - Resistive Layer Alignment

A pattern of 100 µm wide lines and gaps is included in the copper layer, while a double-square marker is printed with the resistive pattern (figure 7.5). The markers are placed congruently on both layers and cover each other as long as the layers are perfectly aligned and their dimensions are in agreement with the design. Each visible line or gap indicates a misalignment of 100 µm. The difference of the misalignment observed on two distant markers sheds light on the relative dilatation of the two patterns.

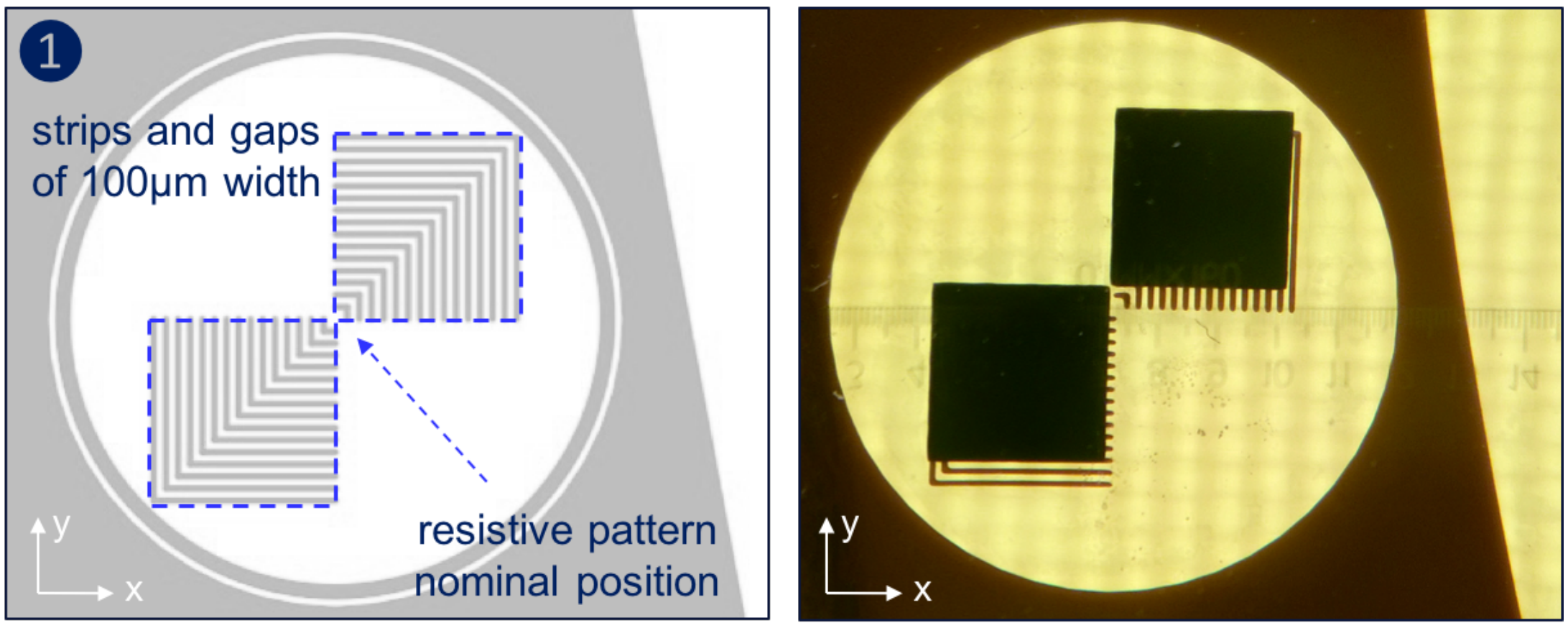

**Figure 7.5.:** Left: Concept of the inter-layer alignment QC marker with the pattern of 100 µm wide lines and gaps on the copper pattern and the nominal position of the covering squares. Right: Exemplary picture showing a misalignment of +400 µm in vertical and -150 µm in horizontal direction. [126]

### 7.4.2. Drilled Hole Positioning on the Copper Pattern

A copper ring of 100 µm thickness is positioned concentric with the nominal drilling position referring to the copper pattern (figure 7.6). The hole radius is 150 µm smaller than the copper ring radius. Once drilled, the hole's position can be validated on ±100 µm precision if the copper ring is untouched by the hole. The board is disqualified if the ring is completely ruptured, proving a position inaccuracy of > 200 µm.

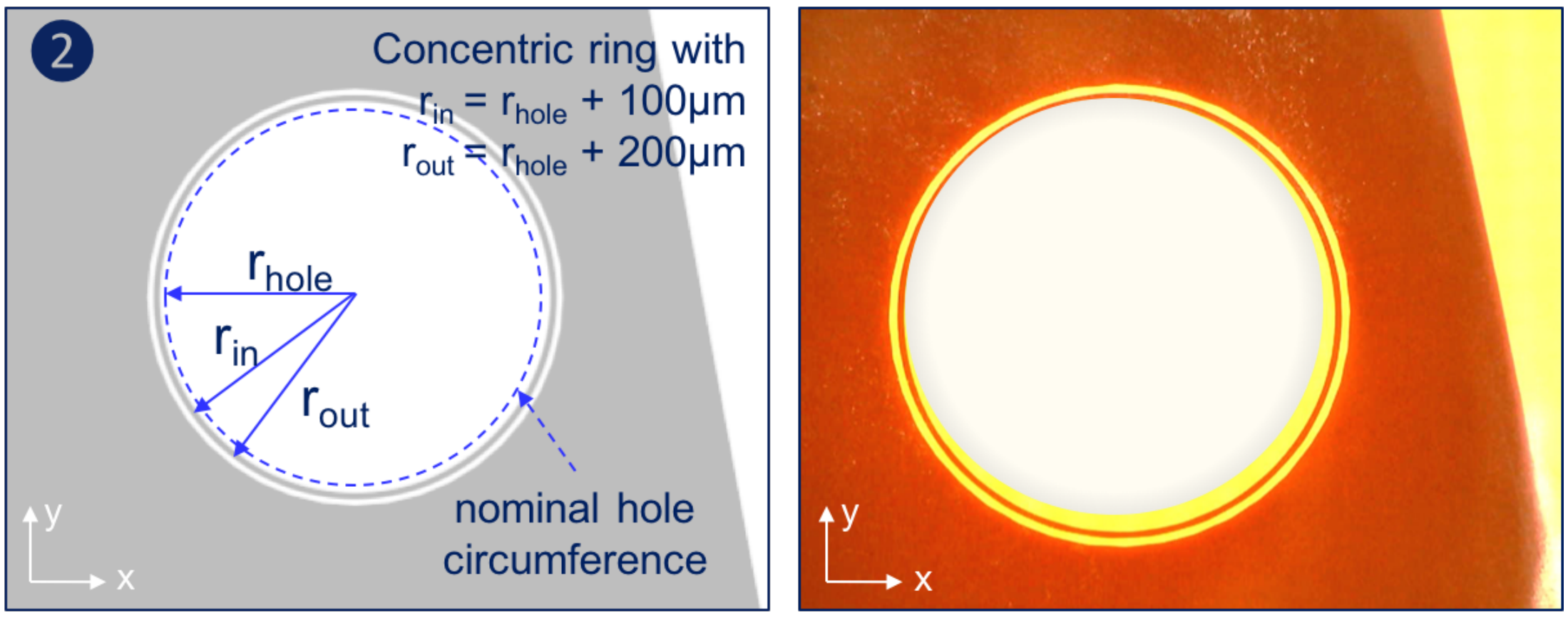

**Figure 7.6.:** Left: Concept of the hole position QC marker with a concentric 100 µm thick copper ring around the nominal hole circumference. Right: Exemplary picture showing a deviation of the hole's position of > 100 µm but < 200 µm referred to the copper pattern. [126]

### 7.4.3. Edge Milling Accuracy referring to the Strips

A set of four ladder-patterns with 100 µm lines and gaps are placed across the milling line and shifted by ±50 µm to each other (figure 7.7). By counting the non-removed lines the milling precision can be assessed on a ≤ 100 µm-level. Combined with a straightness measurement along the edge, performed with a rectified ruler on an illuminated table, two markers on the left and right are sufficient to judge the overall milling accuracy.

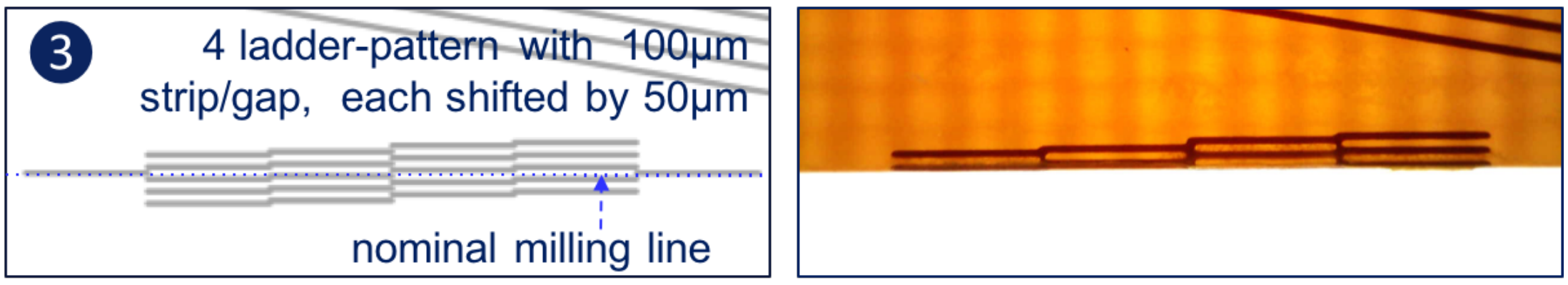

**Figure 7.7.:** Left: Concept of the edge accuracy QC marker comprising four ladder-patterns with 100 µm strips and gaps positioned across the nominal milling line. Right: Exemplary picture showing a cut slightly too far inside the board with ≤ 50 µm deviation from nominal. [126]

## 7.5. QC Tooling and Laboratory Setup for CERN PCB Acceptance

The full certification of each anode PCB to be used in the NSW Micromegas is crucial to the overall detector performance and the certainty of having boards thoroughly checked after delivery contributes to a high level of quality during production. A laboratory for quality control (QC) testing of the anode PCBs has been set up in building 188 on the CERN premises. The conceptual design of the QC work-flow and the involved tooling has been developed as major task of this thesis. However, the realization of the tooling and software has only been possible with the great effort taken by many colleagues throughout the NSW community.

The acceptance QC encompasses a series of testing stations, each equipped towards certain criteria to be tested, as we will discuss in the subsequent sections. The boards pass sequentially through the different survey points and, therefore, a parallel QC on several items is possible. The QC results are stored using the unique PCB identifier and managed with a web browser based interface guiding the operator through each step of the QC procedure. This interface includes the measurement results from the independent QC programs used during the automatized measurement steps and is connected to the NSW logistics database (DB) and the Kobe QC DB for resistive foil production.

To facilitate the localization of defects, irregularities and blemishes, a coarse sector segmentation of the board is applied during the QC, with a quartering along the short direction forming the rows A to D and a segmentation each 10 cm in the long board direction forming the columns -10 to -1 and 1 to 10, referring to the left and right of the central high voltage division line (see figures 7.10 or 7.14).

### 7.5.1. Optical PCB Inspection

A thorough visual inspection of the PCB is, apart from all automatized tests and measurements, of crucial importance to identify localized flaws and irregularities. The optical inspection is structured in three steps:

Starting with a *top-light illumination*, the board's surface is checked for any kind of irregularities including, but not limited to, surface dents, bumps, enclosures of gas or solids, damages of the coverlay or its surface, scratches of and blemishes in the resistive layer as well as structural damages of the FR4 board. Localized dirt and contamination is removed if possible and in case of a surface covering non-cleanliness, it is tested locally if the upcoming cleaning steps will remove this contamination. The correct stacking of the anode board layers and the homogeneity of the plating in the connector regions is checked visually.

Placing the board on an illuminated table, the internal structure of the semitransparent PCB becomes partially visible. Utilizing this *backside illumination* the QC markers, described in the previous section, are controlled and the straightness of the edges is verified. Therefore, the board's edge is pressed against a stratified precision ruler and the illumination profile of the gap is controlled, revealing any steps or bent segments of the supposed to be straight edge. The interference pattern created by the small misalignment between the resistive strips and the readout strips allows to immediately access the alignment quality. Furthermore, enclosures as well as defects in the resistive layer become more clearly visible on the back-light table.

In the last sub-test, the board is inclined towards the source of illumination. With the *reflected light* non-flat structures become apparent. The pillar pattern is checked for completeness, benefiting from the human's eyes' ability to spot pattern irregularities. Bumps, dents and other structures deviating from the nominal board surface can easily be identified and distinguished by their reflection pattern. The flatness of the coverlay can right away be confirmed, since roughness on the Pyralux® surface, resulting from contamination during the coverlay development, causes a diffuse reflection, which is easily distinguishable from the expected bright reflection.

All observed irregularities are localized, utilizing the coarse grid described above, and classified. Therefore, a list of known defects is included in the QC interface. Comments and microscope pictures, possibly including to-scale measurements, can be uploaded to the database, which is of special importance for non-classifiable defects. Besides its type, the operator rates each flaw as a 'minor' or 'major' blemish. Minor defects are defined as either easily recoverable and/or are supposed to have negligible impact on the detector construction and performance. Major irregularities, on the other hand, could cause a threat to the detector and/or their reparation is not a well-established procedure. Depending on the type of defect, a board rejection could be possible and has to be decided on a case-to-case-basis. However, PCBs comprising major blemishes should not be used for Micromegas production unless successfully recovered and retested thereafter.

### 7.5.2. Pattern Dimension Test

The detector's muon reconstruction capability relies on the accuracy of the PCBs copper pattern and the assumption of a strip position accuracy of 30 µm in the precision coordinate. Due to the FR4s sensitivity towards temperature and humidity changes, reflected in an elongation of the PCB, this accuracy requirement is rather challenging. Thus, a QC method yielding sufficient accurate and reliable measurements with a acceptable effort is mandatory.

The QC system we developed is based on the *Rasnik* image analysis, which allows for a very precise (O(1 µm, 2 mrad)) reconstruction of the position of the chess-pattern like mask within an acquired image. Rasniks have been applied with great success in the ATLAS MS alignment system utilizing optics to measure small changes in the camera-mask distance [92] and, therefore, are a technology very well known to the ATLAS Muon collaboration. Differing from the MS alignment, for the PCB QC we opted for a *contact CCD* (c-CCD) based imaging system, where the pattern of the mask is captured distortion-free via a fiber optic plate (FOP) glued on a commercial CCD camera chip.

To access the global pattern dimensions, free from a potential movement of the board, the relative positions of the six Rasnik masks positioned along the left and right rim of each board (see chapter 7.1) are measured simultaneously by six c-CCDs. Since the design positions and distances between the Rasnik masks vary on the 16 different board sizes[1], a flexible but nevertheless highly accurate and repeatable positioning method for the cameras is required. Therefore, each c-CCD camera is mounted permanently on a calibrated *surveyor*, carrying the c-CCD minimal head and the camera head electronics required to acquire the image.

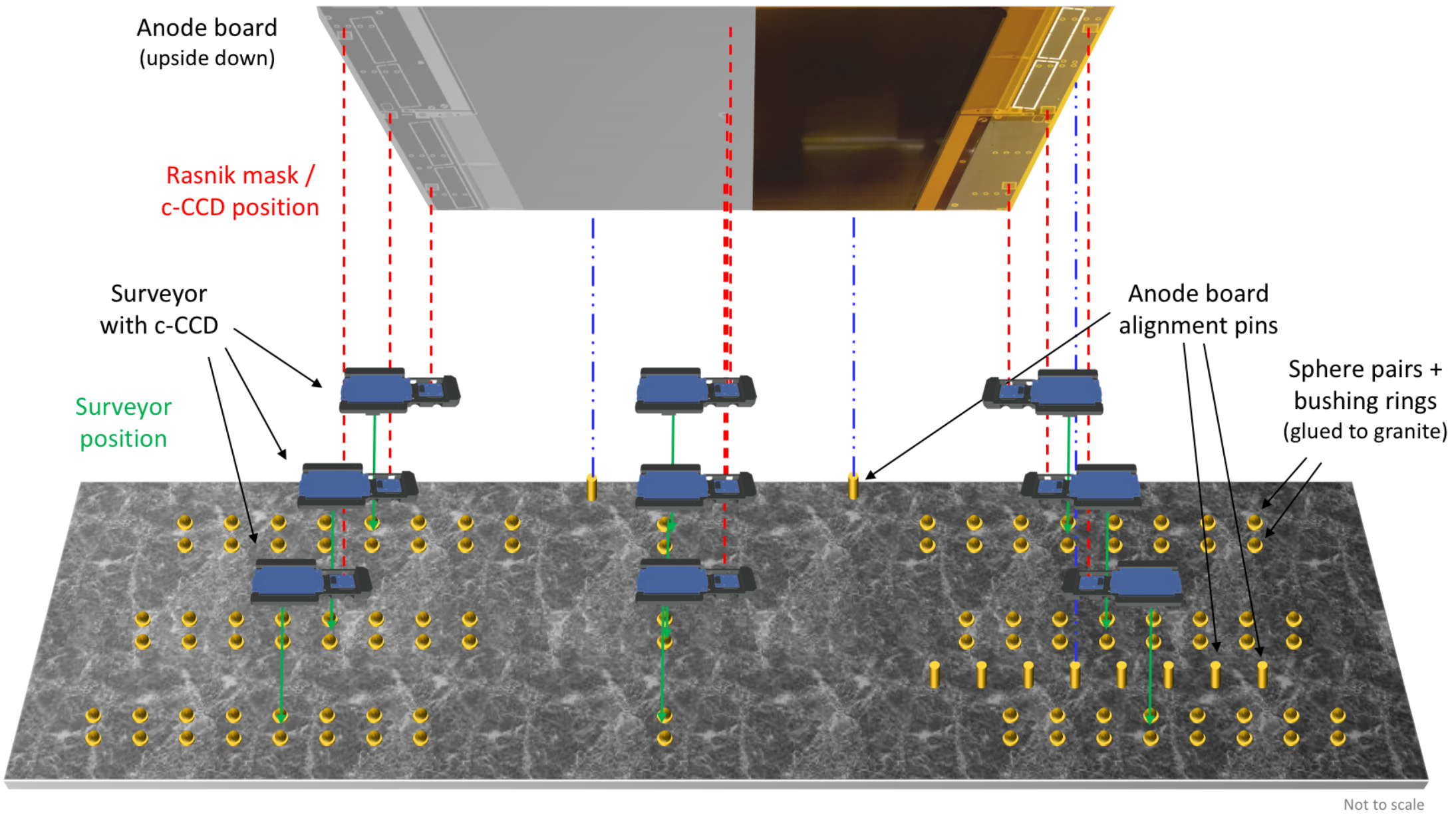

**Figure 7.8.:** Schematic of the setup for PCB dimension measurement. Spheres and position reference cylinders are glued on a marble table and precisely measured for reference. Nine surveyors are kinematically mounted on these spheres, three on either side and three in the center. The positions of the c-CCD on the side-surveyors agree with the Rasnik mask positions on the board once it is placed face down on the setup, referring to pins in the positioning cylinders. (Only the setup part for S-type board measurement is displayed. Spacer material and weights are not shown.)

The surveyors are kinematically placed on a pair of spheres glued to a reference granite table (figure 7.8). The non-occupied table surface is covered with 20 mm thick spacer to ensure a flat surface on one level with the c-CCD-FOP. The board is then placed upside down on the setup and positioned with coarse sub-mm accuracy referring to three position pins, two along the long side and one along the short edge. Applying additional weight on top of the Rasnik mask regions ensures good contact of the PCB with the FOP and flawless image acquisition.

The position $(x, y)_{i,j}$ of mask $i, j$ with $i \in \{left; right\}$ and $j \in \{top; mid; bottom\}$ is determined from a $5 \times 7\,\mathrm{mm}^2$ image of the mask taken by the c-CCD and is accordingly referring to this *frame* ($F$) coordinate system ($(x, y)_{i,j,F}$). Additionally, the relative rotation of the mask in the frame $\beta_{i,j,F}$ is returned. To combine the independent frame measurements and extract the board's dimensions, the mask coordinates must be transformed into a common coordinate system ($(x, y)_{i,j,T}$), provided by the *table* ($T$). Therefore, the mounting of the camera, and thus the frame position, in the *surveyor* ($S$) system must be taken into account, as well as the position of the reference spheres on the table.

The surveyor's coordinate system ($S$) is determined by a cone (origin) and a slot (direction) utilized to position the surveyor on the reference spheres. Being permanently installed on the surveyor, the c-CCD position and rotation can be calibrated yielding a two dimensional translation $T_S^F$ and a rotation $R_S^F$ between the frame and the surveyor coordinate system with:

$$T_S^F = \begin{pmatrix} x_{c-CCD} \\ y_{c-CCD} \end{pmatrix}_S \quad \text{and} \quad R_S^F = \begin{pmatrix} \cos\alpha_S^F & -\sin\alpha_S^F \\ \sin\alpha_S^F & \cos\alpha_S^F \end{pmatrix}. \tag{7.1}$$

Similarly the position of the surveyor on the table is determined by the position of the two spheres. Sphere A will accommodate the cone and, hence, define the surveyors origin, where as sphere B will define the surveyors orientation in the x-y-plane[2]. Precise calibration of the sphere positions with $< 20\,\mu\mathrm{m}$ accuracy yields the spheres coordinates $(x, y)_A$ and $(x, y)_B$, allowing for the calculation of the transformation parameters:

$$T_T^S = \begin{pmatrix} x_A \\ y_A \end{pmatrix}_T, \quad R_T^S = \begin{pmatrix} \cos\alpha_T^S & -\sin\alpha_T^S \\ \sin\alpha_T^S & \cos\alpha_T^S \end{pmatrix} \quad \text{and} \quad \alpha_T^S = \arctan\left(\frac{x_A - x_B}{y_A - y_B}\right). \tag{7.2}$$

The measured coordinates of the Rasnik masks in the table coordinate system can be calculated by:

[1]Eta and stereo boards of one type share the same relative Rasnik positions, only their outer shape differs slightly due to the inclined edges.

[2]The third coordinate is fixed by a third small sphere support on the surveyor touching the granite surface.

$$\begin{pmatrix} x \\ y \end{pmatrix}_{i,j,T} = R_T^S \left[ R_S^F \begin{pmatrix} x \\ y \end{pmatrix}_{i,j,F} - T_S^F \right] - T_T^S. \tag{7.3}$$

Once the measured mask positions $(x, y)_{i,j,T}$ are obtained, they can be compared with their *nominal* design positions $(x, y)_{i,j,N}$:

$$\begin{pmatrix} \Delta x \\ \Delta y \end{pmatrix}_{i,j} = \begin{pmatrix} x \\ y \end{pmatrix}_{i,j,T} - \begin{pmatrix} x \\ y \end{pmatrix}_{i,j,N}. \tag{7.4}$$

The anode board is only coarsely aligned on the table and thus their coordinate systems may be shifted and rotated towards each other, adding a bias to the above comparison. To compensate for the misalignment, before interpretation of the results, the full set of measured coordinates must be corrected with a common three parameter transformation representing the translation $\delta x_{fit}$, $\delta y_{fit}$ and the rotation angle $\gamma_{fit}$. Thus (7.4) is altered to

$$\begin{pmatrix} \Delta x \\ \Delta y \end{pmatrix}_{i,j} = \left[ \begin{pmatrix} \cos\gamma_{fit} & -\sin\gamma_{fit} \\ \sin\gamma_{fit} & \cos\gamma_{fit} \end{pmatrix} \begin{pmatrix} x \\ y \end{pmatrix}_{i,j,T} - \begin{pmatrix} \delta x_{fit} \\ \delta y_{fit} \end{pmatrix} \right] - \begin{pmatrix} x \\ y \end{pmatrix}_{i,j,N}. \tag{7.5}$$

The fit parameters are obtained by minimization of the sum of the distances between the measured and the nominal positions. This ensures an equal weighting of all six masks in the fit.

$$\min_{\delta x_{fit}, \delta y_{fit}, \gamma_{fit}} \sum_{(i,j)} \left| \begin{pmatrix} \Delta x \\ \Delta y \end{pmatrix}_{i,j} \right|^2 \tag{7.6}$$

Once the fit parameters are determined, equation (7.5) yields the deviations of the board's Rasnik masks with respect to their design position. These values can be directly interpreted, for example, $\Delta y$ can be compared to the 30 µm requirement. Alternatively, they can be combined to understand board wide tendencies, such as comparing the mean $\Delta x$ of all masks on the left respectively the right rim to understand the overall board length deviation in strip direction.

To complement this assessment of the global board shape, a second development stage of the tool is planned. Adding three c-CCDs in the center of the PCB, to take an image of the readout strips visible through the Kapton® foil at the interruption of the resistive pattern. This will allow to determine the y-position of the strips modulo the strip pitch, which is much larger than the expected / allowed deviation. Utilizing this additional information a curvature of the board (*'banana-shape'*) or an optical distortion (*'barrel-'* or *'pillow-effect'*) can be reliably detected.

### 7.5.3. Pillar Height Measurement

After investigating an individual pillar measurement, non of these approaches yielded satisfying results. Utilizing laser reflection measurements turned out to be problematic due to the optical transparency of the Pyralux® material and capacitance measurement between a plate pressed on the pillar and the anode lacked the reliability and accuracy to obtain O(1 µm) distance resolution. Therefore, we opted for a mechanical setup to obtain representative height measurements within a well defined pillar area.

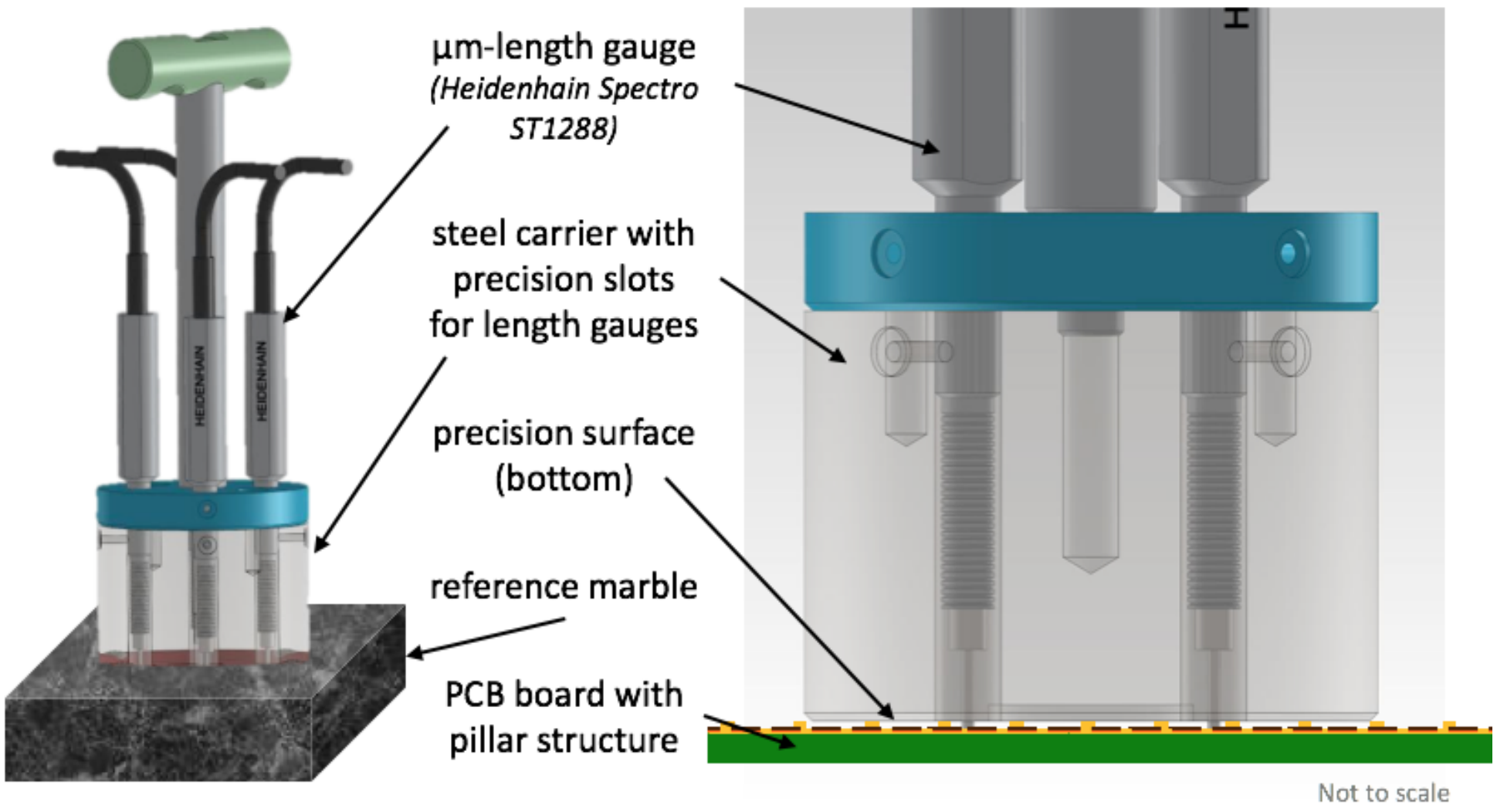

**Figure 7.9.:** Left: Layout of the pillar height tool, placed on a reference marble. Right: Visualization of the measurement principle: length gauges are touching the anode board through the precision surface resting on the pillars.

In our setup four precision length gauges (Heidenhain Spectro ST1288 [130]) are mounted on a stainless steel holder, piercing through a flat surface polished with 1 µm precision (figure 7.9 - left). The zero position of the probes is calibrated on a marble stone. Once placed on the pillar structure, the probe tips touch the anode surface, while the reference plate rests on the top of the pillars, yielding relative measurements between the two planes (figure 7.9 - right). The readout unit (Heidenhain ND 2108G [131]) continuously monitors the measurement and triggers a reading of the stabilized value 1.2 s after the tool has been placed on the PCB. By repeating the measurement once in each sector of the PCB, a full pillar height map is obtained (figure 7.10). Measurements are combined in a root based GUI, guiding the operator through the sector-by-sector measurement and calibration steps.

Since height deviation on a single pillar compared to its neighbors is very unlikely to occur, given the production method (chapter 6.3), this QC is meant to be sensitive to trends in the pillar height thickness across the surface of the board. Therefore, the average height values within each row $\overline{h_i}$ ($i \in \{A, B, C, D\}$) and each pair of neighboring columns $\overline{h_{i,j}}$ ($i \in \{-10, -9, ..., 9\}$; $j = i + 1$) are compared. Boards with a discrepancy larger than 5 µm between these regions are disqualified.

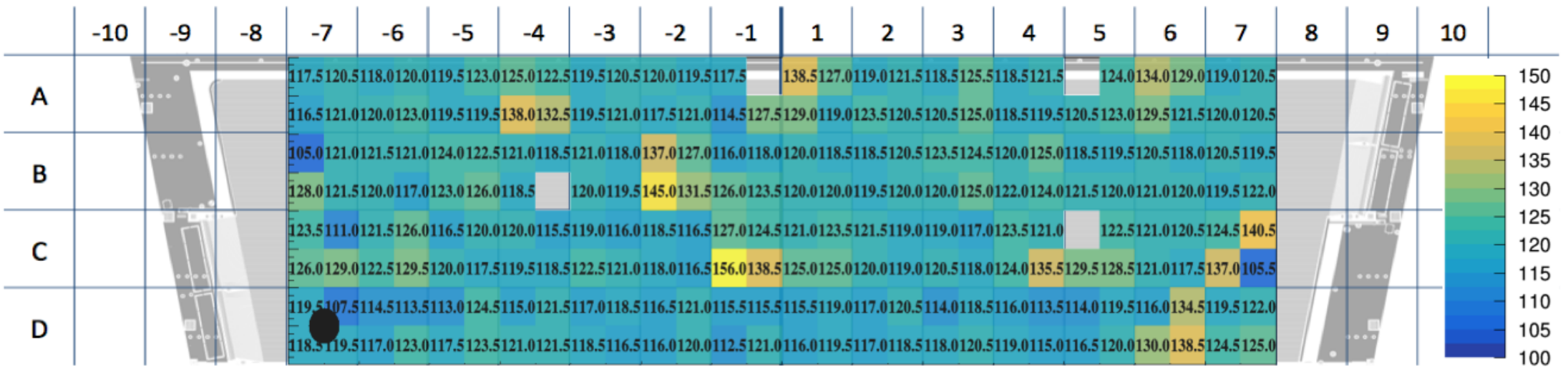

**Figure 7.10.:** Mapping of the pillar height on a NSW anode PCB type SE8. The average distance between anode and top of the pillars is $\overline{h} = 121.3\,\mu\text{m}$ with a slight systematic left-right-increase of $\Delta\overline{h}_{l\leftrightarrow r} = 3.7\,\mu\text{m}$ and a top-bottom shift of $\Delta\overline{h}_{t\updownarrow b} = 3.3\,\mu\text{m}$. Empty spots indicate excluded measurements on pillars, the black dot visualizes the measurement position.

While the length gauges resolution [130], the flatness of the reference surface and the calibration marble contribute with $< 1\,\mu\text{m}$ to the overall resolution and repeatability, the position of the tip on the uneven anode surface (see chapter 6.2) contributes a $\approx 2.5\,\mu\text{m}$ fluctuation (RMS) to the measurements. Dependent on the touching point of the sphere shaped tip, the measurement refers to the anode surface or enters slightly in between two strips. Although this could be avoided by the use of a cylindrical probe tip with a flat surface, all alternatives to the sphere shape increase the risk of damaging the anode surface and have, therefore, been rejected. Probes touching a pillar (figure 7.11 (a)) typically deviate significantly from the overall height average and can be excluded in the data processing, similarly to contamination of $\geq 20\,\mu\text{m}$ filaments or dust particles (figure 7.11 (b)). An additional bias can be caused by localized bumps or dents (figure 7.11 (c) and (d)), or an extended non-flatness of the PCB within the measurement area (figure 7.12). To mitigate the latter effect, the precision surface is limited to a circle of $\varnothing = 7\,\text{cm}$ and the board is placed on a thoroughly cleaned marble table.

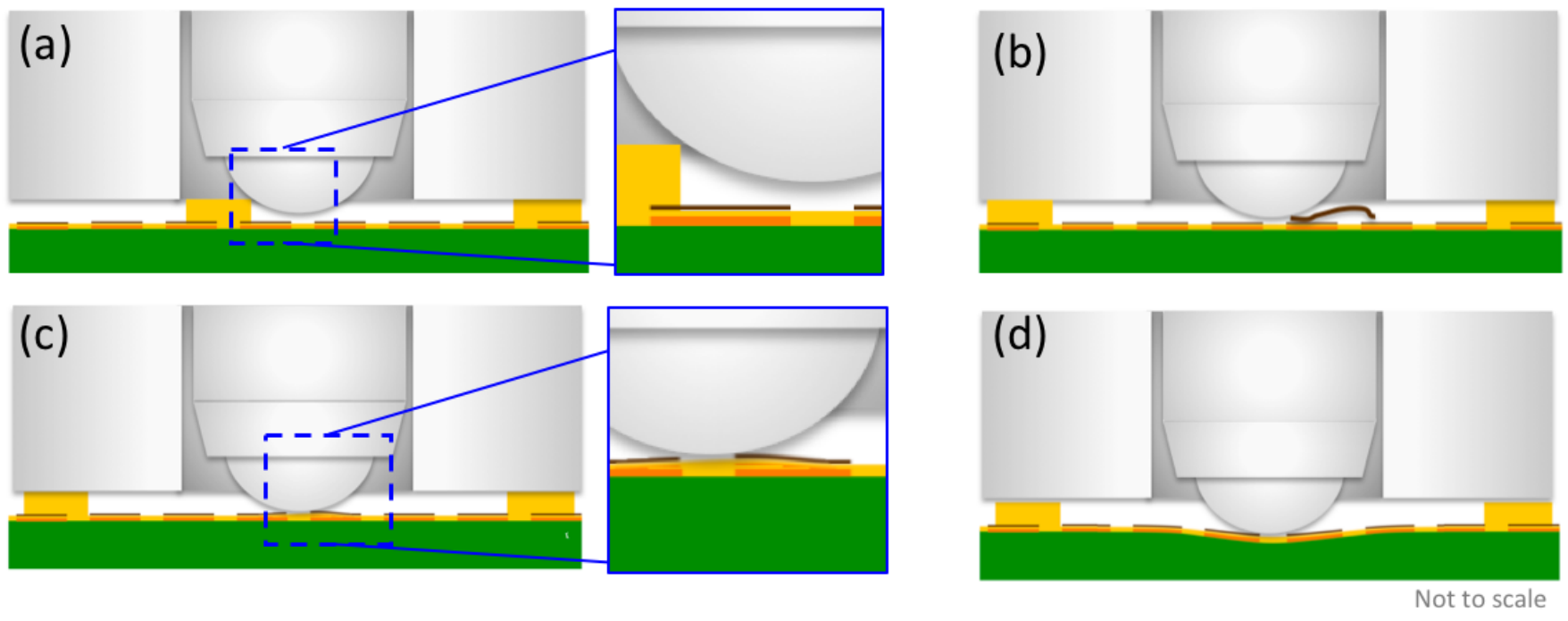

**Figure 7.11.:** Examples for locally biased measurements of the pillar height: probe touches a pillar (a) or the measurement refers to the top of a filament (b), to a bump (c), or dent (d) in the anode surface.

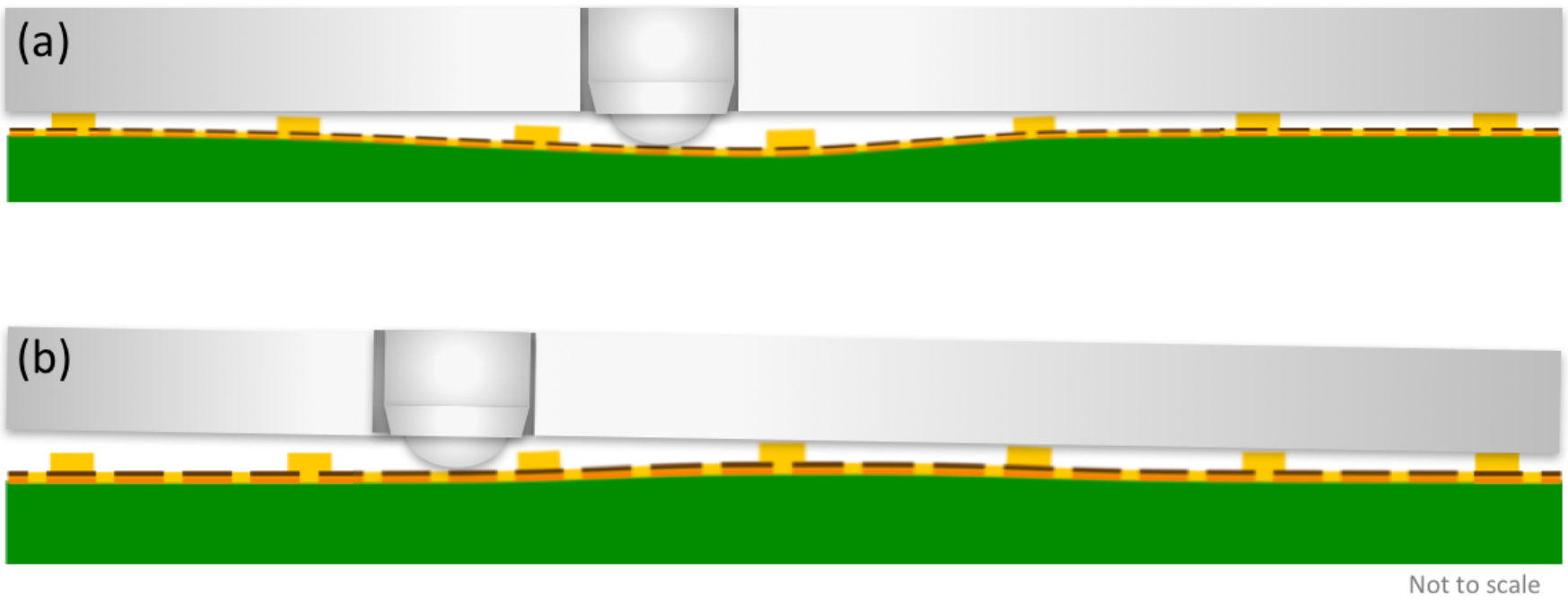

**Figure 7.12.:** Examples for biased measurements of the pillar height by extended PCB non-flatness: while the base plate rests on distant pillars the probe touches into an extended pit (a), or a larger region of reduced PCB thickness (b).

### 7.5.4. Anode Pattern Resistivity Survey

The surface resistivity of the anode determines the detector's behavior under high radiation-rate, demanding an upper limit, as well as the limitation of charge spread, requiring a minimal resistivity. Furthermore, PCB processing conditions like temperature and pressure during gluing and curing steps, and the intensity of surface polishing are reflected in the change of surface resistivity during PCB production. Therefore, a full mapping of the anode's surface resistivity is included in the QC scheme once, directly after the resistive foil production in Kobe (Japan) and again during the acceptance QC at CERN.

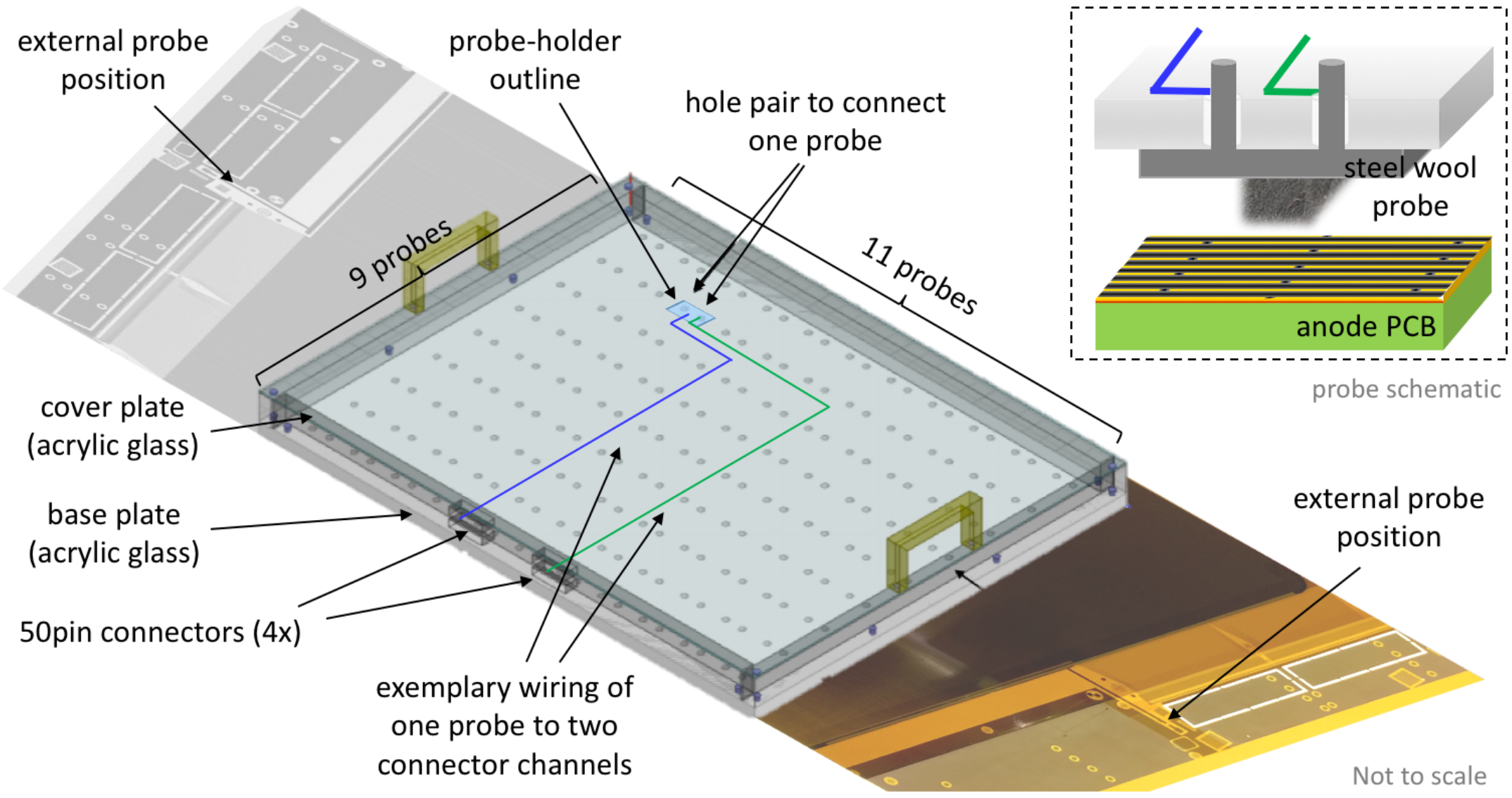

**Figure 7.13.:** Resistivity tool probe carrier plate, positioned on an anode PCB. Only the position and wiring of one probe is indicated. The connection schematic of the steel wool probe is shown in the upper right corner.

The surface resistivity is obtained measuring the impedance between two well-defined probes connecting to the anode surface and converting this probe-to-probe impedance (in units of $\Omega$) to a surface resistivity (in units of $\Omega/\Box$), taking into account the geometry of the probes and the anode. Aiming for a granularity of one measurement every 5 cm in either direction calls for a (semi-)automatized measurement. Therefore, a carrier-plate with $9 \times 11$ probes is positioned on the anode, as shown in 7.13. The probes are $\approx 10 \times 4 \times 3\,\mathrm{mm}^3$ pieces of fine conductive steel wool. This simple solution ensures, due to its softness, good contact to the anode, despite the obstructive pillar structure possibly denying contact to the alternatively considered spring-loaded probes with hard surfaces. The large probe size guarantees connection to $> 20$ anode strips, minimizing the variation in the measured impedance caused by limited positioning accuracy. Each of the probes is connected to two relays, which are closed by default. Opening either one establishes connection of the probe with one pole of a commercial multimeter (Fluke 289 [132]). During a measurement sequence the switcher unit, comprising 200 relays and an Arduino based relay-control, connects two probes to the multimeter, which is read-out via an optical-to-USB connection. Once a stable measurement is received by the root based GUI, the connection is switched to the next pair of probes. Although any probe combination could be realized, the measurement sequence is limited to two modes of operation: either addressing probes neighboring in strip direction (default) or connecting an external probe positioned on the high voltage supply point with a single probe on the anode surface (on demand). Re-positioning of the carrier-plate and appending of the measurement data allows to map the full surface of large PCBs.

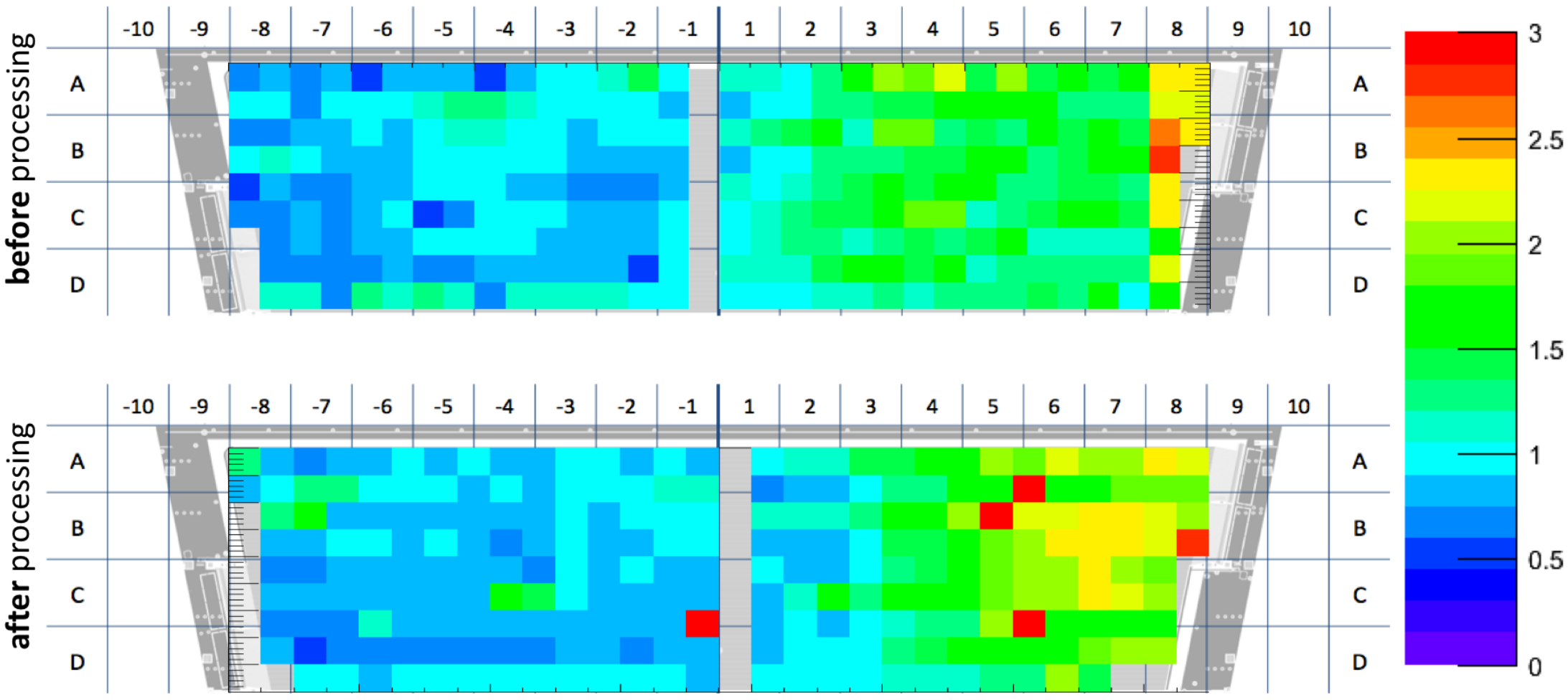

**Figure 7.14.:** Mapping (R in M$\Omega/\Box$) of the surface resistivity of foil SE8_035 before processing (top, measured in Kobe) and on the final PCB (bottom, measured at CERN). While the overall resistivity is slightly increased, the visible left-right asymmetry as well as a high-ohm region (8 B) remain unchanged. Additional single high-ohm spots occurred during processing, possibly due to local damages on the surface. The observable position shift is owed to slightly differing measurement procedures, precisely the positioning of the tool, standardized thereafter.

The values measured during the CERN QC are compared to those obtained in Kobe. An example is shown in figure 7.14. On top of the mapping, the program returns the distributions and static parameters. During initial QC in Kobe, the foils are quality graded (grade A, B or C) based on the average resistivity, the width and shape of the distribution. Grade C foils are rejected while B foils are stored as spares. During final acceptance QC at CERN the change in resistivity is calculated and compared to the allowed shift ($R_{final}/R_{initial} < 3$). A comparison of the mapping will reveal localized irregularities, if they are caused during PCB processing, leading to a possible rejection of the board.

Besides the detailed mapping of the surface resistivity, other electrical attributes of the anode boards are tested, utilizing a commercial multimeter (Fluke 289) and an insulation meter (Megger MIT415 [133]). On each side of the board a template is positioned (figure 7.15), connecting the anode pattern with a set of soft probes (A) and the high voltage supply antenna with a spring-loaded pointy tip (B). The 512 readout strips are shortened by means of a metallic mesh (C) and the Pyralux® coverlay above the silver line is completely covered with a mesh electrode (D). Each probe is accessible with a contact on the template's top side, facilitating the connection to the measurement meter. With the multimeter the conductivity between (A) and (B) is tested ($R < 10\,\mathrm{M}\ \Omega$ requirement). Thereafter, the isolation of the coverlay applied on top of the silver line is verified applying 1 kV between (B) and (D). The insulation of the Kapton® foil is probed by charging up the readout lines (C) with 1 kV to the grounded resistive pattern (B).

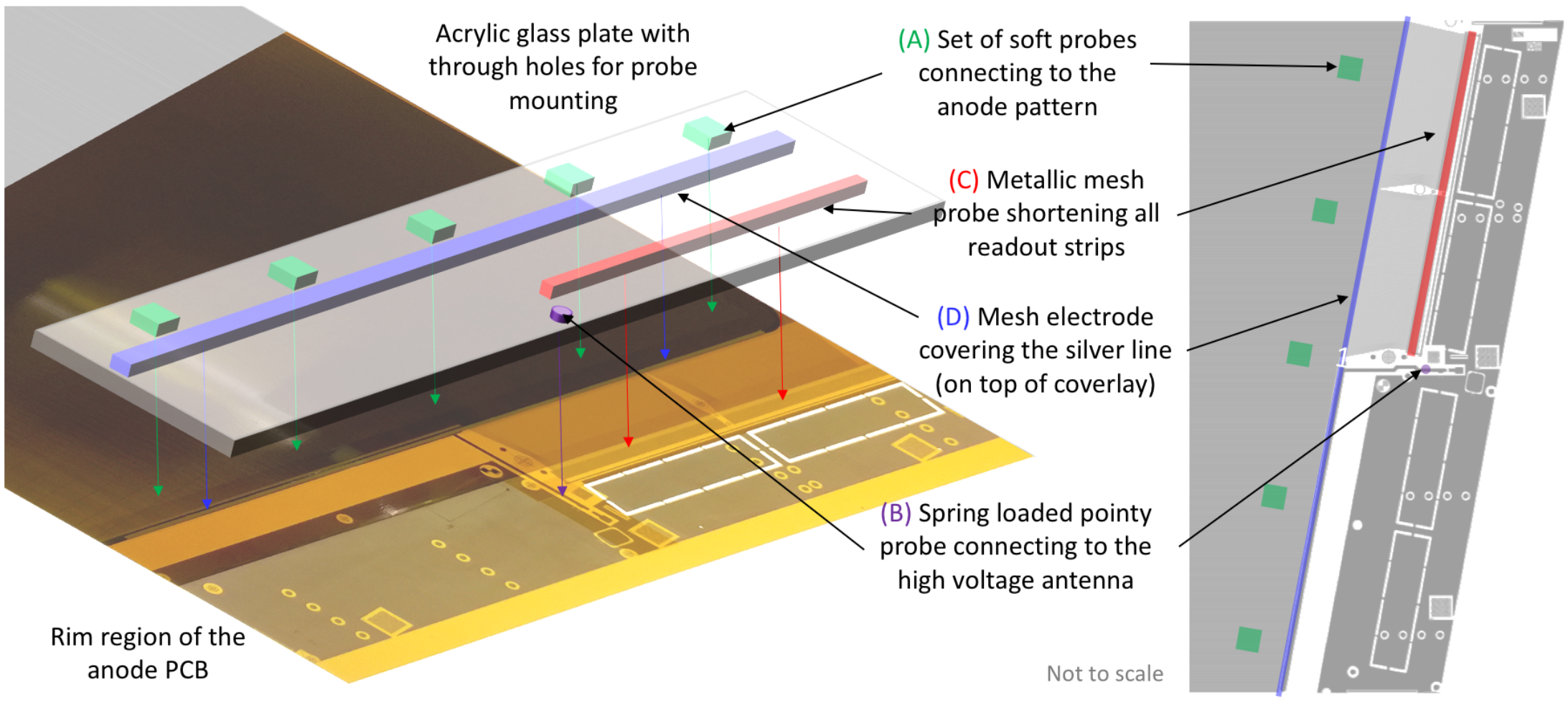

**Figure 7.15.:** Schematic of the template used to test different electrical attributes of the anode PCB. Left: Positioning of the template on the board. The colored arrows point from the probes to the envisaged connection points. Right: Projection of the probe positions on the PCB pattern.

### 7.5.5. Readout Strip Continuity Control

During production cuts of and shorts between the copper strips can be detected optically, utilizing an automatized optical inspection (AOI) machine to compare the image with the layout files. This test of the strip continuity is not any longer possible after covering the readout with the resistive pattern and the Kapton® foil. However, a QC of the readout pattern is important to cross-check the suppliers' conformance statements, to map dead channels and to exclude interruptions of the readout strips occurring during subsequent PCB processing. Such kind of problems have, for example, been observed in the form of etching during the plating step, caused by a chemical reaction with the Akaflex® glue, damaging and partially interrupting many readout strips.

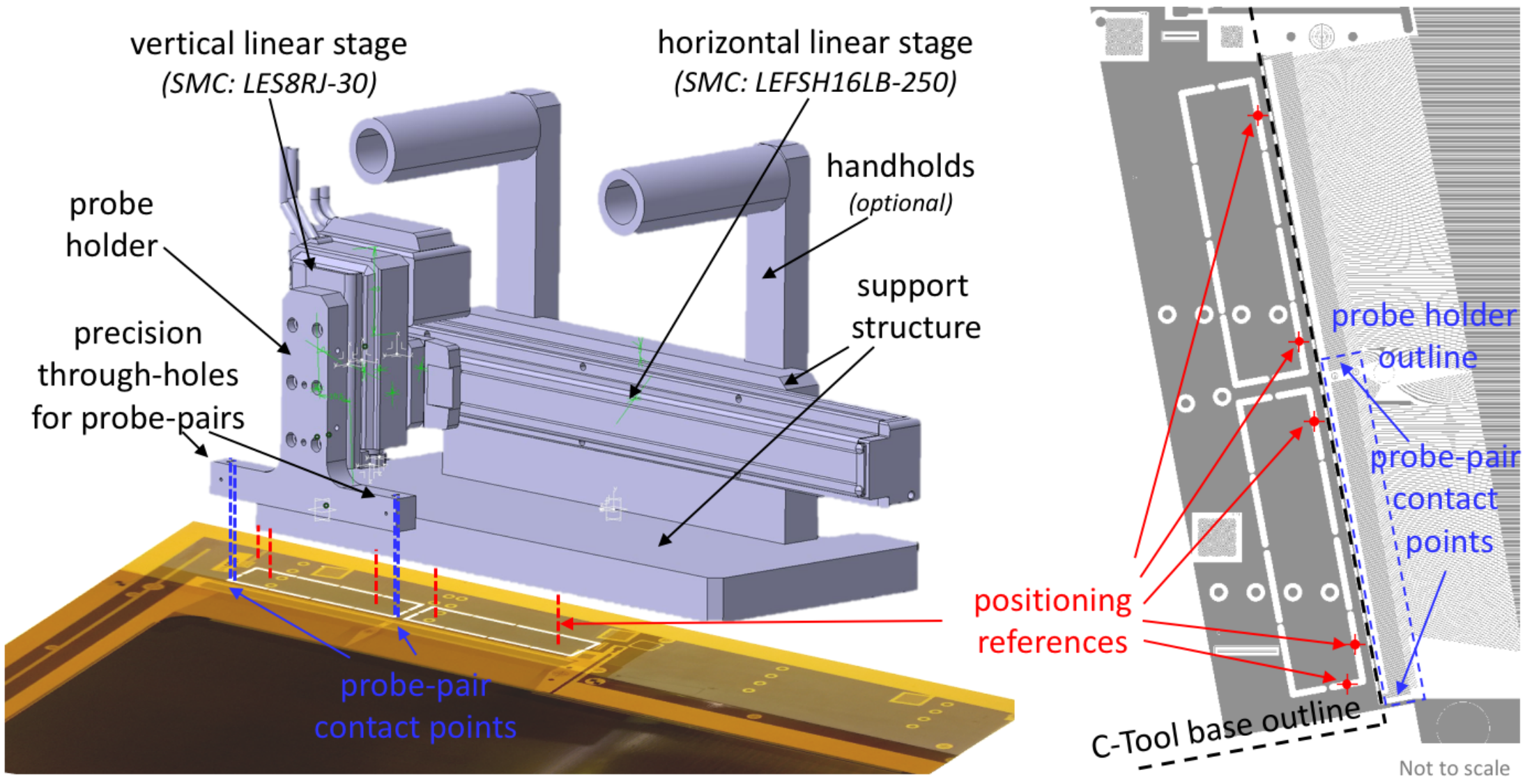

**Figure 7.16.:** Left: Capacitance measurement stage to be positioned above the connector region of the PCB rim. The probe holder carries two pairs of spring-loaded pins which will connect to the readout strips. After capacitance measurement with an LCR meter (not shown) the holder is lifted, displaced horizontally by 400 µm and lowered again to connect to the next strip pairs. Right: Design of the copper structure around the connector region. Positions in the cooling cut-outs used for coarse tool positioning are indicated, as well as the position of the two probe-pairs and their holder in the first measurement position.

A capacitance measurement between neighboring readout strips proved to be an accurate and repeatable method to determine the strips length and, therefore, identify cut strips as well as strips with a direct connection to one of their neighbors. The capacitance is expected to increase linearly by $\Delta C = (1.15 \pm 0.06)\,\mathrm{pF/cm}$ with the length of the readout strip

Being a tedious manual operation, the contacting of the $2 \times 512$ readout strips per board has been automatized using a precision linear stage (SMC LEFSH16) to step across the connector pad and sub-sequentially contact two adjacent strips via spring-loaded probes and a vertical step motor (SMC LES8). To reduce the measurement time, two LCR meter (BK Precision 879B) are used in parallel on two pairs of probes measuring the strips 1-256 or 257-512. The measurement reading and the movement control are combined in a LabView based user interface, automatically detecting strips with low capacitance or unwanted connections.

## 7.6. Summary on the QC/QA Task and PCB Production

The production of resistive anode PCBs at the size, complexity and quality level required for the ATLAS NSW Micromegas posed a great challenge and a non-negligible risk for the success of the whole upgrade project. With no alternative to an industrial production of these core components the proper refinement of the PCB production methods, the technological transfer to industrial producers and in the end the successful production was the top priority of the CERN muon group within the NSW collaboration. Among these tasks the development and implementation of a quality control and quality assurance (QC/QA) scheme, the design and realization of the required tooling and the setup of the QC laboratory was a major task fulfilled in the scope of this thesis.

Based on the first anode PCB QC experience during the construction of the MMSW [115], the set of QC tests has been defined and with ongoing refinement of the PCB requirements and production specifications constantly enhanced. During this process a number of production related, partially severe, blemishes have been identified and solutions to these problems developed in close collaboration with the CERN DT PCB group and the industrial partners. The QC performed at CERN during the module 0 phase of the NSW project contributed immensely to a finalization of the production scheme and a significant increase in production yield. Finally, this three years process resulted the published [125, 126] and herein presented QC/QA scheme which has been accepted by the NSW community and the industrial partners and is now applied during series production.

The tooling and laboratory setups designed during this task have been realized with the help of many colleagues from the NSW community. This allowed for the readiness of the CERN acceptance QC test facilities. With the arrival of the first PCBs for the series production in the first quarter of 2017 both supplier companies ELTOS and ELVIA have finally proven their capability to produce resistive anode PCBs according to the NSW specifications and quality requirements with a satisfying yield. This is a milestone of tremendous importance for the whole NSW upgrade project clearing the path for series detector construction.

# 8. Mesh Selection for the NSW Micromegas

The eponymous component of the Micromegas technology, the micromesh, is the detectors' most precise component. Although a wide range of meshes, mesh geometries and -parameters can be used to build an operational Micromegas, the right choice of this component will permit a wider range in operation parameters and optimize the detector qualities, such as reconstruction efficiency, timing- and energy-resolution, as discussed in chapter 1.3.2 and shown in figure 1.7.

The impact of the micromesh on a Micromegas detector's spatial resolution has already been addressed in early Micromegas R & D [122]. The experimental and simulation studies presented in part I complement these early theoretical discussions and describe in detail the mesh geometry impact on signal formation processes (chapter 3 and 4). Therefore, only the conclusion of mesh selection arguments with respect to electron transparency and the amplification process will be repeated here, referring to the studies in full length in their respective chapters.

Apart from these signal formation related arguments, the large dimensions of the NSW Micromegas call for additional requirements on the mechanical properties of the mesh and add constraints predetermined by the industrial production processes. All these demands have been thoroughly investigated in the scope of this thesis and the 'MM task force 8 on mesh related issues' and concluded in the final choice of the NSW micromesh parameters. The discussion and conclusion presented in the following has been published in [134].

## 8.1. An Ideal Micromesh for a Micromegas Detector

For the construction and operation of $m^2$-size Micromegas detectors with a mechanically floating mesh (chapter 6.3) an ideal micromesh would comprise a multitude of attributes:

(a) The mesh should be **as thin and flat** as possible. Surface structures deviating from a flat plane should be minimal to approximate best to a continuous conductive plate, thus creating a double stage parallel-plate like setup, the very idealization of a Micromegas.

(b) Such a conductive parallel plate idealization is the prerequisite for **homogeneous electrical fields** in both volumes. The homogeneity of the drift field is required for optimal use of a TPC-like track reconstruction, explained in chapter 6.1.1, and a non-homogeneity in the amplification gap contributes to gain fluctuations and eventually effects the energy resolution.

(c) The mesh should provide electrostatic shielding between the two gas volumes and allow for the creation of **independent electrical fields** in the drift- and the amplification region. This is required to fine tune the operation parameters of the detector independently.

(d) Despite dividing the volume electrically, an ideal mesh should be **completely transparent to electrons** traversing from the low drift field to the amplification field, but fully absorbent to ions drifting in the other direction.

(e) The mesh must allow **unhindered gas exchange** between the amplification- and the drift volume, since gas renewal is provided only into the drift gap.

(f) Additionally, the mesh must be **sufficiently robust** to be stretched over large areas and counteract deformations under the electrostatic forces. It must be **mechanically durable** to sustain a sequence of assembly-, transport- and cleaning processes.

(g) Preferably, the mesh production should be simple and **reliable**, ensure flawless meshes and utilize a **quick and cost effective industrial fabrication** method.

Obviously, no material can fully meet all these partially contradicting requirements. Hence, a trade-off between the different demands is necessary, requiring a detailed understanding of each of the above mentioned aspects. In the following, the different mesh production methods and -parameters are discussed and different options for each one are evaluated. Arguments or parameter choices supporting one of the above mentioned criteria are referred to as for example (a+), while being labeled with (d-) if (partially) contradicting the attribute.

## 8.2. Production Technique and Material Options

**Production method -** Meshes or grids can be produced by a variety of techniques. Although photo-lithographical etching [113] or electroforming [135] can yield finer-pitched and flat meshes with O(3 µm) in thickness and O(1 µm) lateral precision (a+), these grid structures typically lack the mechanical robustness to be stretched over several meters (f-). An elaborated approach of etching a copper mesh directly on a support structure of Kapton® pillars was successfully performed at the CERN PCB workshop [136]. It combines high precision with the functionality of integrated spacers, but again lacks mechanical robustness. Additionally to their mechanical drawbacks, these techniques are expensive and difficult to perform on large areas (g-). As already pointed out by Giomataris in [123], weaving of wires to a mesh cloth has proven to be the most reliable method to manufacture accurate and mechanically durable meshes of several $m^2$ size.

**Mesh material -** To be suitable for a Micromegas, the mesh material must first and foremost be conductive and, therefore, metals are the first materials to be considered. While copper or nickel are most commonly used in electroforming, other noble metals can as well be used in electrodeposition processes, yielding favorable electrical attributes but come with an increased cost (g-) and limited mechanical durability (f-). Although less expensive, the thin copper layers most commonly used for photo-lithographical etching suffer from the same mechanical drawbacks. Stainless steel, customarily used to draw wires for fine conductive meshes, combines mechanical robustness (f+) with high dimensional accuracy on µm-level at moderate cost (g+). While austenitic stainless steel type AISI 304 (18 % Cr, 8 % Ni) is the most commonly industrial steel alloy, AISI 316 (18 % Cr, 10 % Ni) features an increased resistance to corrosion. A similar corrosion hardness during high temperature treatment can be obtained by reducing the alloys carbon content below 0.03 %, these compositions are referred to as L-type. Cheaper ferritic stainless steel alloys with typically reduced nickel content suffer from an increased corrosion and, therefore, should be excluded.

Apart from pure metallic meshes, polyamide wires with metallic coating have been considered as an alternative, promising higher flexibility and, therefore, favorable stretching properties. Two types of metalised nylon meshes have been scrutinized using the Exchangeable Mesh (ExMe) detectors (chapter 3.3.1): Coating with a thin aluminum layer via vacuum deposition has proven to maintain the geometrical accuracy of the wires, while causing problems on the electric conductivity (b-). The lack of contact between the wires resulted in an average surface resistivity of $\mathrm{O}(100\,\mathrm{k}\Omega/\square)$ with high non-homogeneity across the surface. A chemically nickel-plated mesh featured good conductivity, but resulted in a degradation of the wire accuracy due to the wet nature of the process. Both coated meshes have undergone the mechanical treatment and cleaning foreseen for the NSW Micromegas production and all samples showed significant damages and delamination of the coating layer (f-), disqualifying the material for an application in the NSW.

## 8.3. Parametrisation of Woven Wire Meshes

**Weaving pattern -** Different weaving patterns, like for example the Dutch weave (figure 8.1), can increase the robustness of the mesh cloth (f+) but cause smaller open areas, reducing the electron transparency (d-). Additionally, they can yield significant asymmetries in warp and weft directions causing asymmetries in the electrical field configuration (b-). Plain weave with identical wires and mesh aperture in warp and weft direction results in the symmetric pattern with the shortest periodicity and represents the best approximation to a flat perforated plane (a+) achievable with wire woven meshes (figure 8.1). Although a symmetric configuration between warp and weft wire is desired to increase the homogeneity of electrical fields (b+), slight variations between the wires on O(1 µm) can significantly facilitate the production process and reduce the risk of broken wires and, therefore, flawed mesh material (g+).

**Mesh parametrisation -** For the symmetrical plain weave, the mesh is completely specified by the wire diameter $d$ and the (mesh) aperture between neighboring wires $a$ (figure 8.1 - right). In an asymmetric case these parameters differ in the warp and weft direction: $d_{weft}$ and $d_{warp}$ respectively $a_{weft}$ and $a_{warp}$. The open area of the mesh is defined as geometrical projection through the mesh plane:

$$O_{sym} = \left[\frac{a}{a+d}\right]^2 \quad \text{repectively} \quad \mathrm{O_{asym}} = \frac{\mathrm{a_{weft}} \times \mathrm{a_{warp}}}{(\mathrm{a_{weft}} + \mathrm{d_{weft}}) \times (\mathrm{a_{warp}} + \mathrm{d_{warp}})}. \tag{8.1}$$

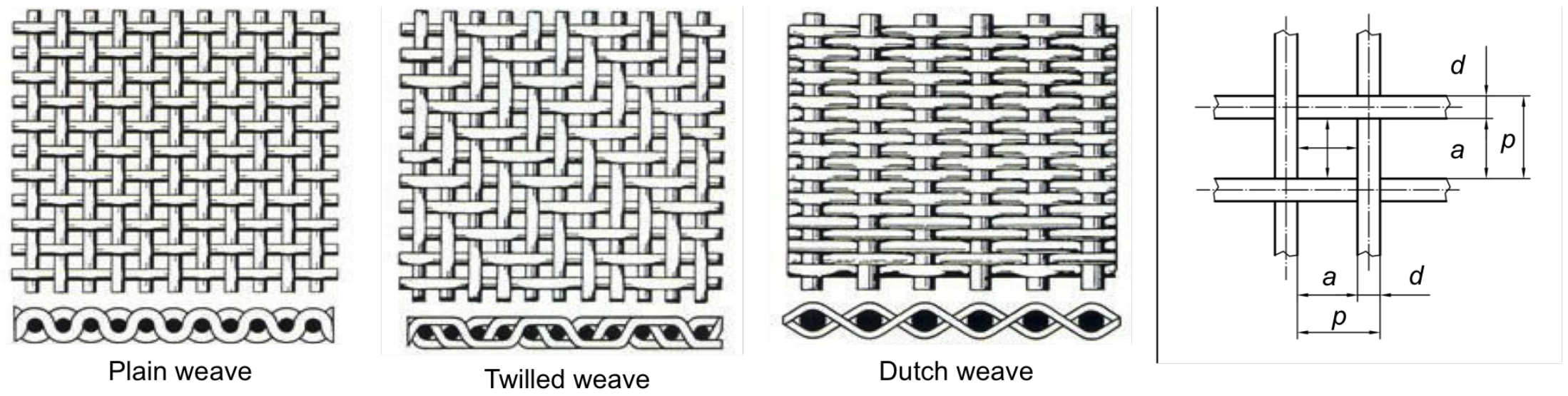

**Figure 8.1.:** Left: Schematic of the three most common weaving patterns: plain weave, twilled weave and (plain) dutch weave (left to right). Right: Defining geometry of a plain weave unit cell. [137].

The mesh pitch or periodicity $p$ is accordingly:

$$p_{warp,weft} = a_{warp,weft} + d_{warp,weft}. \tag{8.2}$$

The naming convention used in mesh weaving industries [137] usually refers to a combination of the wire diameter $d$ with the number or count of lines per inch (lpi), since the first is measured with great care and the latter is a whole-number parameter over the full mesh width and, therefore, most accurate.

**Meshcloth thickness -** The woven wire cloth thickness is mainly determined by the wire diameter and the tension applied on the wires during the weaving process. Without further processing after the weaving the thickness $t$ is limited by

$$t \geq d_{weft} + d_{warp} \approx 2d. \tag{8.3}$$

While an equal tension on warp and weft wires yields the best contact between the wires, a slight deviation reduces significantly the internal tension of the mesh cloth and suppresses the formation of wrinkles in the cloth (g+) by adding only O(2 µm) to the mesh thickness. Although the flatness of the mesh can be increased (a+) by calendering, a process where the mesh cloth is pressed between two precisely parallel rolls, this process is not industrially available for meshes of more than 2 m width.

## 8.4. Choice of the Wire Diameter

Reaching a consensus about weaving the mesh with stainless steel wires in a plain weave pattern, the wire diameter needs to be scrutinized and agreed upon.

**Mechanical constraints -** According to (8.3) thinner wires yield meshes of decreased thickness and increased flatness and, therefore, are better suited to approximate a flat and thin plate-like structure (a+). However, thinner wires are less stable and more prone to wire rapture during the weaving process, causing blemishes in the mesh cloth and reducing the production yield (g-). Additionally, a reduction of the wire diameter requires more wire-length to weave a mesh of comparable open area and thus increases costs per square meter significantly (g-). To avoid production flaws and to keep costs reasonable, a wire diameter of $\geq$ 30 µm is strongly recommended for the NSW Micromegas. Still, the comparison of 30 µm-wire meshes to meshes woven with thinner wires is of interest with regards to future Micromegas applications beyond the NSW project.

**Transparency to electrons -** The mesh's transparency to electrons traversing from the drift into the amplification region has been addressed in detail in the experimental and simulation studies presented in the first part of this thesis. As discussed in chapter 3 and shown in figures 8.2 and 8.3, meshes woven from thinner wires yield a slightly reduced electron transparency compared to thicker wire meshes with comparable open area, and are thus unfavorable (d-).

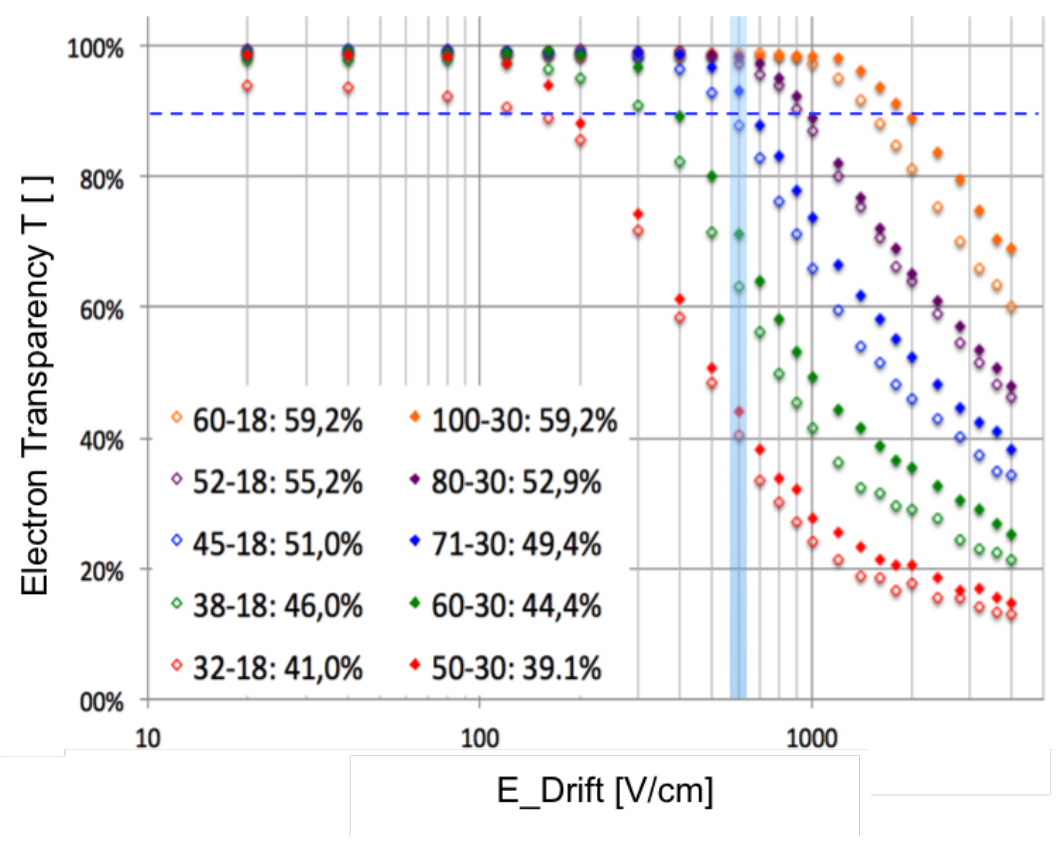

**Figure 8.2.:** Simulated electron transparency of different woven wire meshes ($a - d : O$) as function of the drift field strength. The dashed blue line indicates the NSW Micromegas transparency requirement, the shaded blue region the envisaged drift field strength. [134]

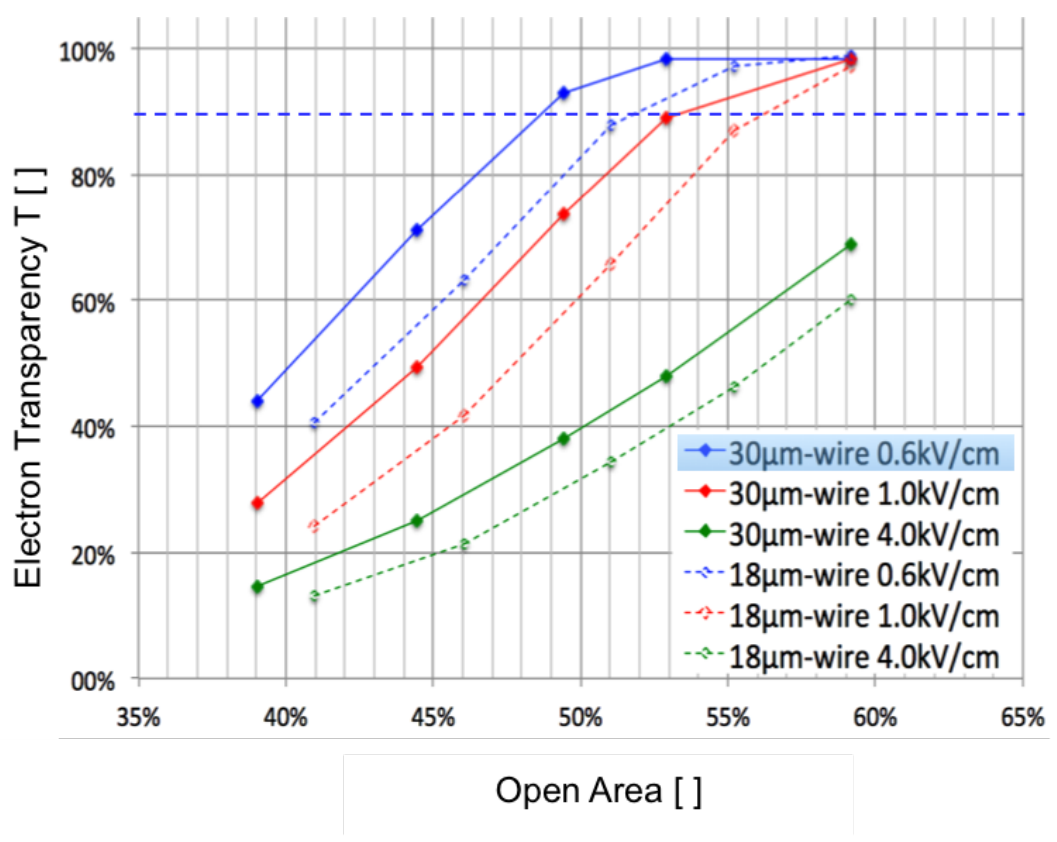

**Figure 8.3.:** Simulated electron transparency as function of the open area for 30 µm- and 18 µm-wire meshes and three drift field settings. The dashed blue line indicates the NSW Micromegas transparency requirement. [134]

**Electrical field strength -** Despite being primarily determined by the potential difference between the cathode, the mesh and the anode, the electrical fields are sensitive to small changes in the mesh's geometry and position. The impact of variation of the mesh parameters has been studied using a finite element study with COMSOL Multiphysics® [63] in a 2-dimension cross-section of the Micromegas at the center of a mesh aperture. Compared to the ANSYS® and Garfield++ Simulation utilized for the majority of the studies presented in part I, this simplification allows for the simulation of a large parameter range and features a high granularity of the finite elements. It neglects, however, the impact of the second wire direction and, therefore, yields slightly biased absolute values in the electrical field strength compared to a full 3-dimensional model. This approach is, therefore, well suited for qualitative argumentation, while requiring the comparison to the full simulation for quantitative statements. An exemplary field map is shown in figure 8.4. The streamlines indicate the most probable electron path, neglecting diffusion in the gas. Electrons created in the drift volume are strongly focused on a path in-between the mesh wires and, accordingly, the electrical field strength along these paths is the main determinant for the Micromegas' gain.

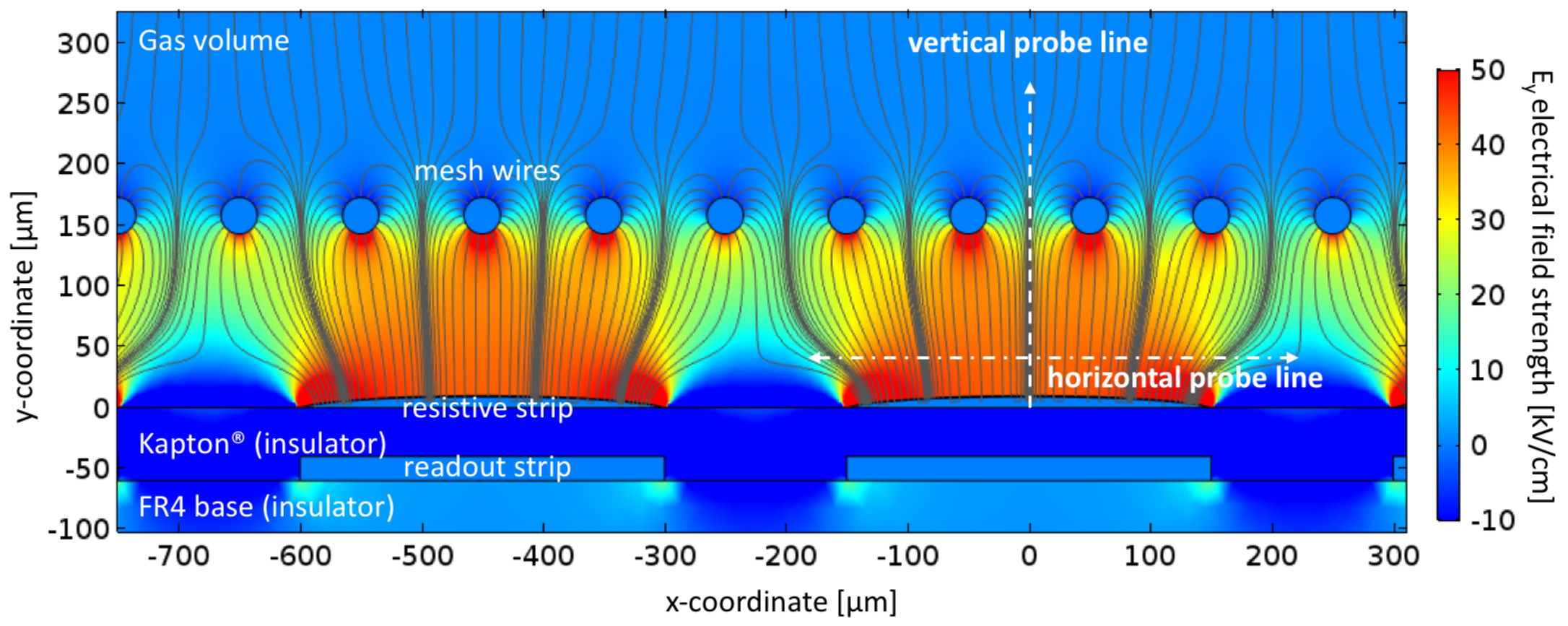

**Figure 8.4.:** Electrical Field strength in the vicinity of the mesh wires. The mesh has a periodicity of 100 µm (diameter = 30 µm, aperture = 70 µm), the strip pitch is 450 µm (strip width = 300 µm). The position and orientation of two probe lines are indicated (dashed). (COMSOL Multiphysics® [63] simulation - the streamline (grey) density is not proportional to the field strength.) [134]

The impact of the wire diameter on the electrical field strength along an electron's most probable path through the amplification gap becomes visible in comparison of figure 8.5 and 8.6, each showing five mesh configurations with different mesh apertures and comparable open areas in the same color. The transition slope between amplification and drift region is systematically steeper for the finer mesh structure, yielding a better approximation to a parallel plate setup (a+). The increased field strength is only partially due to the higher wire density, since the reduced mesh thickness reduces as well the mean distance from the wires to the anode.

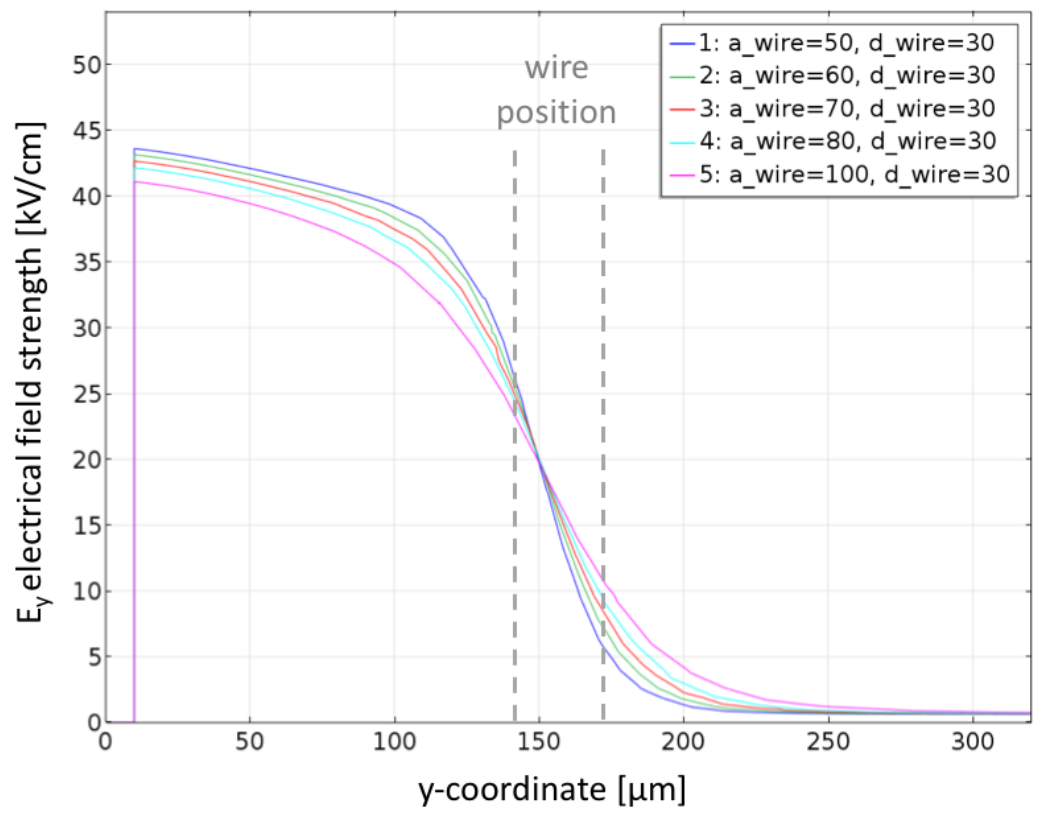

**Figure 8.5.:** Electrical field strength along a vertical probe line in the amplification gap with different apertures of a 30 µm-wire mesh. [134]

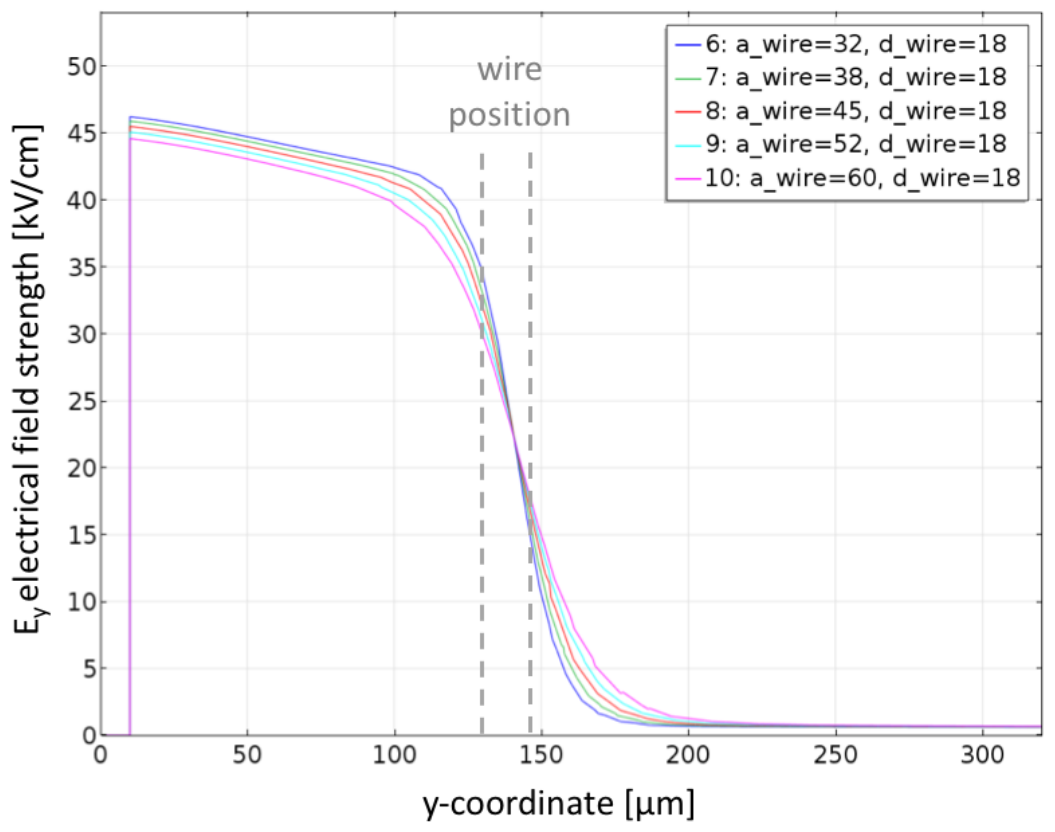

**Figure 8.6.:** Electrical field strength along a vertical probe line in the amplification gap with different apertures of a 18 µm-wire mesh.

In the drift region a decrease of the wire diameter results in an overall reduction of the electrical field strength, as shown in comparison between figure 8.7 and 8.8. The increase of the electromagnetic shielding properties of finer woven meshes results in a reduced bias between the effective field strength $E_{D,\text{eff}}$ and the nominal drift field $E_{D,\text{nom}}$, even at large distances from the mesh as well as in a reduced non-homogeneity along the drift gap.

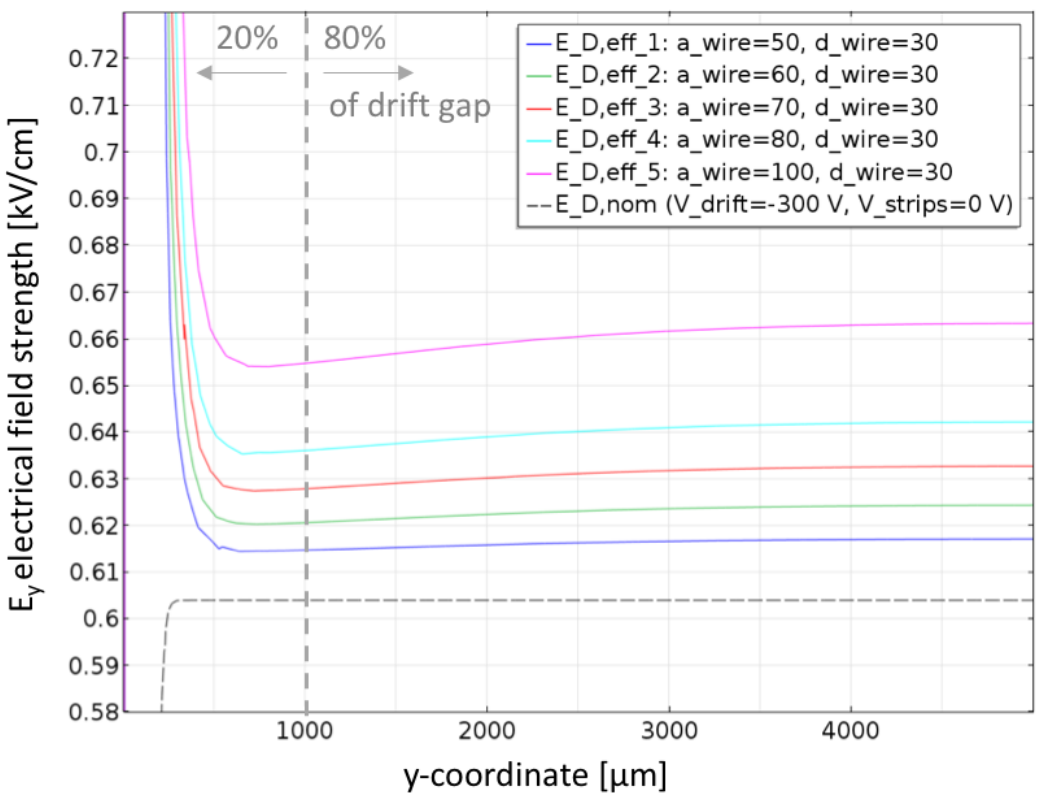

**Figure 8.7.:** Electrical field strength along the drift gap of a Micromegas with different apertures of a 30 µm-wire mesh. [134]

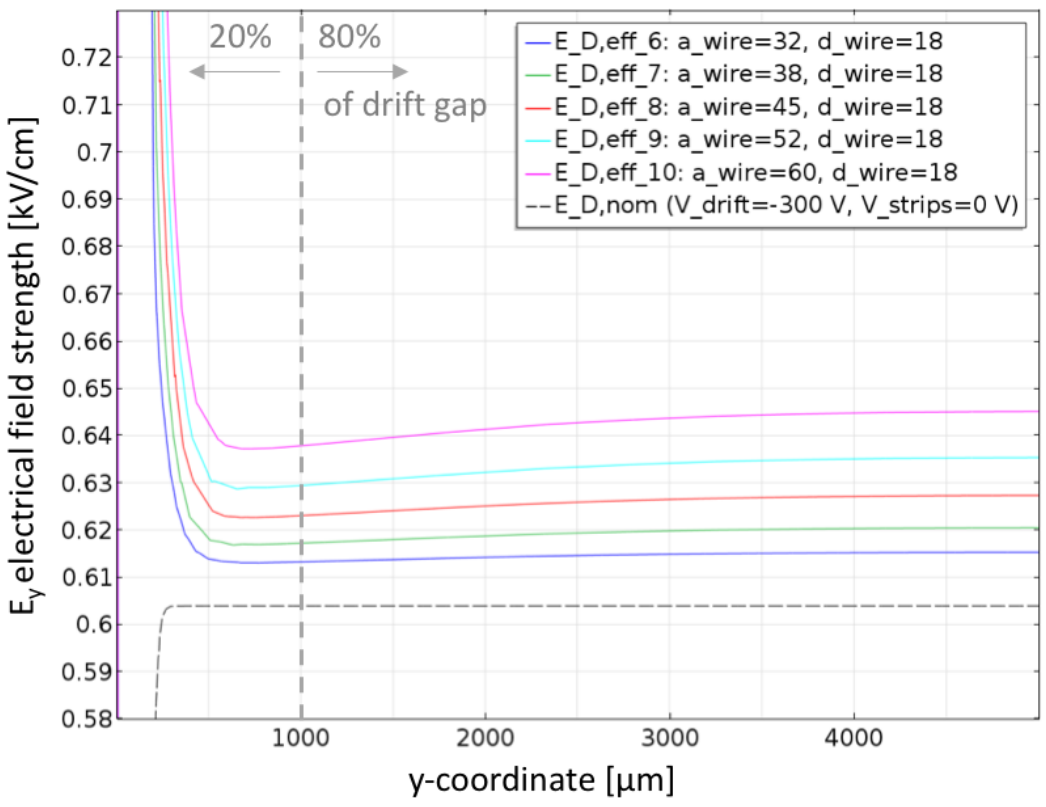

**Figure 8.8.:** Electrical field strength along the drift gap of a Micromegas with different apertures of a 18 µm-wire mesh.

Accordingly, thinner wires are favorable for meshes utilized in a parallel plate-like setup (a+), increasing the field homogeneity along the volumes (b+) and providing a better shielding between them (c+).

## 8.5. Optimization of the Mesh Aperture

**Mechanical considerations -** While for a fixed wire diameter smaller apertures increase the bending angle of the wire during the weaving process and, therefore, the risk of wire rapture (g-), a mesh cloth with larger apertures is less rigid and more prone to damages during handling (f-). This is reflected in a higher Young Module $E$ of a mesh with higher mesh count, or smaller aperture, which effects the sagging of the mesh, stretched with a certain force. Therefore, the aperture of a 30 µm-wire mesh is suggested to be chosen in a range of $50\,\mu\text{m} \leq a \leq 80\,\mu\text{m}$, slightly favoring smaller apertures for their mechanical robustness.

**Transparency to electrons -** The strong dependence of the mesh's electron transparency to its open area is shown in the figures 8.2 and 8.3. With a 30 µm-wire an aperture of $\geq 70\,\mu\text{m}$ (open area $\geq 49\,\%$) is required to reach an electron transparency of $\geq 90\,\%$ (dashed line) with the voltage settings foreseen for the NSW Micromegas (shaded blue region).

**Field strength along the electron's path -** With a decreased distance between the wires, the electrical field strength along the electron's most probable path through the amplification gap (vertical probe line) increases, as shown in figure 8.5. Consequentially, a smaller aperture results in a larger gas gain, as discussed in chapter 4. Simultaneously, the mesh's shielding capability increases with smaller apertures (c+), reducing the continuous offset between the effective drift field strength $E_{D,\mathrm{eff}}$ and its nominal value $E_{D,\mathrm{nom}}$ as well as the region of non-homogeneous field strength in the drift volume (figure 8.7). While a continuous offset can easily be taken into account in a TPC-like reconstruction algorithm, the increase of the non-uniformity within the drift volume is more difficult to cope with. Hence, an aperture not wider than 80 µm between neighboring 30 µm-wires is recommended, to limit the $E_{D,\mathrm{eff}}$ uncertainty in the top 80 % of the drift gap to $\leq 1\,\% \cdot E_{D,\mathrm{nom}}$ .

**Field homogeneity across the regions -** The field homogeneity across the amplification gap, along horizontal probe lines at different distances to the anode, is displayed in figure 8.9. The influence of the wire position on the field homogeneity is clearly visible and most pronounced close to the wires. Alongside the wire periodicity, the non-homogeneity is dominated by the anode strip structure, especially in the lower part of the amplification gap. Figure 8.10 shows the comparably low impact of a mesh aperture variation to the field homogeneity even at a small distance to the mesh of $\leq 40$ µm.

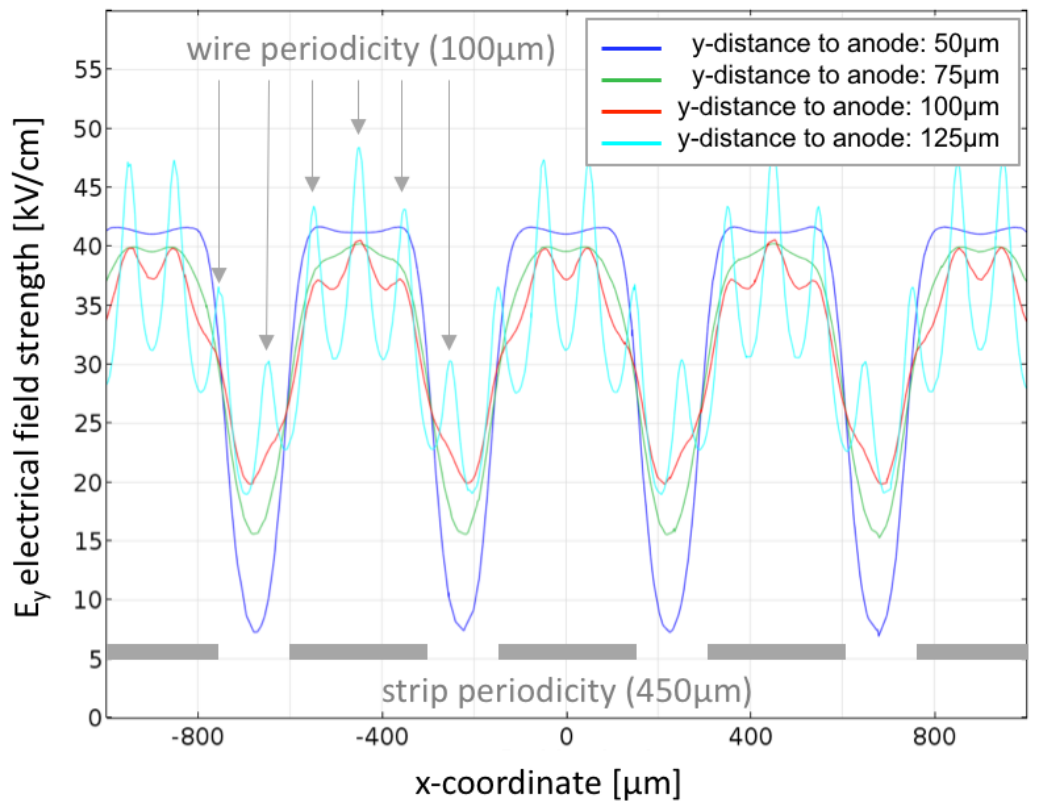

**Figure 8.9.:** Electrical field strength fluctuation along a set of horizontal probe lines parallel to a mesh with 100 µm periodicity. [134]

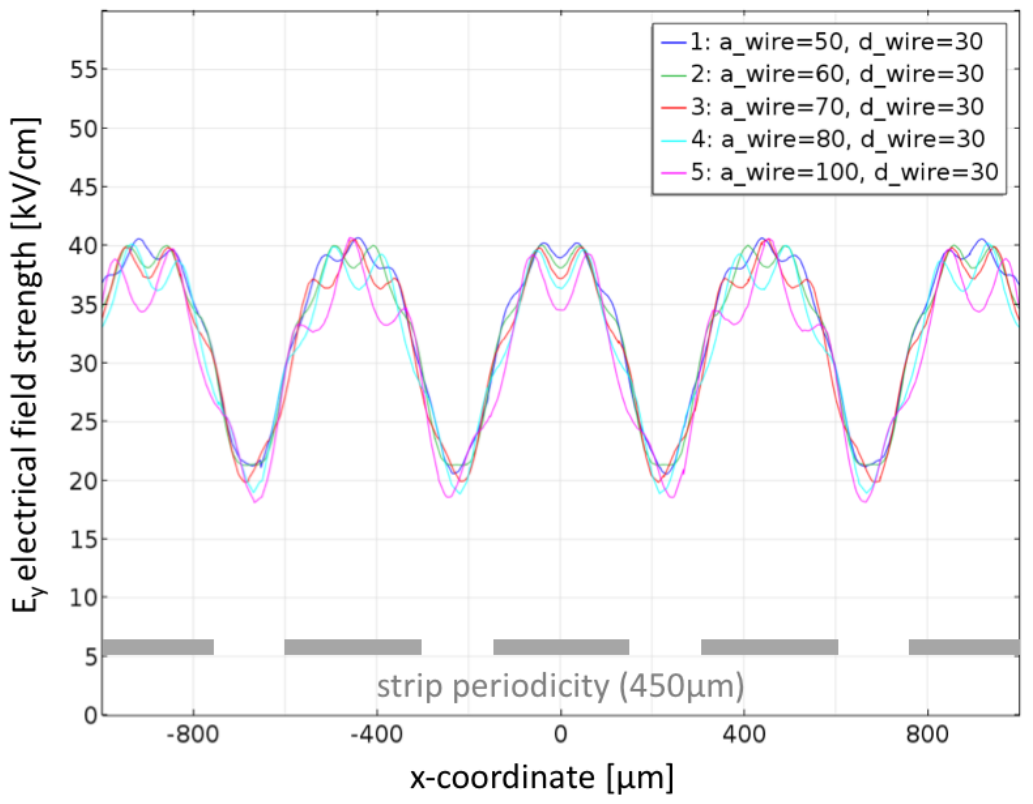

**Figure 8.10.:** Electrical field strength fluctuation along the horizontal probe line 100 µm above the anode, compared for different mesh apertures.

The same periodicity of the field strength in the drift region is visible in figure 8.11. However, the main source of non-homogeneity remains the anode structure. The homogeneity of the drift field strength can be significantly improved (b+) by the reduction of the mesh aperture and, accordingly, increase of the mesh's electrostatic shielding capability (c+), as shown in figure 8.12.

## 8.6. Gas Flow through the Mesh

An effective gas exchange through the mesh is important to keep the gas mixture in the amplification gap pure (e+), since contamination is created there during the ava-lanche formation, primarily by dissociative attachment, like: $e^- + CO_2 \rightarrow O^- + CO$. Gas-flow-simulations with a simplified (flat) mesh model ($d = 30$ µm, $a = 70$ µm) has been performed in collaboration with S. Karentzos to probe the gas transmission behavior through a 70-30 mesh. Figure 8.13 shows a gas exchange time between drift and amplification volume in the order of a few seconds. Given the exchange rate of the full detector (4 Volumes/day), the gas flow hindrance caused by the mesh is negligible.

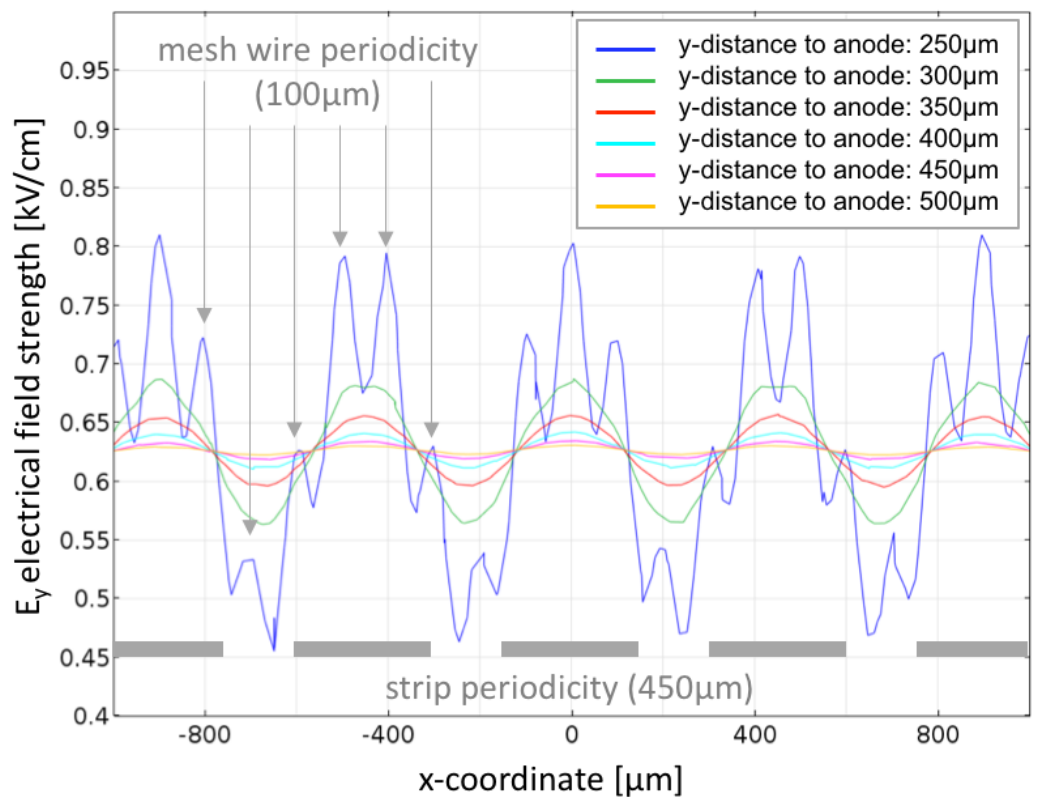

**Figure 8.11.:** Electrical field strength fluctuation along parallel horizontal lines above a mesh with 100 µm periodicity. [134]

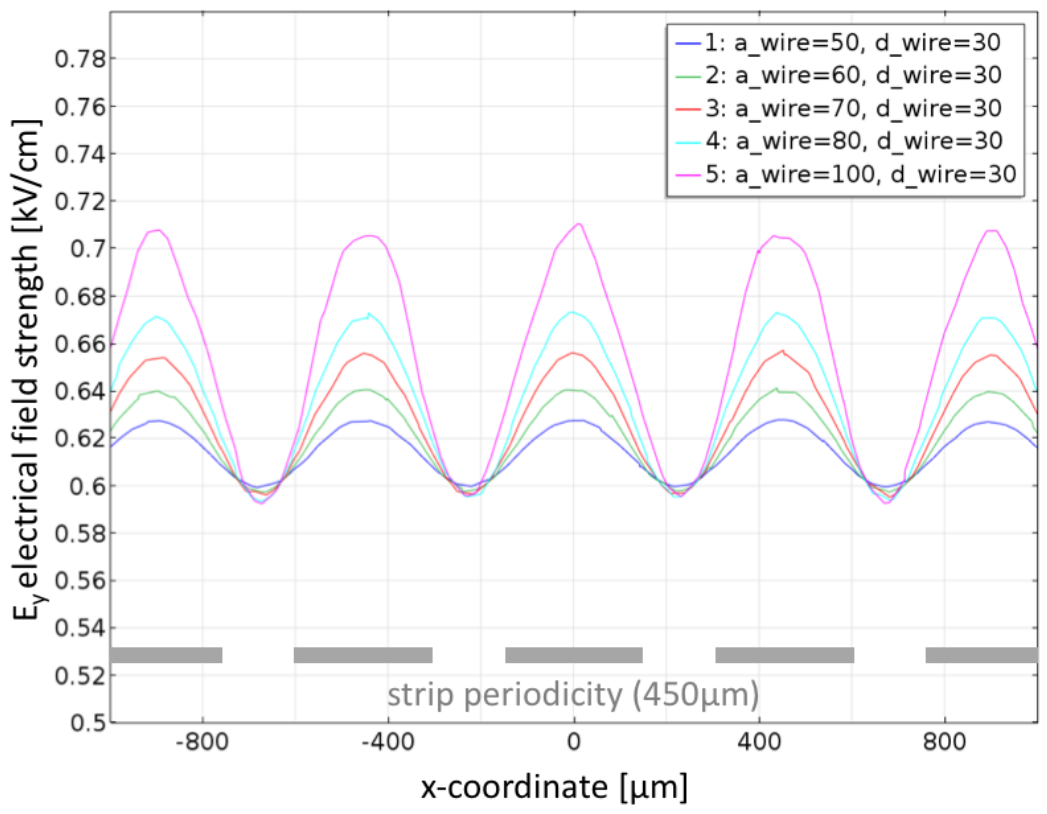

**Figure 8.12.:** Electrical field strength fluctuation along the horizontal probe line 350 µm above the anode, compared for different mesh apertures.

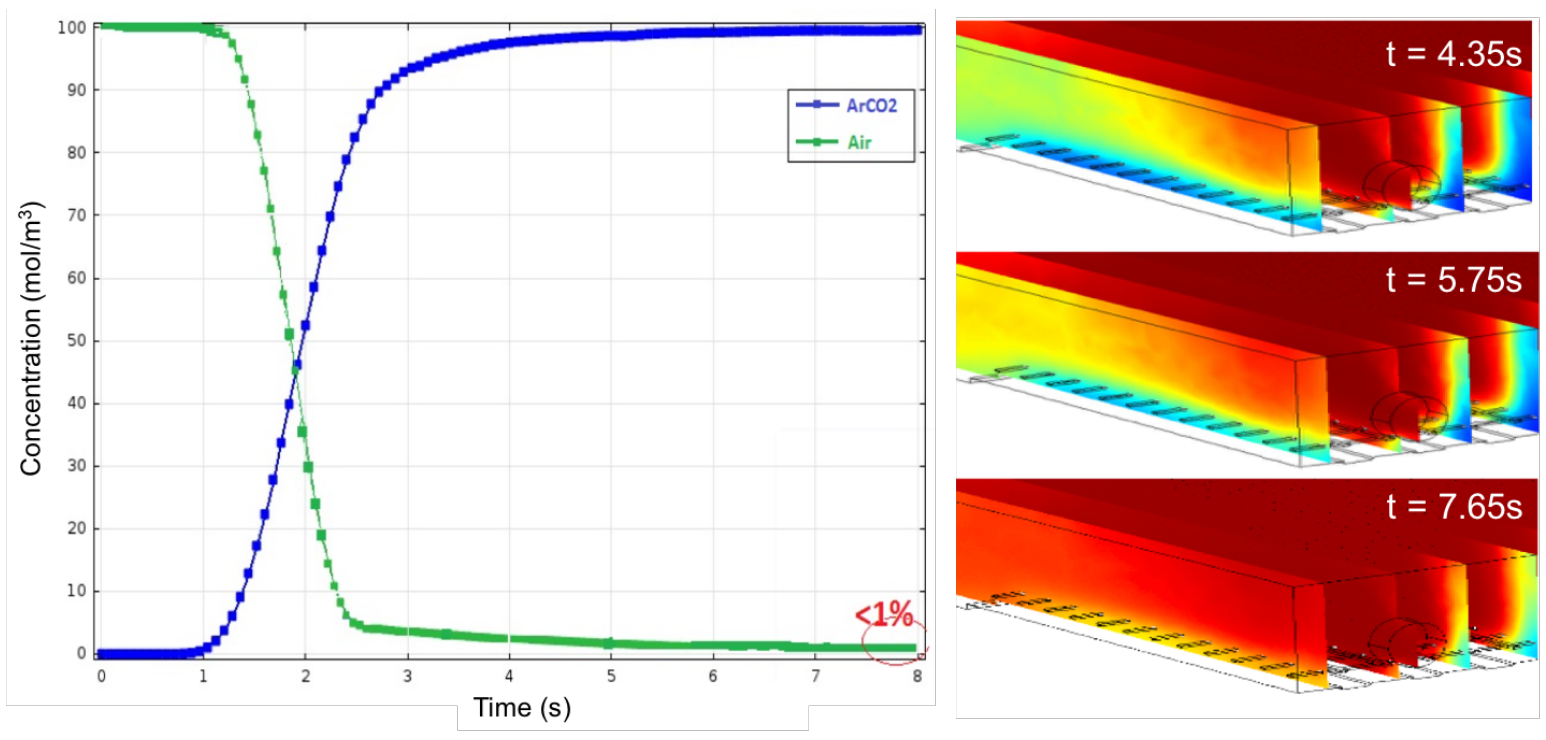

**Figure 8.13.:** Propagation of the gas during the exchange process (blue: old gas; red: new gas) in the drift- and the amplification-region at three sequential instances. [134]

## 8.7. Production Flaws, Blemish Categorization and Quality Assurance

Production flaws and hence blemishes on the product are inevitable to occur during a large scale production. like the weaving of a 2800 $m^2$ mesh cloth. This demands the application of a quality control (QC) and quality assurance (QA) scheme. The international standards document ISO 9044 [137] specifies a wide range of common blemishes and tolerances for their occurrence rate. In addition, most mesh weaving companies follow internal quality regulations based on their experience and quality standards. However, most of these quality criteria are aimed for mesh applications in fields like screen printing or as small-size components of electrical devices and are not well-suited to cope with the needs of an application in Micromegas.

Accordingly, it has been mandatory to scrutinize the possible impact of each of the known types of mesh blemishes to the Micromegas usability and properties. The driving thought behind the categorization of blemishes has been the extent of its impact on the detector. For practicability of the quality requirements we opted for a two-stage differentiation of production flaws: critical blemishes endangering the functionality of a full Micromegas segment or plane and acceptable faults causing only a localized effect on the Micromegas performance. Based on this categorization customized quality criteria have been defined, approved by the NSW collaboration and agreed on with the supplying industries:

- **Type I - critical**: Comprising all kind of free wire ends or movable wires within the cloth. This includes, but is not limited to: bursts, broken wires in either weft or warp direction, kinks within a wire, creepers, pinholes and inclusions or foreign material (for instance wire pieces), if they are trapped in the weaving pattern and cannot be removed without damage of the mesh. In addition, all deviations from the weaving pattern exceeding the double of the maximum tolerance $X_i$ as stated in ISO 9044:1999 chap. 4.2. (Table 1) are considered type I blemishes.

- **Type II - acceptable**: Blemishes yielding a local deviation from the weaving pattern, as long as the continuity of each wire is not violated and the aperture is not exceeding the double of the maximum tolerance $X_i$ as stated in ISO 9044:1999 chap. 4.2 (Table 1). This category of blemishes includes, but is not limited to: double shots, wide shots, close shots, reed marks or tramlines and local variations in warp or weft count.

Each mesh sheet has been subjected to a full surface quality control at the supplying company yielding quality reports for each of the more than 500 mesh sheets. No blemish of Type I were tolerated on any mesh sheet within the sensitive area, later covering the detector's drift panel. The number of sheets comprising any Type II blemishes, up to the ISO 9044 specified maximum quantity per sheet, has been limited to 10 % of the total production.

## 8.8. Mesh Selection and Industrial Production

Choosing the micromesh parameters for the NSW Micromegas required a trade-off between production demands and physics related constrains. The weaving of meshes from stainless steel wires, in the plain weave pattern, has proven to be an affordable and technologically viable choice (g) for large area Micromegas. The requirement of mechanical stability (f) over more than two meter mesh width and cost-considerations (g) exclude wires of less than 30 µm diameter. The requirement on the mesh's transparency to electrons (d) sets a lower limit to the mesh aperture, while on the other hand very large apertures are excluded because of mechanical properties (f), the decrease in amplification-field homogeneity (b) and the increase in interference effects between the electrical fields (c).

In the end a plain weave mesh with 71 µm aperture, 30 µm wire diameter and a mesh count of 250 lines per inch (lpi) is the optimum trade-off between all the above aspects. To strengthen the mesh's mechanical properties, reduce inner tension in the material and increase the production yield a small deviation from the warp-weft-symmetry of $\leq 2$ µm in wire diameter has been allowed for, adjusting the aperture to maintain the 250 lpi mesh count.

The defined specifications for the large-scale production and quality control of 2800 $m^2$ woven wire mesh (71-30, 250 lpi) for the ATLAS NSW Micromegas have been proposed to and accepted by the collaboration. Due to the author's leading contribution to the 'MM task force 8' and the intense interaction with industrial suppliers, the micromesh was the first of the more complex detector components of the NSW Micromegas to have its parameters optimized and agreed upon, well within the original schedule. After a world-wide market survey and tender phase, the full quantity of mesh cloth has been produced between November 2015 and June 2016 and is now available for NSW Micromegas module construction.

# Conclusion

This thesis was dedicated towards the Micromegas technology [36], the study of its fundamental processes during signal formation and the preparation of its large scale application in the ATLAS NSW [9].

The presented methodical studies were based on a systematic assessment of the sub-processes and their dependencies on the detector parameters. This allowed to assess, scrutinize and quantify independently the mechanisms contributing to electron losses by gas attachment and absorption at the mesh. Thereby the comparison between experimental data obtained with dedicated setups, analytic descriptions and simulation of the microscopic processes has been used. These studies, partially published at an earlier stage in [53], have been awarded by the MPGD community with the *George Charpak Young Scientists Award 2015*.

For attachment losses the survey was restricted to the contamination of Ar:$CO_2$ with Oxygen and water vapor, the dominating sources of attachment in laboratory setups. They showed the immense impact of even small levels of Oxygen contamination to the electron yield in low drift fields, as well as an upper limit in drift field strength due to the attachment to $CO_2$. The method can be easily extended to any other gas with known attachment cross sections and we showed that under the condition of small contamination levels the attachment contributions follow a multiplicative combination law.

The study on electron transparency of the micromeshes shed a new light on the electron behavior during mesh transit. An unprecedented level of agreement between simulation prediction and experimental measurements was reached for highly transparent structures with an increasing, but well understood, bias for less transparent meshes. These studies outperform the comparable studies in [66] in terms of the variety of meshes, level of agreement with experimental data and understanding of systematic deviations.

Several tendencies in the gas gain dependence on the fine geometry parameters of the micromesh, the anode structure and the inter-pillar distance have been identified. While micromesh and pillar distance effects followed the expected behavior, the impact caused by two differently produced anodes deviated in a significant and systematic way. Explanation attempts have been discussed, but a conclusive answer can not be given, for lack of a detailed measurement of the µm-scale geometry.

In collaboration with the University Paris-Sud (Orsay) measurements with the SER-setup [87] have been extended and complemented with simulation studies on the gain spectra and gain variance. The study allowed for an experimental extraction of penning transfer rates and yielded an astonishing agreement in the avalanche size distribution between experimental data and simulation prediction without gain scaling. For the comparison of gain fluctuations over a wider gain range a mathematical avalanche ex-trapolation method was introduced and validated, yielding a good agreement to the experimental data as published in [85]. The remaining systematic discrepancy in the gain fluctuation could be traced back to experimentally not distinguishable multiple electron events and a contribution from secondary avalanches.

This unique combination of refined simulation tools with an outstanding experimental method has a huge potential for further studies of other gases or a broadened parameter space. Especially the impact of small Oxygen contamination on the gain, barely experimentally studied in Micromegas, would be of great interest. Similarly, a repeated measurement with $ArCO_2$ mixtures commonly used in large scale experiments like ATLAS would be of high value for these collaborations. The author's effort to extend the SER studies had to be abandoned due to the limited access to the experimental setup. However, we encourage to revive and continue these promising studies. A similar potential for continuation of signal formation studies lies in the developed exchangeable mesh (ExMe) detectors, which are offering a wide parameter space to explore in subsequent studies. To further extend the experimental accessible parameter space two complementary detector R & D projects have been initialized: measurements in a pressure controlled environment at the University of Würzburg and a variable gap Micromegas prototype to be constructed in cooperation with the University of Mainz.

Besides an involvement in the Micromegas R & D leading to the discussed technological choices and the construction and testing of the MMSW detectors [115], the author's impact on the ATLAS NSW upgrade relates primarily to the preparation, production and quality control of the anode PCBs and the micromesh.

For the PCB production a detailed quality control (QC) and quality assurance (QA) scheme has been developed, refined and finally agreed upon with the NSW collaboration, the ATLAS review boards as well as the industrial suppliers. New methods for a quick quality assessment using QC markers have been designed as well as a variety of measurement tools required to set up the CERN acceptance QC laboratory. The development and realization of the latter has been a huge collaboration effort leading to an on-time readiness for the recently started series production of the NSW Micromegas anode PCBs. Summaries of these developments have been published in [125, 126].

The characterization, production preparation and quality assurance (QA) for the micromeshes became the responsibility of the 'Task Force 8' coordinated and lead by the author. Decisively influenced by the studies on electron transparency and additional surveys on the impact of the micromesh geometry [134] the parameters for the woven wire mesh were determined, proposed and accepted by the NSW community. These specifications included the outcome of strong interaction with the world-leading industrial producers of woven wire meshes. They took into account production related constraints, cost optimization and quality assurance aspects. Following a market survey and tender process, a producer has been selected and the full quantity of micromeshes for the NSW Micromegas detectors became available for construction in summer 2016.

The completion of both tasks and the start, respectively the finalization, of the associated production were regarded as milestones on the path to series production of the NSW Micromegas detectors. Accordingly, the work conducted in the scope of this thesis and summarized herein has been a significant contribution to the NSW project.

# List of Terms and Abbreviations

**ADDC** **A**RT **D**ata **D**river **C**ard - Electronics card for processing of the NSW ART data.

**AFP** **A**TLAS **F**orward **P**roton - Upgrade project to the ATLAS far detector (5.2.3).

**AISI** **A**merican **I**ron and **S**teel **I**nstitute - Institution issuing the commonly used norm for steel alloys.

**ALICE** **A** **L**arge **I**on **C**ollider **E**xperiment - Particle physics experiment at the LHC [30].

**AOI** **A**utomatized **O**ptical **I**nspection - Group of machines for automatized optical measurement.

**ART** **A**ddress-in-**R**eal**T**ime - ASIC designed and produced in the scope of the NSW upgrade.

**ASIC** **A**pplication **S**pecific **I**ntegrated **C**ircuit - Microchip designed and utilized for a specific application.

**ATLAS** **A** **T**oroidal **L**HC **A**pparatu**S** - Particle physics experiment at the LHC [6].

**ATLAS ID** **ATLAS** **I**nner **D**etector - Detector subsystem of the ATLAS (5.1.2).

**BIS78** **B**arrel **I**nner layer **S**mall sections **7** and **8** - Upgrade project and chamber type for the ATLAS MS.

**Calo** **Calo**rimeter - Group of detectors to measure a particle's energy, in this context used as abbreviation for the ATLAS calorimeter detector sub-system(s) (5.1.3).

**CAST** **C**ERN **A**xion **S**olar **T**elescope - Particle physics experiment at CERN.

**c-CCD** **c**ontact **C**harge-**C**oupled **D**evice - type of image sensor.

**centroid** - Hit reconstruction scheme or method based on the center of charge in an hit.

**CERN** **C**onseil **E**uropéen pour la **R**echerche **N**ucléaire or European Organization for Nuclear Research in Geneva (Switzerland).

**CMS** **C**ompact **M**uon **S**olenoid - Particle physics experiment at the LHC [31].

**CNC** **C**omputer **N**umerical **C**ontrol - Type of precision milling- / drilling-machine.

**CSC** **C**athode **S**trip **C**hamber - Type of GD technology (1.2.3).

**DAQ** **D**ata **A**c**Q**usition - Process of or system to process and store data from an experimental measurement.

**DB** **D**ata**B**ase.

**DCS** **D**etector **C**ontrol **S**ystem - here: the ATLAS DCS.

**DT** **D**etector **T**echnology - here: Department at CERN.

**ELTOS** ELTOS S.p.A. Elettronica Toscana - PCB producer.

**ELVIA** ELVIA Group - Printed Circuit Boards - PCB producer.

**EM** **E**lectro **M**agnetic - here: abbreviation for ATLAS Calorimeter sub-systems.

**EMEC** **E**lectro **M**agnetic **E**nd-**C**ap - Detector sub-system of the ATLAS Calo (5.1.3).

**ESFRI** **E**uropean **S**trategy **F**orum on **R**esearch **I**nfrastucture.

**ExMe** **Ex**changeable **Me**sh Micromegas - Detector build by the author during the R & D program in the scope of this thesis.

**EYETS** **E**xtended **Y**ear **E**nd **T**echnical **S**top - Period of maintenance at the LHC.

**FCal** **F**orward **Cal**orimeter - Detector sub-system of the ATLAS Calo (5.1.3).

**FEB** **F**ront-**E**nd **B**oards - Electronics card carrying the NSW sTGC front end electronics.

**FELIX** **F**ront-**E**nd **L**ink **I**nterface e**X**change - Central processing unit in the ATLAS TDAQ scheme.

**FEM** **F**inite **E**lement **M**ethod - Approximation method used in nummerical modelling.

**FOP** **F**iber **O**ptic **P**late.

**FPGA** **F**ield **P**rogrammable **G**ate **A**rray - Flexibly configurable integrated circuit.

**FR4** **F**lame **R**etardent class **4** - Classification of fiber glass epoxy material.

**FTk** **F**ast **T**rac**k**er - Upgrade project to the ATLAS detector [101].

**FWHM** **F**ull **W**idth at **H**alf **M**aximum - Parameter utilized to determine the width of a peaked distribution.

**GBT** **G**iga**B**it **T**ransceiver.

**GD** **G**aseous **D**etector - Group of detectors relying on ionization in a gas volume.

**GEM** **G**aseous **E**lectron **M**ultiplier - Type of GD technology (1.2.4).

**GridPix** **In**grid **Pix**el - Type of GD technology combining an InGrid detector with a pixelized readout chip (1.2.6).

**HEC** **H**adronic **E**nd-**C**ap - Detector sub-system of the ATLAS Calo (5.1.3).

**HEP** **H**igh **E**nergy **P**hysics - Scientific field in physics research.

**HL-LHC** **H**igh-**L**uminosity **LHC** - Envisaged upgrade of the LHC accelerator.

**HLT** **H**igh **L**evel **T**rigger - Logic cluster in the ATLAS trigger scheme (5.1.5).

**IBF** **I**on **B**ack **F**low - process of ion drift through a structure in a GD.

**IBL** **I**nsertable **B** **L**ayer - Detector Upgrade to the ATLAS ID [99]).

**ID** - **ID**entification when referring to a particle or **ID**entifier when referring to an item.

**InGrid** **In**tegrated **Grid** - Type of GD technology (1.2.6).

**IP** **I**nteraction **P**oint - Collision point of two protons in the LHC, or rather one of its detector.

**ISO** **I**nternational **S**tandards **O**rganization - Institution issuing norms and standards for industrial processes and products.

**ITk** **I**nner **T**rac**k**er - Upgrade project to the ATLAS ID (5.2.4).

**L0** **L**evel **0** or Level 0 trigger - referring to the ATLAS trigger scheme (5.2.4).

**L1** **L**evel **1** or Level 1 trigger - referring to the ATLAS trigger scheme (5.1.5).

**L1DDC** **L**evel **1** **D**ata **D**river **C**ard - Electronics card for processing of the NSW ROC data.

**L2** **L**evel **2** or Level 2 trigger - referring to the ATLAS trigger scheme (5.1.5).

**LAr** **L**iquid **Ar**gon - here: material or the ATLAS Calorimeter sub-systems or the corresponding technology (5.1.3).

**LEM** **Large** **E**lectron **M**ultiplier - Type of gaseous detector technology similar to the THGEM (1.2.4).

**LHC** **L**arge **H**adron **C**ollider - Particle collider at CERN, Geneva (Switzerland) [89].

**LHCb** Particle physics experiment at the LHC [90].

**LINAC** **LIN**ear **AC**celerator - here: First stage of the LHC pre-faccelerator complex.

**LM1** **L**arge **M**odule **1** - Inner Micromegas module of the large NSW wedge.

**LS** **L**ong **S**hutdown - Upgrade and maintenance period of the LHC.

**LXCAT** - Open-access project for collecting, displaying, and retrieving electron and ion scattering cross sections (lxcat.net).

**MC** **M**onte **C**arlo - here: Random number based simulation technique.

**MDT** **M**onitored **D**rift **T**ubes - Type of GD technology with radial geometry (1.2.1 ).

**Micromegas** **Microme**sh **ga**seous **s**tructures - Type of GD technology (1.2.5).

**MiniFCal** **Mini** **F**orward **Cal**orimeter - Considered upgrade of the ATLAS Calo detector system (5.2.4).

**MIP** **M**inimal **I**onizing **P**article - a charged particle with a momentum corresponding to the energy range of minimal energy losses (figure 2.2).

**MM** **M**icro**M**esh gaseous structure - Occasionally used abbreviation for the Micromegas technology (1.2.5).

**MMFE8** **M**icro**M**egas **F**ront-**E**nd **8** - Electronics card carrying the NSW Micromegas front end electronic components.

**MMSW** **M**icro**M**egas **S**mall **W**heel prototype - Detector module build during the R & D program of the NSW Micromegas [115].

**MPGD** **M**icro **P**attern **G**aseous **D**etector(s) - Group of GD technologies.

**MS** **M**uon **S**pectrometer - Detector sub-system of the ATLAS dedicated to the measurement of muons (5.1.4).

**MWPC** **M**ulti-**W**ire **P**roportional **C**hamber - Type of GD technology (1.2.3).

**NJD** **N**ew **J** **D**isk - Shielding disk applied to the NSW.

**NSW** **N**ew **S**mall **W**heel - Upgrade of the ATLAS Muon detector, sheduled for LS2 [9].

**NTP** **N**ormal **T**emperature and **P**ressure - Experimental conditions corresponding to $T = 20\,°\mathrm{C}$ and $p = 1\,\mathrm{atm} = 1013\,\mathrm{mbar}$.

**PCB** **P**rinted **C**ircuit **B**oard - Integrated circuit commonly used in all kind of electronic devices, here: component of a (Micromegas) particle detector.

**PP** **P**arallel Plate - Setup of two parallel electrodes creating a uniform field in a GD.

**PS** **P**roton **S**yncrotron - Particle collider at CERN, Geneva (Switzerland).

**QA** **Q**uality **A**ssurance - Procedure before, during and after production of an item to ensure its compliance with quality requirements.

**QC** **Q**uality **C**ontrol - Testing process to verify the quality of an item.

**QED** **Q**uantum **E**lectro **D**ynamics - Theory of electromagnetic interaction.

**R & D** **R**esearch and **D**evelopment - here: In the context of detector technology.

**RD51** **R**esearch and **D**evelopment Collaboration **51**: Micro-Pattern Gas Detector Technologies - Research cooperation in the field of detector physics.

**RF** **R**adio **F**requency.

**RMS** **R**oot **M**ean **S**quare - statistical parameter to measure the spread of data in a distribution.

**ROC** **R**ead**O**ut-**C**ontroller - ASIC designed and produced in the scope of the NSW upgrade.

**RoI** **R**egion **o**f **I**nterest - here: Detector region where a trigger signal indicated a trespassing particle.

**RPC** **R**esisitive **P**late **C**hamber - Type of GD technology using a PP geometry (1.2.2).

**SCA** **S**low **C**ontrol **A**dapter - Electronics controlling chip for the NSW.

**SCT** **S**emi**C**onductor **T**racker - Detector subsystem of the ATLAS ID (5.1.2).

**SER** **S**ingle **E**lectron **R**esponse - Experimental method for R & D on GDs.

**sFCal** small-gap **F**orward **Cal**orimeter - Considered upgrade of the ATLAS Calo detector system (5.2.4).

**sMDT** **s**mall **M**onitored **D**rift **T**ubes - Type of GD technology based on the MDT.

**SPS** **S**uper **P**roton **S**yncrotron - Particle collider at CERN, Geneva (Switzerland).

**sSoT** single **S**trip **o**ver **T**hreshold - Hit reconstruction scheme or method.

**sTGC** **s**trip **T**hin **G**ap **C**hambers - Type of GD technology.

**SUSY** **SU**per **SY**mmetry - Group of particle physics theories extending the Standard Model.

**TDAQ** **T**rigger and **D**ata **A**c**Q**usition - Combination of the Trigger and DAQ process or system.

**TDS** **T**rigger **D**ata **S**erializer - ASIC designed and produced for the NSW upgrade.

**TGC** **T**hin **G**ap **C**hambers - Type of GD technology (1.2.3).

**THGEM** **TH**ick **G**aseous **E**lectron **M**ultiplier - Type of gaseous detector technology similar to the LEM (1.2.4).

**Tmm** **T**est **m**icro**m**egas - Series of detector prototypes build during the R & D program of the NSW Micromegas.

**ToT** **T**ime-over-**T**hreshold - Parameter of a signal used in signal processing.

**TPC** **T**ime **P**rojection **C**hamber - Type or operation mode of PP-like GDs (1.2.3).

**µ-TPC** - Hit reconstruction method in a very thin, therefore 'µ', TPC.

**TQF** **T**estchamber '**Q**uadro **F**romagi' - Detector prototype build during the R & D program of the NSW Micromegas, the name refers to the quartered surface structure.

**TRT** **T**ransistion **R**adiation **T**racker - Detector subsystem of the ATLAS ID (5.1.2).

**TtP** **T**ime-to-**P**eak - Parameter of a signal used in signal processing.

**UA** **U**nderground **A**rea - Name of a series of experiments conducted at CERN.

**VMM** - Front end electronics ASIC designed and produced in the scope of the NSW upgrade project in cooperation with the RD51 collaboration.

**VOTAT** **V**ary **O**ne **T**hing at **A** **T**ime - Problem solving strategy for complex multi-parameter systems, involving the explicit variation of a single parameter under preservation of all others [44].